\documentclass{wissdoc}

\usepackage[american]{babel}

\usepackage{verbatim}

\usepackage{graphicx}

\newcommand{\blankpage}{
  \clearpage{\pagestyle{empty}\cleardoublepage}
}
\input{babarsym} 

\begin{document} 
\frontmatter
\titlehead{\large\resizebox{!}{1.2cm}{%
\includegraphics*{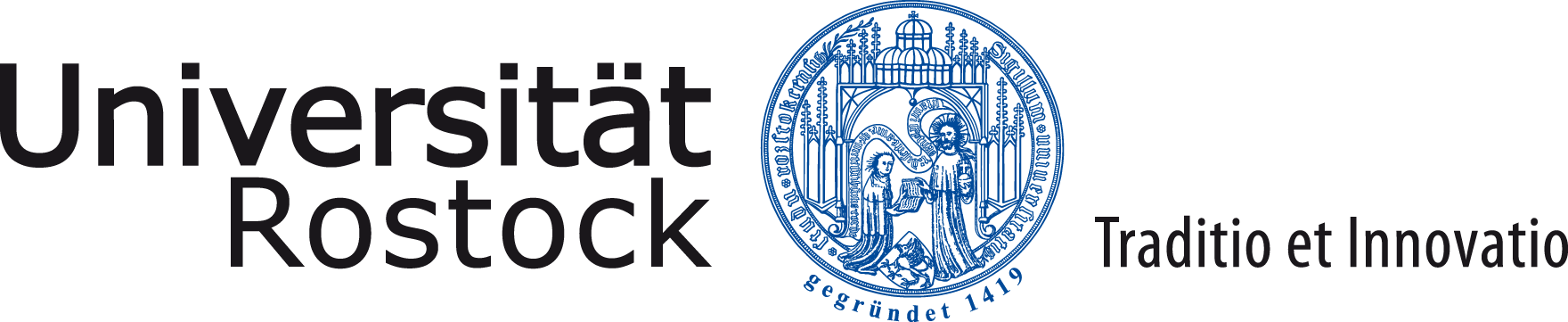}}\hfill
} 
\titlefoot{%
\hfill \resizebox{!}{1cm}{%
\includegraphics*{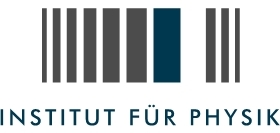}}%
}


\begin{titlepage}
\begin{center}
\hbox{}
\vfill
\boldmath
{\huge\bfseries Search for the decay
		$\Bm \rightarrow \Lambda_c^+ \antiproton \ellm \nulb$\\
		with the \babar Detector \par}
\unboldmath
\vskip 1cm
DISSERTATION\\[2ex]
zur\\[2ex]
Erlangung des akademischen Grades\\[2ex]
Doctor rerum naturalium (Dr. rer. nat.)\\[2ex]
der Mathematisch-Naturwissenschaftlichen Fakult\"at\\[2ex]
der Universit\"at Rostock\\[2ex]
\end{center}
\vskip 7cm
{\bf vorgelegt von}\\[2mm]
Torsten Leddig\\[2mm]
aus Rostock \\[2mm]
 geb. am 19. Februar 1984 in Demmin

\vfill
\end{titlepage}

\blankpage
\newpage
Gutachter: \\[2mm]
\begin{tabular}{c}
  Privat-Dozent Dr. Roland Waldi, Universit\"at Rostock \\
\end{tabular}
\blankpage

\subsection*{Abstract}
This work presents the search for the semileptonic baryonic \B decay $\Bm \ra \LCp \antiproton \ellm \nulb$.
The used data comprises the complete \babar data set of $470 \cdot 10^{6}$ \BBb events, collected at the SLAC National Accelerator Laboratory. Using a pole-model decay simulation, we obtain upper limits with $90\%$ confidence level of
\begin{align*}
    \BR(\Bm \ra \LCp \antiproton \en \nueb)\cdot \frac{\BR(\LCp \ra \proton \Km \pip)}{5\%} &< 1.2 \times 10^{-4},\\
    \BR(\Bm \ra \LCp \antiproton \mun \numb)\cdot \frac{\BR(\LCp \ra \proton \Km \pip)}{5\%} &< 2.5 \times 10^{-4},\\
    \BR(\Bm \ra \LCp \antiproton \ellm \nulb)\cdot \frac{\BR(\LCp \ra \proton \Km \pip)}{5\%} &< 1.0 \times 10^{-4}.
\end{align*}
These results are in slight tension with predictions based on the measurement of $\Bm \ra \antiproton \proton \ellm \nulb$ and $\Bm \ra \LCp \antiproton \pim$.

\subsection*{Kurzfassung}

In der vorliegenden Arbeit wird die Suche nach dem semi-leptonischen, baryonischen \B Zerfall $\Bm \ra \LCp \antiproton \ellm \nulb$ pr\"asentiert. Die genutzten Daten entsprechen dem kompletten \babar Datensatz von $470 \cdot 10^{6}$ \BBb Paaren, welcher am SLAC National Accelerator Laboratory gesammelt wurde. Mittels einer Zerfallssimulation auf Grundlage eines Polmodels  erhalten wir Limits mit $90\%$ Vertrauensniveau von
\begin{align*}
    \BR(\Bm \ra \LCp \antiproton \en \nueb)\cdot \frac{\BR(\LCp \ra \proton \Km \pip)}{5\%} &< 1.2 \times 10^{-4},\\
    \BR(\Bm \ra \LCp \antiproton \mun \numb)\cdot \frac{\BR(\LCp \ra \proton \Km \pip)}{5\%} &< 2.5 \times 10^{-4},\\
    \BR(\Bm \ra \LCp \antiproton \ellm \nulb)\cdot \frac{\BR(\LCp \ra \proton \Km \pip)}{5\%} &< 1.0 \times 10^{-4}.
\end{align*}
Diese Ergebnisse stehen in leichtem Widerspruch zu Vorhersagen basierend auf der Messung der Zerf\"alle $\Bm \ra \antiproton \proton \ellm \nulb$ und $\Bm \ra \LCp \antiproton \pim$.

\blankpage

\tableofcontents

\listoffigures

\listoftables

\mainmatter
\chapter{Introduction}

In 1973 Kobayashi and Maskawa postulated a third quark family, consisting of the top and bottom quark. These new quarks were needed to incorporate CP-violation into the electroweak Standard Model framework. The first measurement of a bound \bbbar state took place in 1977, when the Columbia-Fermilab-Stony Brook collaboration at Fermilab discovered the \OneS resonance. After further data taking, they were able to obtain evidence for two additional resonances, the \TwoS and the \ThreeS in the process $\proton + nucleus \rightarrow \mumu + X$. Later on these results were confirmed by \epem annihilation experiments at DORIS (storage ring at DESY, Hamburg, Germany) and at CESR (storage ring at Cornell University, USA). In addition, experiments at DORIS and CESR were able to establish the $\Upsilon$ as a bound state of a $b$ quark, with charge $-\frac{1}{3}$, and its anti-particle.\\
In order to estimate further quantum numbers the production of hadrons with b-flavor (e.g. \B-mesons: $\B = \b\qbar$) was required. Theoretical models proposed that the yet unseen \FourS and higher resonances should have a mass above the \BBbar production threshold, which would enable the production of \BBbar pairs.
The CLEO and CUSB experiments at CESR observed these higher resonances in 1980. In the next decades experiments like CLEO and ARGUS (Detector at the DORIS storage ring) collected data on the decay of \B-mesons. Therefore, \B-mesons were produced in the reaction \epem~$\rightarrow$~\upsbb. These experiments opened a new field of research known as \B-physics, providing a testing ground for the Standard Model. This new research field enabled physicists to confirm parts of the Standard Model with high precision. In addition, \B-meson decays are a sensitive probe for physics beyond the Standard Model, and allow us to test new theories.\\
In the late 90s of the 20th century the \B-Factories \babar (Detector at SLAC, USA) and {\it Belle} (Detector at KEK, Japan) started data taking and have collected large data samples ever since. These data sets enable studies of rare \B-decays.\\
A substantial fraction of all \B-decays produce baryons in the final state, which was confirmed by the ARGUS collaboration in 1992. They measured the inclusive branching fraction of \B-mesons into baryons \cite{Albrecht:1992he} to be

\begin{equation}
	\BR(\B \rightarrow baryons + X) = (6.8 \pm 0.5_{\rm stat} \pm 0.3_{\rm syst})\%.
	\label{eq:1}
\end{equation}

Up to now little is known about the underlying production mechanisms of baryons in weak \B-decays. One of the major drawbacks is that for most baryonic \B-decays several decay mechanisms have to be taken into account without knowledge of their impact on the total branching fraction. Since theoretical predictions are rare, due to the non-perturbative nature of the describing quantum field theory (quantum chromodynamics, QCD), it is necessary to measure \B-decays proceeding exclusively via one mechanism. This requirement is met by the semileptonic \B-decay $\Bm \ra \LCp \antiproton \ellm \nulb$ investigated in the present work.

\section{The Standard Model}

The Standard Model of Particle Physics was developed in the 20th century and has become one of the most successful theories in physics. According to the Standard Model all matter in the universe consists of a few fundamental particles, the leptons and the quarks. While leptons can occur isolated, quarks can only exist in bound states, the mesons and baryons. Moreover, three of the four fundamental interactions can be described by the standard model. The interactions incorporated into the theoretical framework of the Standard Model are the electromagnetic, the weak and the strong interaction. Up to now it is not possible to describe gravity in terms of the Standard Model since a quantum field theory of gravity is still missing.\\ 
In the next sections the Review of Particle Physics 2012 \cite{PDG:2012} was taken as reference for particle properties unless stated differently.

\subsection{The Fundamental Particles}\label{particles}

In the Standard Model two groups of fundamental particles are used to describe the visible matter in the universe: the quarks and the leptons. Both groups consist of fermions, particles with spin $1/2$. The next sections give an overview of the basic properties of these two groups. Here, we follow the description given in my diploma thesis \cite{Leddig:Diplom}.

\subsubsection{The Leptons}

The first member of this group of particles, the electron, was discovered in the year 1897 by J.J. Thomson. The muon followed in 1937, and the the tauon $\tau$, as last charged member of this group, was discovered in 1975. The neutral partner of these charged leptons was detected in 1956 by Cowan and Reines. In 1962 it was shown by Lederman {\it et al.} that there is a substantial difference between the electron and the muon neutrino. Today the existence of three types of neutrinos, corresponding to the charged leptons, is established. \\
In Table \ref{tab:1} the basic properties of the three charged and the three uncharged leptons are shown. Here the electric charge is given in units of the electron charge $e$.
	\begin{table}[H]
		\centering
		\caption{The leptons and their basic properties}
		\begin{tabular}{ccc}\toprule
						& mass in \mevcc		& electric charge	\\
			\toprule
			electron 		& $0.510998928 \pm 0.000000011$	& $-1$ 			\\
			electron-neutrino 	& $< 2 \cdot 10^{-6}$		& $0$  			\\
			\midrule
			muon 			& $105.6583715 \pm 0.0000035$	& $-1$ 			\\
			muon-neutrino 		&  $< 2 \cdot 10^{-6}$		& $0$  			\\
			\midrule
			tauon 			& $1776.82 \pm 0.16$		& $-1$ 			\\
			tauon-neutrino 		& $< 2 \cdot 10^{-6}$		& $0$  			\\
			\bottomrule
		\end{tabular}
		\label{tab:1}
	\end{table}
As can be seen in Table \ref{tab:1}, the leptons can be arranged in three families, each family consisting of a charged particle and the corresponding neutrino.\\
While the masses of the charged leptons are well known it has only been possible to estimate upper limits for the neutrino masses. Up to now only the mass differences between the different neutrino types have been measured with high precision. But neither the mass hierarchy of the neutrinos nor an absolute mass value for one of the neutrinos has been measured.

\subsubsection{The Quarks}\label{quarks}

The abundance of discovered hadrons, particles bound by the strong interaction, required a new ordering scheme. This scheme was introduced by M. Gell-Mann, who arranged the known hadrons into several multiplets. Later on this classification could be explained by a new group of fundamental particles which are the building blocks of the known hadrons. This new group of particles was coined quarks.\\ 
Today we know six quarks of different flavor (up, down, charm, strange, top and bottom), arranged in three families, analogous to the three lepton families. The three quark-families, together with their basic properties, can be seen in Table \ref{tab:2}.
	\begin{table}[H]
		\centering
               	\caption{The quarks and their basic properties}
		\begin{tabular}{ccc}\toprule
					& current mass in \mevcc& electric charge	\\
			\toprule 
			up 		& $2.3^{+0.7}_{-0.5}$	& ${+\frac{2}{3}}$ 	\\
			down	 	& $4.8^{+0.7}_{-0.3}$	& ${-\frac{1}{3}}$ 	\\
			\midrule
			charm		& $1275 \pm 25$		& ${+\frac{2}{3}}$ 	\\
			strange		& $95 \pm 5$		& ${-\frac{1}{3}}$	\\
			\midrule
			top		& $173500 \pm 600 \pm 800$	& ${+\frac{2}{3}}$	\\
			bottom 		& $4180 \pm 30$		& ${-\frac{1}{3}}$	\\
			\bottomrule
		\end{tabular}
		\label{tab:2}
	\end{table}
As it can be seen from the table, quarks exist with two different charges, ${+\frac{2}{3}}$ and ${-\frac{1}{3}}$. Further their masses range from $\approx 1 \mevcc$ to $174 \gevcc$, a multitude of the proton mass. 

\subsection{The Fundamental Forces}\label{forces}

These two groups of fundamental particles interact with each other via four fundamental forces, listed in Table \ref{tab:7} in the order of their relative strength.
\begin{table}[H]
	\centering
        \caption{The four fundamental forces, listed in the order of their relative strength, at $Q^2 = 0$}
	\begin{tabular}{llll}\toprule
		force		&	strength		&	theory	&	gauge-bosons \\\midrule
		strong		&	$1 \cdot 10$		&	QCD	&	8 gluons \gluon \\
		electromagnetic	&	$1 \cdot 10^{-2}$	&	QED	&	photon \g \\
		weak		&	$1 \cdot 10^{-13}$	&	QFD	&	\W and \Z \\
		gravitational	&	$1 \cdot 10^{-42}$	&	GTR	&	graviton \\
		\bottomrule
	\end{tabular}
	\label{tab:7}
\end{table}
In the context of the SM only the first three interactions can be described by a quantum field theory (QFT). Gravity can only be described in terms of the general theory of relativity which cannot be combined with the QFT of the SM. Hence, gravity is no part of the SM of Particle Physics.

\subsubsection{The electromagnetic interaction}

Quantum electrodynamics, which describes the electromagnetic interaction, is the oldest and most successful theory of the dynamic theories. It describes the interaction between charged particles by the exchange of a massless, electrically neutral boson (particle with integer spin): the photon \g. In the limit of strong fields QED passes into classical electrodynamics, described by Maxwell's equations.\\ 
Up to now QED predictions meet experiments with an extremely high degree of accuracy: currently about $10^{-12}$ \cite{wiki}.


\subsubsection{The weak interaction}

In contrast to the other interactions described in the Standard Model the weak interaction distinguishes between left and right. It only affects left handed particles (particles with a spin antiparallel to their momentum) and right handed anti-particles, which means that it violates parity symmetry ({\bf P}). Moreover the weak interaction violates {\bf CP}-symmetry, i.e. particles and anti-particles behave differently under the weak interaction. This is a contribution to the dominance of matter in the universe. It is also the only interaction that is able to change the flavor of a particle, i.e. it can change an up-type quark into a down-type quark and vice versa.\\
The weak interaction acts on all quarks and leptons, and is described by the exchange of heavy bosons, the $\W^{\pm}$ and $Z^0$ which have a mass of $(80.385 \pm 0.025) \gevcc$ and $(91.1876 \pm 0.0021) \gevcc$ \cite{PDG:2012}, respectively. The large mass of the force carriers leads to a range smaller than the diameter of a nucleus, in contrast to the electromagnetic interaction which has an infinite range.\\
While the weak interaction conserves the family in the lepton sector, i.e. the number of particles from a specific family is conserved, this does not hold true for weak processes in the quark sector. In order to explain this discrepancy Cabibbo suggested 1963 that the quark generations are ``rotated'' for the weak interaction, consequently the weak eigenstates are different from the strong eigenstates. The eigenstates of the weak interaction are
\begin{center}
	\begin{equation}
		\left(
			\begin{array}{c}
				u \\
				d^\prime
			\end{array}
		\right),
		\left(
			\begin{array}{c}
				c \\
				s^\prime
			\end{array}
		\right),
		\left(
			\begin{array}{c}
				t \\
				b^\prime
			\end{array}
		\right)
		\label{eq:2}
	\end{equation}
\end{center}
in contrast to the strong eigenstates
\begin{center}
	\begin{equation}
		\left(
			\begin{array}{c}
				u \\
				d
			\end{array}
		\right),
		\left(
			\begin{array}{c}
				c \\
				s
			\end{array}
		\right),
		\left(
			\begin{array}{c}
				t \\
				b
			\end{array}
		\right)
		\label{eq:3}
	\end{equation}
\end{center}
Here $d^\prime$, $s^\prime$ and $b^\prime$ are linear combinations of the physical quarks $d$, $s$ and $b$. The relation between the ``twisted'' and the physical quarks is given  by the {\it Cabibbo-Kobayashi-Maskawa matrix} (CKM matrix) $V_{CKM}$:
\begin{center}
	\begin{equation}
		\left(
			\begin{array}{c}
				d^\prime \\
				s^\prime \\
				b^\prime
			\end{array}
		\right) = 
		\left(
			\begin{array}{ccc}
				V_{ud} & V_{us} & V_{ub} \\
				V_{cd} & V_{cs} & V_{cb} \\
				V_{td} & V_{ts} & V_{tb} 
			\end{array}
		\right) \cdot
		\left(
			\begin{array}{c}
				d \\
				s \\
				b
			\end{array}
		\right)
		\label{eq:4}
	\end{equation}
\end{center}
At this point $V_{ab}$ names the relative coupling strength between $a$ and $b$. Experiments have delivered the following magnitudes of all nine CKM matrix elements \cite{PDG:2012}:
\begin{center}
	\begin{equation}
		\left| V_{CKM} \right|= 
		\left(
			\begin{array}{ccc}
				0.97427 \pm 0.00015 		& 0.22534 \pm 0.00065 		& 0.00351^{+0.00015}_{-0.00014}	\\
				0.22520 \pm 0.00065 		& 0.97344 \pm 0.00016 		& 0.0412^{+0.0011}_{-0.0005} 	\\
				0.00867^{+0.00029}_{-0.00031} 	& 0.0404^{+0.0011}_{-0.0005}	& 0.999146^{+0.000021}_{-0.000046} 
			\end{array}
		\right)
		\label{eq:5}
	\end{equation}
\end{center}
As it can be seen from eq. \eqref{eq:5} transitions inside one family (e.g. $c \rightarrow s$) are much more likely than transitions between different families (e.g. $c \rightarrow d$) which are called ``Cabibbo-suppressed''.

\subsubsection{The strong interaction}\label{strong}

\begin{figure}[H]
	\centering\includegraphics[width=.55\textwidth]{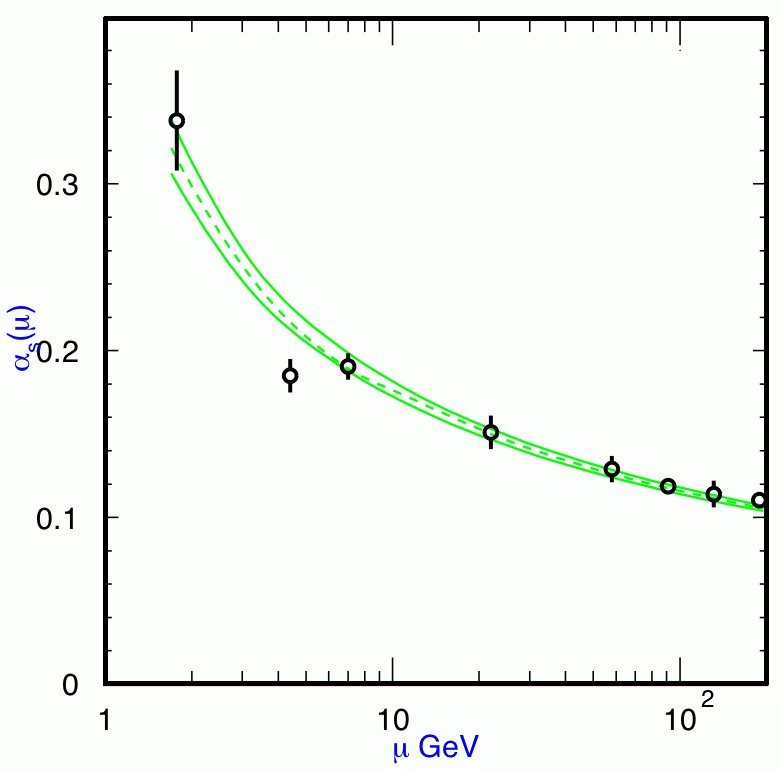}
	\caption{Summary of the values of $\alpha_s(\mu)$ at the energy scale of $\mu$ where they are measured. The lines show the central values and the $\pm 1 \sigma$ limits of the average \cite{pdg}.}
	\label{fig:2}
\end{figure}

The strong interaction is responsible for the coupling of the quarks in bound states. While the electromagnetic interaction couples to the electric charge, and the weak interaction couples to the weak charge, the strong interaction couples to the color-charge of the quarks. Quarks exist in three colors (red, green and blue) and anti-quarks in three anti-colors (anti-red, anti-green and anti-blue). But it is only possible to observe the colorless bound states. In mesons the color of the quark and the anti-color of the anti-quark compensate (e.g. red and anti-red), while in baryons all three colors have to appear which leads to a colorless particle. \\
The strong interaction is mediated by eight gluons, massless particles carrying a color- and an anti-color charge. In contrast to the electromagnetic and weak force, the strong force can not be described in terms of a simple $1/r$ law ($r$ is the distance between the interacting particles), since the strong coupling constant is a running constant, as shown in Figure \ref{fig:2}. At high energies the quarks are close together and the interaction is weak (asymptotic freedom), while at low energies the distance between them is large and the interaction is strong (confinement). \\
Asymptotic freedom can be explained by the self-interacting nature of the gluons. This self-interaction leads to a weak interaction between the quarks at small distances, and hence to asymptotic freedom. To understand confinement the strong interaction can be interpreted as a string, if two quarks are separated the energy density stored in the connecting string rises. At sufficient high densities the string breaks up forming a quark-antiquark pair, which are bound to the primary quarks. Consequently quarks can not be referred to as free particles at low energies.

\subsection{Bound States}

As described before the strong interaction binds the quarks into colorless states, the hadrons.
Today two colorless bound states of quarks are well established: the mesons and the baryons. This two groups will be explained in more detail in the next two sections.\\
Recent measurements by Belle \cite{Choi:2007wga} and LHCb \cite{Aaij:2014jqa} seem to point to a new group of bound states, so called tetraquarks. But, since the nature of the observed tetraquark candidate $Z(4430)$ is not confirmed, this group of hadrons will be omitted.

\subsubsection{Mesons}\label{meson}

Mesons, consisting of a quark and an antiquark, pose the most simple combination of quarks into a colorless bound state. Meson ground states can be divided into pseudoscalar and vector mesons, depending on the spin orientation of the quark and antiquark. If the spins are aligned antiparallel $\uparrow \downarrow$ the meson is called a pseudoscalar meson, while a parallel $\uparrow \uparrow$ alignment is called a vector meson. Similar to atomic spectroscopy, excited states of mesons are possible by different values of the angular momentum $L$.\\
However, the picture, that a meson only consists of a quark and an antiquark, is too simple. Since the quarks are bound by gluons which are self-interacting and can fluctuate into quark-antiquark pairs the inner structure of a meson is much more complicated. The current model of the structure of a meson is that it consists of the valence quarks, the sea quarks and the sea gluons. The meson type is determined by the valence quarks alone.\\
The basic properties of the mesons relevant for this work are shown in Table \ref{tab:3}. Here the width is an equivalent value to the lifetime for short-lived particles.

\subsubsection{Baryons}\label{baryon}

Baryons, as second type of a bound state of quarks, consist of 3 valence quarks. The naming scheme relates to the isospin, as well as the quark content. For example baryons containing only two \u or \d quarks is called a $\Lambda$ (isospin $0$)  or $\Sigma$ (isospin $1$). If the third quark is a charm or bottom quark it is given as index. For different mass states of the same quark content and isospin configuration the mass is given, e.g. $\Sigma_c(2520)$. Consisting of three valence quarks they follow the Fermi statistics. Well known baryons are the proton and the neutron as constituents of the nucleus. Table \ref{tab:4} shows the basic properties of the baryons relevant for this work.

\begin{table}[h]
\setlength{\tabcolsep}{.5em}
	\centering
        \caption{Mesons relevant for this work and their basic properties}
	\begin{tabular}{ccccc}\toprule
		Meson		& quarks	&	mass in \mevcc		& lifetime			& width $\Gamma$ in MeV	\\
		\midrule 
		\pip	& $\u \dbar$ & $139.57018 \pm 0.00035$	& $(2.6033 \pm 0.0005) \cdot 10^{-8}$ 	& $-$			\\
		\Kp	& $\u\sbar$	& $493.677 \pm 0.016$		& $(1.2385 \pm 0.0024) \cdot 10^{-8}$ 	& $-$			\\
		\Bz	& $\d \bbar$ & $5279.4 \pm 0.5$		& $(1.530 \pm 0.009) \cdot 10^{-12}$	& $-$			\\
		\FourS		& $\b \bbar$		& $10579.4 \pm 1.2$		& $-$					& $20.5 \pm 2.5$	\\
		\bottomrule
	\end{tabular}
	\label{tab:3}
	\vspace{0.5cm}
	\centering
        \caption{Baryons relevant for this work and their basic properties}
	\begin{tabular}[H]{ccccccc}\toprule
		Baryon				& quarks	&	mass in \mevcc		& lifetime			& width $\Gamma$ in MeV	\\
		\midrule 
		\proton, \antiproton		& $\u \u \d$	& $938.27203 \pm 0.00008$	& $> 10^{31}$ years		& $-$			\\
		$\Lambda_c^+$	& $c \u \d$	& $2286.46 \pm 0.14$		& $(200 \pm 6) \cdot 10^{-15}$s 	& $-$			\\
		$\Sigma_c(2455)^{++}$		& $c \u \u$	& $2453.98 \pm 0.16$		& $-$				& $2.26 \pm 0.25$	\\
                $\Sigma_c(2455)^{+}$		& $c \u \d$	& $2452.9 \pm 0.4$		& $-$				& $<4.6$		\\
                $\Sigma_c(2455)^{0}$		& $c \d \d$	& $2453.74 \pm 0.16$		& $-$				& $2.16 \pm 0.26$	\\
		\bottomrule
	\end{tabular}
	\label{tab:4}
\end{table}
\clearpage


\boldmath
\chapter{Baryonic \B decays}
\unboldmath
Baryons are the main constituent of the visible matter in our universe, but despite their large importance for our understanding of the universe little is known about their production mechanisms. 
A possibility is the production in decays of heavy mesons. Here, \B mesons are the first known mesons heavy enough to decay into a large variety of baryonic final states. In addition, decays into baryons make up a significant part of the overall branching fraction of \B mesons.

In the next sections a summary of the most striking results is given.


\section{Multiplicity hierarchy}

\B-decays to baryons are not as well studied as mesonic decays. Previous measurements show a strong hierarchy of the branching fractions depending on the final state multiplicity, as shown in Table \ref{tab:LambdaC} for \B decays to a $\LCp$ baryon and in Table \ref{tab:Dbranch} for decays to a charmed meson, accompanied by a baryon-antibaryon pair. 
\begin{table}[b]
  \begin{center}
    \caption{Branching fraction results for $B \rightarrow  \Lambda_c^+ \bar{p} \; {\rm m}\cdot \pi$ with $ {\rm m}= 0,1,2,3$, ordered according to their multiplicity\cite{PDG:2012}.}
    \begin{tabular}{lr}\hline
      decay mode & ${\cal B} \pm \sigma (10^{-4})$  \\\hline
      $\bar{B}^0 \rightarrow \Lambda_c^{+}\bar{p}$ & $0.20 \pm 0.04$  \\\hline
      $\bar{B}^{0} \rightarrow \Lambda_c^{+}\bar{p}\pi^0$         & $1.9 \pm 0.5$  \\
      $B^{-} \rightarrow \Lambda_c^{+}\bar{p}\pi^-$               & $2.8 \pm 0.8$  \\\hline
      $\bar{B}^{0} \rightarrow \Lambda_c^{+}\bar{p}\pi^-\pi^+$ & $11.2 \pm 3.2$ \\\hline
      $B^{-} \rightarrow \Lambda_c^{+}\bar{p}\pi^-\pi^-\pi^+$     & $22 \pm 7$  \\
      \hline
    \end{tabular}
    \label{tab:LambdaC}
  \end{center}
\end{table}
For \B decays to a charmed baryon the branching fraction increases by an order of magnitude comparing the two-body decay $\bar{B}^0 \rightarrow \Lambda_c^{+}\bar{p}$ with the three-body decays $\bar{B}^{0} \rightarrow \Lambda_c^{+}\bar{p}\pi^0$ and $B^{-} \rightarrow \Lambda_c^{+}\bar{p}\pi^-$. This increase is in contrast to the mesonic \B decays, where the three-body branching fraction is at the same order of magnitude as for the corresponding two-body mode. A possible explanation for the strong increase from the two-body to the three-body decay comes from the resonant substructure. The additional pion allows for $\Sigma_c$ and nucleon resonances, thus increasing the number of possible decay paths. Further, according to \cite{Suzuki:2007pw} the production of a two-body baryonic final state requires a hard gluon, introducing a strong suppression factor. Adding a light meson reduces the invariant mass of the remaining system, allowing for a soft gluon in the baryon production.
For the decay into a charmed meson accompanied by a baryon-antibaryon pair the branching fraction reaches its maximum for multiplicities of four, as can be seen in Table \ref{tab:Dbranch}.
\begin{table}[h]
  \begin{center}
    \caption{Branching fraction results for $B \rightarrow D \proton \antiproton \; {\rm m}\cdot \pi$ with $ {\rm m}= 0,1,2$ \cite{PhysRevD.85.092017}, ordered according to their multiplicity.}
    \begin{tabular}{lc}\toprule
      \B decay					& $\BR \pm \sigma_{\rm stat} \pm \sigma_{\rm syst}$ $(10^{-4})$\\\midrule
      $\Bzb \ra D^0 \proton \antiproton$		& $1.02 \pm 0.04 \pm 0.06$\\
      $\Bzb \ra D^{*0} \proton \antiproton$		& $0.97 \pm 0.07 \pm 0.09$\\\midrule
      $\Bzb \ra D^+ \proton \antiproton \pim$		& $3.32 \pm 0.10 \pm 0.29$\\
      $\Bzb \ra D^{*+} \proton \antiproton \pim$		& $4.55 \pm 0.16 \pm 0.39$\\
      $\Bm \ra D^0 \proton \antiproton \pim$		& $3.72 \pm 0.11 \pm 0.25$\\
      $\Bm \ra D^{*0} \proton \antiproton \pim$		& $3.73 \pm 0.17 \pm 0.27$\\\midrule
      $\Bzb \ra D^0 \proton \antiproton \pim \pip$	& $2.99 \pm 0.21 \pm 0.45$\\
      $\Bzb \ra D^{*0} \proton \antiproton \pim \pip$	& $1.91 \pm 0.36 \pm 0.29$\\
      $\Bm \ra D^{+} \proton \antiproton \pim \pim$	& $1.66 \pm 0.13 \pm 0.27$\\
      $\Bm \ra D^{*+} \proton \antiproton \pim \pim$	& $1.86 \pm 0.16 \pm 0.19$\\\bottomrule
    \end{tabular}
    \label{tab:Dbranch}
  \end{center}
\end{table}

\section{Threshold enhancement}\label{sect:thresh}

\begin{figure}
  \subfigure[]{
    \includegraphics[width=.46\textwidth]{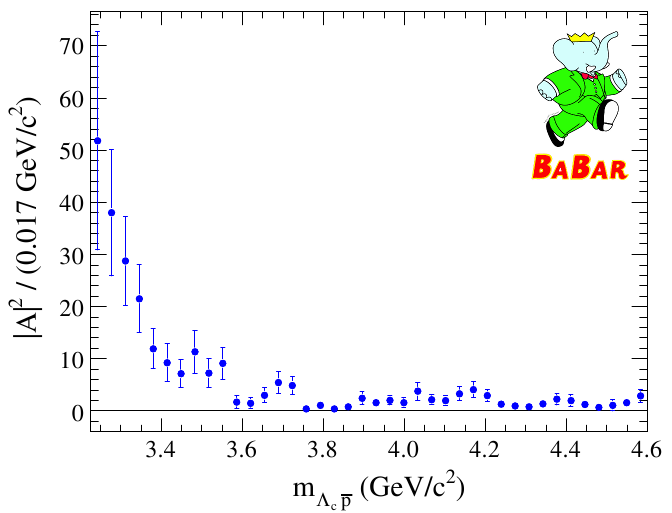}
    \label{subfig:thresh_Lcppi}
  }
  \subfigure[]{
    \includegraphics[width=.5\textwidth]{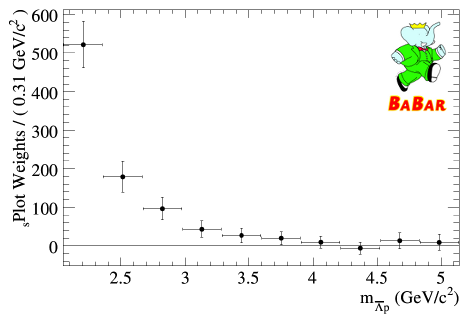}
    \label{subfig:thresh_Lppi}
  }
  \subfigure[]{
    \includegraphics[width=.58\textwidth]{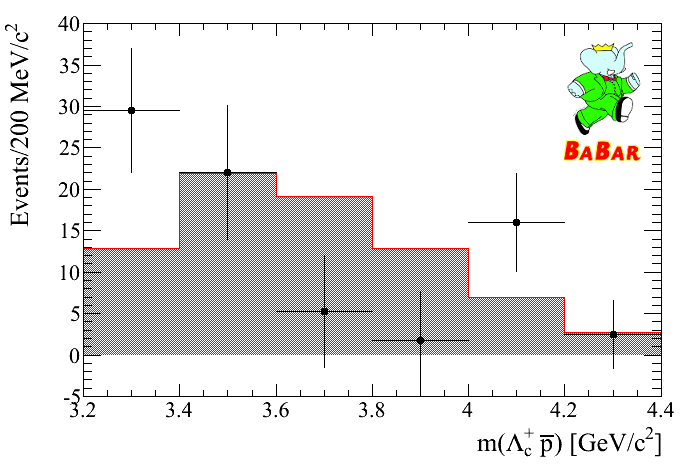}
    \label{subfig:thresh_LcpKpi}
  }
  \subfigure[]{
    \centering\includegraphics[width=.38\textwidth]{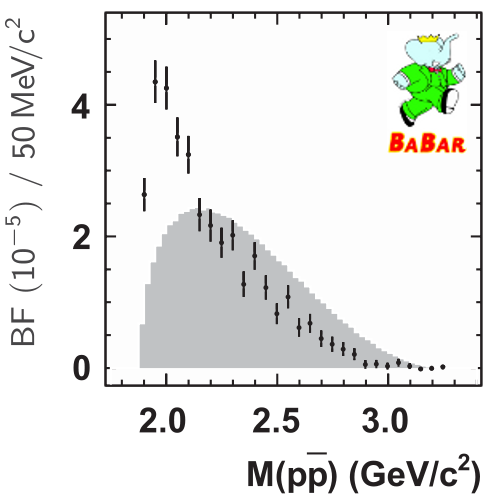}
    \label{subfig:thresh_Dpppi}
  }
  \caption{The invariant baryon antibaryon mass distributions for \subref{subfig:thresh_Lcppi} $\Bm \ra \LCp \antiproton \pim$ \cite{Aubert:2008ax}, \subref{subfig:thresh_Lppi} $\Bzb \ra \Lambda\antiproton \pip$ \cite{PhysRevD.79.112009}, \subref{subfig:thresh_LcpKpi} $\Bzb \ra \LCp \antiproton \Km \pip$ \cite{PhysRevD.80.051105,PoS.EPS-HEP.2009.215} and \subref{subfig:thresh_Dpppi} $\Bzb \ra D^+ \proton \antiproton \pim$ \cite{PhysRevD.85.092017}. For the upper row the signal distribution was divided by the expectation from a simple phase space model. The lower row shows the signal distribution, with the phase space expectation represented as a histogram for $\Bzb \ra D^+ \proton \antiproton \pim$.}
  \label{fig:thresh}
\end{figure}

A feature of baryonic \B decays observed quite frequently is an enhancement at low invariant baryon-antibaryon masses. Examples are shown in Fig. \ref{fig:thresh}. Common to all the given examples is a deviation from a simple phasespace model at the invariant-mass threshold. 

Explanations for this feature come from different sides. For decays of the type $\B \ra K+X$, with the fundamental subprocess $\bbar \ra \sbar+ g +g$ a strong contribution from a flavor-singlet penguin is expected. In terms of baryonic \B decays like $\Bp \ra \proton \antiproton \Kp$ a dominant contribution of a $\proton \antiproton$ bound state with $J^{PC} = 0^{\pm +}$ can provide a fair fraction of the observed final state \cite{Rosner:2003bm}.

\begin{figure}[t]
  \subfigure[]{
    \includegraphics[width=.48\textwidth]{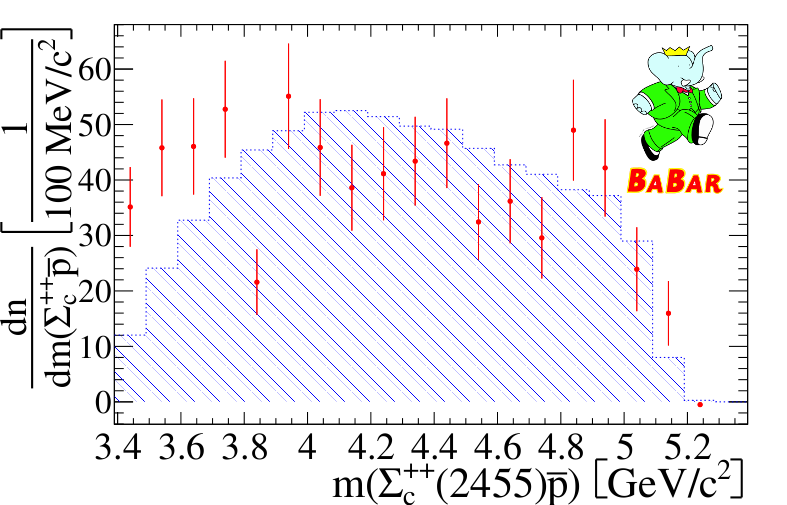}
    \label{subfig:SigmaCplpl2455p}
  }
  \subfigure[]{
    \includegraphics[width=.48\textwidth]{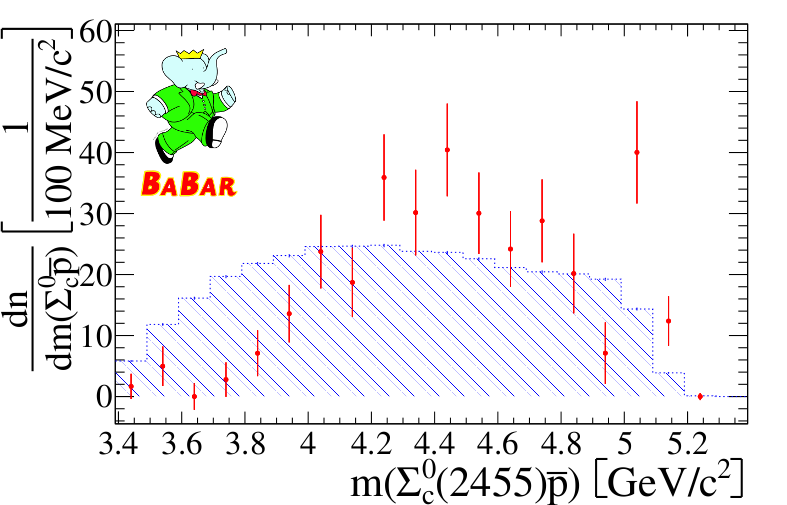}
    \label{subfig:SigmaC02455p}
  }  
  \caption{Comparison of phase space distributed signal Monte Carlo (histogram) with $_sP$lotted data (data points), \subref{subfig:SigmaCplpl2455p} shows the invariant $\Sigma_c(2455)^{++}\antiproton$ mass, and \subref{subfig:SigmaC02455p} the invariant $\Sigma_c(2455)^{0}\antiproton$ mass.}
  \label{fig:SigmaC2455p}
\end{figure}

For decays not dominated by flavor-singlet penguins a possible explanation for the threshold enhancement comes from the fragmentation into hadrons. If the baryons are neighbors in the fragmentation chain one expects their invariant mass to be low. A pole-model related rule of thumb, given in \cite{hartmann2011study}, is to check if the decay could proceed via an initial meson-meson or baryon-antibaryon configuration. In the first case one of the mesons decays into a baryon antibaryon pair, giving rise to the low mass enhancement. In the latter case no such enhancement is expected. This model is quite successful in explaining the results in $\Bzb \ra \LCp \antiproton \pim \pip$. In this analysis \cite{hartmann2011study} an enhancement is seen for the resonant subdecay $\Bzb \ra \Sigma_c(2455)^{++}\antiproton \pim$, while no enhancement is visible for $\Bzb \ra \Sigma_c(2455)^0 \antiproton \pip$. A comparison of phase space simulation and experimental data is shown in Fig. \ref{fig:SigmaC2455p}. Comparing the two distributions evidence for a low mass enhancement in $m(\Sigma_c(2455)^{++}\antiproton)$ is visible, while the low mass region in $m(\Sigma_c(2455)^{0}\antiproton)$ is unpopulated.
\begin{figure}[b]
  \subfigure[]{
    \includegraphics[width=.31\textwidth]{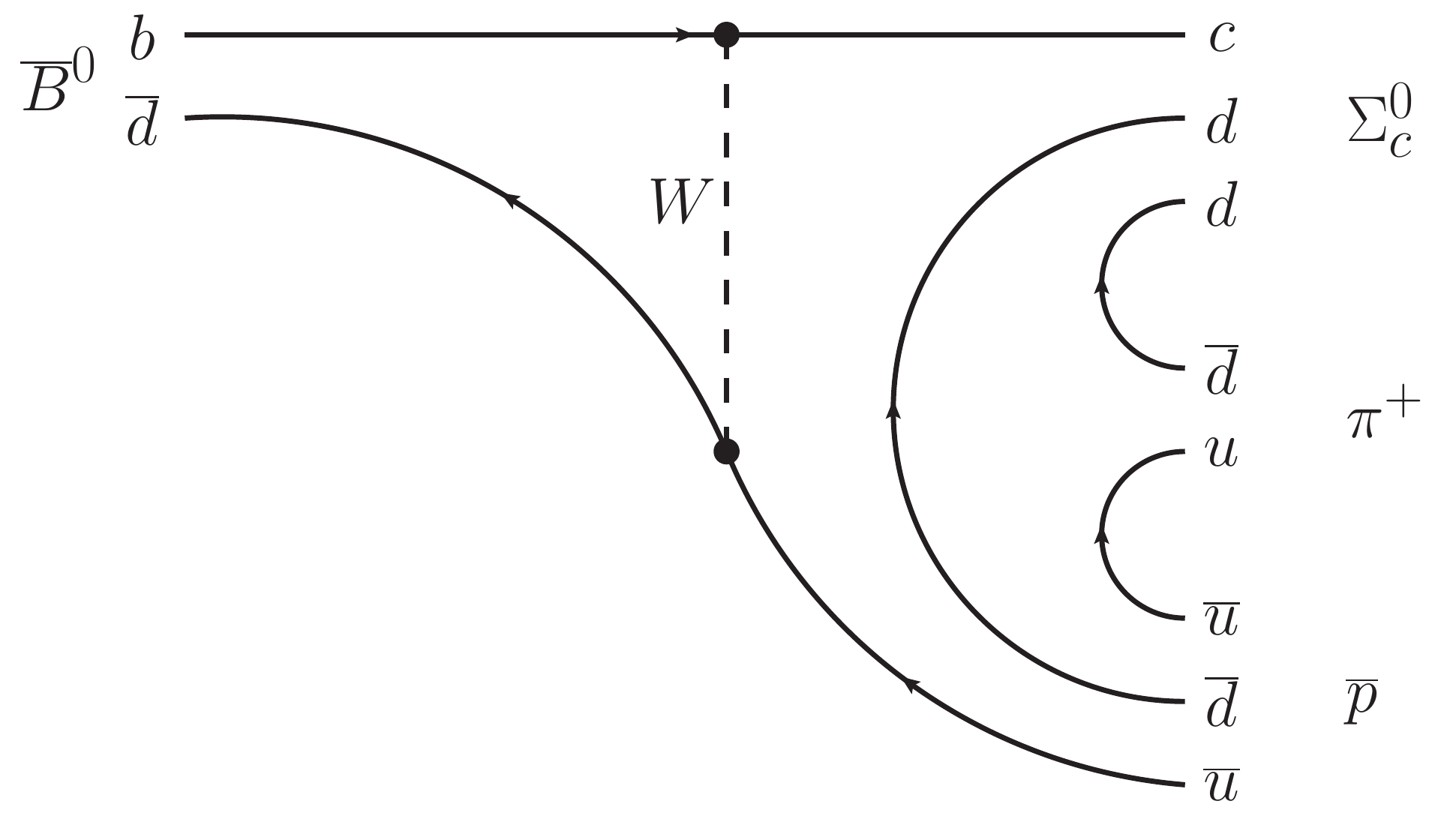}
    \label{subfig:SigmaC0_A}
  }
  \subfigure[]{
    \includegraphics[width=.31\textwidth]{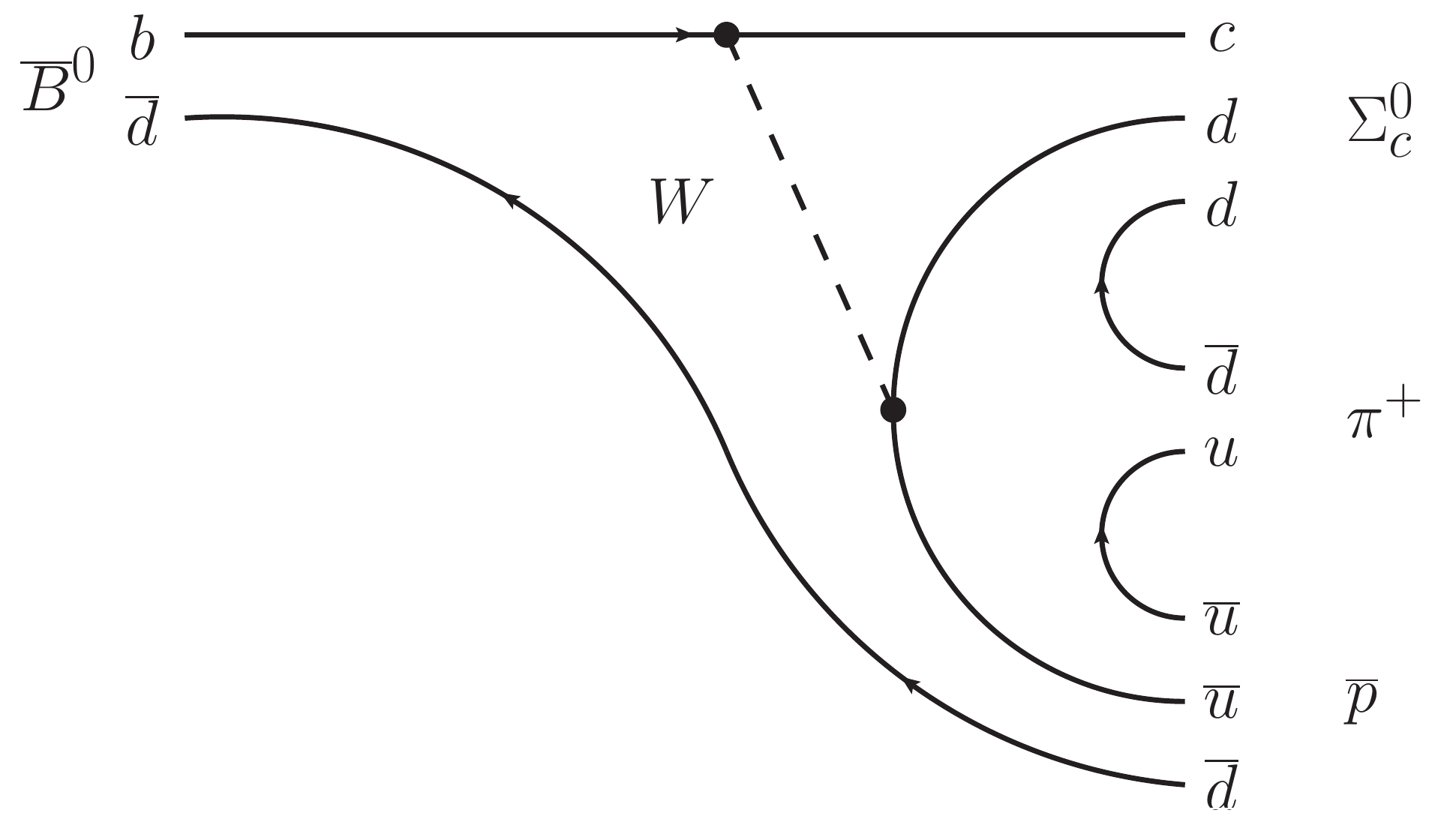}
    \label{subfig:SigmaC0_2a}
  }  
  \subfigure[]{
    \includegraphics[width=.31\textwidth]{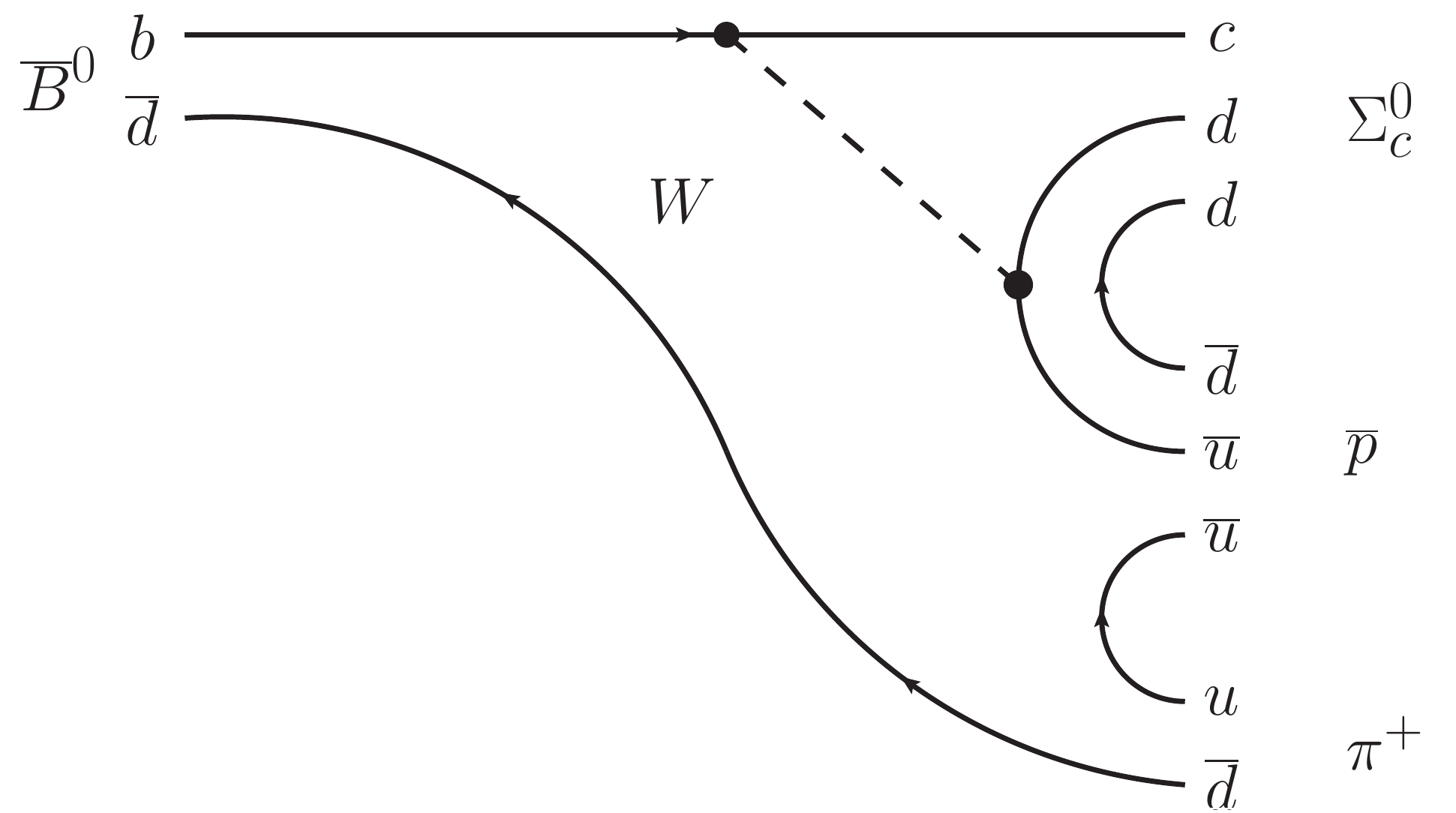}
    \label{subfig:SigmaC02455p_feyn_2b}
  }
  \caption{Comparison of the Feynman graphs responsible for the decay $\Bzb \ra \Sigma_c(2455)^{0}\antiproton \pip$.}
  \label{fig:SigmaC0}
\end{figure}
\begin{figure}[h]
  \subfigure[]{
    \includegraphics[width=.31\textwidth]{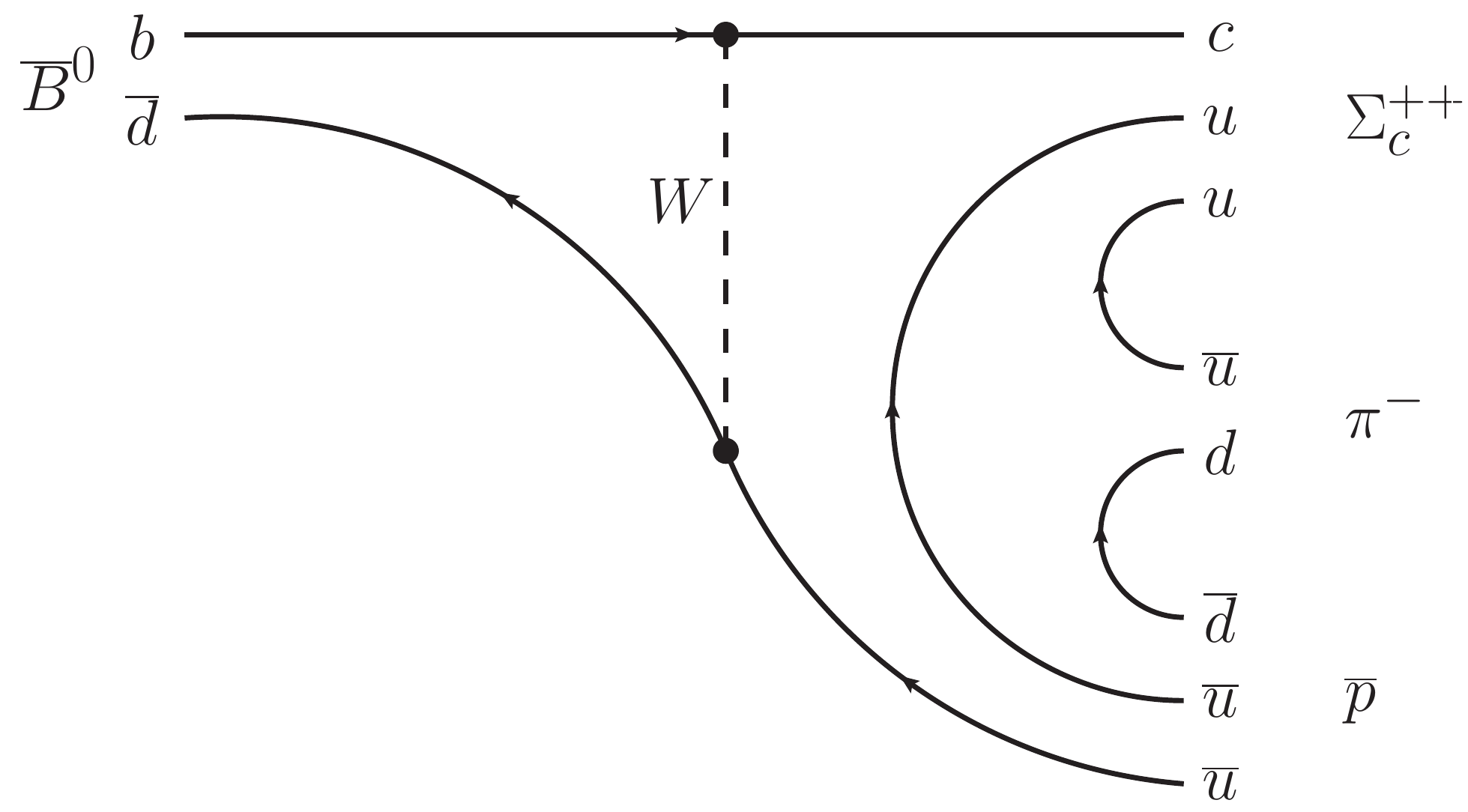}
    \label{subfig:SigmaCplpl_A}
  }
  \subfigure[]{
    \includegraphics[width=.31\textwidth]{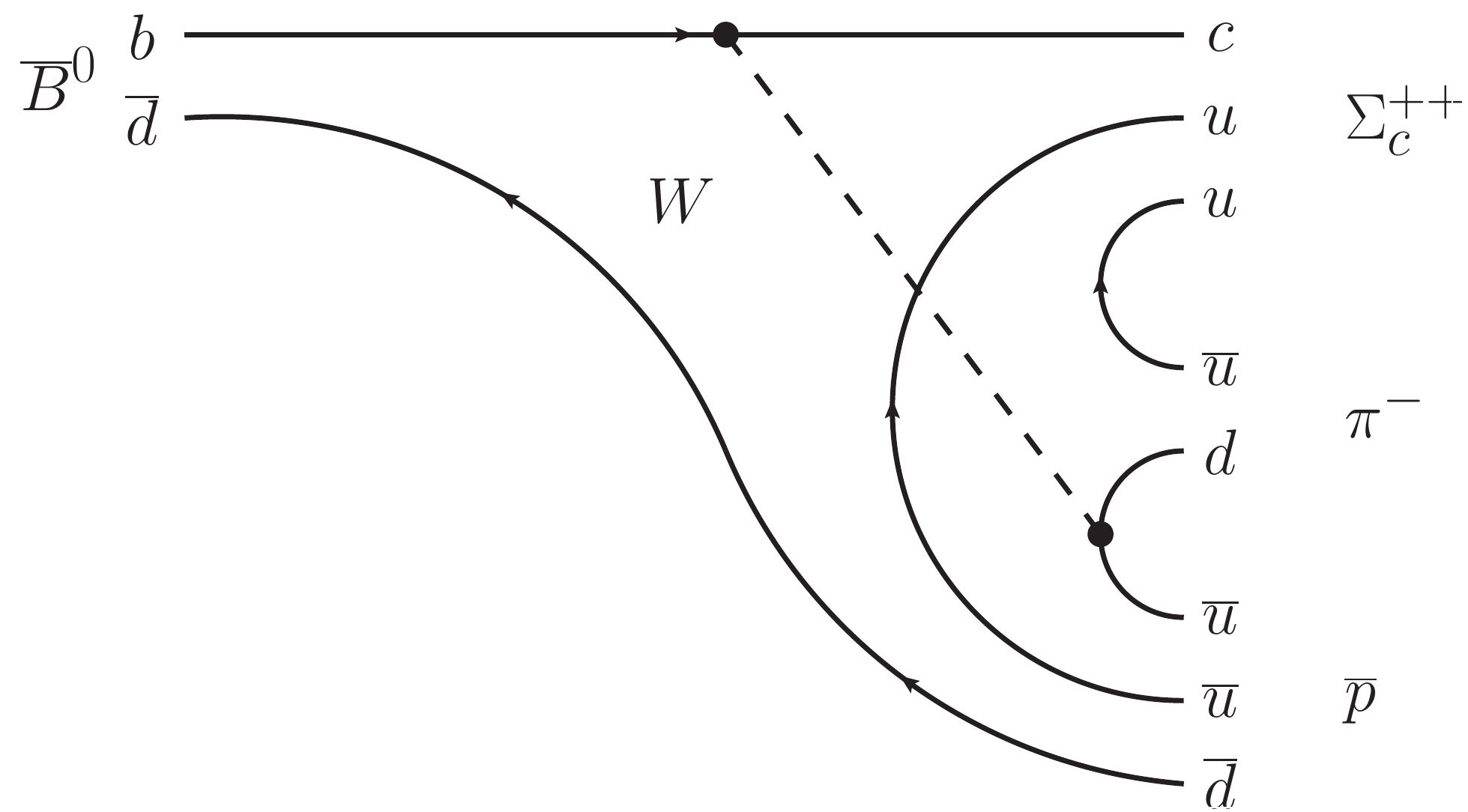}
    \label{subfig:SigmaCplpl_2g}
  }  
  \subfigure[]{
    \includegraphics[width=.31\textwidth]{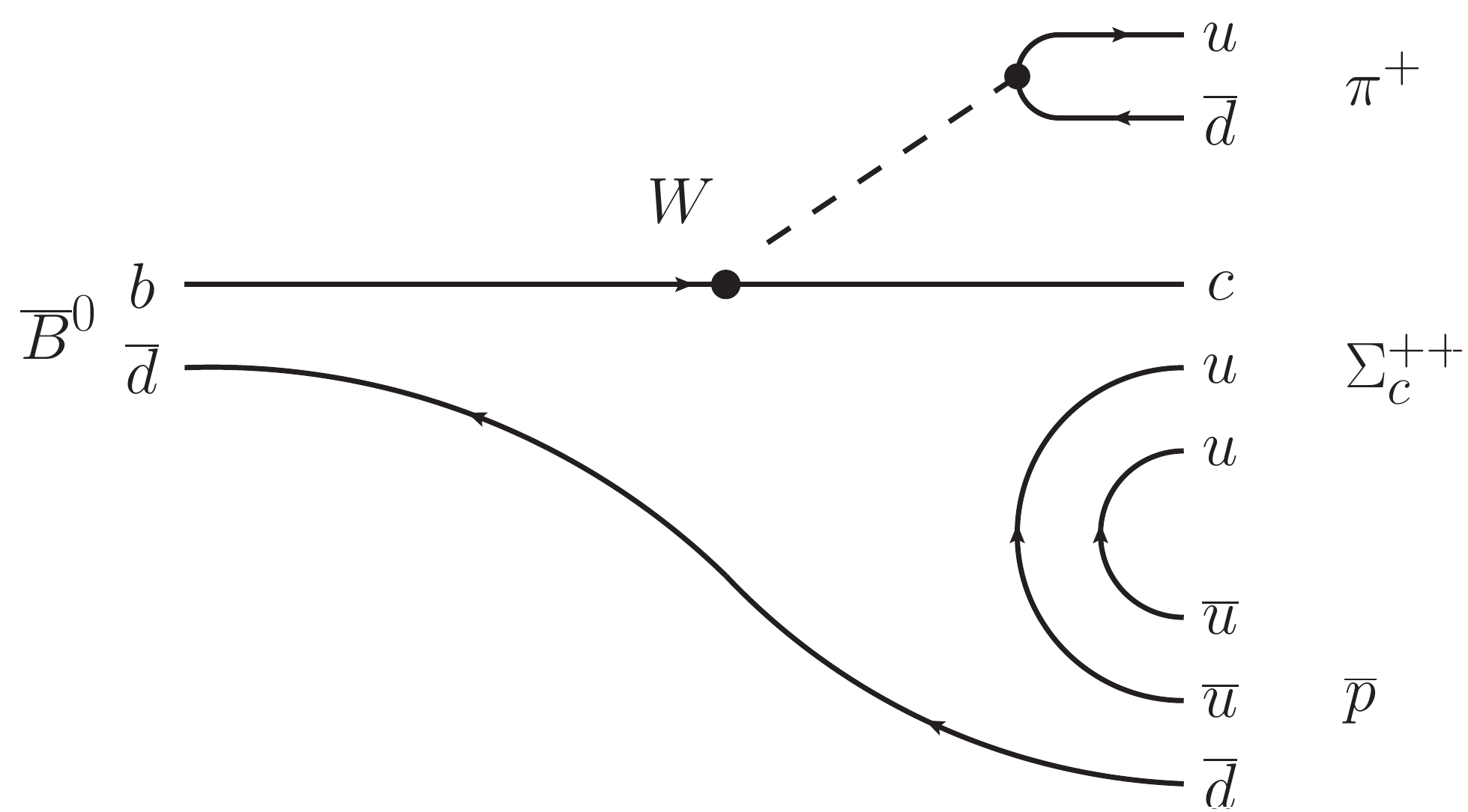}
    \label{subfig:SigmaCplpl_1}
  }
  \caption{Comparison of the Feynman graphs responsible for the decay $\Bzb \ra \Sigma_c(2455)^{++}\antiproton \pim$. }
  \label{fig:SigmaCplpl}
\end{figure}
In \cite{hartmann2011study} a simple model to explain the absence of an enhancement in $m(\Sigma_c(2455)^0\overline{p})$ is suggested. In this model the Feynman diagrams are categorized into two classes, according to their quark configuration after the weak decay.
\begin{itemize}
  \item {\bf meson-meson} configuration, i.e., before quark fragmentation, the quarks are arranged in two (virtual) mesons.
  \item {\bf diquark-diquark} configuration, i.e., before quark fragmentation, the quarks are arranged in a diquark ($q_1 q_2$) and an anti-diquark state.
\end{itemize}
In the {\bf meson-meson} configuration one of the mesons fragments into a baryon anti-baryon pair, while the second meson carries away momentum, leading to a threshold enhancement. In the {\bf diquark-diquark} configuration such an enhancement is not possible, since we already start from a baryon anti-baryon configuration. Comparing the Feynman graphs shown in Fig. \ref{fig:SigmaC0} and \ref{fig:SigmaCplpl} only the external graph in Fig. \ref{subfig:SigmaCplpl_1} is in the {\bf meson-meson} configuration, which explains the absence of an enhancement for $\Bzb \ra \Sigma_c(2455)^{0}\antiproton \pip$.

This approach seems to be valid for other \B meson decays into baryons as well, and is equivalent to the explanation via the fragmentation mechanism given in \cite{Rosner:2003bm}.

\section{Semileptonic \B decays into baryons}\label{semileptBdecay}

\begin{figure}[h]
  \begin{center}
    \includegraphics[width=.6\textwidth]{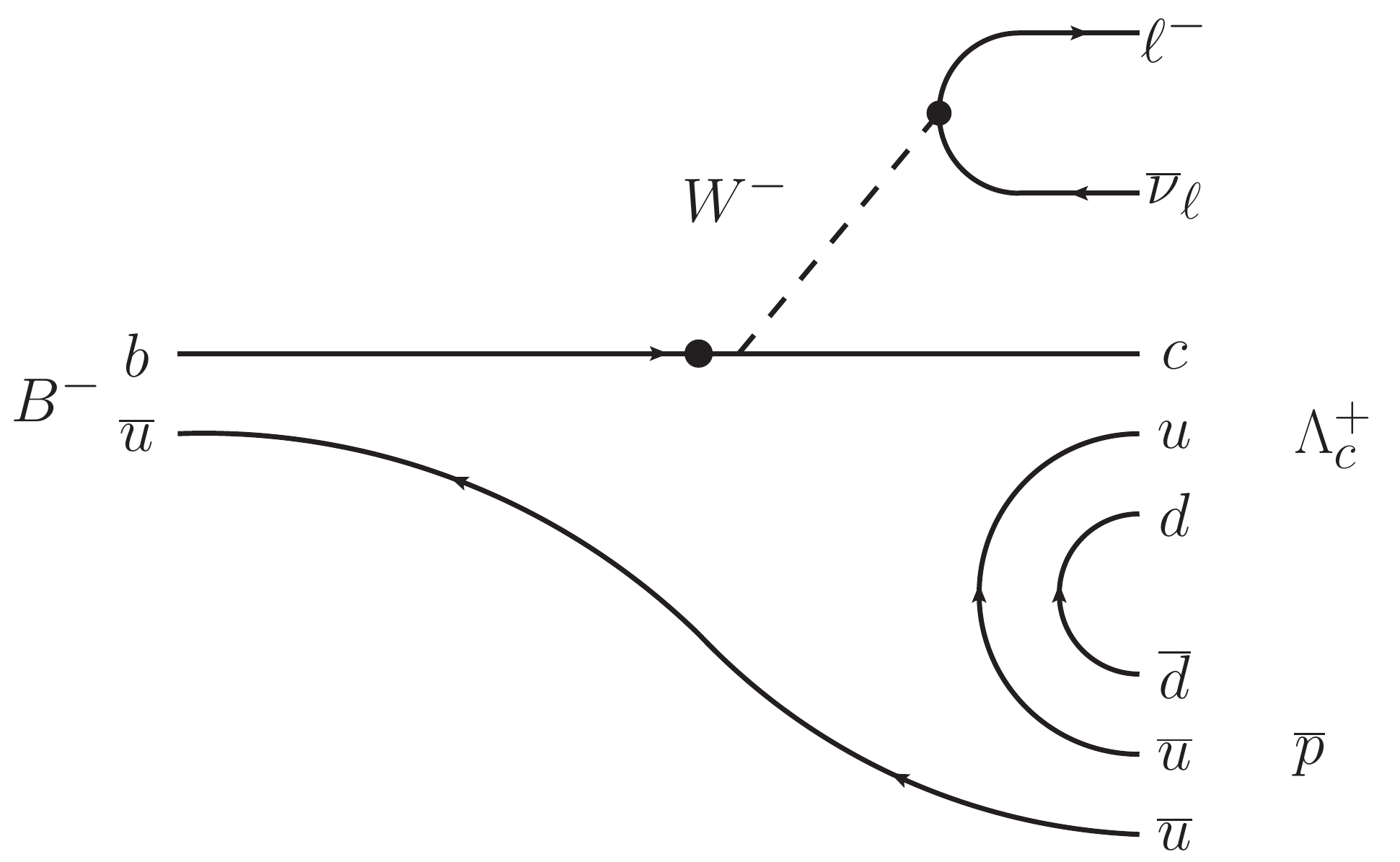}
  \end{center}
  \caption{Feynman graph for the decay $\Bm \ra \LCp \antiproton \ellm \nulb$.}
  \label{fig:feyn:BtoLcplnu}
\end{figure}

Starting from the Feynman graphs shown in \ref{fig:SigmaC0} and \ref{fig:SigmaCplpl} we can assume that the threshold enhancement is caused by the external Feynman graph. In order to assess the relative influence of the external graph we have to investigate a decay, that can proceed only via an external graph. The ideal object for such a study is the semileptonic decay $\Bm \ra \LCp \antiproton \ellm \nulb$, with $\ellm = (\en, \mun)$ and $\nulb = (\nueb, \numb)$ (Charge conjugation is implied throughout this work). The corresponding Feynman graph is shown in Fig. \ref{fig:feyn:BtoLcplnu}.

Up to now, there are only upper limits for its relative strength available. The CLEO collaboration showed that the ratio of $\B \ra \LCp X \en \nueb$ to $\B \ra \LCp X$ is smaller than $5\%$ at $90\% \; {\rm C.L.}$\cite{Bonvicini:1997fn}. 
But this result has two caveats. First, the lepton momentum is required to be greater than $0.6\gevc$, which reduces background from fake and secondary electrons, but the signal efficiency as well. Second, uncorrelated $\Bbar \ra \LCm X$ events pose a large source of systematic uncertainty. To circumvent these caveats \babar uses a tagged approach, by reconstructing a \B meson in a hadronic mode and looking for the signal in its recoil \cite{Lees:2011yh}. With this approach \babar determines the before mentioned ratio to be
\begin{equation}
  \frac{\BR(\Bbar \ra \LCp X \ellm \nulb)}{\BR(\Bbar \ra \LCp X)}<3.5\%
\end{equation}
at the $90\%$ confidence level.

The first direct measurement of a semileptonic baryonic \B decay is a recent publication by the Belle collaboration \cite{Tien:2013nga}. Like the previously mentioned \babar measurement they use a tagged approach, and found evidence for the decay $\Bm \ra \proton \antiproton \ellm \nulb$. For this measurement Belle studied $615$ exclusive hadronic decays of charged \B mesons. Paired with the large dataset of $772$ million \BBbar pairs they measured a branching fraction of
\begin{equation}
  \BR(\Bm \ra \proton \antiproton \ellm \nulb) = (5.8^{+2.4}_{-2.1}({\rm stat}) \pm 0.9({\rm syst}))\times 10^{-6}
\end{equation}
with a significance of $3.2\sigma$. The corresponding upper limit at $90\% {\rm C.L.}$ is
\begin{equation}
  \BR(\Bm \ra \proton \antiproton \ellm \nulb) < 9.6\times 10^{-4}.
\end{equation}

Starting from the measurement of $\Bm \ra \proton \antiproton \ellm \nulb$ a rough estimate of the branching fraction for $\Bm \ra \LCp \antiproton \ellm \nulb$ can be obtained. Neglecting phase space differences, the only difference is the CKM matrix element in the \b quark decay. For the former $|V_{ub}|$ has to be considered, while the latter depends on $|V_{cb}|$. The ratio $|V_{cb}/V_{ub}|$ equals $9.9$ \cite{PDG:2012}, and hence we can expect
\begin{equation}
  \BR(\Bm \ra \LCp \antiproton \ellm \nulb) \approx (5.7 \pm 2.9)\times 10^{-4}.
  \label{eq:pred:CKM}
\end{equation}

Another approach to obtain a branching fraction estimate is the fully hadronic decay $\Bm \ra \LCp \antiproton \pim$ which is measured to be $(2.8 \pm 0.8) \times 10^{-4}$\cite{PDG:2012}. Neglecting any influence of internal $W$ emission diagrams we can obtain a branching fraction by assuming
\begin{equation}
  \frac{\BR(\Bm \ra \LCp \antiproton \ellm \nulb)}{\BR(\Bm \ra \LCp \antiproton \pim)} = \frac{\BR(\tau^- \ra \ellm \nulb \nut)}{\BR(\tau^- \ra \pim \nut)}.
\end{equation}
$\BR(\tau^- \ra \pim \nut)$ is measured to $(10.83\pm 0.06)\%$, $\BR(\tau^- \ra \en \nueb \nut)$ to $(17.83 \pm 0.04)\%$, $\BR(\tau^- \ra \mun \numb \nut)$ to $(17.41 \pm 0.04)\%$ \cite{PDG:2012}. Neglecting the small difference between the latter two the ratio is roughly $1.8$, which leads to
\begin{equation}
  \BR(\Bm \ra \LCp \antiproton \ellm \nulb) \approx (4.7 \pm 1.3)\times 10^{-4}.
  \label{eq:pred:LCpppi}
\end{equation}
An estimate for the semileptonic branching fraction could be obtained in the isospin analysis of $\Bzb \ra \LCp \antiproton \piz$ and $\Bzb \ra \LCp \antiproton \eta$ \cite{ebertPHD}. Based on isospin relations the author predicts the relative strength of the external Feynman graph in $\Bm \ra \LCp \antiproton \pim$ of $(0.20 \pm 0.45)\times 10^{-4}$. Combined with the ratio of $\tau$ decays this leads to a $90\%$ confidence level upper limit for the semileptonic decay of
\begin{equation}
  \BR(\Bzb \ra \LCp \antiproton \en \nueb) < 1.4 \times 10^{-4}.
\end{equation}
This result shows a slight tension with the predictions given in eq. \eqref{eq:pred:CKM} and \eqref{eq:pred:LCpppi}. But given the limitations of these predictions the isospin prediction might be the most reliable one.

\chapter{The \babar experiment}\label{aims}
\begin{figure}[H]
	\centering\includegraphics[width=.75\textwidth]{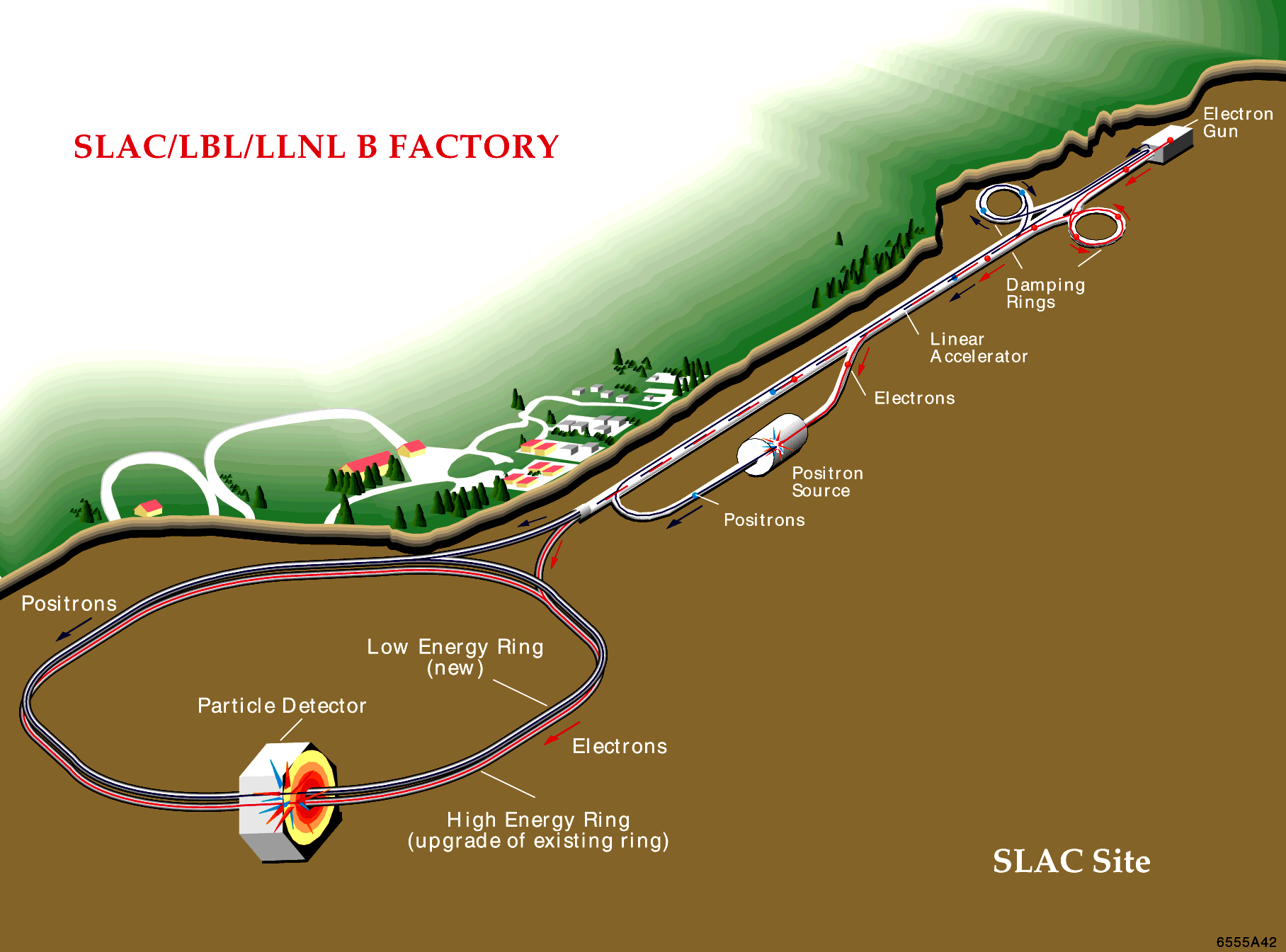}
	\caption{A schematic view of the SLAC site with the \epem accelerator and the \babar detector}
	\label{fig:3}
\end{figure}

The \babar experiment, operated from 1999 to 2008, was designed and built to perform a systematic study of \CP-asym\-metries in the decays of neutral \B-mesons. Furthermore, \babar allowed a sensitive measurement of the CKM matrix element $|V_{ub}|$ and observations of rare $D$, \B and $\tau$-decays. Together these results are capable of putting constraints on fundamental parameters of the Standard Model.
In addition a wide spectrum of physics topics, like baryonic \B-decays or charm- and tau-physics, could be studied with the \babar detector.
For the investigation of \B-mesons \babar was operated on the energy of the \FourS ($10.58\gev$) resonance, which decays into a \BBb pair with a probability of more than $96 \%$\cite{PDG:2012}. The data set collected on this resonance is the starting point for the presented analysis.\\
The experimental setup,  shown in Fig. \ref{fig:3}, is described in detail in the next sections, following the description given in \cite{Leddig:Diplom}.

\section{The \epem accelerator}

Electrons and positrons were accelerated in the $3.2\km$ long Linac up to energies of several \gev and injected into the \pep2 storage rings, where they were brought to collision inside the \babar detector.\\
Inside \pep2, electrons were stored in the high energy ring (HER) with an energy of $9 \gev$, while the positrons were stored in the low energy ring (LER) with an energy of approx. $3.1 \gev$. These energies meet the requirement of a center-of-mass energy $\sqrt{s}$ equal to the mass of the \FourS resonance, which decays into \BBb pairs, half of the time into a \BzBzb- and the other half into a \BpBm-pair.\\
The cross-sections for the different reactions possible in the \epem collision at $\sqrt{s} = 10.58 \gev$ can be seen in table \ref{tab:8}. {\it Bhabha}-scattering ($\epem \rightarrow \epem$) has by far the highest cross-section, while the other reactions are near $1 \nb$. Consequently, in most of the collisions, no \bbbar-pair is produced.\\
The unique chance \babar offered were the asymmetric beam energies, leading to a boost of the center-of-mass system into the direction of the electron beam. The boost is crucial for studies of \CP asymmetries, by allowing a measurement of the difference in the decay times of the two \B mesons. Therefore, the decay vertices of the \B-mesons have to be measured. The decay time difference can now be estimated by measuring the distance between the two vertices.
	\begin{table}[hb]
		\caption{Production cross-sections at $\sqrt{s} = 10.58 \gev$. The \epem cross-section is the effective cross-section within the experimental acceptance \cite{bpb}.}
		\centering
		\begin{tabular}{cc}\toprule
			$\epem \rightarrow$ 	& cross-section ($\nb$)	\\\midrule
			\bbbar			& $1.05$			\\
			\ccbar			& $1.30$			\\
			\ssbar			& $0.35$			\\
			\uubar			& $1.39$			\\
			\ddbar			& $0.35$			\\
			\tautau			& $0.94$			\\
			\mumu			& $1.16$			\\
			\epem			& $\approx 40$		\\\bottomrule
		\end{tabular}
		\label{tab:8}
	\end{table}

\section{The \babar detector}

\begin{figure}[h]
	\subfigure[]{
		\centering\includegraphics[width=.9\textwidth]{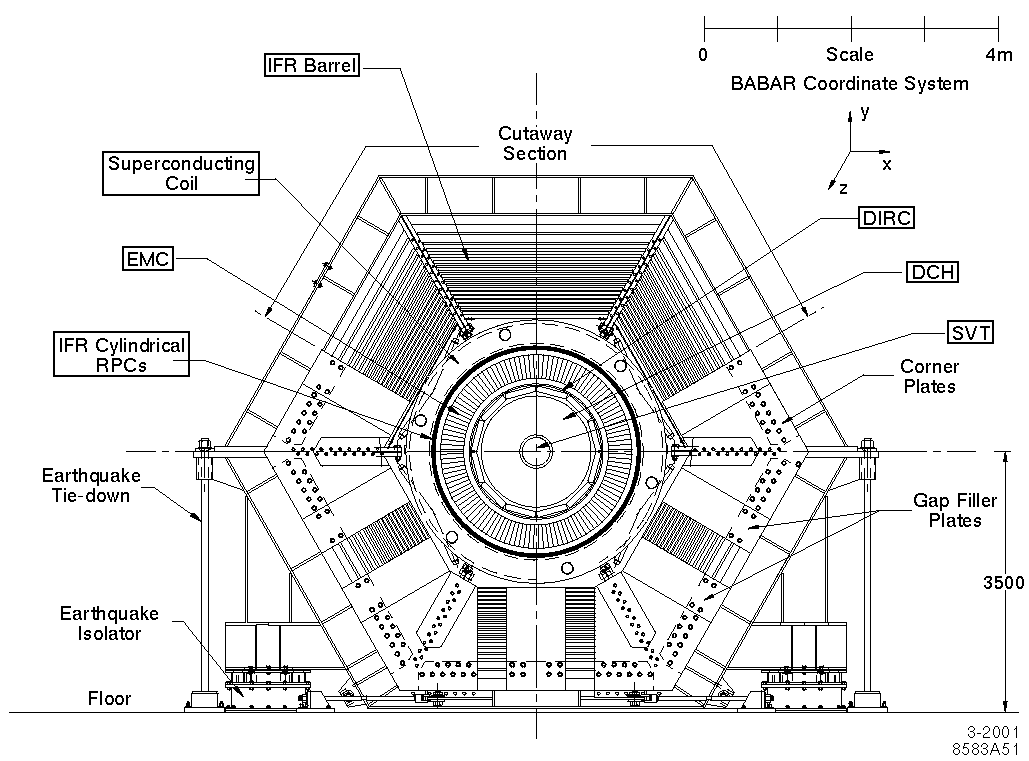} 
		\label{pic:detfront}
	}
	\subfigure[]{
		\centering\includegraphics[width=.9\textwidth]{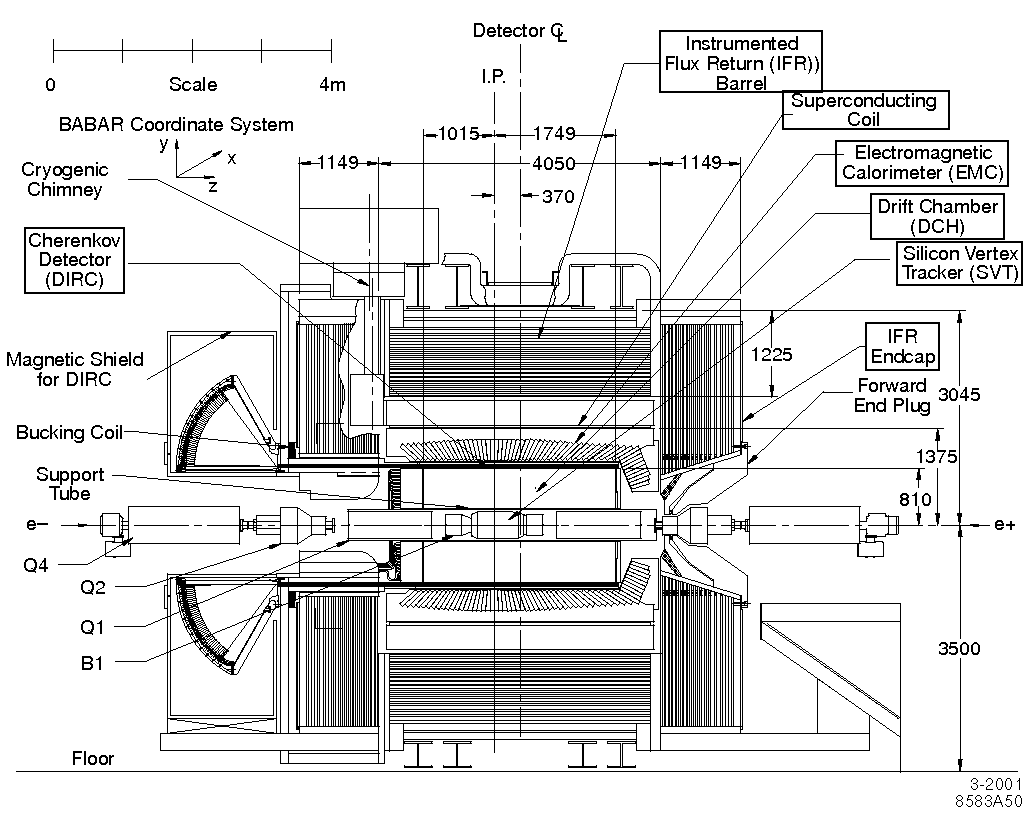} 
		\label{pic:detside}
	}
	\caption{The \babar detector: In the upper right corner the \babar coordinate system is shown, $z$ is the direction of the electron beam, $y$ points upwards while $x$ points horizontally away from the center of the \pep2 ring. \subref{pic:detfront} shows the detector end view (looking into the direction of the HER beam), while \subref{pic:detside} shows a longitudinal cut of the detector.}
	\label{pic:det}
\end{figure}

The foremost requirement on the detector was a maximal acceptance in the center-of-mass system. Due to the asymmetric beam energies the decay products were boosted in the forward direction of the laboratory frame. Thus, an asymmetric detector was necessary to optimize the detector acceptance. Furthermore, an excellent vertex resolution, as well as a good discrimination between \electron, \mmu, \pion, \kaon, and \proton over a wide kinematic range and the capability to detect and identify neutral particles were necessary.\\
\babar has been designed to meet all of these requirements. Picture \ref{pic:det} shows a schematic view of the \babar detector. The detector was arranged cylindrically around the beam pipe and consisted from the center outwards of the following subsystems:
\begin{itemize}
	\item The Silicon Vertex Tracker (SVT), was providing precise position information on charged tracks, is the only tracking device for very low-energy charged particles,
	\item the Drift Chamber (DCH) provided the main momentum measurement for charged particles, and helped in particle identification by measuring the energy loss of traversing particles,
	\item the Detector of Internally Reflected Cherenkov light (DIRC), responsible for the identification of charged hadrons,
	\item the Electromagnetic Calorimeter (EMC), was providing information about neutral particles as well as a good electron identification,
	\item the superconducting coil, was providing a $1.5 \rm T$ solenoidal magnetic field for the momentum measurement in the DCH, and
	\item the Instrumented Flux Return (IFR) was used for muon and neutral hadron identification.
\end{itemize}
The different parts will be explained in the following section. A more detailed description of the detector can be found in \cite{bpb}, \cite{Aubert:2001tu} and \cite{tdr}.
\clearpage
\subsection{Silicon Vertex Tracker}

\begin{figure}[H]
        \centering\includegraphics[width=.9\textwidth]{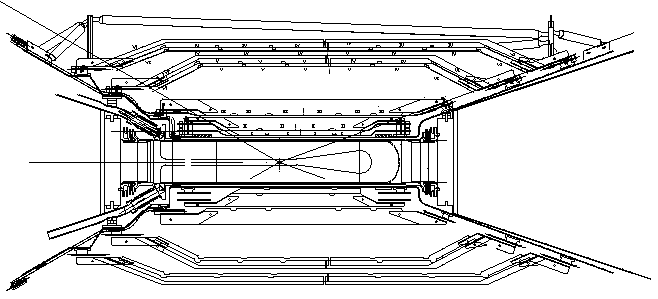}
        \caption{Longitudinal view of the SVT}
        \label{fig:svt}
\end{figure}

\begin{figure}[H]
        \centering\includegraphics[width=.7\textwidth]{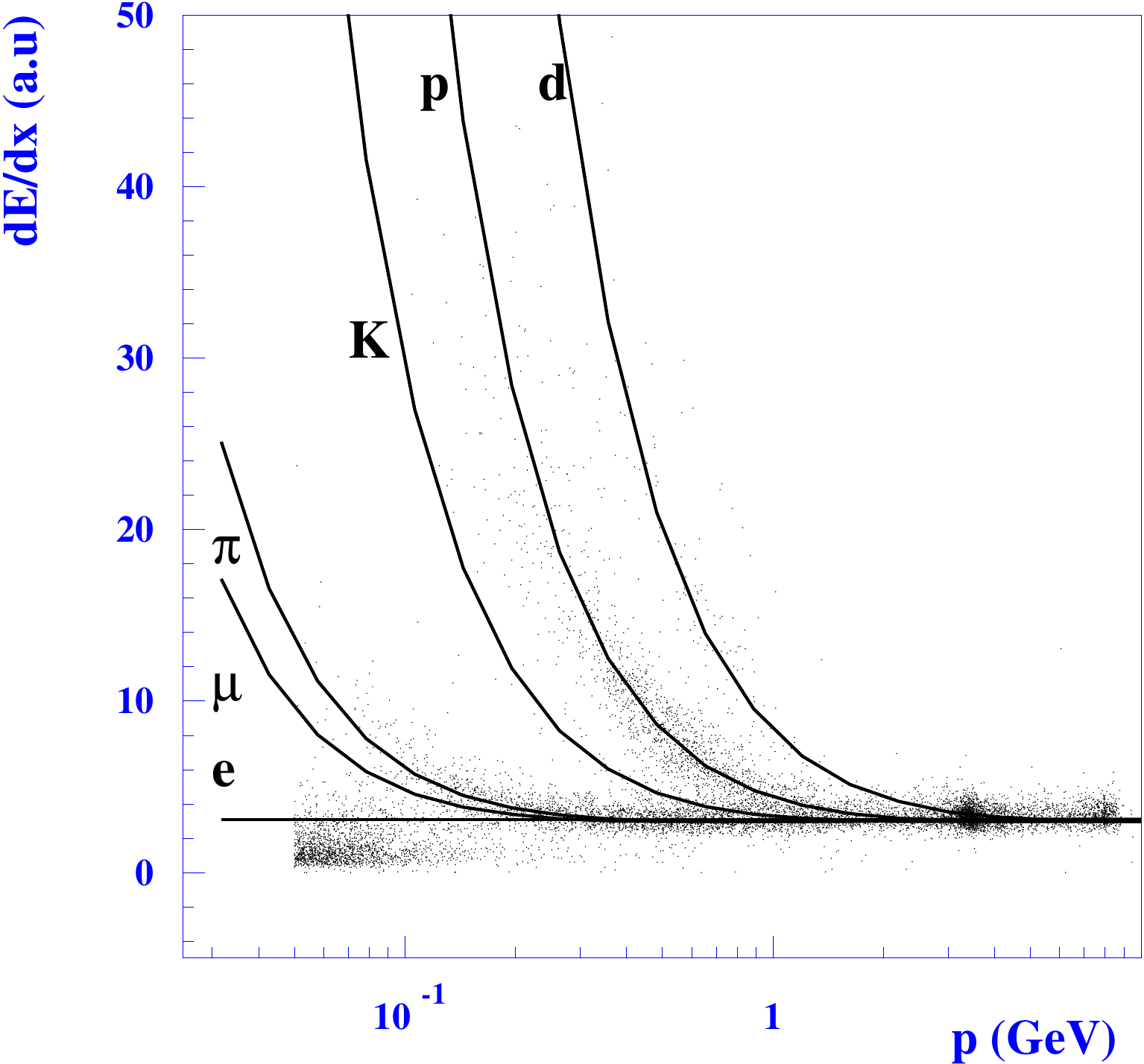}
        \caption{Energy loss per travelled path versus momentum in the SVT for different particles. $e$ - electron, $\mu$ - muon, $K$ - kaon, $p$ - proton, $d$ - deuteron}
        \label{fig:svtdedx}
\end{figure}

The Silicon Vertex Tracker (SVT) was the detector component with the smallest distance to the interaction point (IP). Consequently, it was the first component providing information on the flight path (track) of the particles emerging from the interaction point. Furthermore, it was the only source of information for low momentum particles that did not reach the drift chamber due to their deflection caused by the magnetic field.\\
The SVT, shown in Fig. \ref{fig:svt}, was located inside a $4.5 \m$ long support tube. It consisted of five concentric cylindrical layers of double-sided silicon detectors. These five rings had radii of $3.3 \cm$ up to $14.6 \cm$. If a charged particle crossed the SVT it generated electron-hole pairs leading to a signal. Due to the segmentation of the SVT, the signals in the different segments provided information about the path of the particle. Besides, the SVT also provided information for the particle identification of charged particles with momenta less than $700 \mevc$ by measuring the rate of energy loss \dedx \cite{jepson}. Figure \ref{fig:svtdedx} shows the energy loss per travelled path against the momentum of the particle for different particles in the SVT.\\
The three rings closest to the IP delivered data about the position and the angle of a track for the track reconstruction with a spatial resolution of $10 - 15 \mum$ \cite{Re:2004da}. The two outer rings had a resolution of $30 - 40 \mum$ \cite{Re:2004da} and were important for the measurement of the momentum of particles with a small transversal momentum as well as for the separation of geometrically close tracks.
\subsection{Drift Chamber}

\begin{figure}[H]
        \centering\includegraphics[width=.7\textwidth]{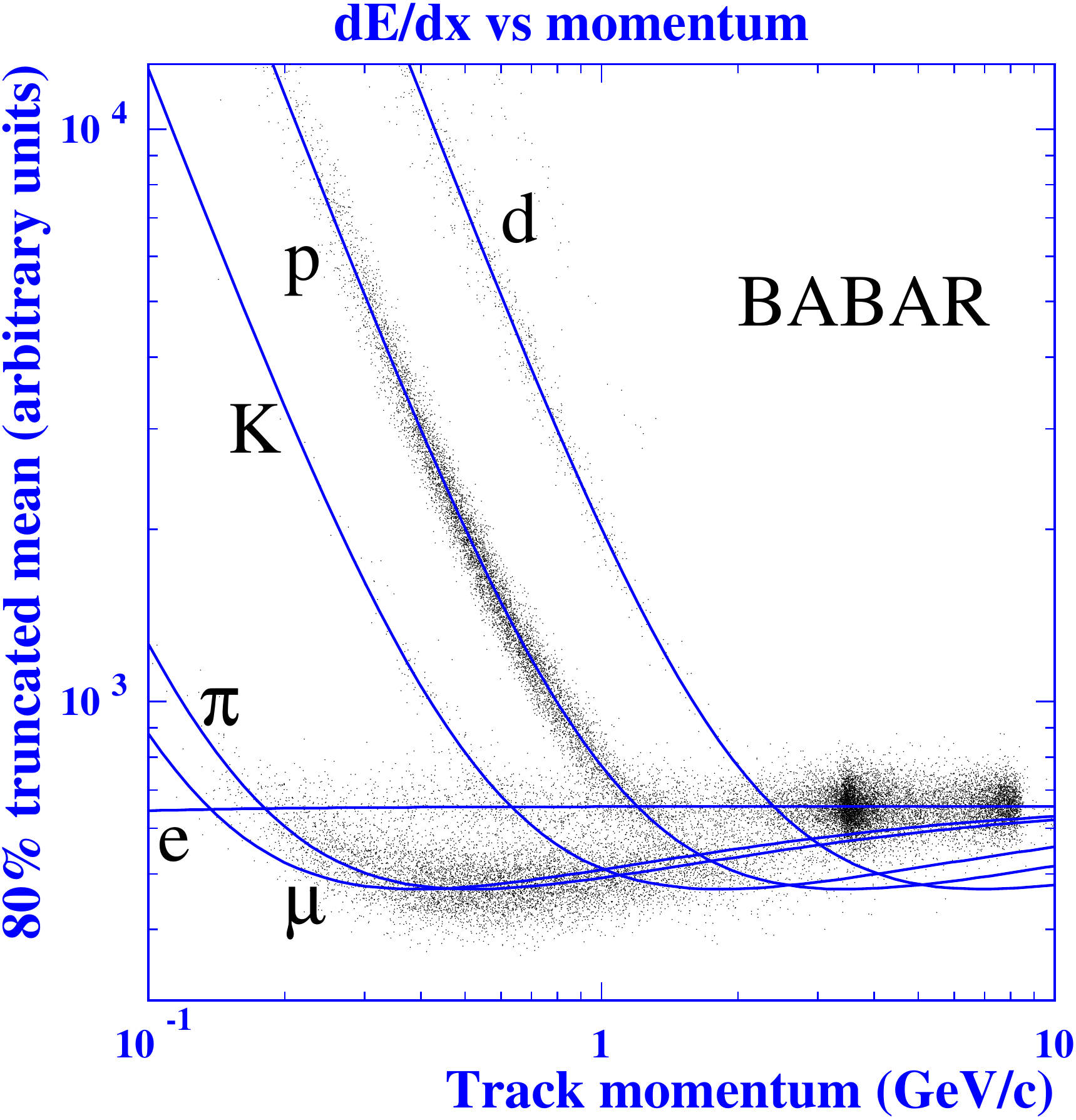}
        \caption{energy loss per travelled path versus momentum in the DCH for different particles ($e$ - electron, $\mu$ - muon, $K$ - kaon, $p$ - proton, $d$ - deuteron)}
        \label{fig:dchdedx}
\end{figure}

\begin{figure}[H]
        \centering\includegraphics[height=.6\textwidth]{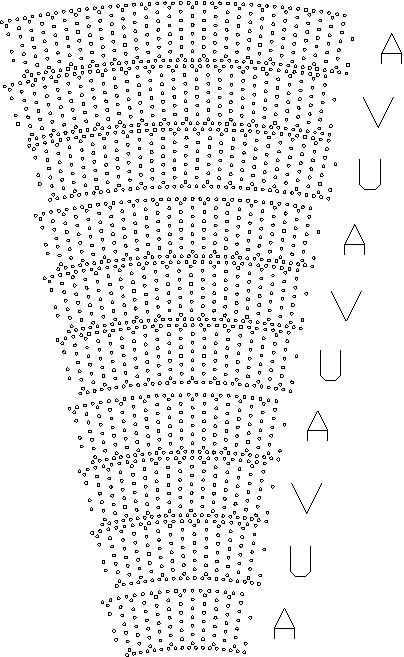}
        \caption{Cell layout in the \babar drift chamber}
        \label{fig:wiresdch}
\end{figure}

\begin{figure}[H]
	\centering\includegraphics[width=.7\textwidth]{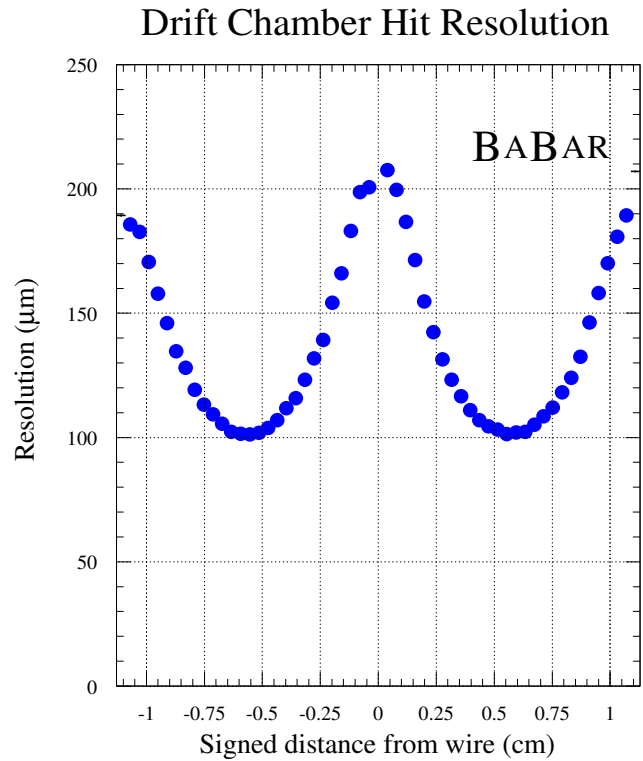}
	\caption{Spatial resolution of the drift cells}
        \label{fig:driftres}
\end{figure}

The drift chamber (DCH) was the main tracking device of the \babar detector. It measured at least 40 space coordinates per track in the central region, ensuring a high reconstruction efficiency for tracks with transverse momentum greater than $100 \mevc$. Further, the drift chamber contributed to the particle identification. Therefore, the rate of energy loss was measured, this rate is characteristical for the different types of particles. Figure \ref{fig:dchdedx} shows the rate of energy loss versus momentum.\\
The drift chamber was a $280 \cm$ long cylinder, with an inner radius of $23.6 \cm$ and an outer radius of $80.9 \cm$. It consisted of $7104$ hexagonal cells, which were formed by $6$ field wires in the corners of the hexagon and a sense wire in the center of the cell. The cells were arranged in $10$ superlayers of $4$ layers each. Axial (A) and stereo (U,V) superlayers alternate as it can be seen in Fig. \ref{fig:wiresdch}. This led to a $z-$coordinate resolution of $700 \mum$ \cite{Kelsey:2004an}.\\
If a particle crossed the drift chamber it ionized the gas mixture ($80 \%$ Helium, $20 \%$ Isobutane) inside the chamber. The generated ions and electrons drifted towards the field wires or to the sense wires, respectively, due to a difference in the voltage of $1960 \rm V$ between the sense and field wires. The generated signal could be used for tracking as well as for particle identification by the measurement of \dedx. Figure \ref{fig:driftres} shows the spatial resolution of the drift cells.

\subsection{Superconducting Solenoid}

Although the superconducting solenoid, together with the instrumented flux return, was the outermost detector component its function was closely related to the drift chamber. The solenoid, a superconducting coil, created a magnetic field of $1.5$ Tesla inside the drift chamber leading to a deflection of the track of a charged particle. This deflection was used to measure the charge as well as the momentum of a charged particle with a transverse momentum resolution of $\sigma(p_T)/p_T \approx 0.45   \% + 0.13 \% \cdot p_T$ \cite{Kelsey:2004an}.\\

\subsection{Detector for internally reflected Cherenkov light}

\begin{figure}[h]
        \centering\includegraphics[width=.7\textwidth]{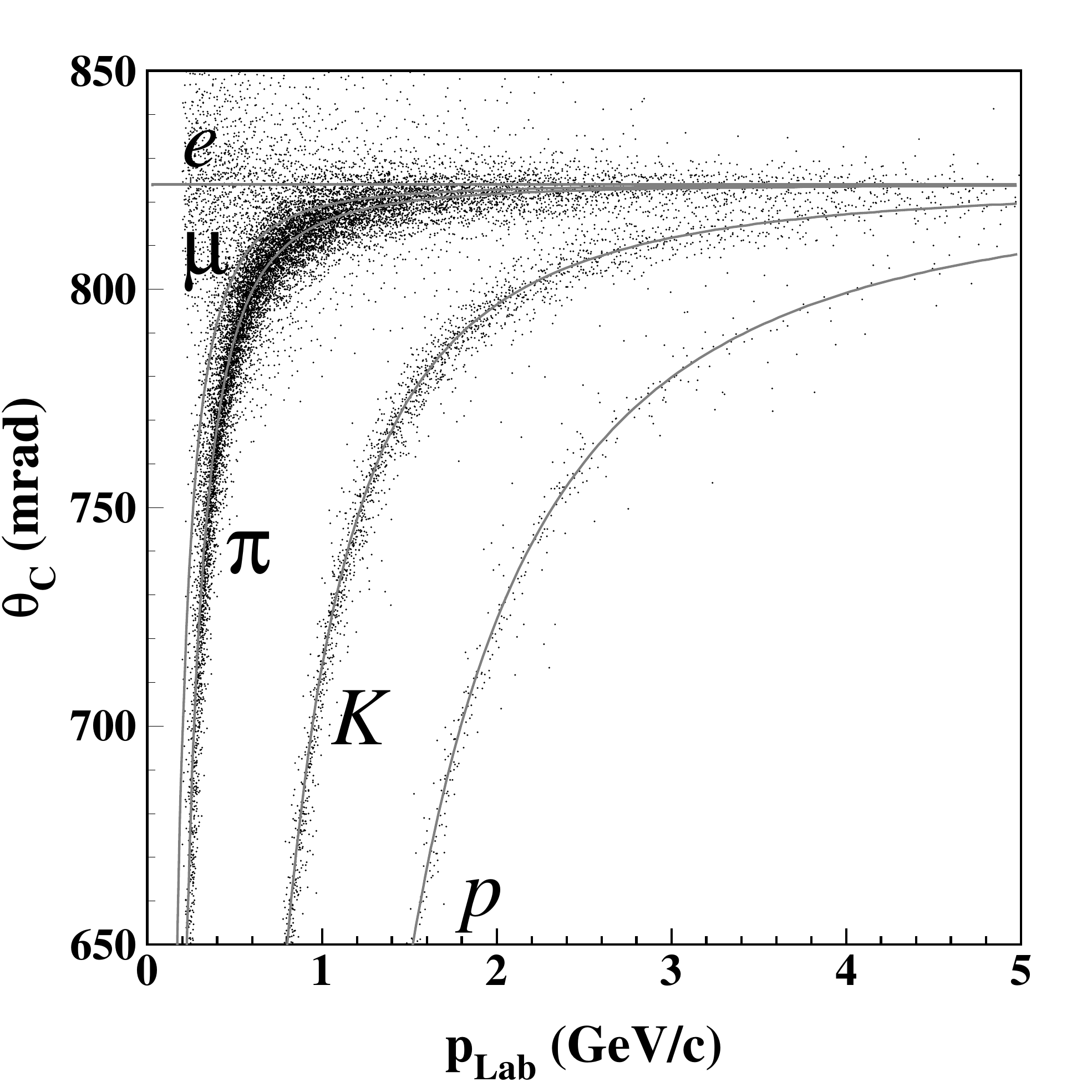}
        \caption{Cherenkov angle against momentum for different particles}
        \label{fig:dirctheta}
\end{figure}

The Detector for internally reflected Cherenkov light (DIRC) was capable of identifying pions and kaons with momenta greater than $0.7 \gevc$ as well as protons with momenta between $1.3 \gevc$ and $4 \gevc$. This was complementary to the drift chamber, which could identify particles with lower momenta.\\
\begin{figure}[H]
	\subfigure[]{
		\centering\includegraphics[width=.48\textwidth]{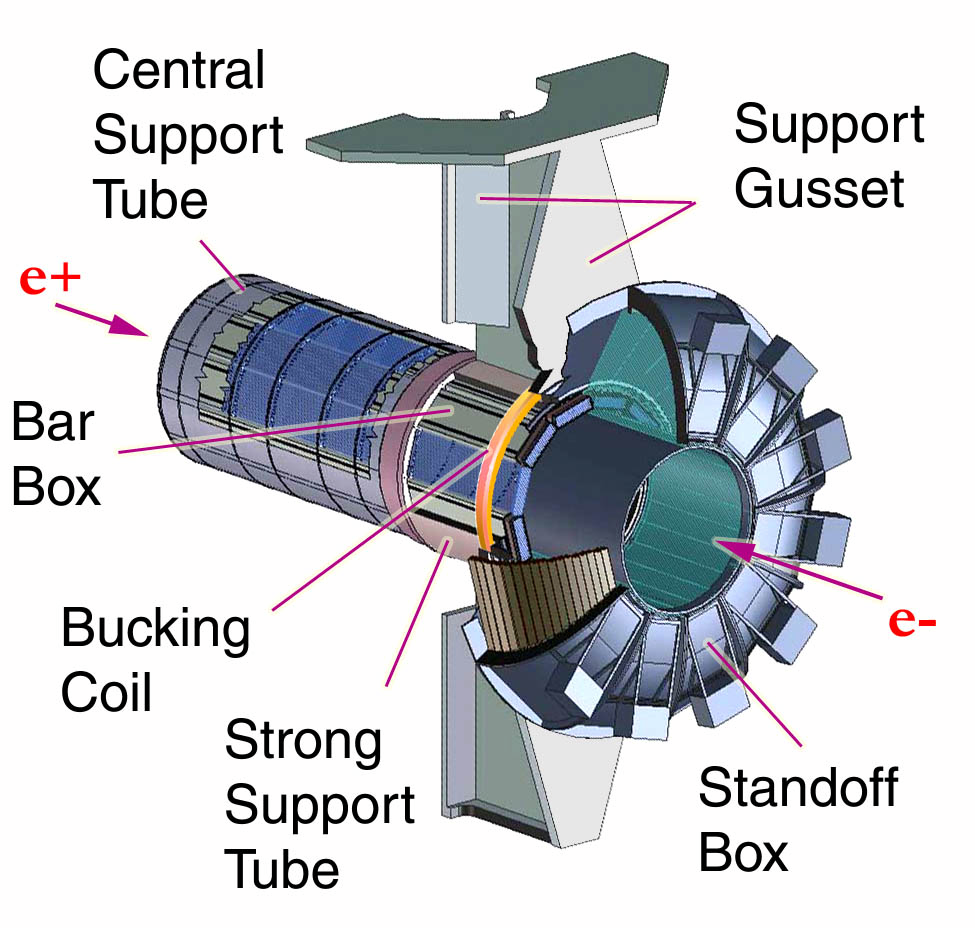} 
		\label{pic:dirc1}
	}
	\subfigure[]{
		\centering\includegraphics[width=.48\textwidth]{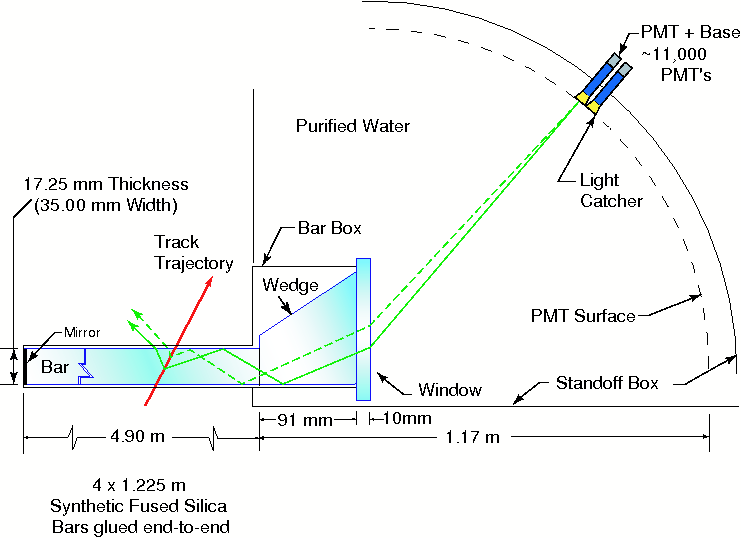} 
		\label{pic:dirc2}
	}
	\caption{Fig. \subref{pic:dirc1} shows a schematic view of the DIRC, the bar box containing the quartz bars and the standoff box are clearly visible. Fig. \subref{pic:dirc2} illustrates the mode of operation of the DIRC. The emitted Cherenkov light is guided by total internal reflection to the standoff box where the Cherenkov angle is measured with the help of photomultiplier tubes.}
	\label{fig:dirc}
\end{figure}
Fig. \ref{fig:dirc} shows a schematic view of the DIRC. It consisted of $144$ bars of synthetic quartz with a refractive index of $n \approx 1.474$ surrounding the drift chamber. If a particle crossed these bars with a velocity larger than the speed of light inside the quartz it produced Cherenkov light. The Cherenkov photons were emitted at an angle $\theta_c$ relatively to the direction of the particle. This angle depends on the mass and momentum of the particle.
\begin{equation}
	\cos \theta_c = \frac{1}{\beta n} = \frac{\sqrt{1 + \left(\frac{m}{p}\right)^2}}{n}
\end{equation}
Inside the bars the Cherenkov photons were reflected many times until they entered the standoff box at the rear side of the detector. This box was filled with about $6000$ liters of purified water and was equipped with nearly $11000$ photo multiplier tubes to detect the Cherenkov angle. Fig. \ref{fig:dirctheta} shows the discrimination power of the Cherenkov angle. The angular resolution of the DIRC for a single photon was $7 \mrad$ \cite{Adam:2004fq}.
\clearpage

\subsection{Electromagnetic Calorimeter}

\begin{figure}[H]
        \centering\includegraphics[width=.8\textwidth]{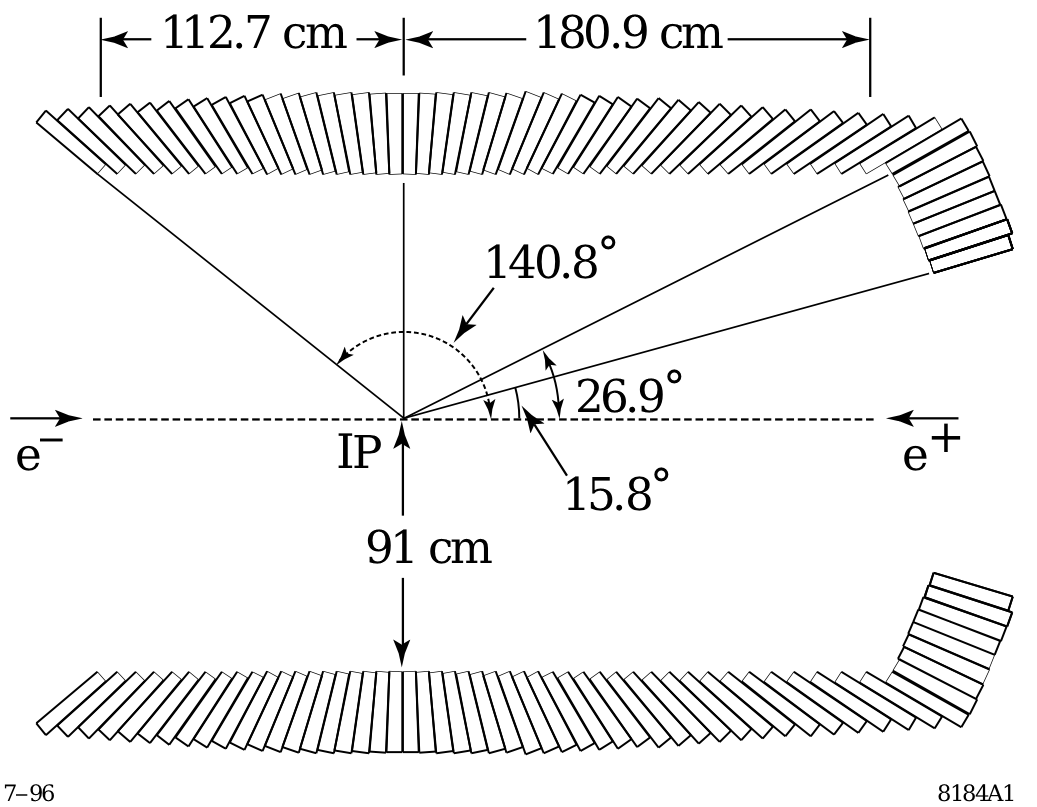}
        \caption{schematic view of the EMC, showing the calorimeter barrel and the forward End Cap.}
        \label{fig:emc}
\end{figure}

The \babar electromagnetic calorimeter (EMC), as shown in Fig. \ref{fig:emc}, was a hollow cylinder surrounding the drift chamber. The calorimeter barrel consisted of $5760$ thallium-doped CsI crystals arranged in $48$ rings of $120$ crystals, while the End Cap was composed of $820$ crystals.
The calorimeter was designed for excellent efficiency as well as good energy ($\sigma_E/E$) and angular resolution of the energy range $20\mev$ to $9\gev$. A detailed study \cite{Ruland:2009zz} gave
\begin{align}
  \frac{\sigma_E}{E} &= \frac{(2.30 \pm 0.03 \pm 0.3)\%}{\sqrt[4]{E(\gev)}} \oplus (1.35 \pm 0.08 \pm 0.2)\% \\
  \sigma_\theta &= \sigma_\phi = \frac{(4.16\pm 0.04)\mrad}{\sqrt{E(\gev)}}.
\end{align}
Here, the $\oplus$ denotes a quadratic summation.
Due to the asymmetric design the calorimeter covered a solid angle of $-0.775 \le \cos \theta \le 0.962$ in the laboratory frame and $-0.916 \le \cos \theta \le 0.895$ in the center-of-mass frame \cite{bpb}.\\
High energetic photons entering a crystal were converted into \epem pairs by interacting with the crystal. The created electrons and positrons emitted bremsstrahlung photons which could convert into \epem pairs again. An electromagnetic shower developed. The generated photons could be absorbed by the scintillator material leading to an excitation of the crystal atoms. When the atom returned into its groundstate it emitted the absorbed energy as light. This light could be detected by photo diodes glued to the rear end of the crystals. Here the amount of measured light was proportional to the energy lost inside the crystals by a particle crossing the EMC.\\
In addition to the detection of photons the calorimeter was an important detector component for the discrimination between electrons and other particles. Details on the electron identification with the help of the EMC are given in section \ref{sect:pid:electron}.

\subsection{Instrumented Flux Return}

\begin{figure}[h]
  \begin{center}
    \includegraphics[width=.8\textwidth]{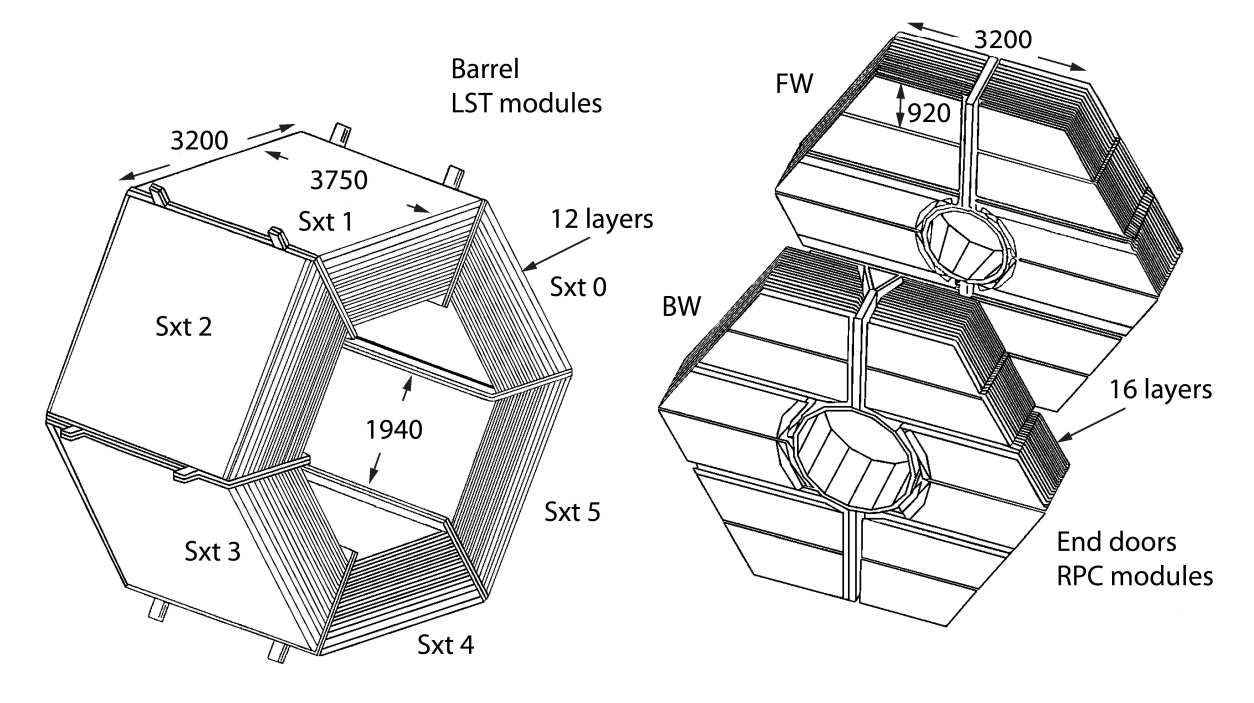}
  \end{center}
  \caption{Final layout of the IFR with barrel sextants, forward (FW) and backward (BW) end caps, the dimensions are given in $\mm$.}
  \label{fig:det:IFR}
\end{figure}

The outermost layer of the \babar detector was the instrumented flux return (IFR). Its main purpose as a detector component was the identification of muons as well as neutral hadrons like the $K_L^0$ mesons. Consisting of a central part (Barrel) and two End doors the IFR covered a solid angle range down to $300\mrad$ in the forward, and $400\mrad$ in the backward direction \cite{bpb}. In the initial setup the IFR was instrumented with $806$ Resistive Plate Chambers (RPCs), arranged in $19$ layers in the barrel and $18$ layers in the End Caps. Already in the first year of data taking the RPCs showed serious aging problems and had to be replaced by second generation RPCs and Limited Streamer Tubes (LSTs). The final configuration of the IFR consisted of $12$ layers of LSTs in the barrel and 16 layers of second generation RPCs in the forward End Cap. In the backward End Cap the original RPCs were retained. During the upgrade to LSTs the remaining flux return slots were filled with brass absorber plates, and some external steel plates were added. Thereby the pion rejection ability of the muon identification algorithm, described in section \ref{sect:pid:muon}, was improved \cite{TheBABAR:2013jta}. A sketch of the final IFR configuration can be seen in Fig. \ref{fig:det:IFR}.

\subsection{Trigger}

The \babar trigger consisted of two levels: a hardware level (called Level 1) and a software level (called Level 3) and had to decide which events observed by the \babar detector were interesting enough to be kept and recorded for later analysis.\\
The Level 1 trigger system consisted of four subsystems: the charged particle trigger (DCH trigger - DCT), the neutral particle trigger (EMC trigger - EMT), the cosmic trigger (IFR trigger - IFT) and the Global Level Trigger (GLT).\\
The DCT and EMT received information from the Drift Chamber and the Calorimeter, respectively, processed it and sent the condensed information to the Global Trigger. This trigger tried to match the angular locations of calorimeter towers and drift chamber tracks and generated Level 1 trigger signal, which were passed to the Level 3 trigger. In addition to the information from the DCT and EMT the Global Trigger also uses IFT information to trigger on cosmic rays and $\mu$-pair events.\\
After the Level 1 trigger the Level 3 trigger, operating on an online farm, analyzed the event data from the DCH and EMC in conjunction with the Level 1 trigger information for further background reduction. Besides the physics filter this trigger stage also performed Bhabha veto and the selection of calibration events as well as critical online monitoring tasks \cite{trigger}.

\section{Event reconstruction}\label{lists}

The events that have been selected to be kept were passed to the reconstruction software which reconstructed the tracks in an event and assigned particle hypotheses to the tracks.\\
During the reconstruction neutral and charged particles had to be considered separately. Charged particles interacted with every detector component and left a track inside the detector that could be reconstructed. In contrast, neutral particles only interacted with the EMC and the IFR. Consequently, their reconstruction relied on information delivered by these two detector components like their shower shape in the EMC and the energy deposit in the detector.

\subsection{Neutrals}

All neutrals in the \babar framework are based on \texttt{CalorNeutral}, a standard list created during online production. This list contains all EMC clusters that are not assigned to any track in the event. For all these entries the photon mass-hypothesis is applied.

Based on \texttt{CalorNeutral} \babar uses two photon lists, namely \texttt{GoodPhotonLoose} and \texttt{GoodPhotonDefault} \cite{ref:Neutrals}. The first one contains all candidates from \texttt{Calor\-Neutral} satisfying the following criteria:
\begin{itemize}
  \item Minimum raw energy of $0.03\gev$
  \item Maximum Lateral Moment $LAT$ of $0.8$.
\end{itemize}
The \texttt{GoodPhotonDefault} list contains all candidates from \texttt{GoodPhotonLoose} with a minimum raw energy of $0.1\gev$.

\subsection{Track reconstruction}\label{tracklists}

For charged particles a multi-step track reconstruction takes place.
The first step is to search for track points in both tracking devices (SVT and DCH) separately. 
Due to the multi-layer design of the SVT and DCH it is possible to reconstruct the path a particle took inside these detector components, therefore a curve is fitted to the track points in the respective subdetector.
Afterwards the reconstruction program tries to assign partial tracks inside the SVT and in the DCH to each other. After a successful assignment the combined track is fitted again to provide a good momentum measurement.\\
The least stringent requirement for an accepted track is a successful reconstruction of the track as well as a successful momentum measurement. A track fulfilling these conditions is assigned to the \texttt{ChargedTracks} list under a pion hypothesis.\\ 
Table \ref{tab:tracking} shows the requirements for the tracking lists used at \babar, as they can be found in \cite{trackingTF}. The different selection criteria shown in table \ref{tab:tracking} are:
\begin{itemize}
	\item transverse momentum $p_t$ $\left[\gevc\right]$
	\item momentum $p$ $\left[\gevc\right]$
	\item distance of closest approach (DOCA) to the $z$-axis in the $xy$-plane $d_{0}$ $\left[cm\right]$
	\item $z$-coordinate of the point of closest approach $z_0$ $\left[cm\right]$
	\item fit probability $P(\chi^2)$
\end{itemize}
\begin{table}[H]
	\caption{Tracking lists used at \babar and the selection criteria for these lists}
	\begin{center}
		\begin{tabular}{lccccc}\toprule
			tracking list			& $p_t$		& $p$	& $|d_0|$	& $|z_0|$	& $P$	\\\midrule
			\texttt{ChargedTracks}		& -		& -	& -		& -		& $>0$	\\
			\texttt{GoodTracksVeryLoose}	& -		& -	& $<1.5$	& $<2.5$	& $>0$	\\
			\texttt{GoodTracksLoose}	& $>0.05$	& $<10$	& $<1.5$	& $<2.5$	& $>0$	\\
              \bottomrule
		\end{tabular}
	\end{center}
	\label{tab:tracking}
\end{table}
For the latter neutrino reconstruction we have to rely on \texttt{ChargedTracks} since identified $V_0$ tracks, including $K_s$, $\Lambda$ and converted $\gamma$ are not included in the more stringent lists \texttt{GoodTracksVeryLoose} and \texttt{GoodTracksLoose}.

\subsection{Charged Particle Identification}\label{pidlists}

The identification of charged particles is crucial for many analyses performed at \babar. For a high quality of the identification all \babar subdetectors contributed in a complementary way to charged particle identification. The SVT and DCH provides \dedx measurements, the DIRC provides a velocity measurement, the EMC discriminates electrons, muons and hadrons according to their energy deposit and their shower shape while the IFR characterizes muons and hadrons according to their different interaction pattern. 

\subsubsection{Electron identification with the EMC}\label{sect:pid:electron}

For the electron identification with the calorimeter we have to distinguish between two methods.
\begin{enumerate}
  \item Electron identification using $E/p$
  \item Electron identification using the shower shape
\end{enumerate}
For the first method the ratio $E/p$ is measured, where $E$ is the energy of a shower in the calorimeter, and $p$ is the measured three-momentum of the corresponding charged track. When an electron enters the calorimeter it produces an electromagnetic shower, depositing its energy in the calorimeter. Thus, the ratio $E/p$ is expected to be close to unity for an electron. In contrast muons and charged hadrons deposit only a fraction of their energy in the calorimeter, leading to smaller values of $E/p$. In addition to the good separation of electrons this method can also be used for a discrimination of muons against hadrons. While muons, as single minimum-ionizing particles, have a well-defined peak in the $E/p$ distribution, hadrons have additional tails at higher $E/p$ values. A caveat of the $E/p$ method is that hadrons interact electromagnetically as well as hadronically with the calorimeter. In the latter case they initiate a hadronic shower, depositing a large fraction of their energy in the calorimeter. Since the resulting $E/p$ values are rather large a discrimination against electrons by $E/p$ alone is difficult. An improvement of the electron identification can be achieved by using the different shower shape for electrons and hadrons. While electrons deposit most of their energy in two or three crystals hadronic showers are more extended. A quantity reflecting this difference is the lateral moment $LAT$
\begin{equation}
  LAT = \frac{\sum_{i=3}^NE_ir_i^2}{\sum_{i=3}^NE_ir_i^2+E_1r_0^2+E_2r_0^2}.
\end{equation}
Here, the energies $E_i$ are ordered according to their value ($E_1>E_2>\hdots >E_N$), with $N$ being the number of crystals associated with the shower. $r_i$ and $\phi_i$ are the polar coordinates of the crystal in the plane perpendicular to the line pointing from the interaction point to the center of the shower. $r_0$ is the average distance between two crystals ($\approx 5\cm$). Since the summation omits the  two crystals with the highest energy deposit this quantity is expected to be small for electromagnetic showers. An even better discrimination between electrons and hadrons is achieved by including information about the azimuthal distribution of the shower. In general electromagnetic showers are expected to be isotropic in $\phi$, while hadronic showers are far more irregular in $\phi$.

\subsubsection{Muon identification with the IFR}\label{sect:pid:muon}

The first step for muon identification with the IFR is the track reconstruction. Due to the low occupancy this is an uncritical task, since only a tiny fraction of tracks overlap each other. First, tracks are reconstructed in each sector using a clustering algorithm. In the second step the clusters in different sectors are merged, using the extrapolation of the charged tracks measured by the tracking systems (SVT and DCH) into the IFR. Such composite clusters are considered as candidates for muons or charged hadrons. \\
For the muon identification the algorithm has to decide, whether the detected track has been produced by a muon or a pion. The other hadron candidates, kaons and protons, are identified by the other particle identification sub-systems since they show a similar signature like the pion in the IFR. For the discrimination of muons against charged hadrons different IFR variables are used, e.g. the number of strips in IFR cluster, the number of measured interaction lengths, and the continuity of IFR hits in the 3-D IFR cluster \cite{bpb,muonPID}.

\subsubsection{BDT selectors}\label{sect:pid:BDT}
For muons we apply the BDT selector, which is based on a Bagged Decision Tree (also Bootstrap Aggregating Decision Tree). A decision tree splits an $n$ dimensional input set (with $M$ entries) into rectangular subsets (nodes), where for each split all variables are considered and the split that leads to the largest increase in the figure of merit (e.g. the muon efficiency) is selected. This split is repeated recursively for each of the resulting nodes. The recursion ends when the number of entries in a node hits a pre-defined minimum $m$, or if the figure of merit doesn't change significantly. If the majority of entries in this node are signal entries this node is classified as a signal node, otherwise it is a background node.
To enhance the reliability and reduce the variance of this procedure many trees are trained on bootstraped replicas of the training set. The bootstraped replica is obtained by sampling with replacement from the original training set, until the size of the replica equals the size of the original training set. For an improvement of the performance compared to a single tree the classifier used has to be sensitive to small changes in the training set. This is obtained by setting a small value for $m$, and thus overtraining the trees. While each tree has a poor predictive power the final vote, a majority vote of all trained trees, has a high predictive power \cite{Narsky:2005hn}.
The major goal of the BDT selectors is to provide a constant muon efficiency over time, momentum $\vec{p}$ and polar angle $\theta$. While the last two are easy to understand the constant efficiency over time incorporates the aging of the detector components for the muon identification, as well as upgrades of the IFR. The muon efficiencies for the different PID lists can be found in Tab. \ref{tab:BDT_eff}. Details on the BDT input variables and performance can be found in \cite{Lit:BAD1853}. 
\begin{table}[h]
  \caption{The target efficiencies for the muon BDT lists. While in the first four lists a constant muon efficiency was targeted, the following four lists aimed at a constant pion mis-ID. The last two lists are specially designed for muons with momenta in the range $0.3$ to $0.7\gevc$.}
  \begin{center}
    \begin{tabular}{lcc}\\\toprule
      List name			& muon eff. 	& pion mis-ID 	\\\midrule
      muBDTVeryLoose   		& $90.0\%$   	& variable	\\
      muBDTLoose 		& $80.0\%$  	& variable	\\
      muBDTTight   		& $70.0\%$   	& variable	\\
      muBDTVeryTight   		& $60.0\%$   	& variable	\\
      muBDTVeryLooseFakeRate   	& variable   	& $5.0\%$  	\\
      muBDTLooseFakeRate 	& variable  	& $3.0\%$  	\\
      muBDTTightFakeRate   	& variable   	& $2.0\%$ 	\\
      muBDTVeryTightFakeRate   	& variable   	& $1.2\%$  	\\
      muBDTLoPLoose   		& $70.0\%$   	& variable	\\
      muBDTLoPTight   		& $60.0\%$   	& variable	\\
      \bottomrule
    \end{tabular}
  \end{center}
  \label{tab:BDT_eff}
\end{table}

\subsubsection{KM selectors}
The KM lists are based on Error Correcting Output Code, that combines multiple binary classifiers to form a multiclass classifier. Here, the binary classifiers are trained differently. In the case of the KM selectors seven Bootstrap Aggregate Decision Trees $t_i$ were trained according to Table \ref{tab:KM_train}.
\begin{table}[h]
  \caption{The used indicator matrix in the KM selector. Each entry indicates whether the given training sample should be treated as signal (1) or background (-1).}
  \begin{center}
    \begin{tabular}{crrrrrrr}\\\toprule
      Class & $t_0$ & $t_1$ & $t_2$ & $t_3$ & $t_4$ & $t_5$ & $t_6$ \\\midrule
      $K$   & $1$   & $1$   & $1$  & $1$   & $1$   & $1$  & $1$    \\
      $\pi$ & $-1$  & $1$   & $-1$ & $1$   & $-1$  & $1$  & $-1$    \\
      $p$   & $1$   & $-1$  & $-1$ & $1$   & $1$   & $-1$ & $-1$    \\
      $e$   & $1$   & $1$   & $1$  & $-1$  & $-1$  & $-1$ & $-1$    \\
      \bottomrule
    \end{tabular}
  \end{center}
  \label{tab:KM_train}
\end{table}
For the classification of a given track each classifier $t_i$ is asked to give an output between $-1$ and $1$, resulting in a string of seven real values between $-1$ and $1$. This string is then compared to the individual codeword for class $k$. For a kaon the codeword is the first row in Table \ref{tab:KM_train}, for a $\pi$ the second and so on. The comparison is here done by calculating the generalized Hamming distance $H_k$, the sum of squared differences, to the codeword for each class $k$. The track is then assigned to the class with the lowest Hamming distance. In order to obtain multiple tightness levels we can use the distance $H_k$ itself as well as ratios of distances. For a kaon list we can use $H_K$, $H_{\pi}/H_K$, $H_p/H_K$ and $H_e/H_K$. For the other particle classes we use the analogue ratios. This method allows us to control the probability for a given class as well as the misidentification rate.
More details on these selectors, especially on the chosen input variables for the seven classifiers can be found in \cite{Lit:BAD2199}.

For high flexibility special selectors for the pre-selection of recorded data, were introduced. These \texttt{CombinedSuperLoose} selectors combine the least restrictive selectors for a given particle type to allow the user to switch to his selector of choice in the event reconstruction. 

\clearpage
\chapter{Software and Datasets}

\section{Software}

For the reconstruction of the decay we use the following packages inside the \babar framework
\begin{itemize}
  \item \texttt{analysis-52}
  \item \texttt{BetaMiniUser V00-04-05}
  \item \texttt{PDT V00-07-00}
  \item \texttt{workdir V00-04-21}
\end{itemize}
The event selection and background reduction uses the data analysis package \texttt{ROOT} \cite{cern} and \texttt{iPython} \cite{iPython} with \texttt{SciPy} \cite{SciPy}.

In the background reduction we apply a random forest, provided by \texttt{SciPy}, to distinguish between signal and background candidates. A random forest is a modification of the bagging method, as described in sect. \ref{sect:pid:BDT}, where a large collection of trees is build, and the response is averaged over all trees. The benefit is that one can use models with a small predictive power, thus unbiased, for classification in this method \cite{hastie}. For bagged trees the bias is the same as for the individual tree. In consequence the only improvement can be obtained in terms of the variance of the average
\begin{equation}
  \rho \sigma^2 + \frac{1-\rho}{B}\sigma^2,
\end{equation}
which depends on the pairwise tree correlation $\rho$, the individual tree variance $\sigma^2$ and the number of trees $B$. While the second term vanishes for large $B$ the first term stays unchanged.

The idea behind the random forest algorithm is to reduce the correlation $\rho$ without increasing the variance $\sigma^2$. This is achieved by a random selection of the input variables during the tree-growing process. In the following we will concentrate on a bootstrapped dataset. For bootstrapping the dataset $\bf Z$, containing $N$ candidates described by $p$ variables, is divided into $B$ datasets. For the division we draw from $\bf Z$ randomly with replacement $B$ datasets $Z_1, \hdots , Z_B$ containing the same number of events $N$ as the original dataset $\bf Z$. 

The general algorithm of the training of a random forest is shown in algorithm \ref{alg:rf_train}.
\begin{algorithm}
\caption{Training of a random forest \cite{hastie}}
\label{alg:rf_train}
\begin{enumerate}
  \item for $b=1$ to $B$:
    \begin{enumerate}
      \item Draw a bootstrap sample $Z_b$ of size $N$ from the training data
      \item Grow a tree $T_b$ to $Z_b$, by recursively repeating the following for each terminal node, until minimum node size $n_{\rm min}$ (number of candidates) is reached
        \begin{enumerate}
          \item select $m$ variables at random from the $p$ variables
          \item pick the best variable/split-point among the $m$
          \item split the node into two daughter nodes
        \end{enumerate}
    \end{enumerate}
  \item output the ensemble of trees $\{ T_B \} ^B_1$.
\end{enumerate}
\end{algorithm}

The split is done by selecting the variable most suitable to discriminate signal from background, and dividing the dataset into two hemispheres. Subsequently each hemisphere is handed over to one of the two daughter nodes. Finally, each node $l$ represents a subregion $R_l$, with $N_l$ observations of the whole input data set. The proportion of class $k$ in this region is thus given by
\begin{equation}
  \hat{p}_{lk} = \frac{1}{N_l}\sum_{x_i \in R_l} I(y_i=k),
\end{equation}
where $x_i$ denotes the candidate to be classified and $y_i$ the class of $x_i$. The function $I(y_i=k)$ returns $1$ if $y_i=k$ and $0$ otherwise. For final nodes all observations in the node are classified as the class with the highest value for $\hat{p}_{lk}$.

In the application phase each tree returns a classification prediction $\hat{C}_b(x)$. The resulting classification prediction from the random forest is then the majority vote of $\hat{C}_b(x)$, with $b=1\hdots B$.

The relevant tuning variables here are the number of trees $B$, the minimal node size $n_{\rm min}$ and the number of randomly to choose variables $m$.

\section{Datasets}

Data taking on the \FourS resonance at \babar took place in six run periods between February 2000 and August 2007. The data set used in this analysis comprises the complete data collected by \babar on the \FourS resonance, with an integrated luminosity of
\begin{equation}
  \lum_{\rm OnPeak} = (424\,000 \pm 1900_{\rm syst})\invpb.
\end{equation}
This corresponds to a total number of \BBb pairs of
\begin{equation}
  N_{\BBb} = 470\,960\,000 \pm 116\,000_{\rm stat} \pm 2\,826\,000_{\rm syst}.
  \label{eq:Nbb}
\end{equation}

In addition to the \texttt{OnPeak} data set various Monte Carlo modes have been used to study background behavior and for the training of the random forest. The used background modes are listed in Table \ref{tab:SPmodes}.
\begin{table}
  \caption{Monte Carlo modes used for background studies. Given are their SP number in \babar bookkeeping and the contained number of events.}
  \begin{center}
  \begin{tabular}{lcr}\toprule
    mode 				& SP number 	& number of events 	\\\midrule
    $\epem \ra \qqbar$, $q=u,d,s$ 	& 998 		& $2\,190\,254\,000$	\\
    $\epem \ra \ccbar$ 			& 1005 		& $1\,127\,360\,000$ 	\\
    $\epem \ra \BpBm$			& 1235		& $707\,282\,000$	\\
    $\epem \ra \BzBzb$			& 1237		& $716\,219\,000$	\\
    \bottomrule
  \end{tabular}
  \end{center}
  \label{tab:SPmodes}
\end{table}

\begin{figure}[h]
  \begin{center}
    \includegraphics[width=.6\textwidth]{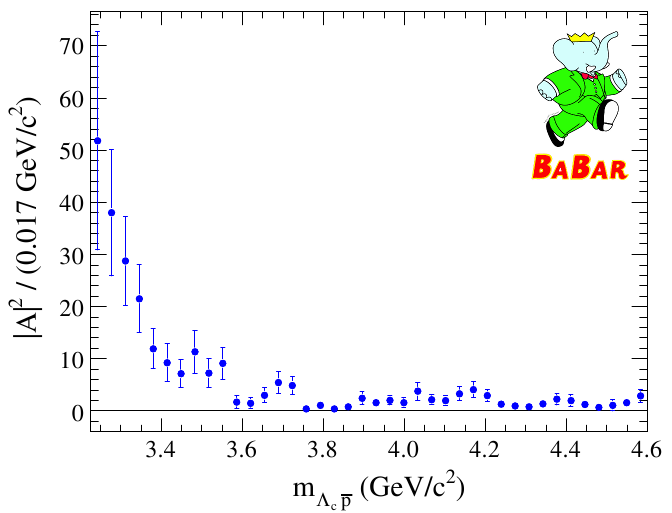}
  \end{center}
  \caption{Invariant $\LCp \antiproton$ mass for the decay $\Bm \ra \LCp \antiproton \pim$, divided by the expectation from a phase space model.}
  \label{fig:MCmodel:thresh}
\end{figure}

For the signal description we chose two different approaches, a simple phase space model, and a simulation reproducing the threshold enhancement seen in various baryonic \B-decays (see sect. \ref{sect:thresh}).
For the first one, the Phase Space (PS) model, we simulate a \B-meson decaying directly into the final state particles $\LCp$, $\antiproton$, $\ellm$ and $\nulb$. Here, the four-momenta are distributed uniformly in phase space. This model is intended for studies of the systematic uncertainties arising from the chosen decay model. 
For the second model, the Weak Interaction (WI) model, we assume that the hadronic form factor is given by the decay $\Bm \ra \LCp \antiproton \pim$. For this purpose we defined a meson pole $M$, decaying into the baryon anti-baryon pair $\LCp \antiproton$.
\begin{align}
  \Bm \ra M &\ellm \nulb \notag\\
  M \ra &\LCp \antiproton \notag\\
  \LCp &\ra \proton \Km \pip
\end{align}
Mass and width of the $M$ are chosen to reflect the properties of the enhancement known from other decay modes, especially from $\Bm \ra \LCp \antiproton \pim$\cite{Aubert:2008ax}. The mass is chosen in such a way that the mass peak is slightly above threshold, at $3.225\gevcc$ and the width is derived from the $m(\LCp \antiproton)$ distribution in \cite{Aubert:2008ax} (Fig. \ref{fig:MCmodel:thresh}) to be $200\mevcc$. 
The semi-leptonic decay $\Bm \ra M \ellm \nulb$ is modeled according to the weak matrix element
\begin{equation}
  |{\cal M}|^2 = (p_{\B}p_{\nu})(p_{\ell}p_{M}),
\end{equation}
under the assumption that the spectator system collapses into one particle.
The subsequent decay $M \ra \LCp \antiproton$ is simulated according to phase space.

\begin{figure}[h]
  \subfigure[]{
    \includegraphics[width=.48\textwidth]{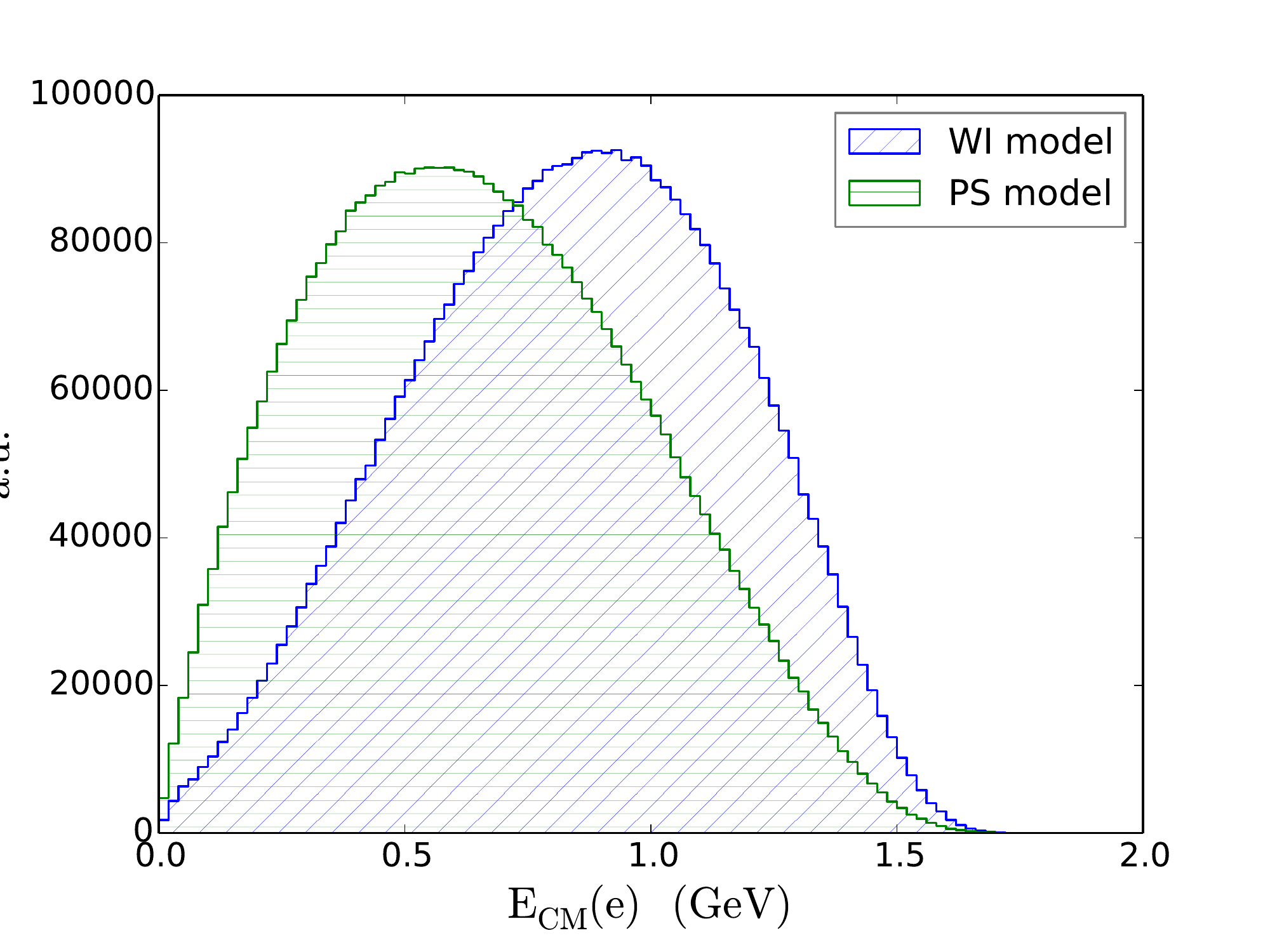}
    \label{subfig:PS:WI:leptonEnergy:e}
  }
  \subfigure[]{
    \includegraphics[width=.48\textwidth]{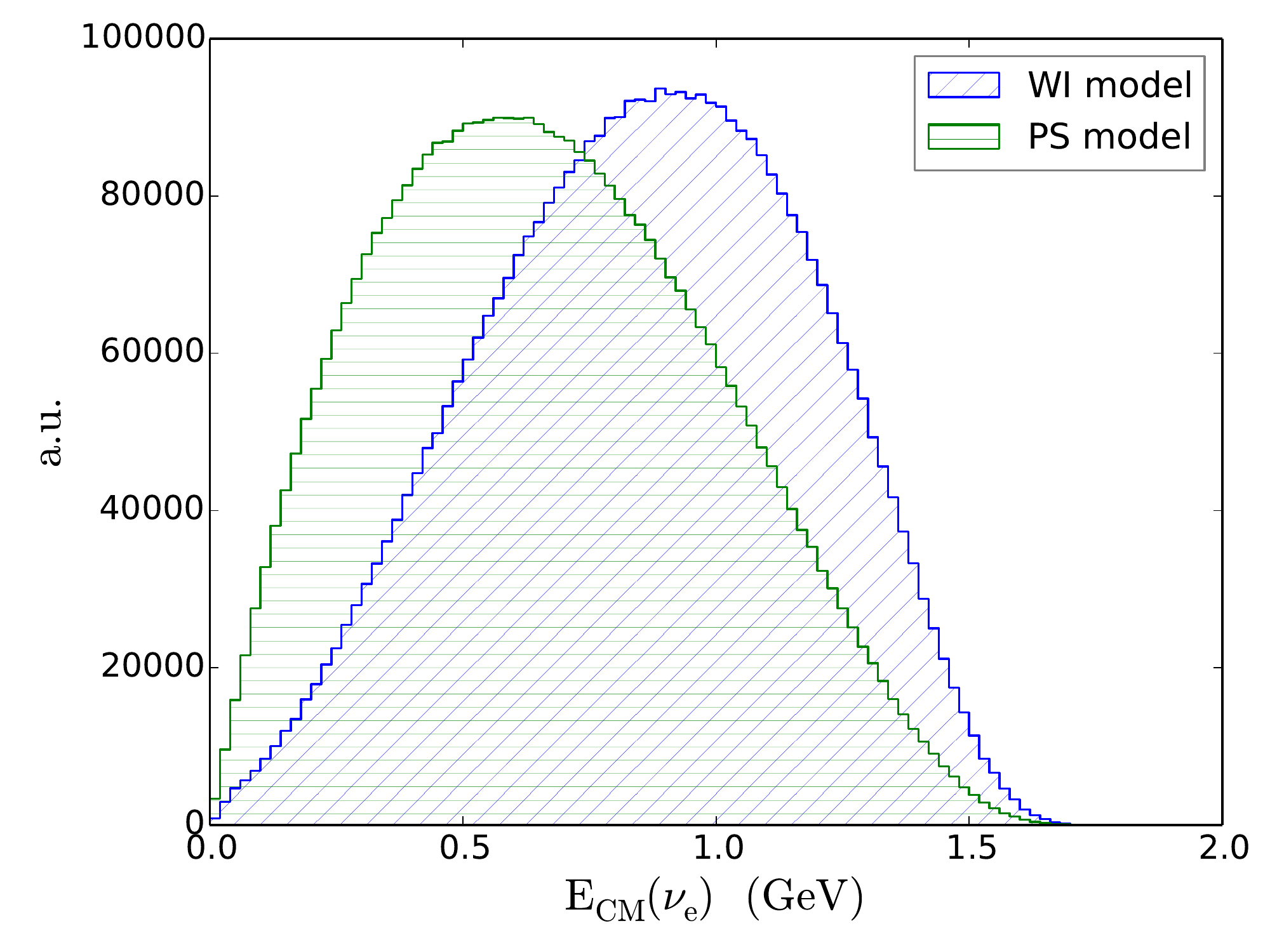}
    \label{subfig:PS:WI:neutrinoEnergy:e}
  }
  \subfigure[]{
    \includegraphics[width=.48\textwidth]{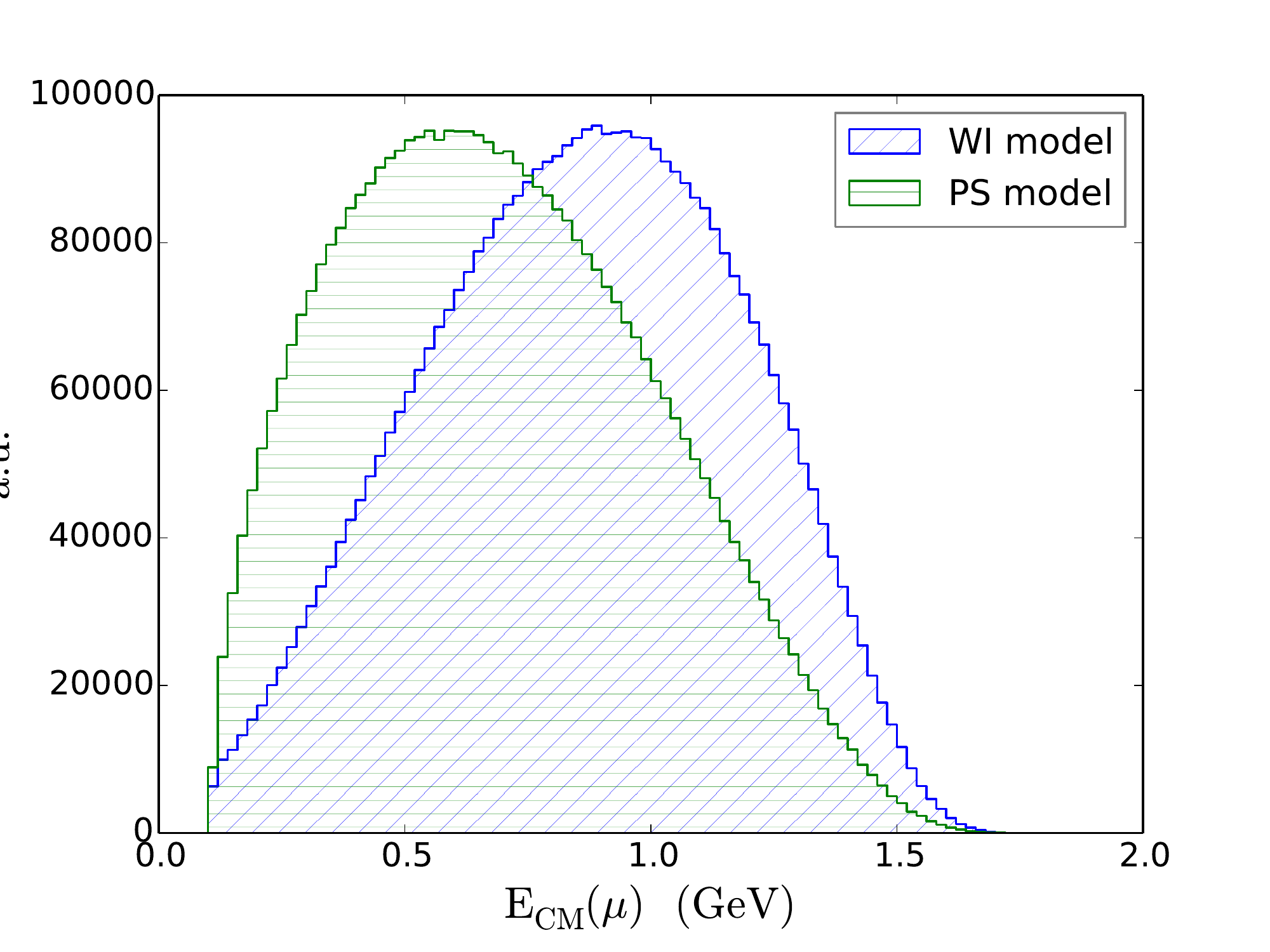}
    \label{subfig:PS:WI:leptonEnergy:mu}
  }
  \subfigure[]{
    \includegraphics[width=.48\textwidth]{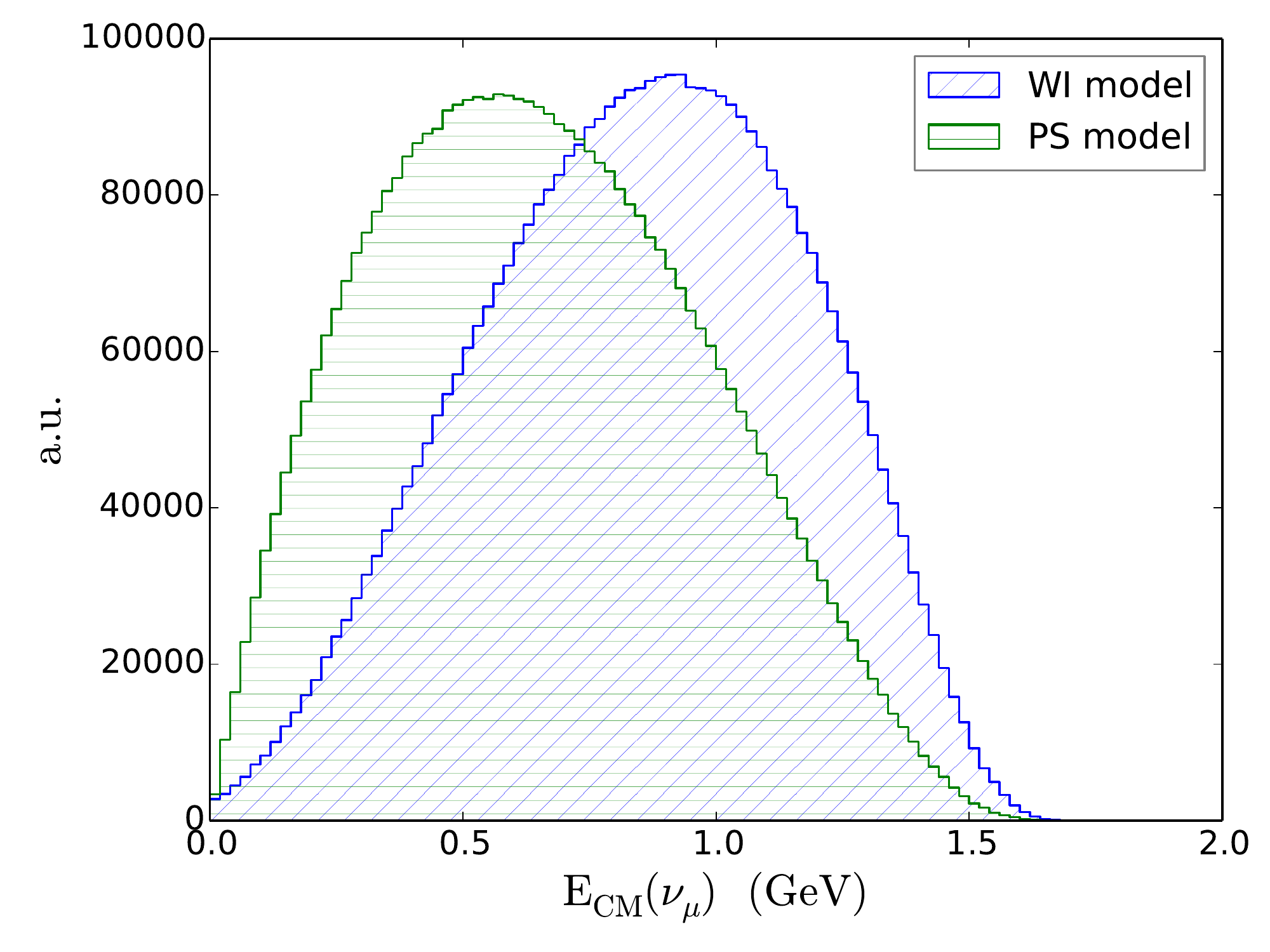}
    \label{subfig:PS:WI:neutrinoEnergy:mu}
  }
  \caption{Comparison of the charged lepton (left) and neutrino (right) energy for the PS and WI model, scaled to the same integral. The first row shows the electron, and the last row the muon channel.}
  \label{fig:PS:WI:compEnergy}
\end{figure}

\begin{figure}[h]
  \subfigure[]{
    \includegraphics[width=.48\textwidth]{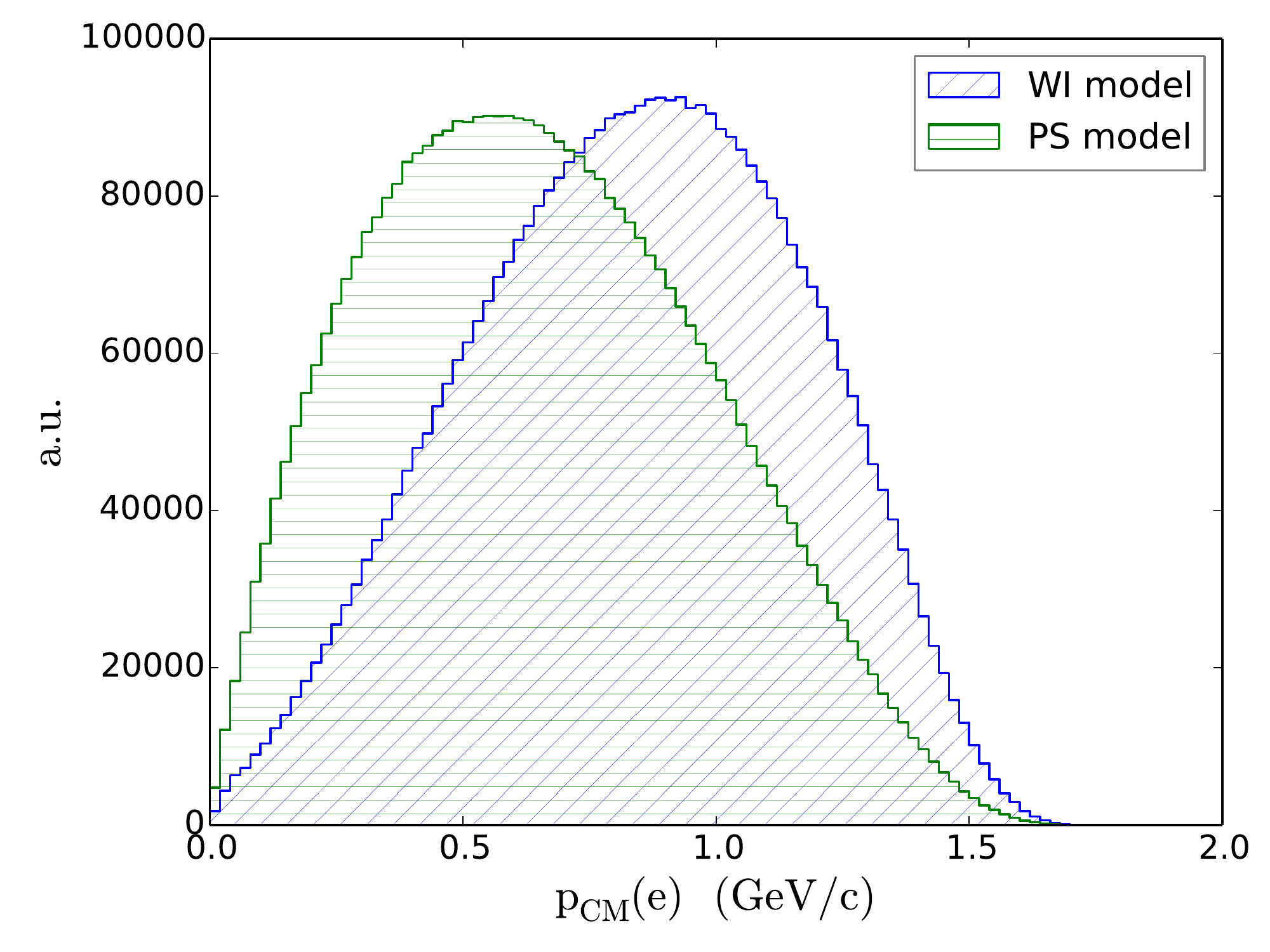}
    \label{subfig:PS:WI:leptonMomentum:e}
  }
  \subfigure[]{
    \includegraphics[width=.48\textwidth]{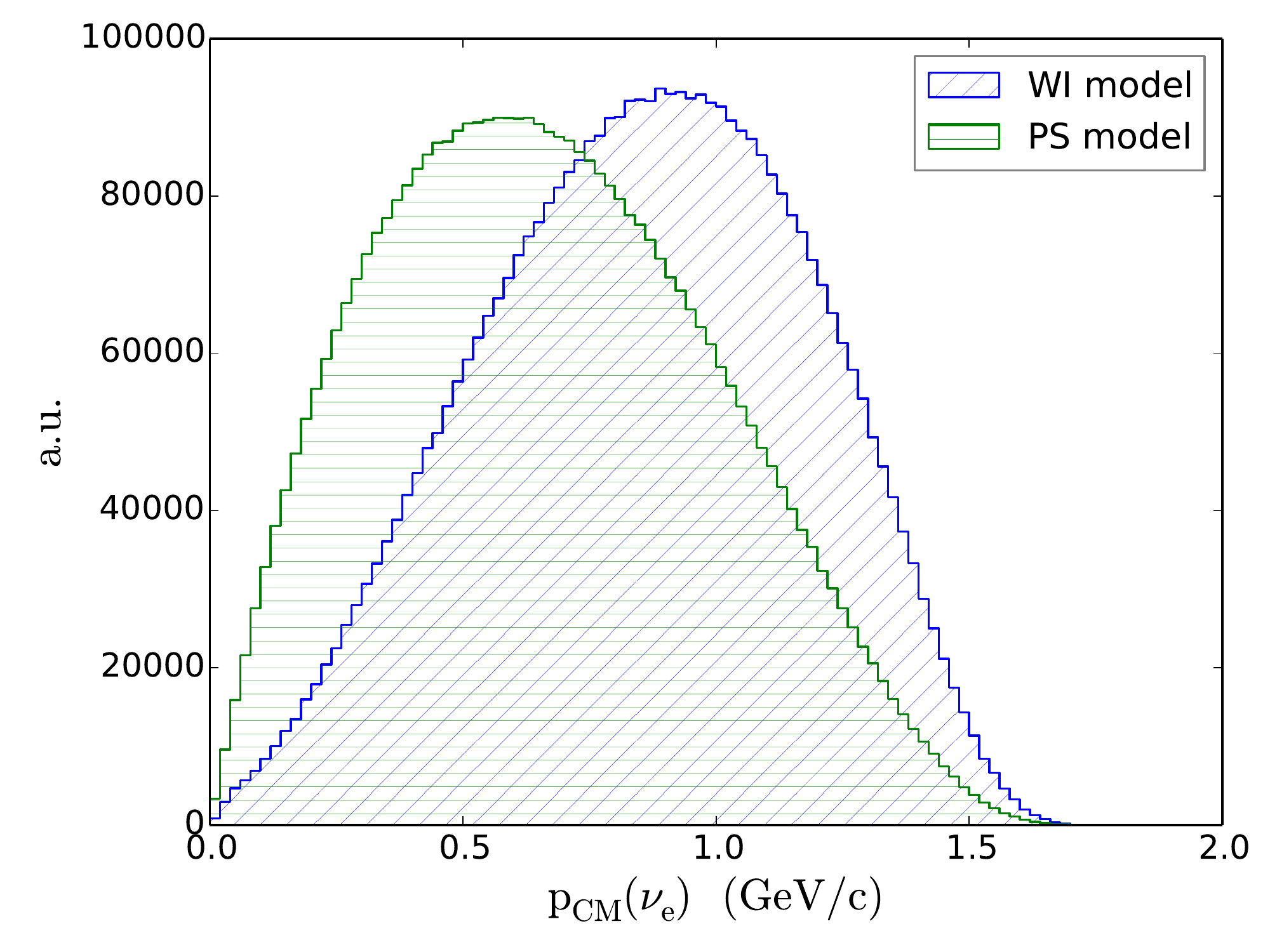}
    \label{subfig:PS:WI:neutrinoMomentum:e}
  }
  \subfigure[]{
    \includegraphics[width=.48\textwidth]{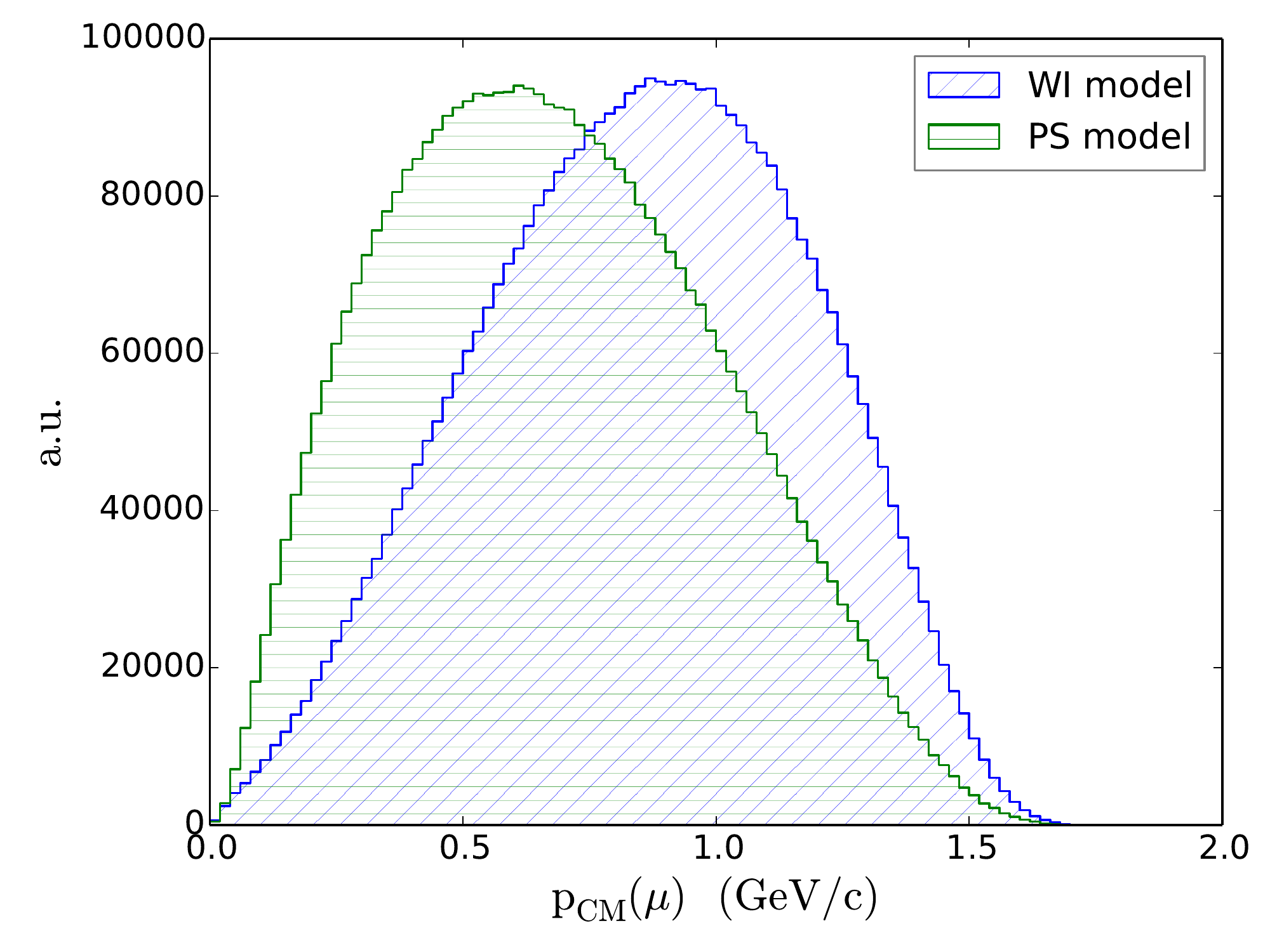}
    \label{subfig:PS:WI:leptonMomentum:mu}
  }
  \subfigure[]{
    \includegraphics[width=.48\textwidth]{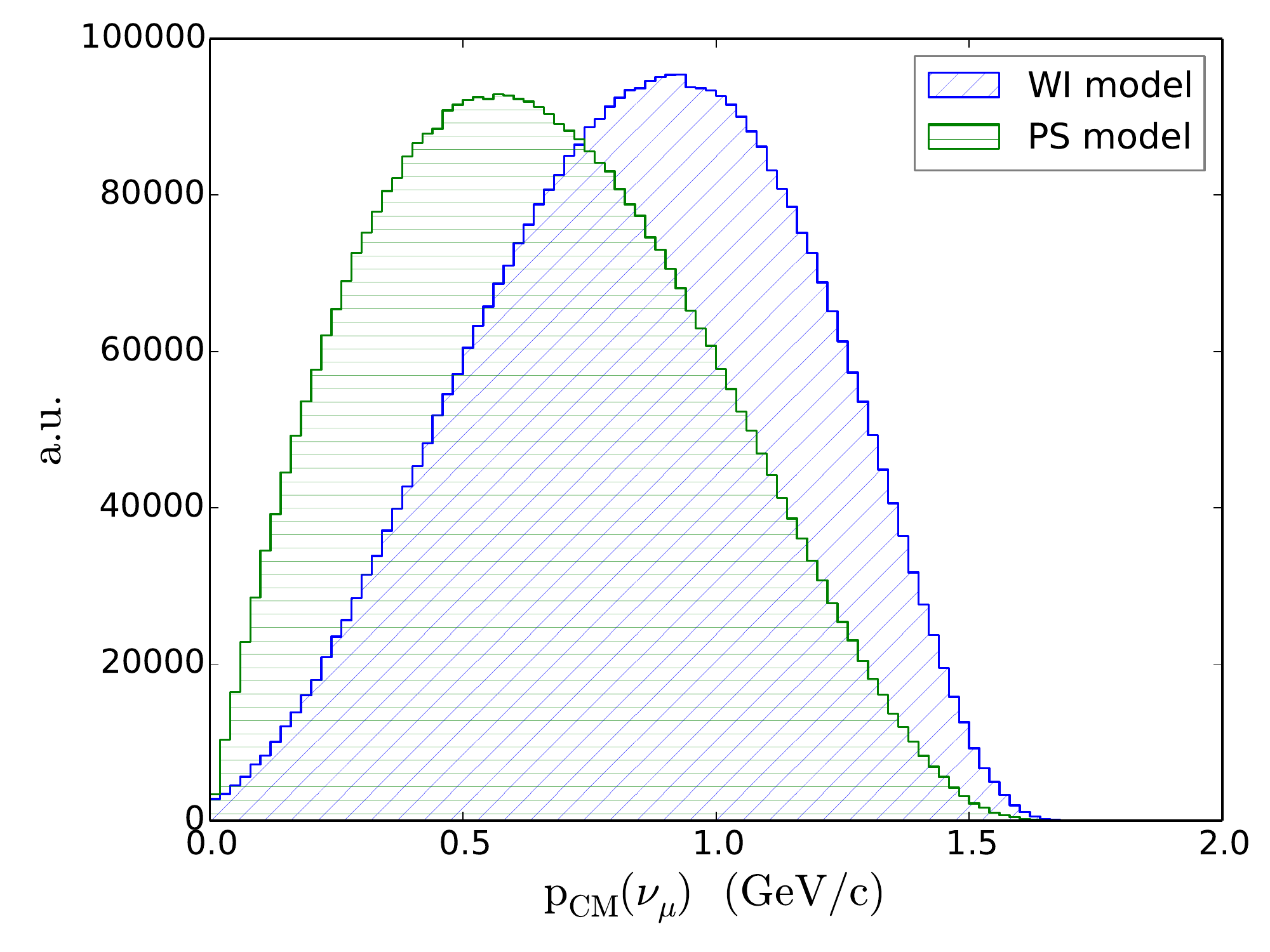}
    \label{subfig:PS:WI:neutrinoMomentum:mu}
  }
  \caption{Comparison of the charged lepton (left) and neutrino (right) momentum for the PS and WI model, scaled to the same integral. The first row shows the electron, and the last row the muon channel.}
  \label{fig:PS:WI:compMomentum}
\end{figure}

\begin{figure}[h]
  \subfigure[]{
    \includegraphics[width=.48\textwidth]{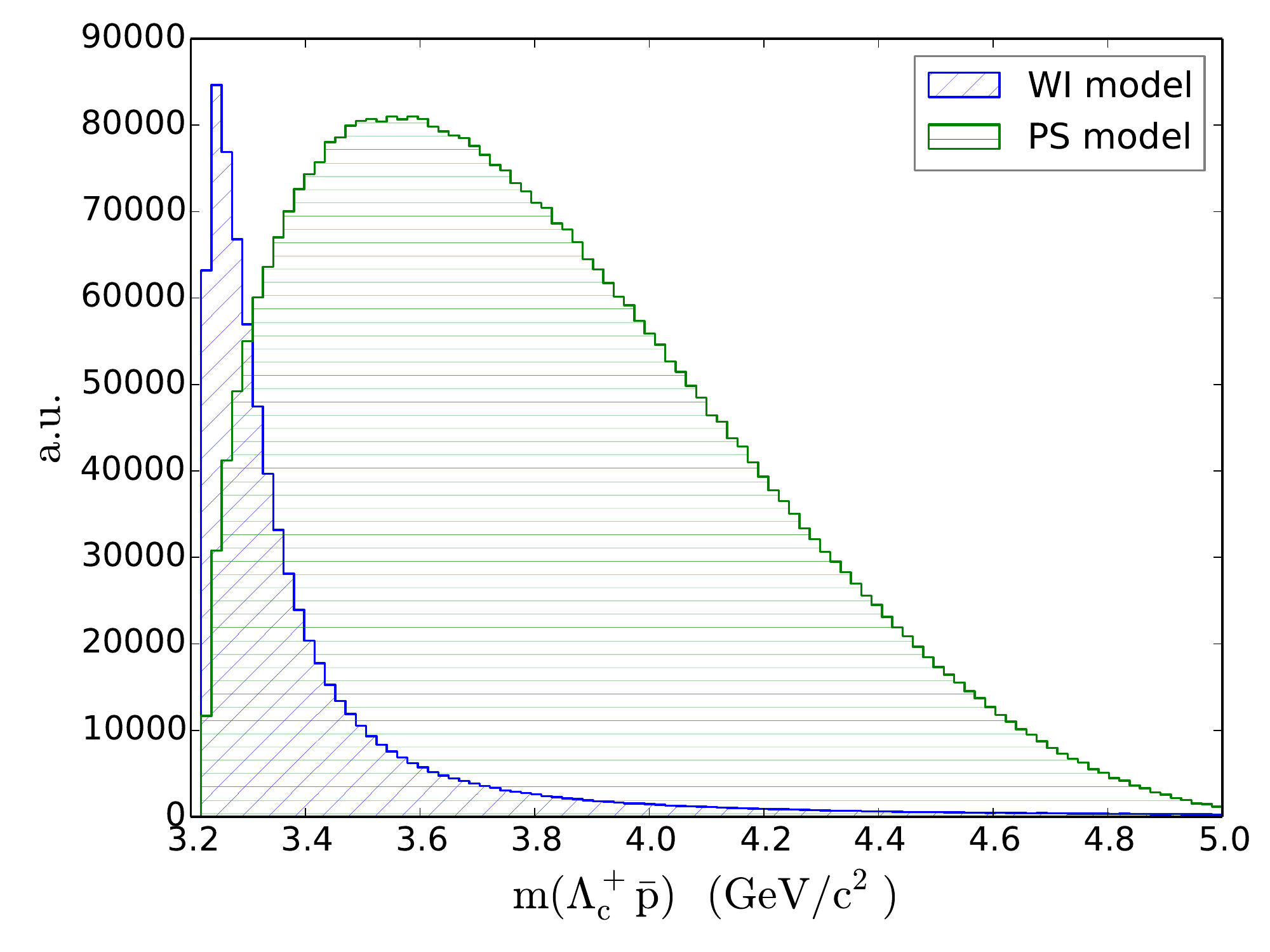}
    \label{subfig:PS:WI:mLambdaCproton:e}
  }
  \subfigure[]{
    \includegraphics[width=.48\textwidth]{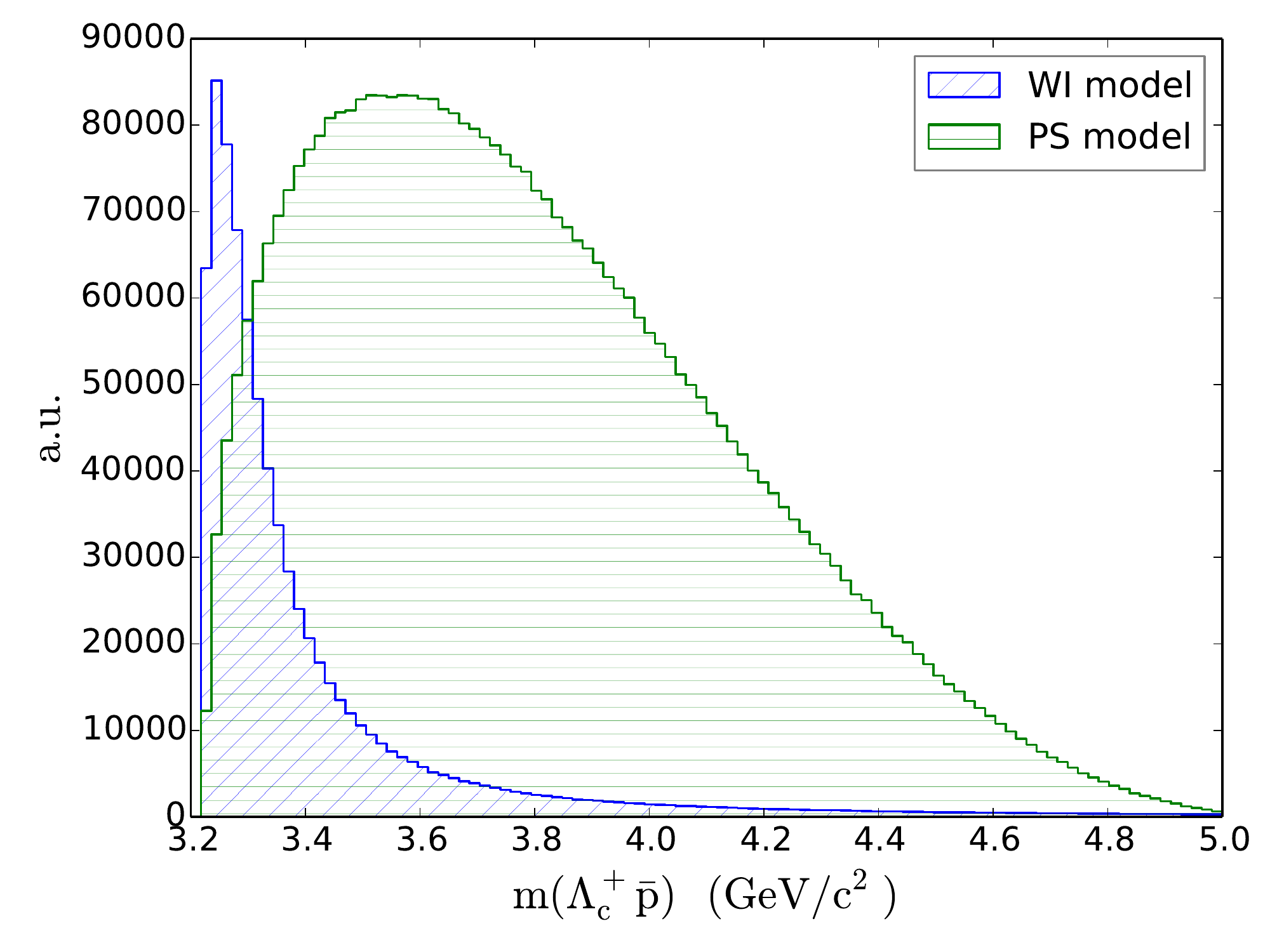}
    \label{subfig:PS:WI:mLambdaCproton:mu}
  }
  \caption{Comparison of the invariant $\LC\ \antiproton$ mass for the PS and WI model, scaled to the same height. Fig. \subref{subfig:PS:WI:mLambdaCproton:e} shows the electron, and \subref{subfig:PS:WI:mLambdaCproton:mu} the muon channel.}
  \label{fig:PS:WI:compMomentummLambdaCproton}
\end{figure}

The WI model is expected to give a more realistic momentum and energy spectrum for the charged lepton and the neutrino as well as a more realistic $m(\LCp \antiproton)$ distribution. Fig. \ref{fig:PS:WI:compEnergy}, \ref{fig:PS:WI:compMomentum} show a comparison of energy and momentum for both models on generator level in the center-of-mass frame.
For the invariant $\LCp \antiproton$ mass a comparison on generator level, as shown in Fig. \ref{fig:PS:WI:compMomentummLambdaCproton}, shows a peak above threshold for the Weak Interaction model, while the Phase Space model shows a broad spectrum.

All signal modes, generated using \evtgen \cite{Lange:2001uf} for the decay simulation and \geant \cite{Agostinelli:2002hh} for modelling the detector response, are given in Table \ref{tab:SPmodesSignal}.
 \begin{table}
   \caption{Monte Carlo modes used for signal studies and the number of generated events. The upper half shows the direct decay of the \B into the final state, described by a simple phase space model. The lower half gives the modes reproducing the threshold enhancement.}
   \begin{center}
     \begin{tabular}{lr}\toprule
       mode 							& number of  signal events	\\\midrule
       $\Bm \ra \LCp \antiproton \en \nueb$ 			& $4\,180\,000$	\\
       $\Bm \ra \LCp \antiproton \mun \numb$ 			& $4\,220\,000$	\\
       $\Bzb \ra \LCp \antiproton \pip \en \nueb$ 		& $4\,230\,000$	\\
       $\Bzb \ra \LCp \antiproton \pip \mun \numb$ 		& $4\,260\,000$	\\\midrule
       $\Bm \ra M \en \nueb$, $M \ra \LCp \antiproton$	 	& $2\,867\,841$ \\
       $\Bm \ra M \mun \numb$, $M \ra \LCp \antiproton$		& $3\,098\,301$ \\
       \bottomrule
     \end{tabular}
   \end{center}
   \label{tab:SPmodesSignal}
 \end{table}

\clearpage
\chapter{Event reconstruction}

Subject of the present work is the study of the two baryonic \B-decay modes $\Bm \ra \LCp \antiproton \en \nueb$ and $\Bm \ra \LCp \antiproton \mun \numb$.
The reconstruction of the \B candidate is done in four separate steps. In the first step a \LCp candidate is reconstructed which is used in the second step to reconstruct the visible part of the \B decay, namely $\LCp \antiproton\ellm$, henceforth denoted as $Y$ system. 
\begin{align}
     \Bm \ra \underbrace{\LCp \antiproton \ellm}_{\displaystyle Y} \nulb
\end{align}
The third step is the reconstruction of the neutrino as missing energy and momentum in the event. In the last step the reconstructed neutrino is combined with the $Y$ to form a \B candidate.

\boldmath
\section{\LCp reconstruction}
\unboldmath
The \LCp is reconstructed in its dominant decay mode $\LCp \ra \proton \Km \pip$, which has a branching fraction of $(5.0 \pm 1.3)\%$ \cite{PDG:2012}. The $\proton$, $\Km$ and $\pip$ candidates are combined and fitted to a common vertex to form a \LCp candidate. A \LCp candidate is accepted if the vertex fit with the \texttt{TreeFitter} algorithm is successful and the invariant mass of the candidate is within the interval from $2.235$ to $2.332\gevcc$. The particle identification (PID) requirements for the three input particles are listed in Table~\ref{tab:PIDLambdaC}.
\begin{table}[h]
	\begin{center}
		\begin{tabular}{cc}\toprule
			particle	& PID list			\\\midrule
			\proton		& \texttt{pCombinedSuperLoose}	\\
			\Km		& \texttt{KCombinedSuperLoose}	\\
			\pip		& \texttt{piCombinedSuperLoose}	\\
			\bottomrule
		\end{tabular}
		\caption{PID lists used for the \LCp reconstruction.}
		\label{tab:PIDLambdaC}
	\end{center}
\end{table}

\boldmath
\subsection{\LCp mass constraint}\label{sect:LCmassconstr}
\unboldmath

\begin{figure}
  \centering\includegraphics[width=.6\textwidth]{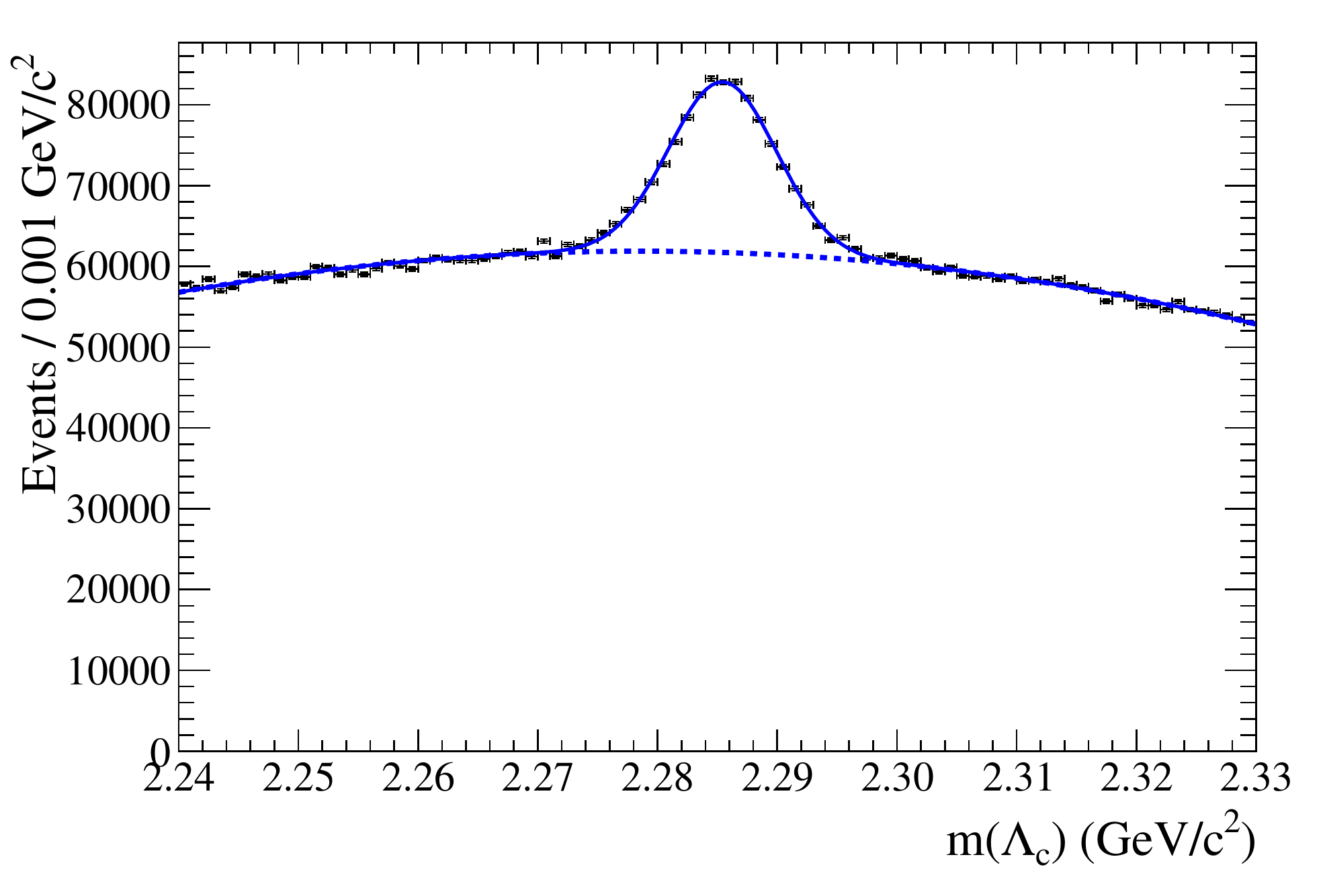}
  \caption{Fitted \LCp mass distribution for OnPeak data.}
  \label{fig:mLambdaC_data_fit}
\end{figure}

 In order to improve the resolution of the reconstructed events a mass constraint on the nominal \LCp mass is applied prior to the $Y$ reconstruction (see section \ref{sect:Yreco}). The standard mass value for the constraint is the one used for the Monte Carlo production which is $2.2849\gevcc$. We use this value in the reconstruction of Monte Carlo events. For data a precise measurement of the \LCp mass shows a momentum dependence of the mass mean value as well as a bias introduced by the SVT material density used in the reconstruction \cite{PhysRevD.72.052006}. 
 To determine the optimal mass value for the constraint we perform an extended maximum-likelihood fit to the $m(\proton \Km \pip)$ distribution in data. As fit function we use a second order Chebychev polynomial for background and a Gaussian for the signal description. The fit can be seen in Fig. \ref{fig:mLambdaC_data_fit}, while the fit parameters are given in Table \ref{tab:mLambdaC_data_fit}.
\begin{table}
  \begin{center}
    \begin{tabular}{cc}\toprule
      Parameter & fitted value\\\midrule
      $N_{\rm bkg} $ & $  5348820\pm 2969$\\
      $N_{\rm sig} $ & $  237499\pm 1924$\\
      $\mu $ & $  2.285446\pm 0.000029$\\
      $\sigma $ & $  0.004489\pm 0.000034$\\
      $b_1 $ & $ -0.033696\pm 0.00076$\\
      $b_2 $ & $ -0.059378\pm 0.00095$\\
      \bottomrule
    \end{tabular}
  \end{center}
  \caption{Fit parameters for the fit to the \LCp mass in OnPeak data.}
  \label{tab:mLambdaC_data_fit}
\end{table}
We decide to constrain the \LCp mass in data to $2.2854\gevcc$.

\boldmath
\section{Reconstruction of the visible decay products}\label{sect:Yreco}
\unboldmath
For the reconstruction of the $Y$ system the previously reconstructed \LCp candidate is combined with an \antiproton, and an \en or \mun candidate. The particle identification at this stage is based on the PID lists given in Table \ref{tab:PIDYreco}. To make sure that the lepton lies within the acceptance region of the SVT, DCH and EMC (excluding the not well calibrated part of the EMC in the backward region) the polar angle $\theta_{\ell}$ of the lepton has to be within $0.41 \hdots 2.37\, {\rm rad}$.
The resulting $Y$ candidate is discarded if a fit of the daughters to a common vertex is not successful. To select those $Y$ candidates with a four-momentum $p_Y$ consistent with a $\Bm \ra Y \nu \ra \LCp \antiproton \ell \nu$ decay we use the angle between the \B meson and the $Y$ candidate $\theta_{\B Y}$ (defined in the center-of-mass system).

\begin{figure}[h]
  \subfigure[]{
    \includegraphics[width=.48\textwidth]{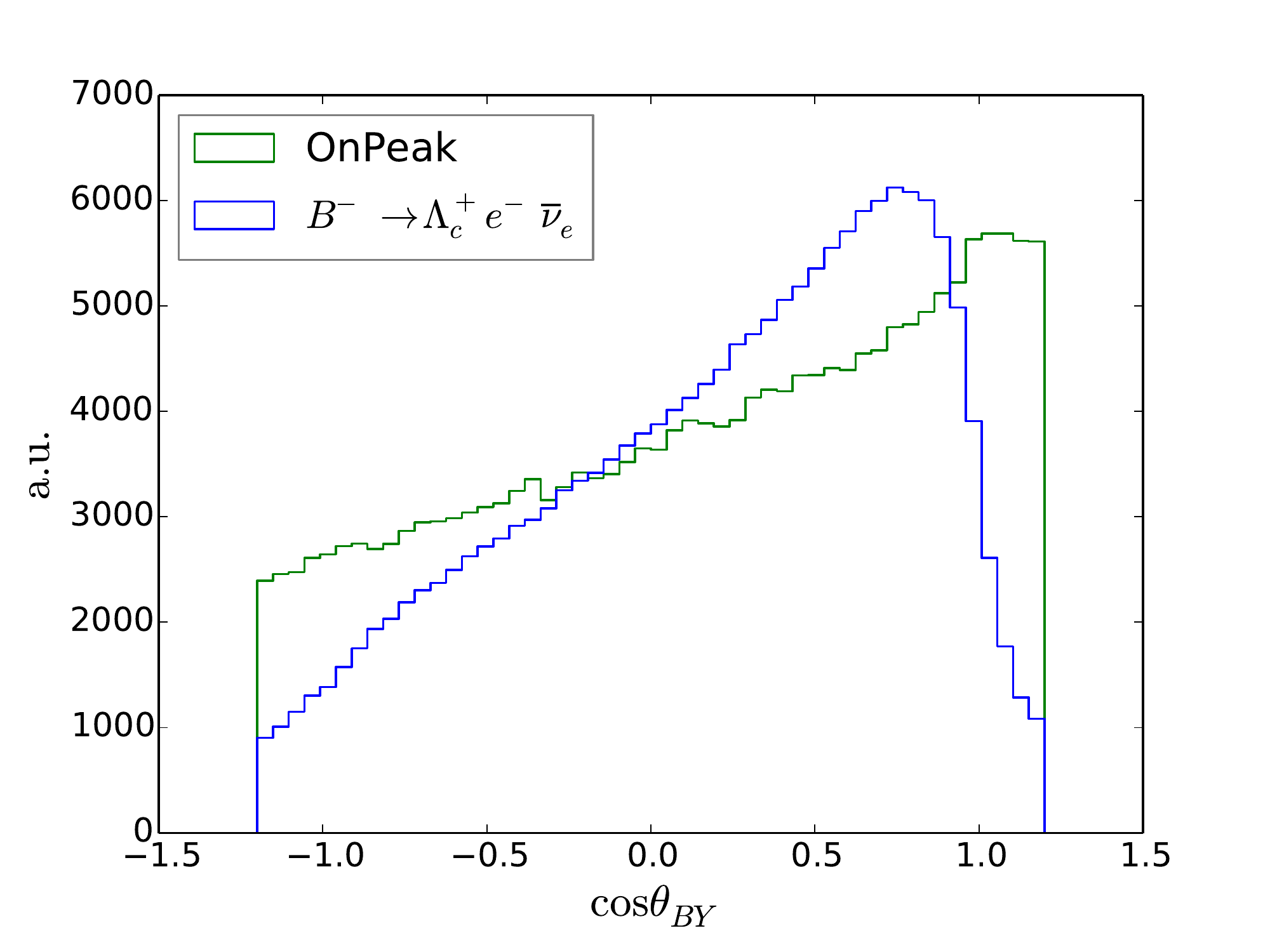}
    \label{cosBY:electron}
  }
  \subfigure[]{
    \includegraphics[width=.48\textwidth]{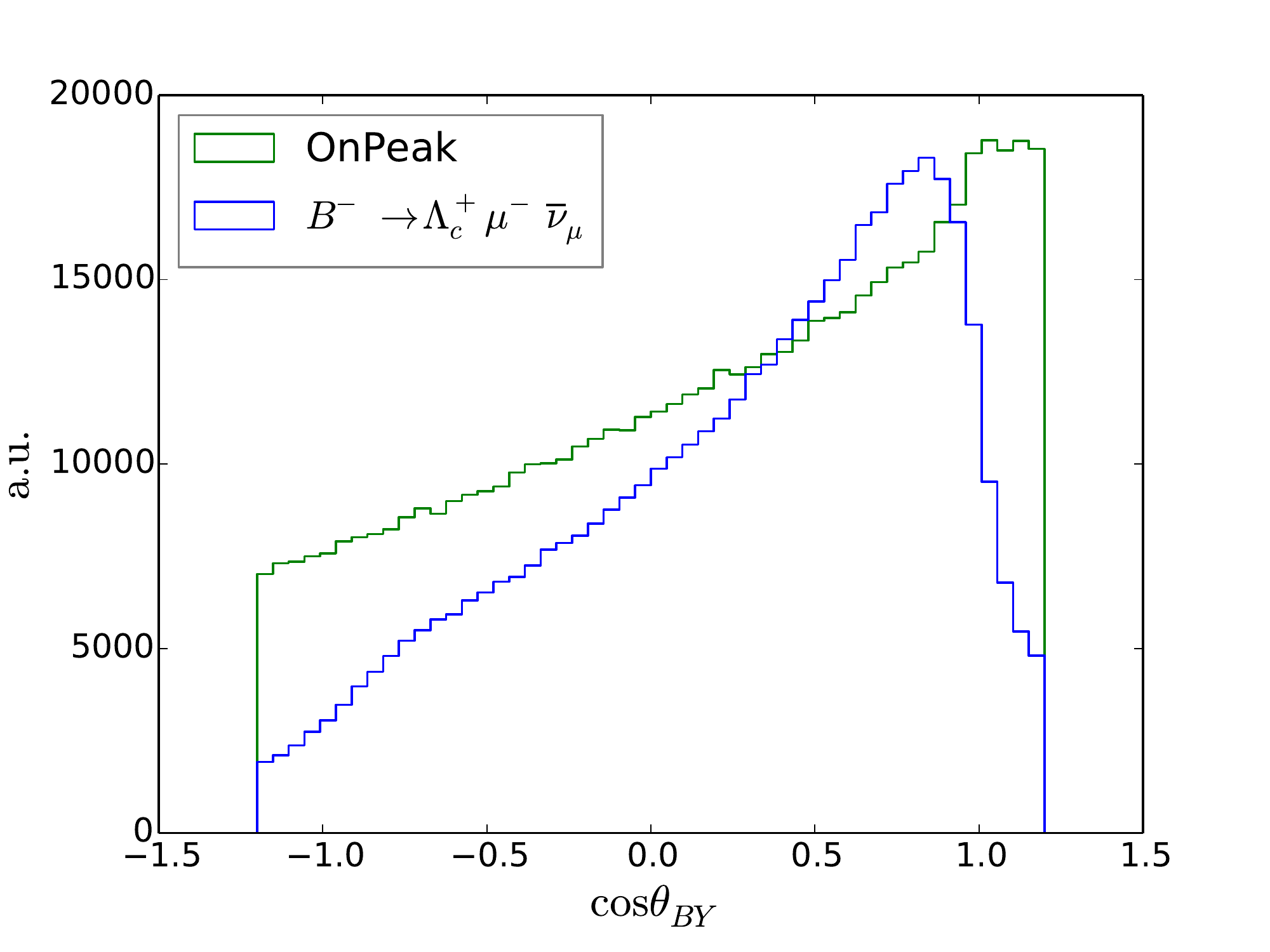}
    \label{cosBY:muon}
  }
  \caption{Comparison of $\cos \theta_{\B Y}$ in OnPeak data and WI signal Monte Carlo for the electron \subref{cosBY:electron} and the muon \subref{cosBY:muon} channel. For both plots a subset of all reconstructed candidates was used, requiring the \mes, \DeltaE and $m(\LCp)$ cuts described later on.}
  \label{fig:comp_cosBY_OnPeak_WI}
\end{figure}

In semileptonic \B decays the four-momentum of the neutrino can be expressed as
\begin{equation}
	p_{\nu}^2 = 0 = (p_{\B} - p_Y)^2 = M_{\B}^2 + M_{Y}^2 - 2(E_{\B}E_{Y} - |\vec{p}_{\B}||\vec{p}_Y|\cos \theta_{\B Y}).
\end{equation}
Here, $E_{\B}$ and $|\vec{p}_{\B}|$ can be derived from the center-of-mass energy. For $M_{\B}$ the world average value can be used. Under the assumption that we have a perfectly reconstructed semileptonic decay, we can determine $\cos \theta_{\B Y}$
\begin{equation}
	\cos \theta_{\B Y} = \frac{2E_{\B}E_Y - M_{\B}^2 - M_{Y}^2}{2|\vec{p}_{\B}||\vec{p}_Y|}.
        \label{eq:cosBY}
\end{equation}
To retain only physical values of $\cos \theta_{\B Y}$ and to suppress background from wrongly reconstructed $Y$ candidates we require $|\cos \theta_{\B Y}|<1.2$, allowing for resolution effects in the reconstruction of this quantity. A comparison of OnPeak data and the WI signal Monte Carlo is shown in Fig. \ref{fig:comp_cosBY_OnPeak_WI}. For light hadronic final states we would expect a strong enhancement at positive values for signal events, while background events (which make up most of the \texttt{OnPeak} data) should show no such peak. This strong difference is smeared out when considering a heavy final state like in the case of $\Bm \ra \LCp \antiproton \ellm \nulb$ where the \B direction is dominated by the $\LCp \antiproton$ system. In consequence the peak for signal Monte Carlo is broadened, while background events tend to prefer values near $+1$ as well.
\begin{table}[h]
	\begin{center}
		\caption{Particle ID lists used for the $Y$ reconstruction.}
		\begin{tabular}{cc}\toprule
			particle	& PID list				\\\midrule
                        \proton		& \texttt{pCombinedSuperLoose}		\\
			\en		& \texttt{eCombinedLoose}		\\
			\mun		& \texttt{muCombinedVeryLooseFakeRate}	\\
			\bottomrule
		\end{tabular}
		\label{tab:PIDYreco}
	\end{center}
\end{table}

\section{Neutrino reconstruction}\label{sect:nu_reco}
Under the assumption that there are no undetected particles in the event the neutrino can be reconstructed as missing energy and missing momentum of the event. In an \epem experiment like \babar the energy and momentum of the CM system are well known for each interaction. To determine the missing energy and momentum in principle all seen particles ($E_i$,$\vec{p}_i$) have to be subtracted from the CM system ($E_{CM}$, $\vec{p}_{CM}$). For neutral particles, mainly photons, the \texttt{GoodPhotonLoose} list is used. For charged particles the \texttt{ChargedTracks} list has to be used since the higher quality list \texttt{GoodTracks\-VeryLoose} does not contain neutral particles that are converted into charged particles inside the detector, e.g. $\gamma \ra \ep \en$, or $K_s^0 \ra \pip \pim$.
\begin{equation}
	(E_{\nu}, \vec{p}_{\nu}) = (E_{\rm miss}, \vec{p}_{\rm miss}) = (E_{CM}, \vec{p}_{CM}) - (\sum_i E_i, \sum_i \vec{p}_i)
\end{equation}
The major problem is to assign the correct particle hypothesis to the charged tracks. In the \texttt{ChargedTracks} list the default hypothesis for each track is the pion hypothesis, i.e. the mass for each track is set to the pion mass.
In order to circumvent this potential bias we assign more realistic particle hypotheses to the tracks for the neutrino reconstruction.
Therefore each element of \texttt{ChargedTracks} is compared in a well defined order with stringent particle identification lists, given in Table \ref{tab:nu_PID}. If the track is not used inside the $\LCp$ or $Y$ the according particle hypothesis is applied to the track, otherwise the same hypothesis as in the $\LCp$ or $Y$ has to be used. The comparison with the particle identification lists is done in the following order:
\begin{enumerate}
	\item \texttt{if} track $\in$ electron list \ra electron PID is assigned
	\item \texttt{else if} track $\in$ kaon list \ra kaon PID is assigned
	\item \texttt{else if} track $\in$ muon list \ra muon PID is assigned
	\item \texttt{else if} track $\in$ proton list \ra proton PID is assigned
	\item \texttt{else} pion PID is assigned.
\end{enumerate}
\begin{table}[h]
  \caption{Particle ID lists used for the neutrino reconstruction.}
  \begin{center}
    \begin{tabular}{ll}\toprule
      particle type & PID list \\\midrule
      electron & \texttt{eKMTight} \\
      kaon & \texttt{KKMTight} \\
      muon & \texttt{muBDTVeryLoose} \\
      proton & \texttt{pKMTight} \\
      \bottomrule
    \end{tabular}
  \end{center}
  \label{tab:nu_PID}
\end{table}

\boldmath
\section{\B reconstruction}
\unboldmath
Subsequently the reconstructed $Y$ and $\nu$ candidate are combined to a \B candidate. Kinematic consistency of the candidate with a \B decay is checked using two variables, the beam energy substituted mass, \mes, and the difference between the reconstructed and expected energy of the candidate, \DeltaE. They are defined as
\begin{align}
	\DeltaE &= E_{\B}^{*} - \sqrt{s}/2 \label{eq:DeltaE}\\
	\mes &= \sqrt{\frac{(s/2 + \vec{p}_{\B} \cdot \vec{p}_{\rm beams})^2}{E^2_{\rm beams}} - \vec{p}^2_{\B}}, \label{eq:mES}
\end{align}
where $s$ refers to the total energy squared of the CM system, $\vec{p}_{\B}$ and $\vec{p}_{\rm beams}$ to the momentum of the \B and the $\epem$ in the laboratory frame and $E_{\B}^{*}$ to the energy of the \B candidate in the CM frame. For the sake of readability we set $c = \hbar =1$.

\begin{figure}
	\subfigure[$\Bm \ra \LCp \antiproton \en \nueb$ Signal Monte Carlo]{
		\includegraphics[width=.48\textwidth]{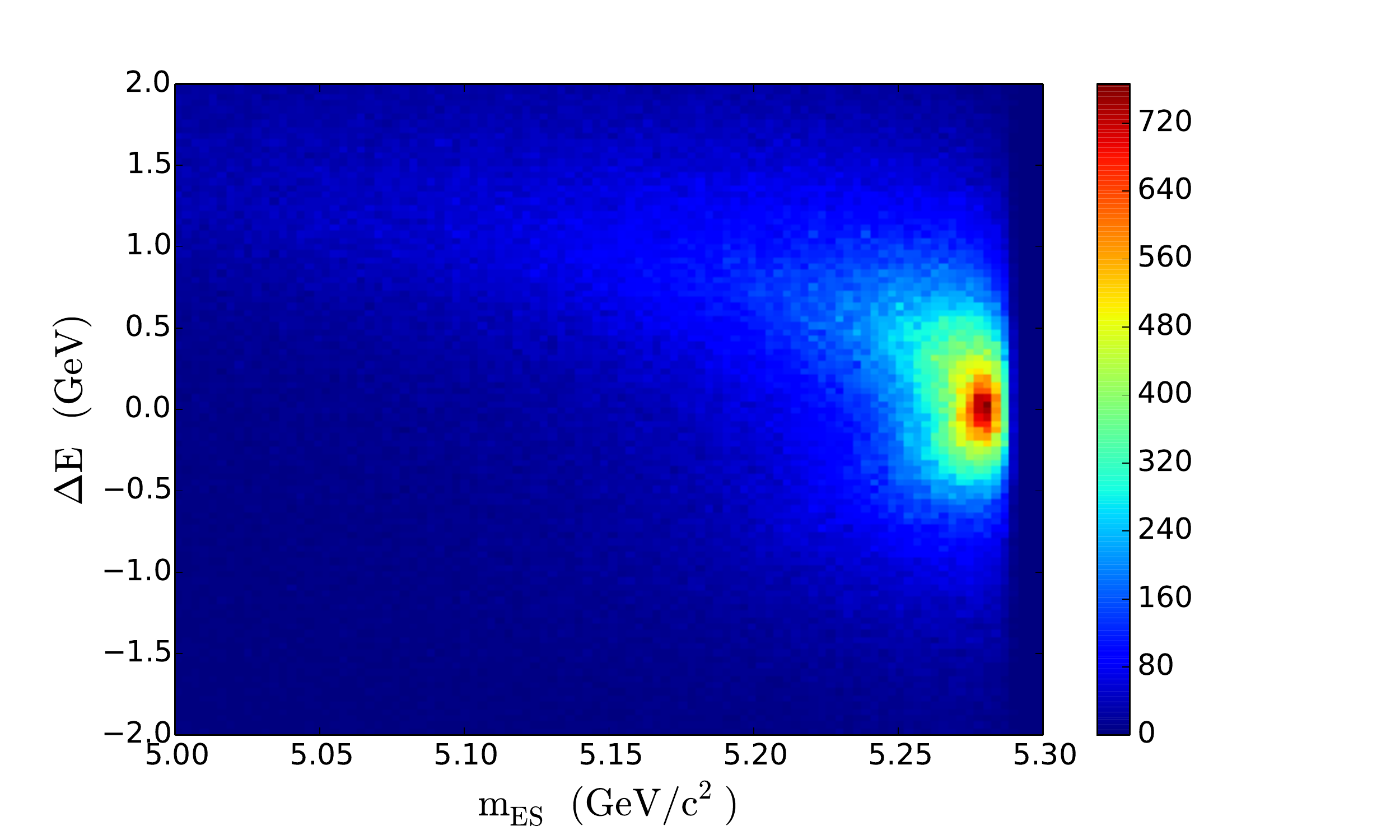}
	}
	\subfigure[$\Bm \ra \LCp \antiproton \mun \numb$ Signal Monte Carlo]{
		\includegraphics[width=.48\textwidth]{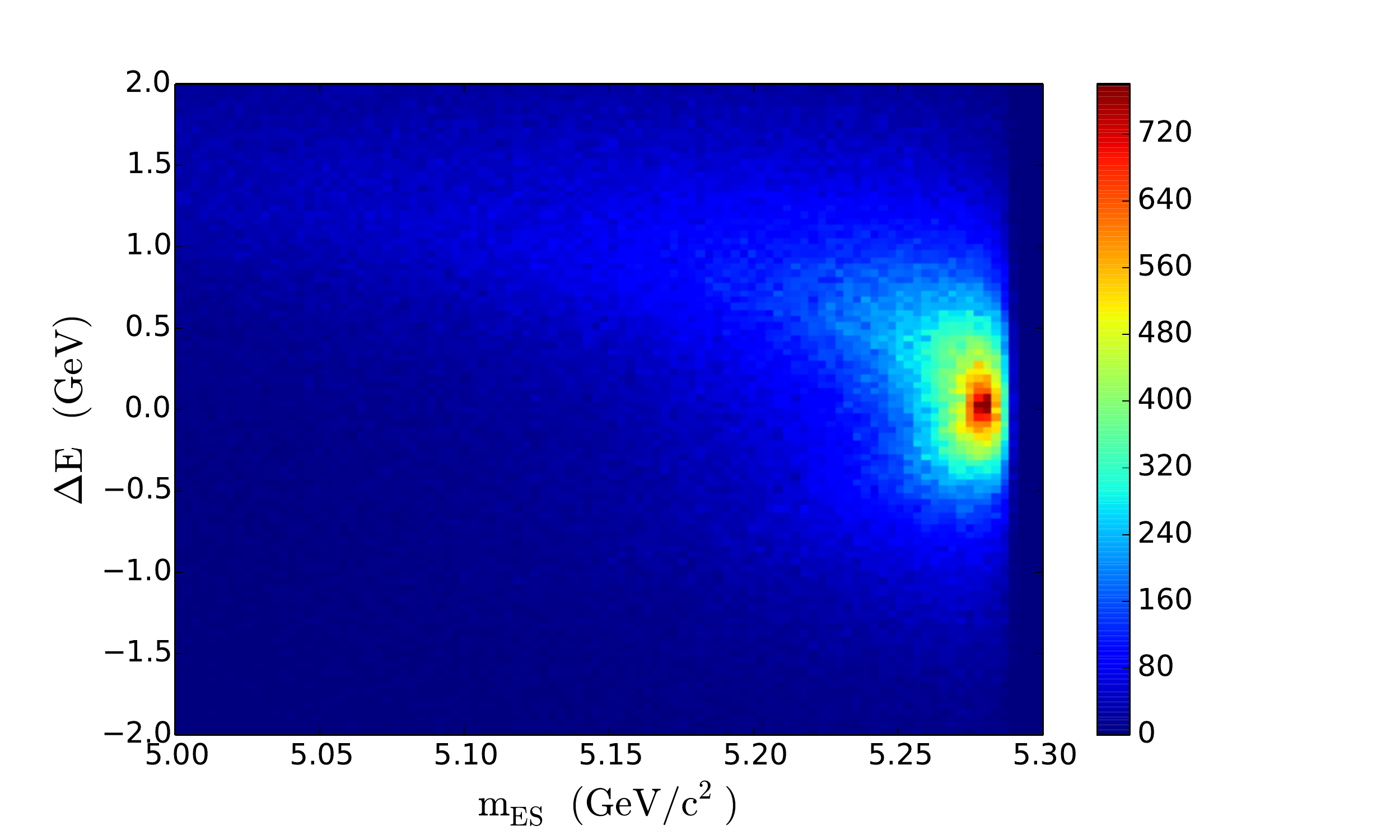}
	}
	\caption{The \DeltaE-\mes plane in WI signal Monte Carlo for the two $\Bm \ra \LCp \antiproton \ellm \nu_{\ell}$ signal modes.}
	\label{fig:DeltaEmESplaneSignalMC}
\end{figure}

The resulting \B candidate has to pass loose cuts on \mes and \DeltaE.
\begin{align}
	-2.0 \gev < \DeltaE < 2.0 \gev \; \mbox{and} \; \mes > 5.0 \gevcc
\end{align}
The resulting two-dimensional \DeltaE-\mes plane in WI signal Monte Carlo (via the \texttt{JETSET} model) for the two signal modes can be seen in Figure \ref{fig:DeltaEmESplaneSignalMC}.
A clear signal peak is visible, making a measurement of these decay channels plausible if the branching fraction is reasonably large.
Furthermore, the distributions show that the signal has large tails in $\mes$ as well as in $\DeltaE$. 

\clearpage
\chapter{Background suppression}

For a good separation of signal and background a detailed background study is necessary. Therefore, we identified two different classes of background.
\begin{enumerate}
	\item continuum events from $\epem \ra \qqbar$ ($q = u, d, s, c$)
	\item combinatorial background from other \B decays
\end{enumerate}
\begin{figure}[h]
  \centering
    \subfigure[]{
      \includegraphics[width=.45\textwidth]{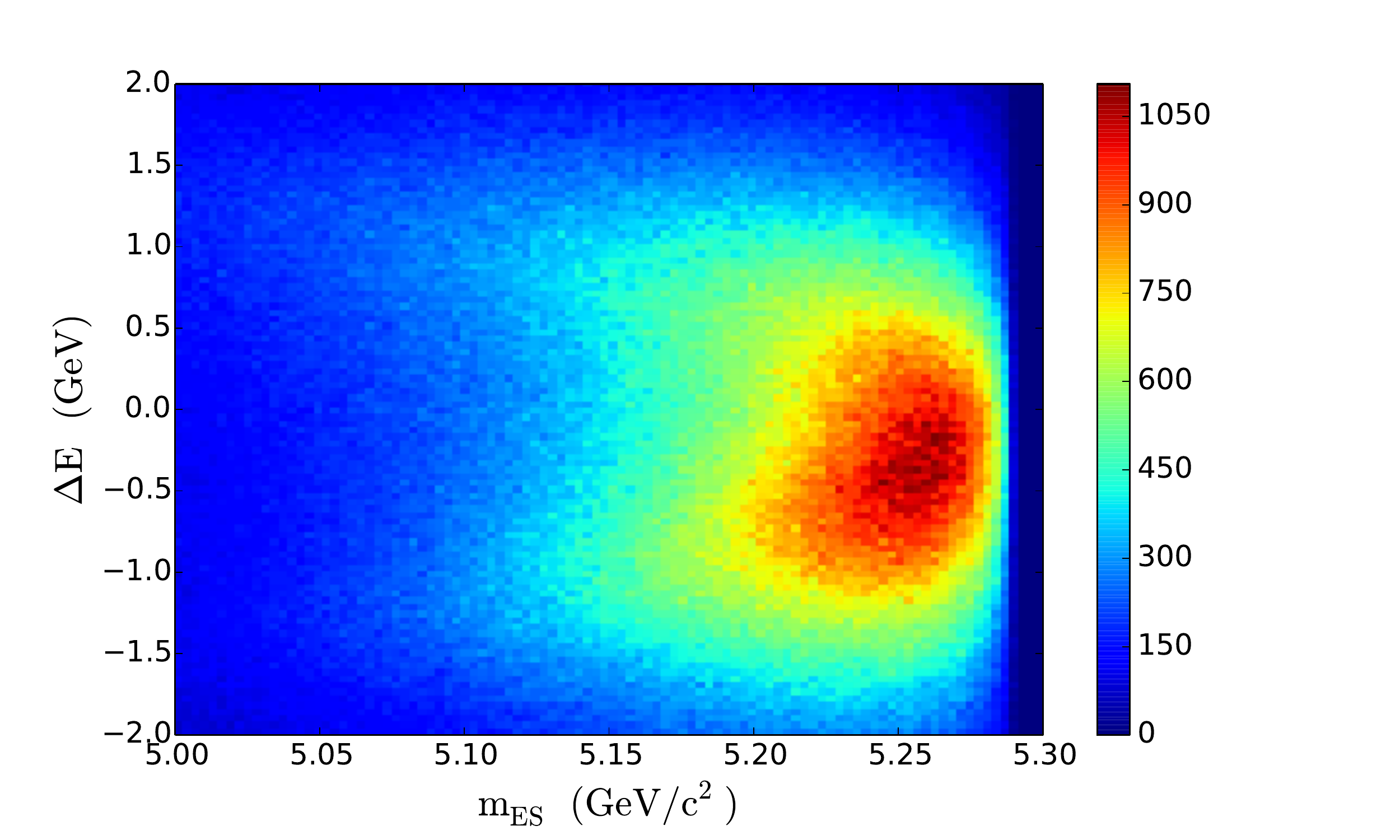}
      \label{subfig:uds}
    }
    \subfigure[]{
      \includegraphics[width=.45\textwidth]{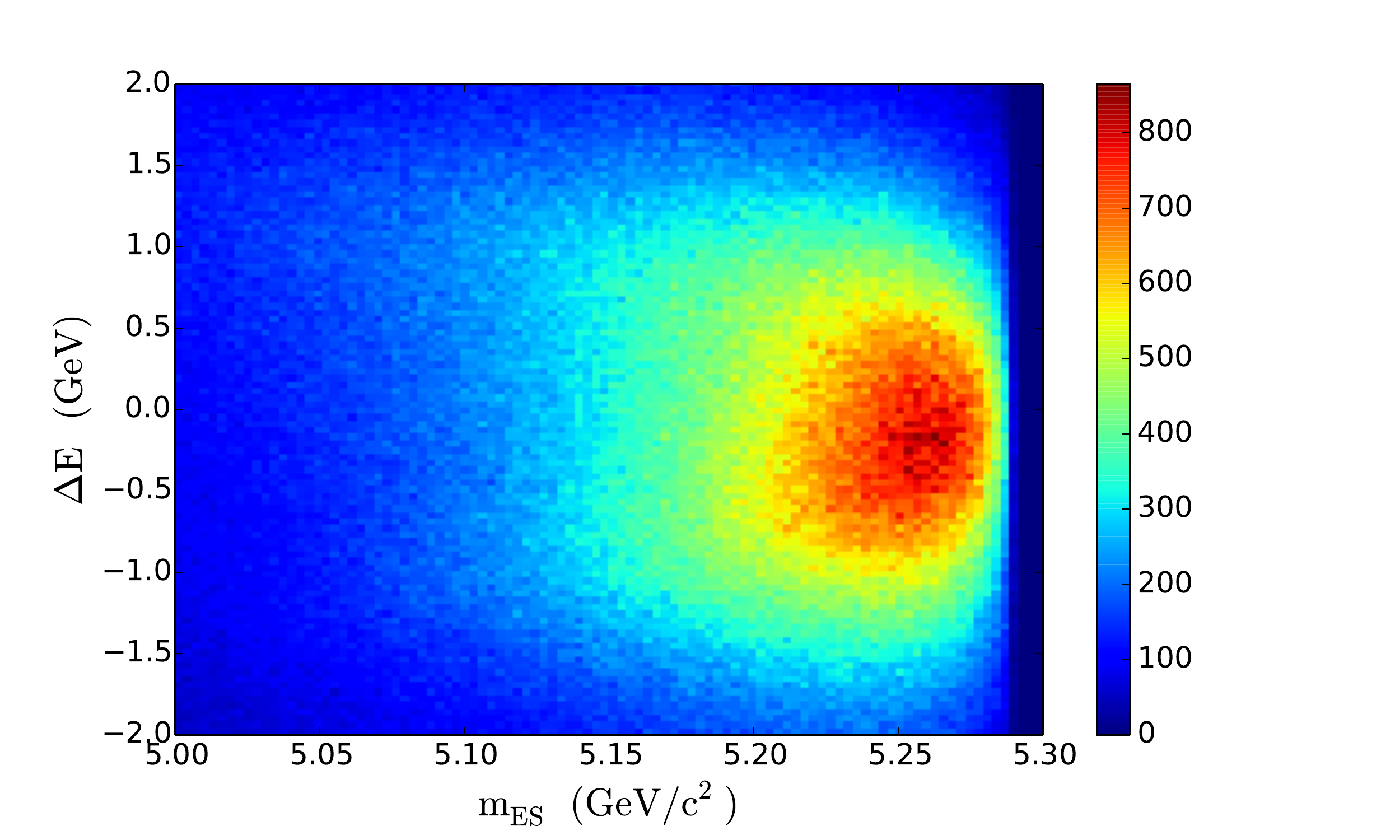}
      \label{subfig:ccbar}
    }
  \caption{The \DeltaE-\mes plane for $uds$ \subref{subfig:uds} and $c \cbar$ \subref{subfig:ccbar} Monte Carlo.}
  \label{fig:2Dback:qqbar}
\end{figure}
\begin{figure}[h]
  \centering
  \subfigure[]{
    \includegraphics[width=.45\textwidth]{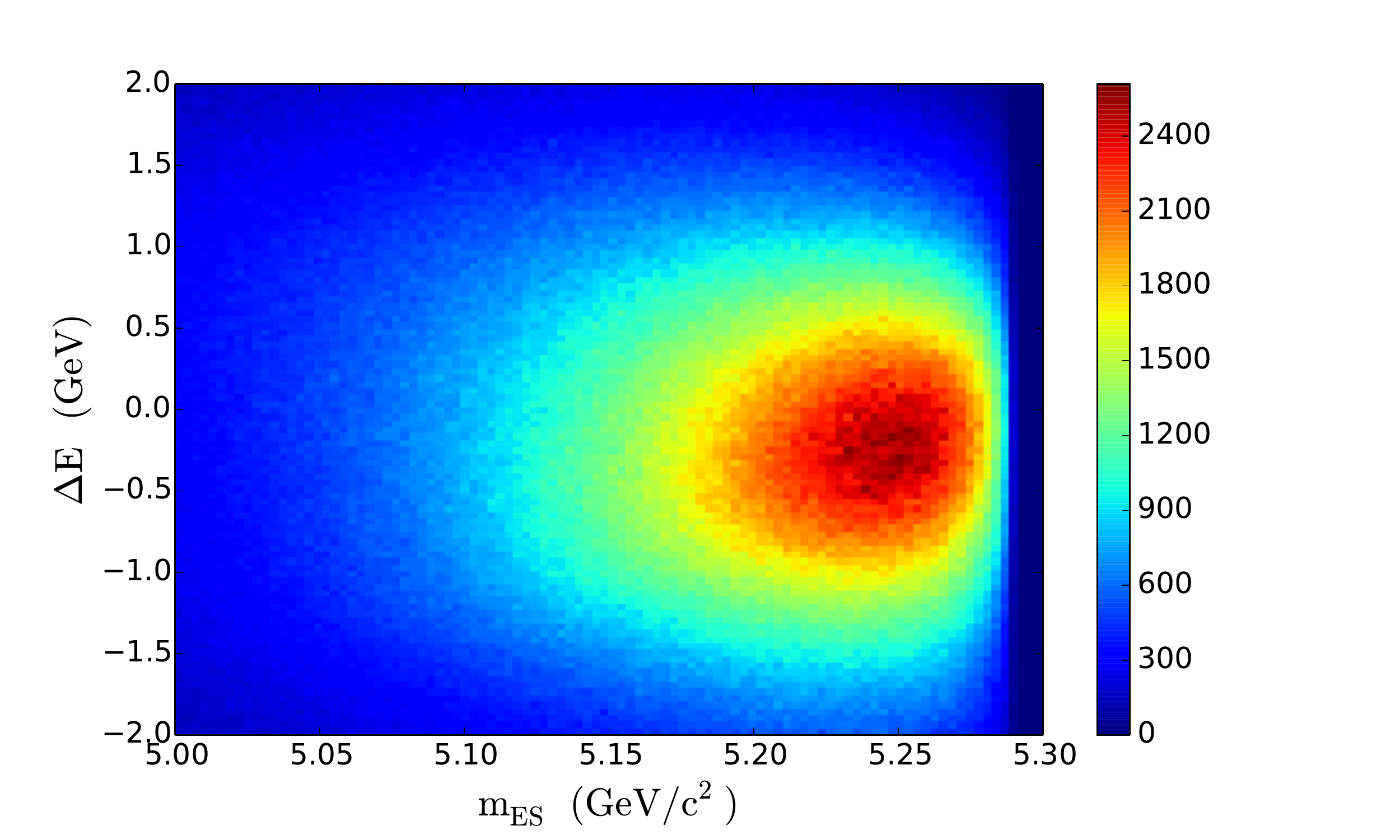}
    \label{subfig:BpBm}
  }
  \subfigure[]{
    \includegraphics[width=.45\textwidth]{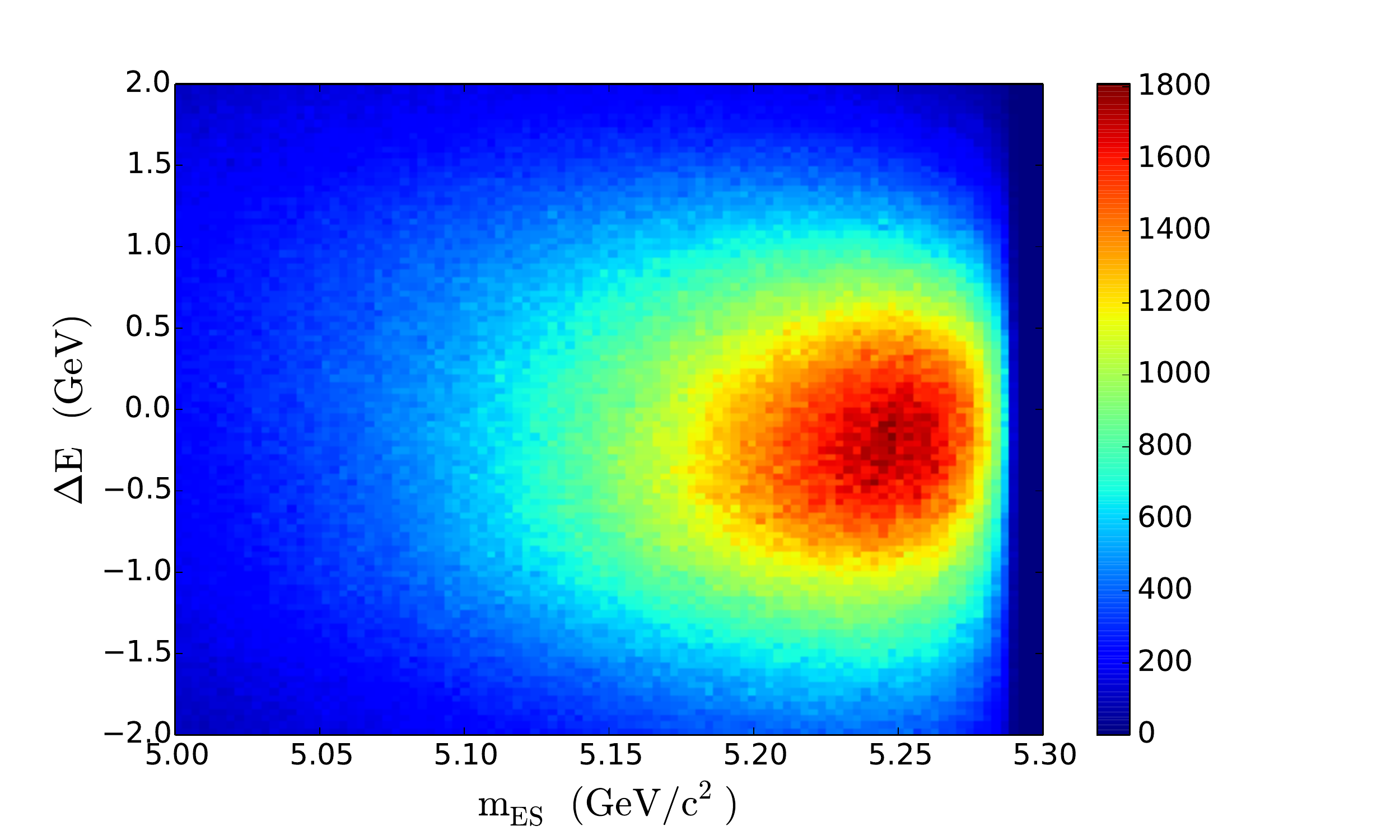}
    \label{subfig:BzBzb}
  }
  \caption{The \DeltaE-\mes plane for $\BpBm$ \subref{subfig:BpBm} and $\BzBzb$ \subref{subfig:BzBzb} Monte Carlo.}
  \label{fig:2Dback:BBbar}
\end{figure}
The two-dimensional \DeltaE-\mes distribution for these two background classes after the decay reconstruction can be found in Figures \ref{fig:2Dback:qqbar} and \ref{fig:2Dback:BBbar}. In comparison to the signal Monte Carlo distribution shown in Figure \ref{fig:DeltaEmESplaneSignalMC} these two background classes are smeared out over a large phase space region. In addition, the $uds$ background shown in Fig. \ref{subfig:uds} is shifted to smaller \mes values. Note, that the concentration of background events close the the signal region is caused by the large invariant mass of the $Y$ system.

\section{\LCp mass cut}

In order to suppress background from generic $\proton$, $\Km$ and $\pip$ combinations in the \LCp reconstruction we decide to use a $\pm 3\sigma$ mass selection region around the central mass value of the \LCp. Due to the observed difference between OnPeak data and Monte Carlo events the cut has to be determined separately for both data sets. Therefore, we fit the $m(\LCp)$ distribution in data and signal Monte Carlo with a Gaussian $g(m; \mu, \sigma)$ for signal and a second order Chebychev polynomial $p(m; b_1, b_2)$ for background. The fit results are given in Fig. \ref{fig:fitLambdaC_data}, \ref{fig:fitLambdaC_SP9938} and Table \ref{tab:fitLambdaC_data}, and \ref{tab:fitLambdaC_SP9938}. 
Due to the high statistics of the signal Monte Carlo sample a better fit would require additional terms for signal description. But for the purpose of a mass cut the shown description suffices.

\begin{tabular}{cc}
  \begin{minipage}{.48\textwidth}
    \begin{figure}[H]
      \includegraphics[width=\textwidth]{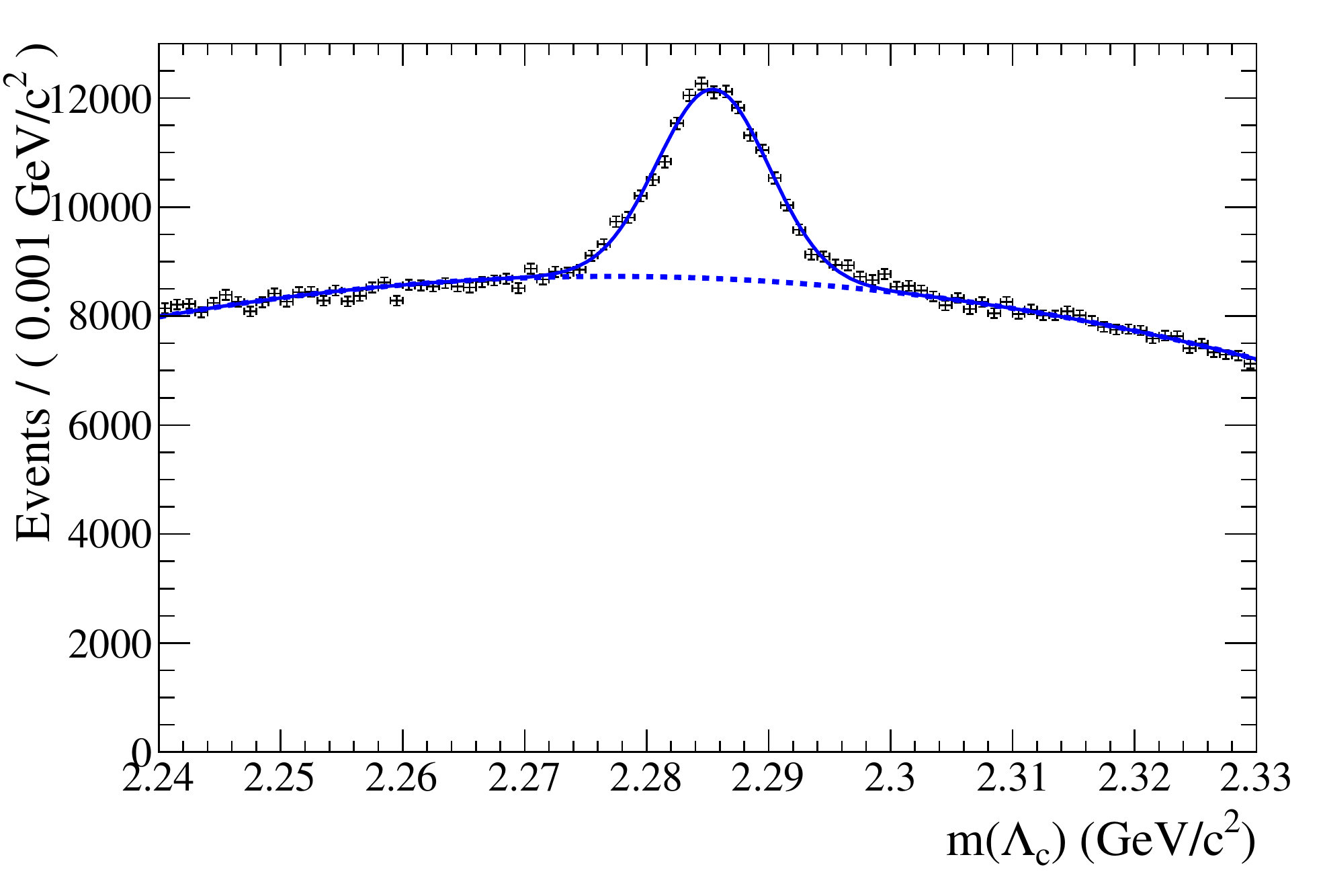}
      \caption{$m(\LCp)$ fit in OnPeak data.}
      \label{fig:fitLambdaC_data}
    \end{figure}
  \end{minipage} &
  \begin{minipage}{.48\textwidth}
    \begin{figure}[H]
      \includegraphics[width=\textwidth]{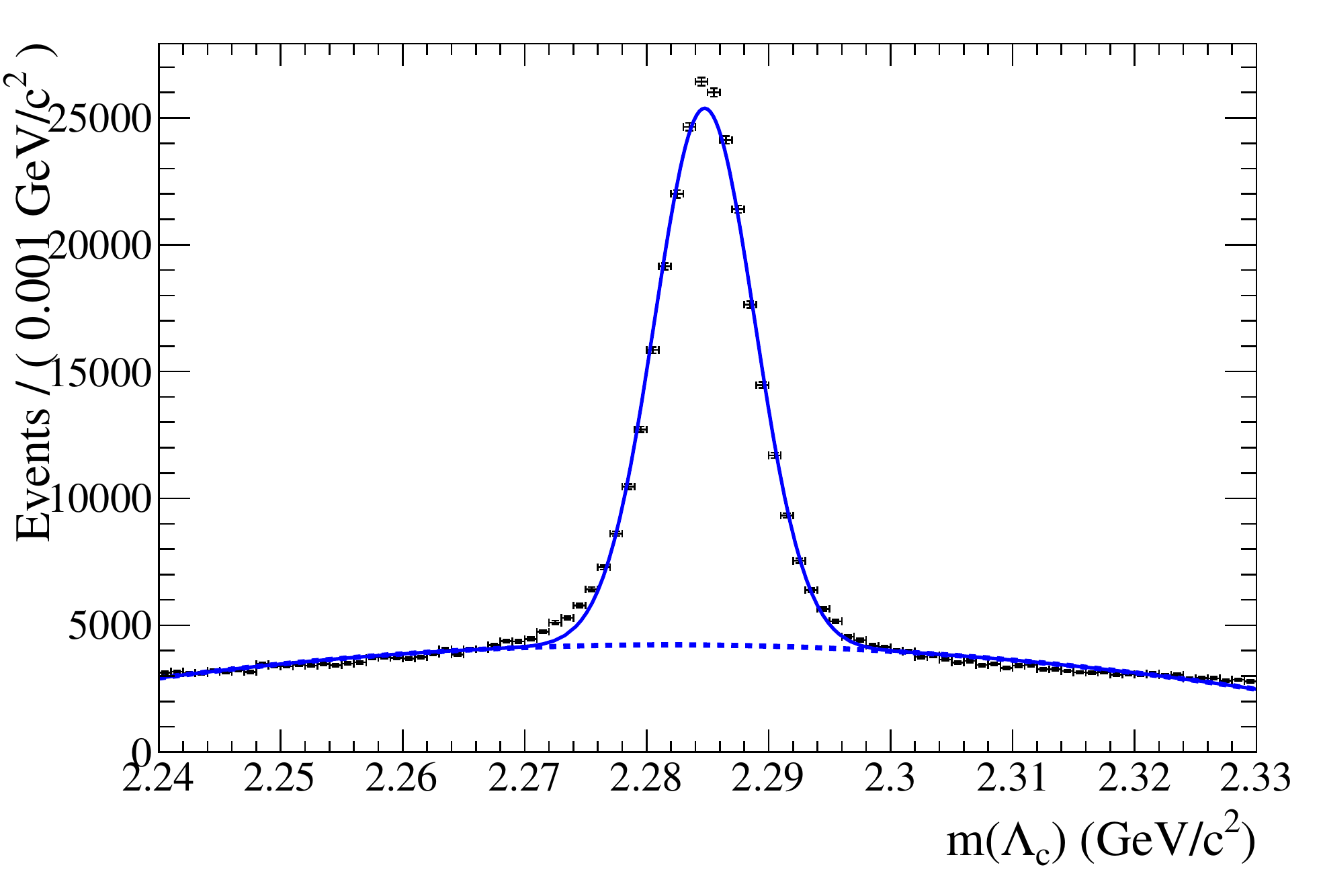}
      \caption{$m(\LCp)$ fit in $\Bm \ra \LCp \antiproton \en \nueb$ MC events.}
      \label{fig:fitLambdaC_SP9938}
    \end{figure}
  \end{minipage} \\
  \begin{minipage}{.48\textwidth}
    \begin{table}[H]
      \caption{Fit parameter for a fit to $m(\LCp)$ in data events.}
      \begin{center}
      \begin{tabular}{lc}
        \toprule
        parameter & value \\\midrule
        $N_{bkg} $ & $  749530\pm 1120$\\
        $N_{sig} $ & $  39835\pm 739$\\
        $\mu $ & $  2.285453\pm 0.000068$\\
        $\sigma $ & $  0.004578\pm 0.000081$\\
        $b_1 $ & $ -0.04754\pm 0.0020$\\
        $b_2 $ & $ -0.06730\pm 0.0026$\\
        \bottomrule
      \end{tabular}
      \end{center}
      \label{tab:fitLambdaC_data}
    \end{table}
  \end{minipage} &
  \begin{minipage}{.48\textwidth}
    \begin{table}[H]
      \caption{Fit parameter for a fit to $m(\LCp)$ in Monte Carlo events.}
      \begin{center}
        \begin{tabular}{lc}\toprule
          parameter & value \\\midrule
          $N_{bkg} $ & $  334006\pm 744$\\
          $N_{sig} $ & $  217413\pm 661$\\
          $\mu $ & $  2.284760\pm 0.000012$\\
          $\sigma $ & $  0.004099\pm 0.000013$\\
          $b_1 $ & $ -0.06399\pm 0.0030$\\
          $b_2 $ & $ -0.21995\pm 0.0038$\\
          \bottomrule
        \end{tabular}
      \end{center}
      \label{tab:fitLambdaC_SP9938}
    \end{table}
  \end{minipage}
\end{tabular}
The fit returns a $\sigma$ of $4.58 \pm 0.09 \mevcc$ for data and $4.10 \pm 0.02 \mevcc$ for Monte Carlo. The larger value for data events is plausible taking into account the observed momentum dependence of the mean value of the \LCp mass for data \cite{PhysRevD.72.052006}. Since we use \LCp candidates from a relatively large momentum range this effect leads to a broader distribution.

For convenience we decide to use the same width for the cut for both, OnPeak data and Monte Carlo, only shifted according to the different mean values, measured in section \ref{sect:LCmassconstr}. Therefore, we decide to use the following cut on the \LCp candidate mass for data
\begin{equation}
  2.2716 < m(\LCp) < 2.2992 \gevcc,
\end{equation}
corresponding to $\pm 3\sigma$, and for Monte Carlo
\begin{equation}
  2.2711 < m(\LCp) < 2.2987 \gevcc,
\end{equation}
corresponding to $\pm 3.4\sigma$.

\section{Particle identification}

For the reconstruction of our signal candidates we use the PID lists used in the \babar internal data reprocessing. While the \texttt{CombinedSuperLoose} PID lists provide a high probability to correctly identify the detected tracks they have a high mis-identification probability as well. In order to reduce the number of mis-identified particles we decide to use the more stringent PID lists, given in Table \ref{tab:PIDlists2}.
\begin{table}[h]
\begin{center}
  \caption{Particle ID lists used for the signal extraction.}
  \begin{tabular}{cc}
    \toprule
    particle & PID list \\\midrule
    proton (\LCp) & \texttt{pKMLoose} \\
    kaon (\LCp) & \texttt{KKMLoose} \\
    pion (\LCp) & \texttt{piKMLoose} \\
    proton   & \texttt{pKMTight} \\
    electron & \texttt{eKMTight} \\
    muon     & \texttt{muBDTTight} \\
    \bottomrule
  \end{tabular}
  \label{tab:PIDlists2}
\end{center}
\end{table}





\section{Continuum background suppression}\label{sect:rf:continuum}

For the reduction of continuum background ($\qqbar$, $q = u, d, s, c$) we decide to use a random forest. The number of split variables is set to $\sqrt{p}$, where $p$ is the total number of discriminating variables. The training data sets for signal and background equal each other in size.

For continuum background we settle on three event shape variables, defined as:
\begin{itemize}
  \item The ratio of the second and zeroth Fox-Wolfram moment $R_2$ \cite{Sjostrand:2006za} for all charged tracks, defined as
    \begin{equation}
      R_2 = \frac{\sum_{ij}^{tracks}|p_i||p_j|P_2(\cos\theta_{ij})}{\sum_{ij}^{tracks}|p_i||p_j|},
    \end{equation}
    with the Legendre polynomial $P_2$ and the angle $\theta_{ij}$ between the momenta $p_i$ and $p_j$.
  \item The sphericity $S$ of the event \cite{Sjostrand:2006za}, derived from the sphericity tensor
    \begin{equation}
      S^{\alpha\beta} = \frac{\sum_ip_i^{\alpha}p_i^{\beta}}{\sum_i|\vec{p}_i|^2},
    \end{equation}
    where $\alpha, \beta=1,2,3$ correspond to the $x$, $y$ and $z$ components.
  \item The cosine of the angle between the thrust axis of the $Y$ candidate and the thrust axis \cite{Sjostrand:2006za} of the rest of the event $\cos\Delta\theta_{\rm thrust}$. The thrust axis is defined as the unit vector $\vec{n}$ that maximizes the thrust $T$
    \begin{equation}
      T = \max_{|\vec{n}|=1}\frac{\sum_i^{tracks}|\vec{n}\cdot\vec{p}_i|}{\sum_i|\vec{p}_i|}.
    \end{equation}
\end{itemize}
A comparison of these three variables between signal and background for the electron channel is given in Fig. \ref{fig:qqbarRF_input}, and for the muon channel in the appendix, Fig. \ref{fig:qqbarRF_input:mu}.
\begin{figure}[h]
  \begin{center}
  \subfigure[]{
    \includegraphics[width=.48\textwidth]{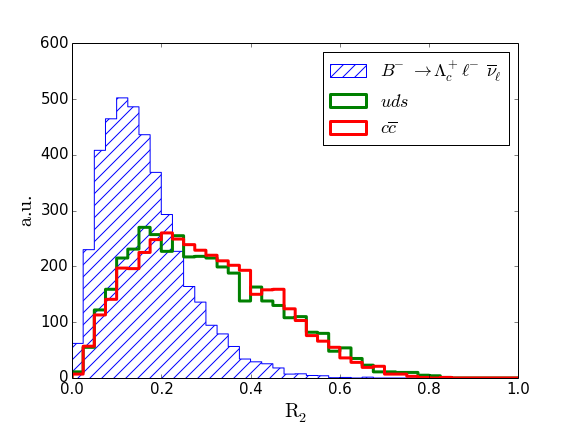}
    \label{subfig:e:qqbar:R2}
  }
  \subfigure[]{
    \includegraphics[width=.48\textwidth]{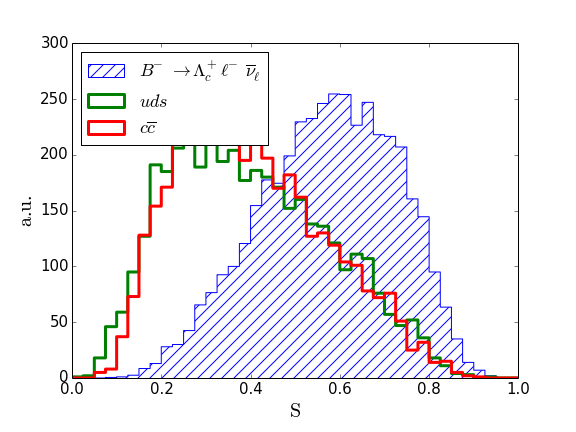}
    \label{subfig:e:qqbar:S}
  }
  \subfigure[]{
    \includegraphics[width=.48\textwidth]{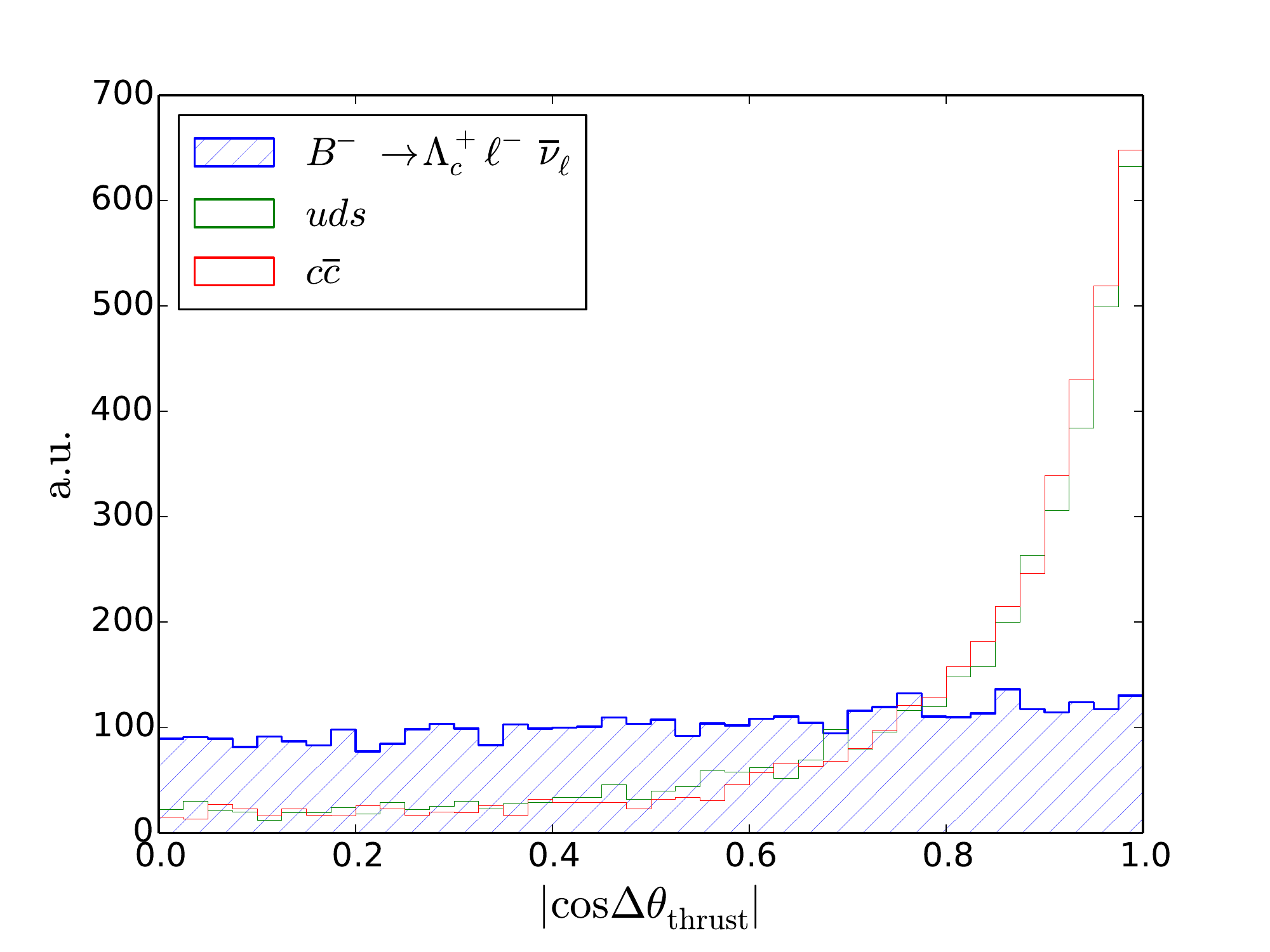}
    \label{subfig:e:qqbar:cosDTh}
  }
  \end{center}
  \caption{Comparison of WI signal Monte Carlo (blue histogram) and $\qqbar$ background Monte Carlo (green and red lines) for the three Random Forest input variables, \subref{subfig:e:qqbar:R2} $R_2$, \subref{subfig:e:qqbar:S} $S$, and \subref{subfig:e:qqbar:cosDTh} $cos\Delta\theta_{\rm thrust}$.}
  \label{fig:qqbarRF_input}
\end{figure}

\begin{figure}[h]
  \begin{center}
    \includegraphics[width=.7\textwidth]{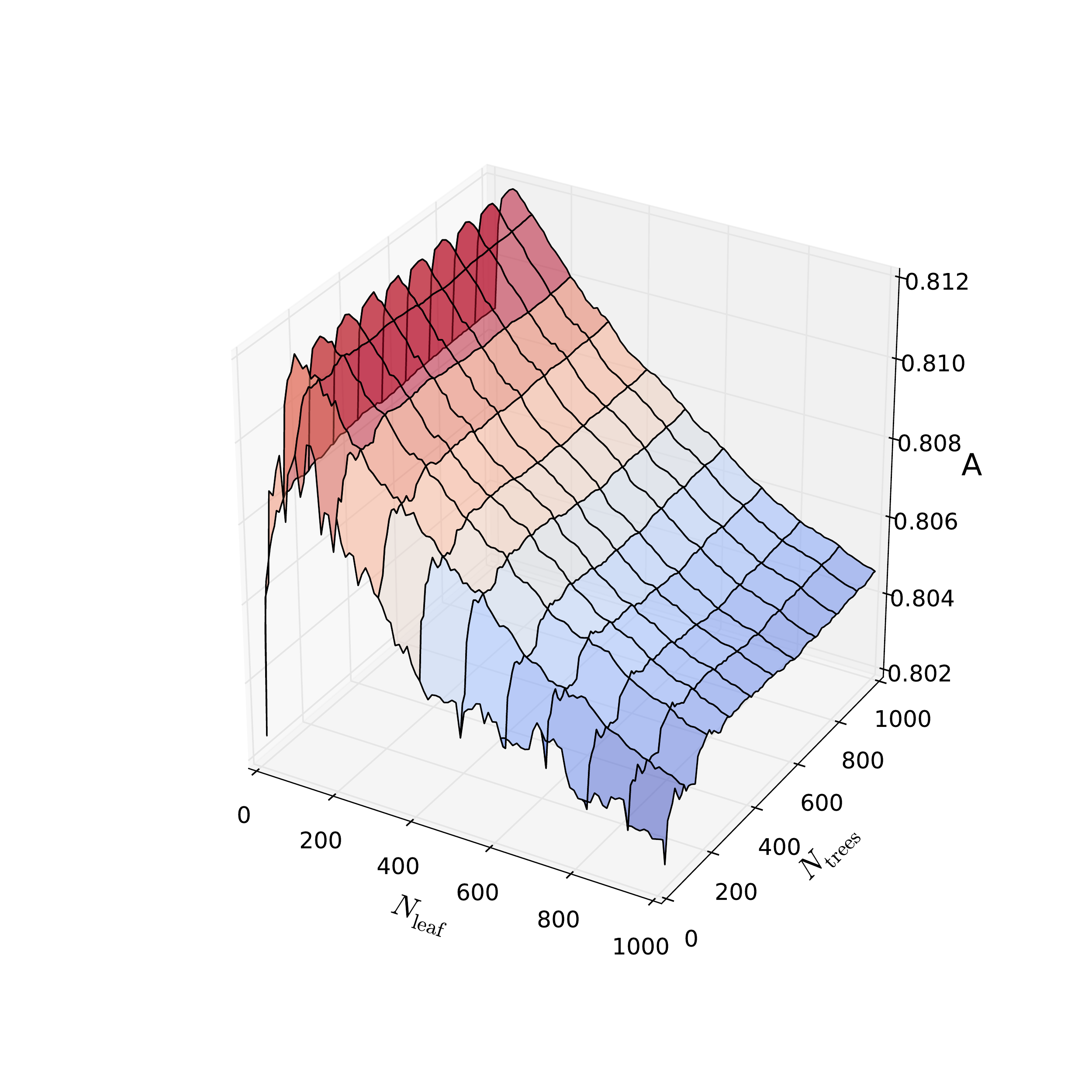}
  \end{center}
  \caption{Area below the ROC curve in dependence of $N_{\rm trees}$ and $N_{\rm leaf}$.}
  \label{fig:qqbar:optRF}
\end{figure}

The random forest (RF) has two major variables to be adjusted for optimal performance, the number of trees $N_{\rm trees}$ and the minimum number of events per leaf $N_{\rm leaf}$. We optimise these two variables by training random forests with both variables ranging from $10$ to $1000$ and calculating the area $A$ below the receiver operating characteristic (ROC) curve. A large value here points to a good performance. A plot of the area in dependence of $N_{\rm trees}$ and $N_{\rm leaf}$ can be seen in Fig. \ref{fig:qqbar:optRF}. The optimum for the RF configuration is at $380$ trees and $60$ events per leaf, but as can be seen from Fig. \ref{fig:qqbar:optRF} the variation in $A$ is not very large (except close to $N_{\rm trees} = 0$), and in consequence the performance deficits by using a not optimal configuration are negligible.

\begin{figure}[h]
  \subfigure[]{
    \includegraphics[width=.48\textwidth]{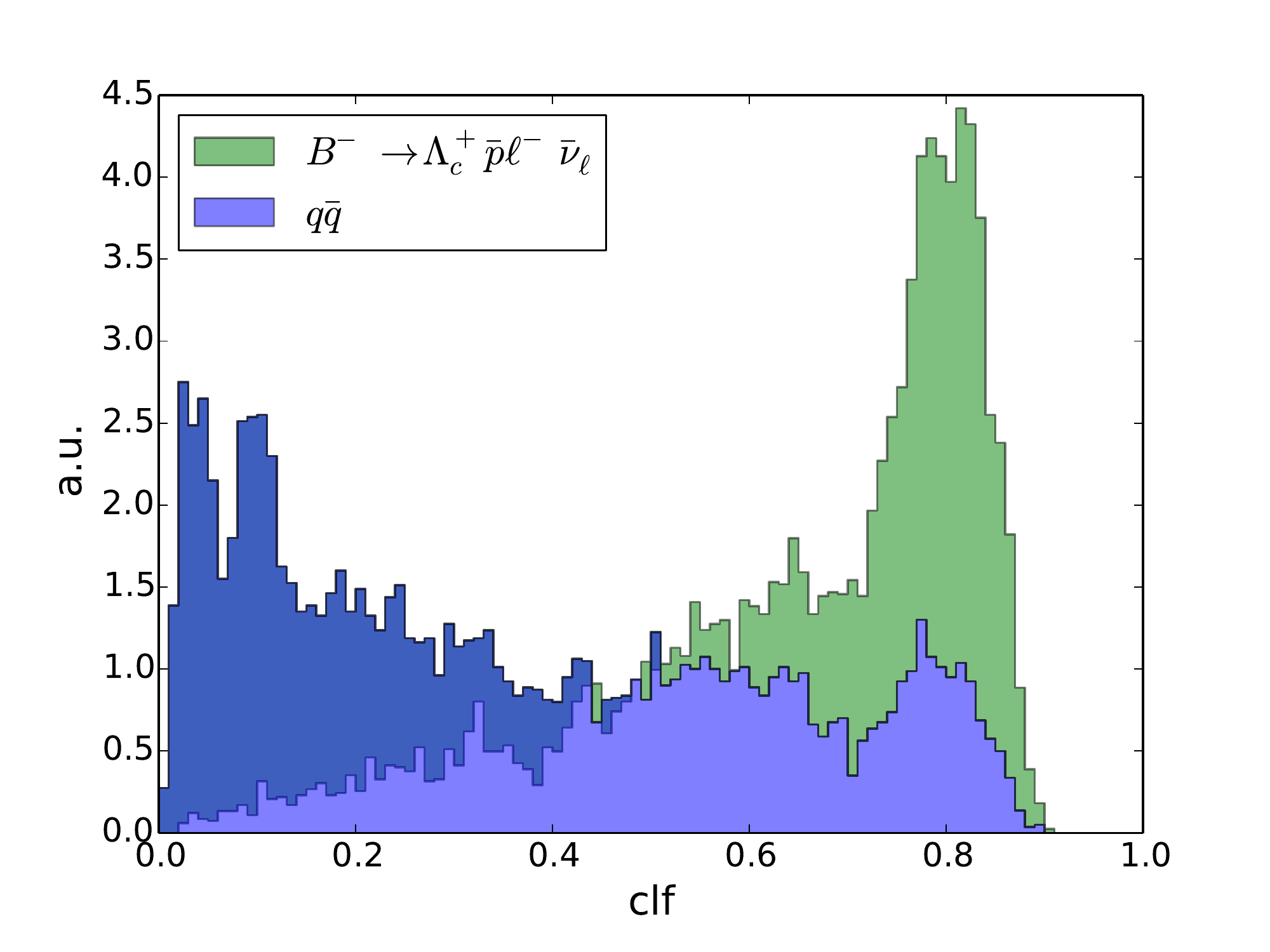}
    \label{subfig:clf:qqbar:electron}
  }
  \subfigure[]{
    \includegraphics[width=.48\textwidth]{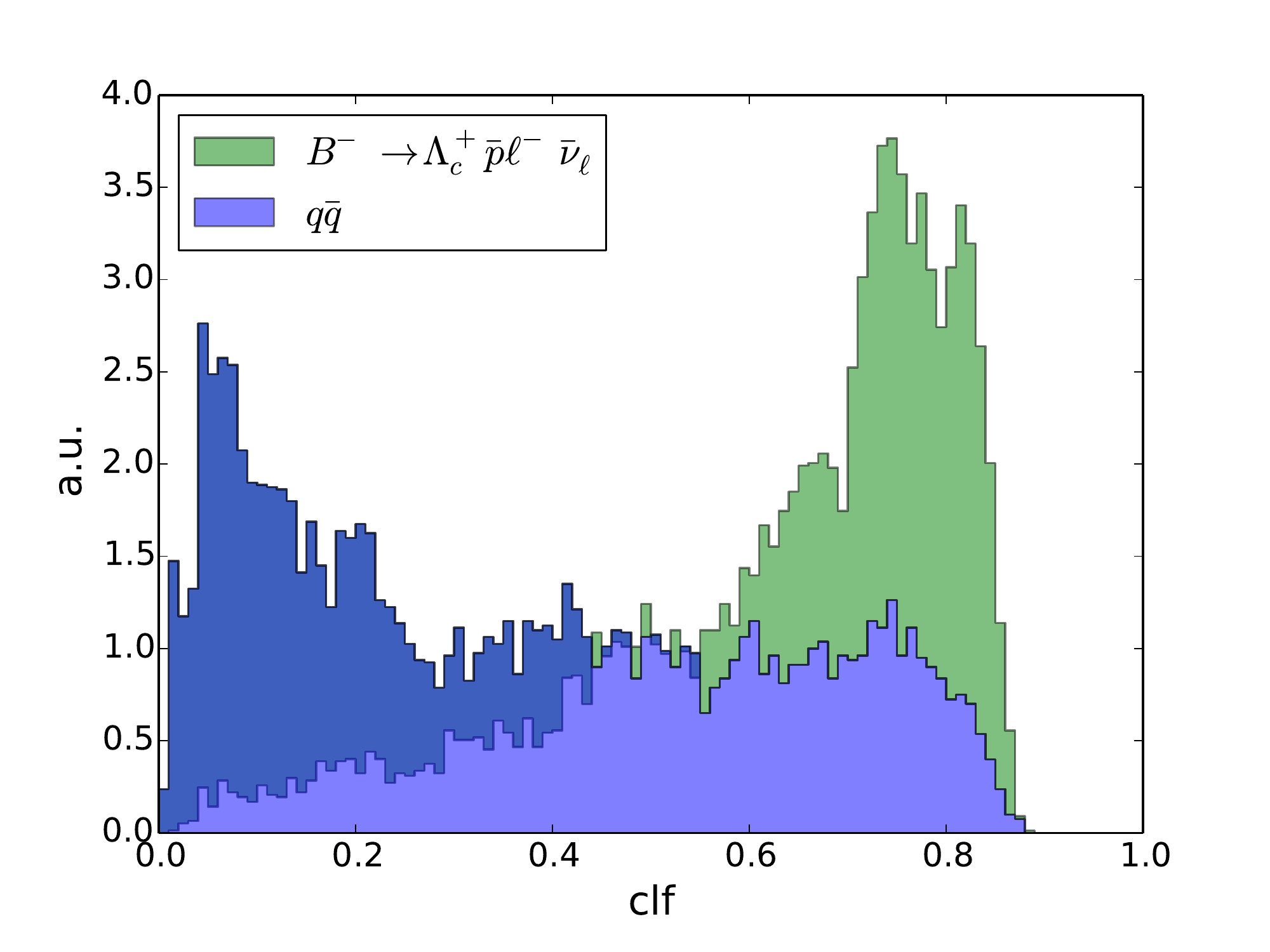}
    \label{subfig:clf:qqbar:muon}
  }
  \caption{The classifier distributions for signal and background after training of the random forest: \subref{subfig:clf:qqbar:electron} $\Bm \ra \LCp \antiproton \en \nueb$, \subref{subfig:clf:qqbar:muon} $\Bm \ra \LCp \antiproton \mun \numb$.}
  \label{fig:clf:qqbar}
\end{figure}

\begin{figure}[h]
  \subfigure[]{
    \includegraphics[width=.48\textwidth]{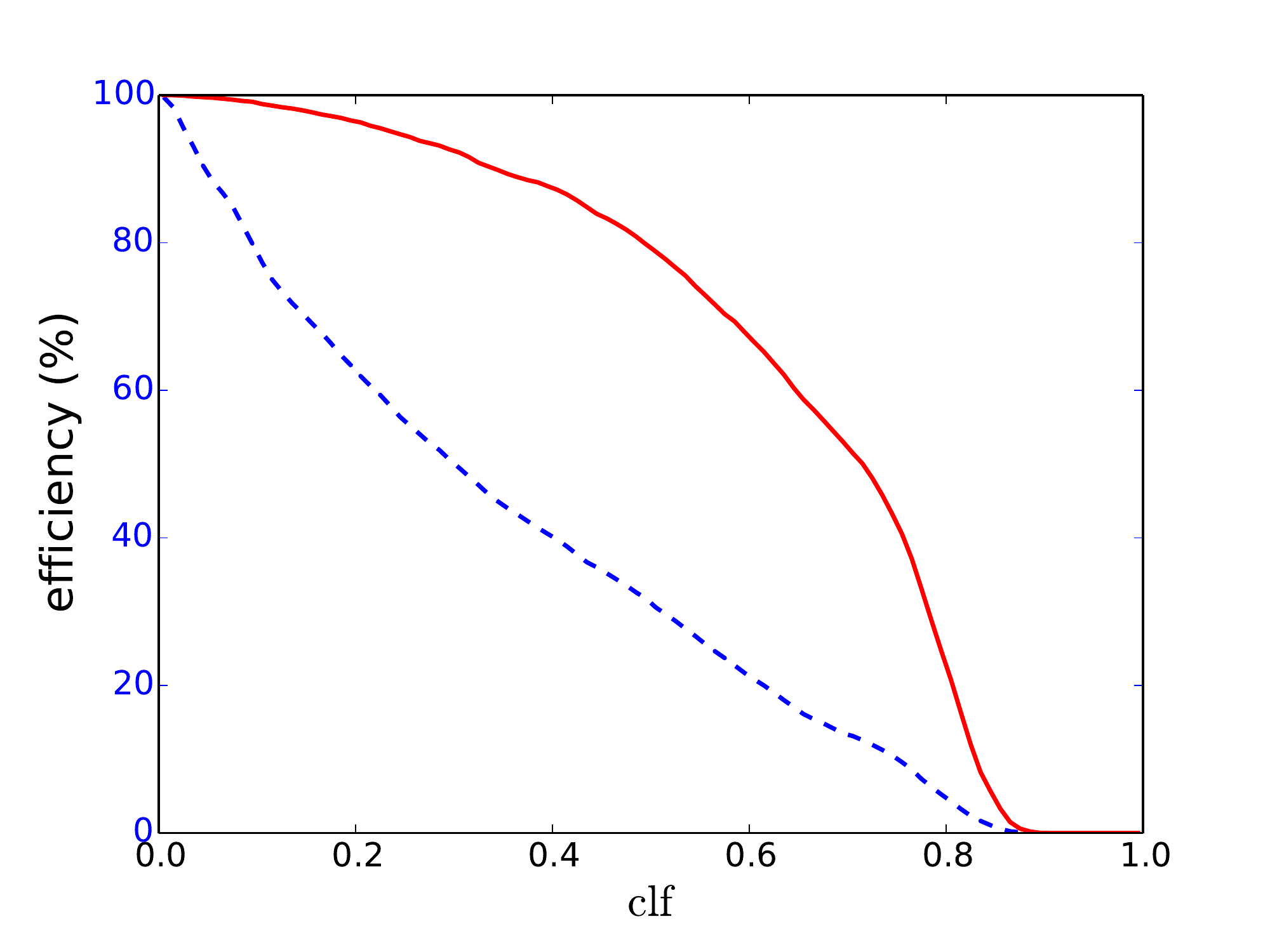}
    \label{subfig:Eff:qqbar:electron}
  }
  \subfigure[]{
    \includegraphics[width=.48\textwidth]{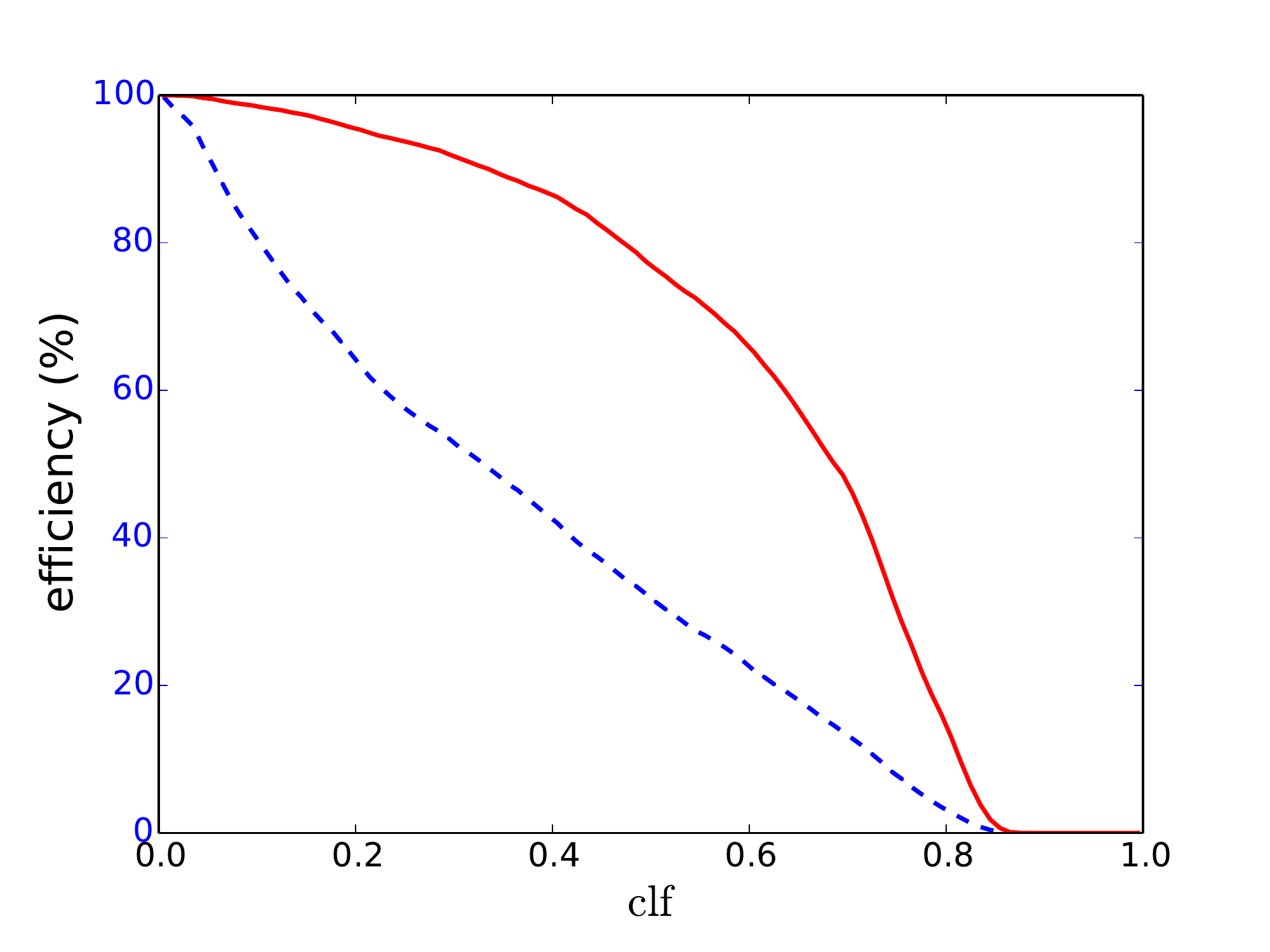}
    \label{subfig:Eff:qqbar:muon}
  }
  \caption{Signal (red, solid line) and background (blue, dashed line) efficiencies in dependence of the cut on the random forest classifier: \subref{subfig:clf:qqbar:electron} $\Bm \ra \LCp \antiproton \en \nueb$, \subref{subfig:clf:qqbar:muon} $\Bm \ra \LCp \antiproton \mun \numb$.}
  \label{fig:Eff:qqbar}
\end{figure}

For continuum background we decide to use the optimal set of $N_{\rm trees}=380$ and $N_{\rm leaf} = 60$ for the electron and the muon channel.
The classifier distribution for signal and background after training is shown in Fig. \ref{fig:clf:qqbar}.
Maximizing $S/\sqrt{S+B}$, with the signal yield $S$ scaled to an expected branching fraction of $10^{-4}$ and the background yield $B$ (including \BBbar background) scaled to data luminosity, returns an optimal cut of $>0.4$ for the electron case, preserving $88\%$ of the signal, while reducing continuum background to $41\%$. For the muon channel we obtain an optimal cut of $>0.4$ as well, preserving $87\%$ of the signal and reducing background to $42\%$. 
A comparison of signal and background efficiency in dependence of the classifier cut is shown in Fig. \ref{fig:Eff:qqbar}.

\clearpage
\boldmath
\section{\B background suppression}
\unboldmath

\begin{figure}[h]
  \begin{center}
    \includegraphics[width=.7\textwidth]{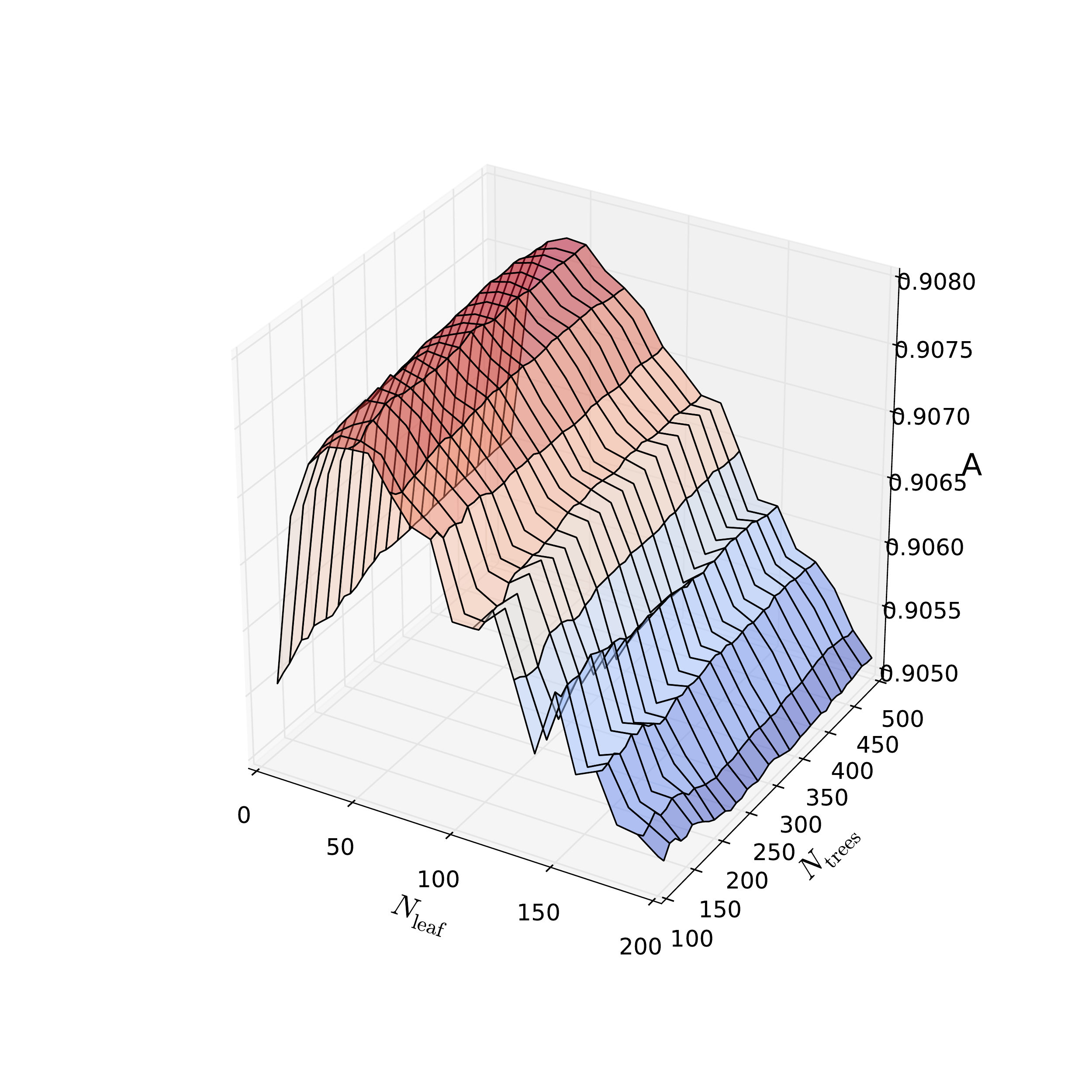}
  \end{center}
  \caption{Area below the ROC curve in dependence of $N_{\rm trees}$ and $N_{\rm leaf}$.}
  \label{fig:BBbar:optRF}
\end{figure}

\begin{figure}[h]
  \subfigure[]{
    \includegraphics[width=.48\textwidth]{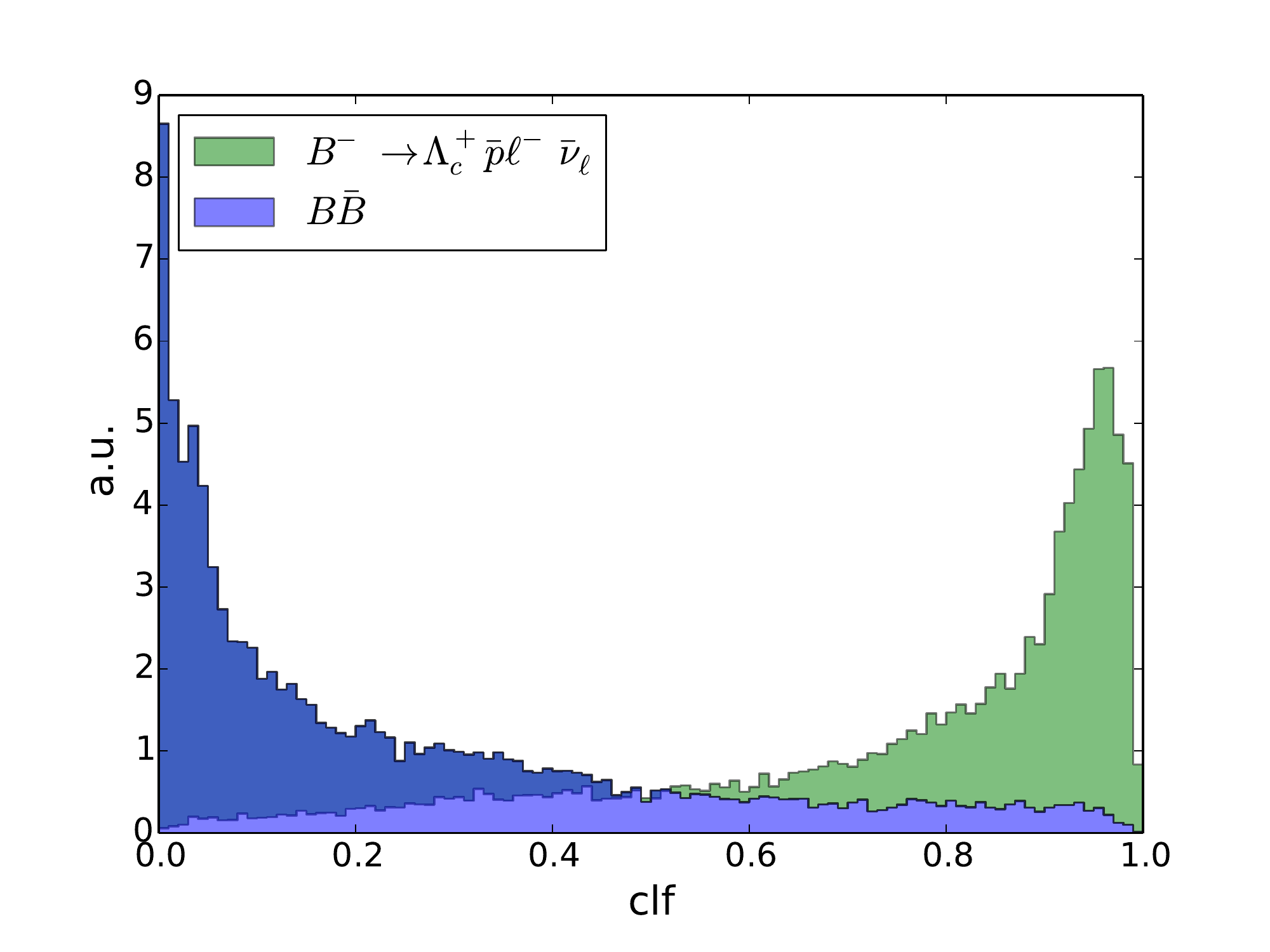}
    \label{subfig:clf:BBbar:electron}
  }
  \subfigure[]{
    \includegraphics[width=.48\textwidth]{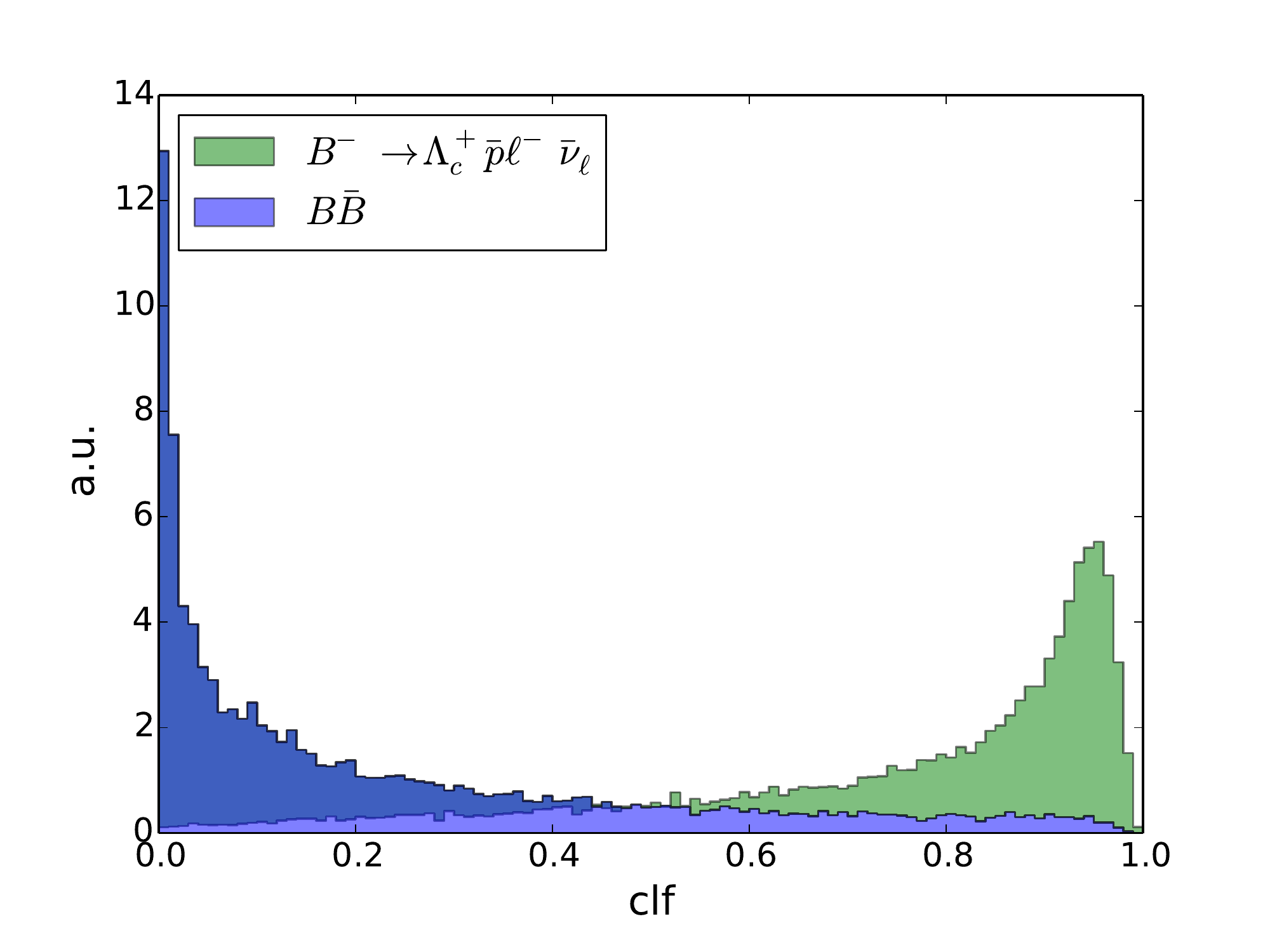}
    \label{subfig:clf:BBbar:muon}
  }
  \caption{The classifier distributions for signal and background after training of the \BBbar random forest: \subref{subfig:clf:BBbar:electron} $\Bm \ra \LCp \antiproton \en \nueb$, \subref{subfig:clf:BBbar:muon} $\Bm \ra \LCp \antiproton \mun \numb$.}
  \label{fig:clf:BBbar}
\end{figure}

In the next step after continuum background reduction we have to reduce generic \BBbar background as well. Therefore, we train an additional random forest, based on four input variables:
\begin{itemize}
  \item The lepton momentum in the CM system,
  \item the angle between the lepton momentum and the momentum of the $\LCp \antiproton$ system in the CM frame,
  \item the vertex probability for the $Y$ candidate, and
  \item the transverse neutrino momentum in the CM system.
\end{itemize}
A comparison of WI signal Monte Carlo and generic \B Monte Carlo data is shown in Fig. \ref{fig:e:BBbar_RF_vars} for the electron and in the appendix (Fig. \ref{fig:mu:BBbar_RF_vars}) for muons. For the vertex probability signal as well as background contain events with a vertex probability of zero. But, the fraction of these candidates in the background is much larger. As shown in the inset the number of events with a probability larger $0.1$ stays nearly constant for the signal simulation, while it converges to zero for background events.
\begin{figure}[h]
  \subfigure[]{
    \includegraphics[width=.48\textwidth]{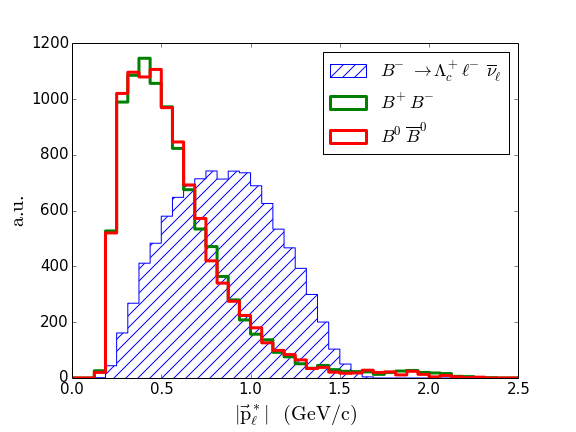}
    \label{fig:e:BBbar_RF_vars:p_lT}
  }
  \subfigure[]{
    \includegraphics[width=.48\textwidth]{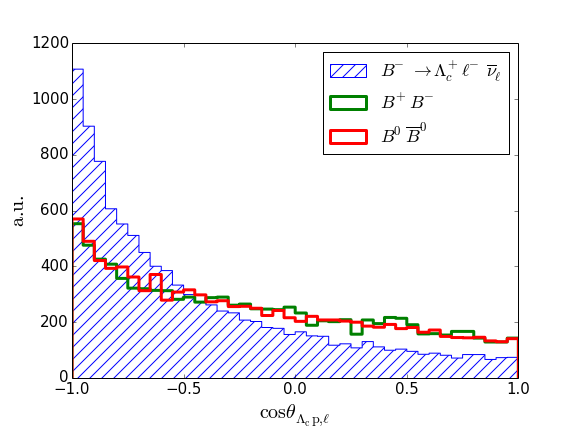}
    \label{fig:e:BBbar_RF_vars:cosLcpl}
  }
  \subfigure[]{
    \includegraphics[width=.48\textwidth]{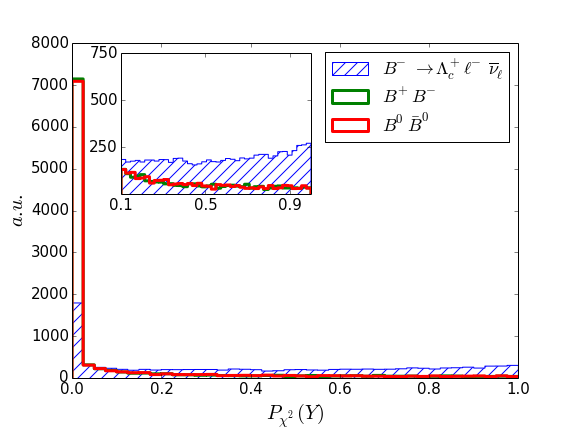}
    \label{fig:e:BBbar_RF_vars:PY}
  }
  \subfigure[]{                                
    \includegraphics[width=.48\textwidth]{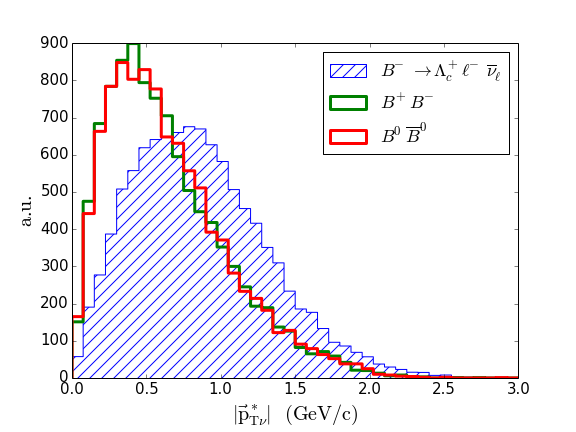}    
    \label{fig:e:BBbar_RF_vars:pNu}
  }
  \caption{Comparison of WI signal simulation events and generic \B decays: \subref{fig:e:BBbar_RF_vars:p_lT} $|\vec{p}_{\ell}^{*}|$, \subref{fig:e:BBbar_RF_vars:cosLcpl} $\cos\theta_{\Lambda_c p, \ell}$, \subref{fig:e:BBbar_RF_vars:PY} $P_{\chi^2}(Y)$, and \subref{fig:e:BBbar_RF_vars:pNu} $p_{T\nu}$.}
  \label{fig:e:BBbar_RF_vars}
\end{figure}
As for the $\qqbar$ background we optimize the \BBbar random forest as well. For performance reasons we decide to vary the number of trees between $100$ and $500$ and the number of events per leaf between $10$ and $200$. The resulting distribution of the area below the ROC curve can be seen in Fig. \ref{fig:BBbar:optRF}. The optimal RF performance is achieved with $450$ trees and a minimum number of $40$ events per leaf. As for the \qqbar RF the variation of the area below the ROC curve is rather small, for $N_{\rm trees}>>100$.
The classifier distributions for signal and background are shown in Fig. \ref{fig:clf:BBbar}.

\boldmath
\subsection{\BBbar Classifier as fit variable}
\unboldmath

\begin{figure}[h]
  \begin{center}
    \includegraphics[width=.48\textwidth]{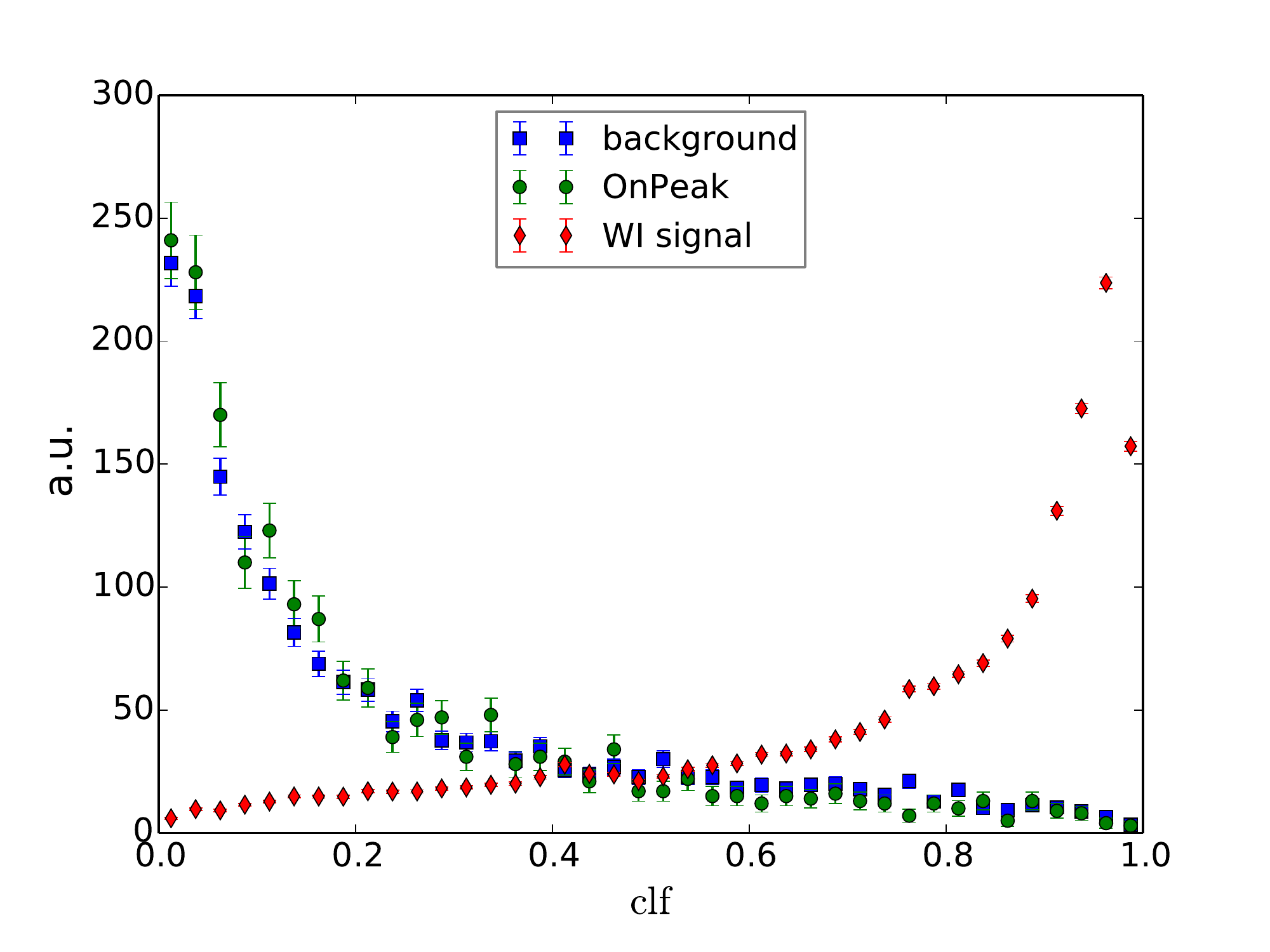}
  \end{center}
  \caption{Comparison of background Monte Carlo (blue squares), \texttt{OnPeak} data (green dots) and WI signal Monte Carlo (red diamonds) for the \BBbar Random Forest classifier.}
  \label{fig:target:classifier}
\end{figure}

\begin{figure}[h]
  \subfigure[]{
    \includegraphics[width=.48\textwidth]{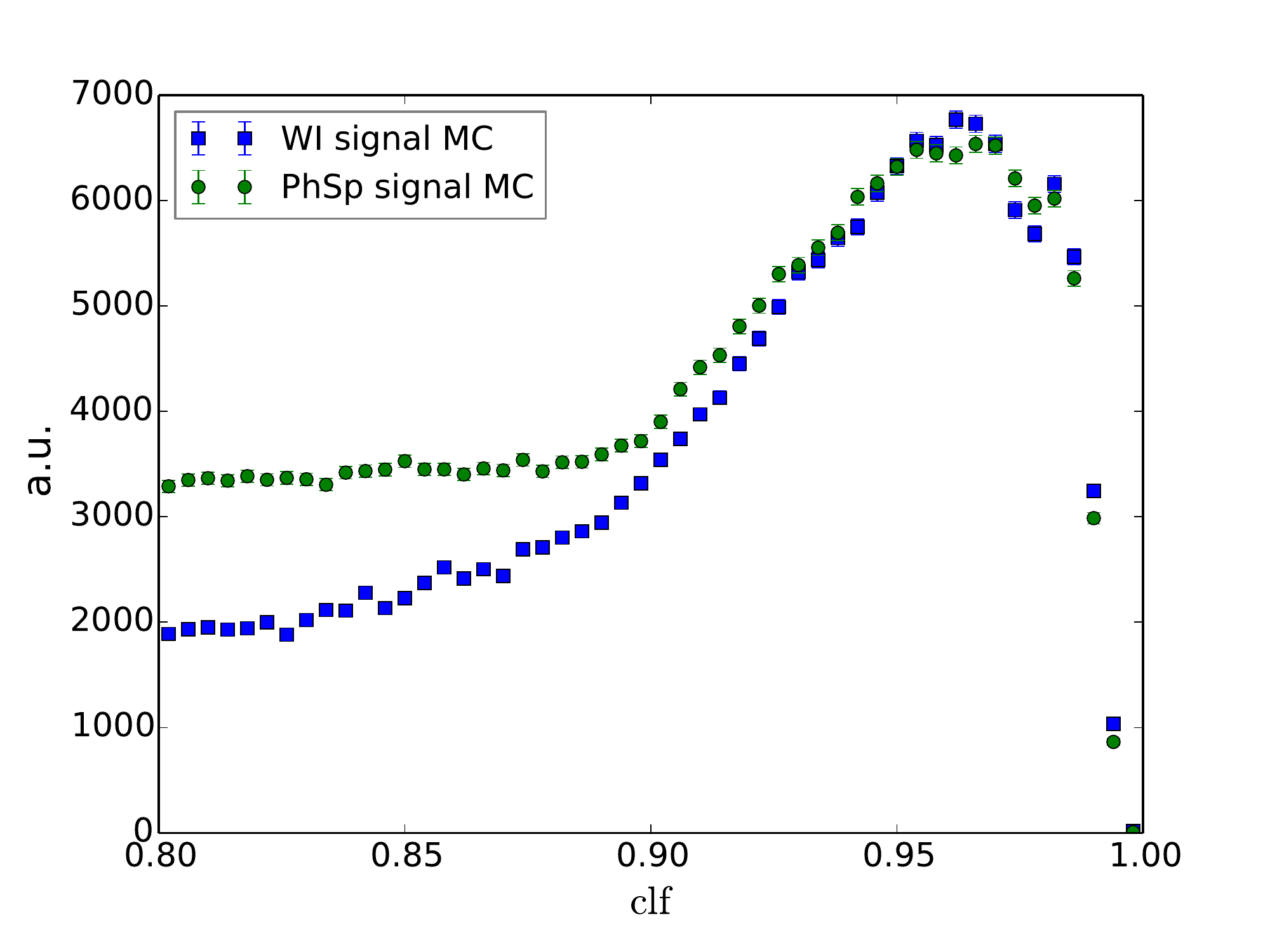}
    \label{fig:clf:Closeup:electron}
  }
  \subfigure[]{
    \includegraphics[width=.48\textwidth]{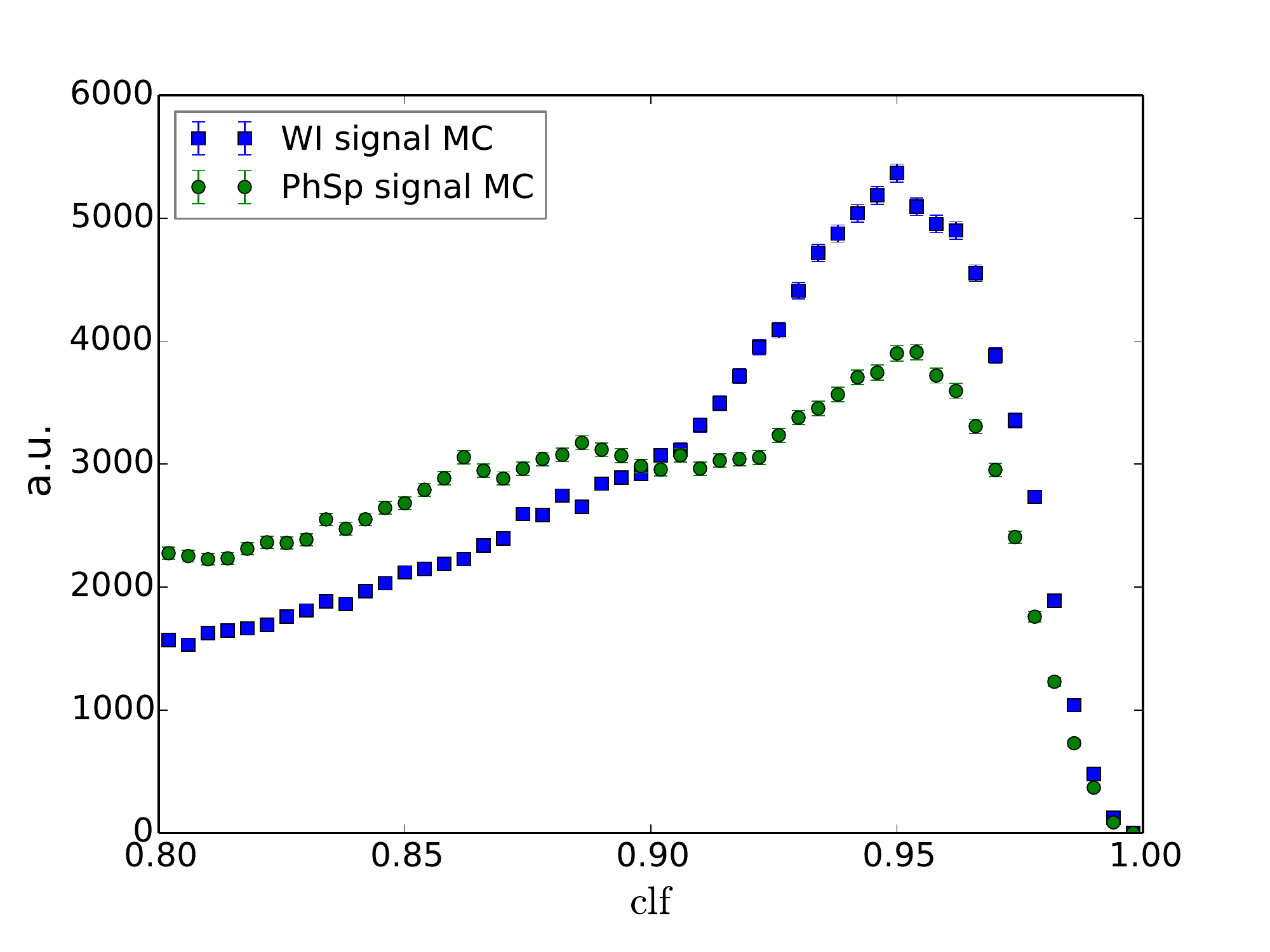}
    \label{fig:clf:Closeup:muon}
  }
  \caption{The \BBbar classifier distribution for the WI electron \subref{fig:clf:Closeup:electron} and muon \subref{fig:clf:Closeup:muon} signal Monte Carlo. The blue squares show the WI model, and the green dots the phase space model. Shown is the range from $0.8$ to $1.0$.}
  \label{fig:target:clf_CloseUp}
\end{figure}

A possible fit variable is the Random Forest output for the \BBbar background itself. This variable offers by construction a good separation of signal and background, as can be seen in Fig. \ref{fig:target:classifier}. In addition the agreement between the background Monte Carlo and the data distribution within the uncertainties is reasonable for low values, although it seems to underestimate background a little. For large values the background Monte Carlo simulation overestimates data background.
Another problem arises when it comes to find a functional description of the signal distribution. A closer look at large classifier ($clf$) values, as shown in Fig. \ref{fig:target:clf_CloseUp} for the electron channel, shows a shoulder on the right hand side of the signal peak, as well as some structure on the left side. This would require a rather complex fit function, which strongly depends on the specific decay model. Thus, we decide not to use the classifier as a fit variable, but rather select candidates above a certain threshold in the classifier.

\clearpage
\chapter{Analysis}

\begin{figure}[h]
  \subfigure[]{
    \includegraphics[width=.48\textwidth]{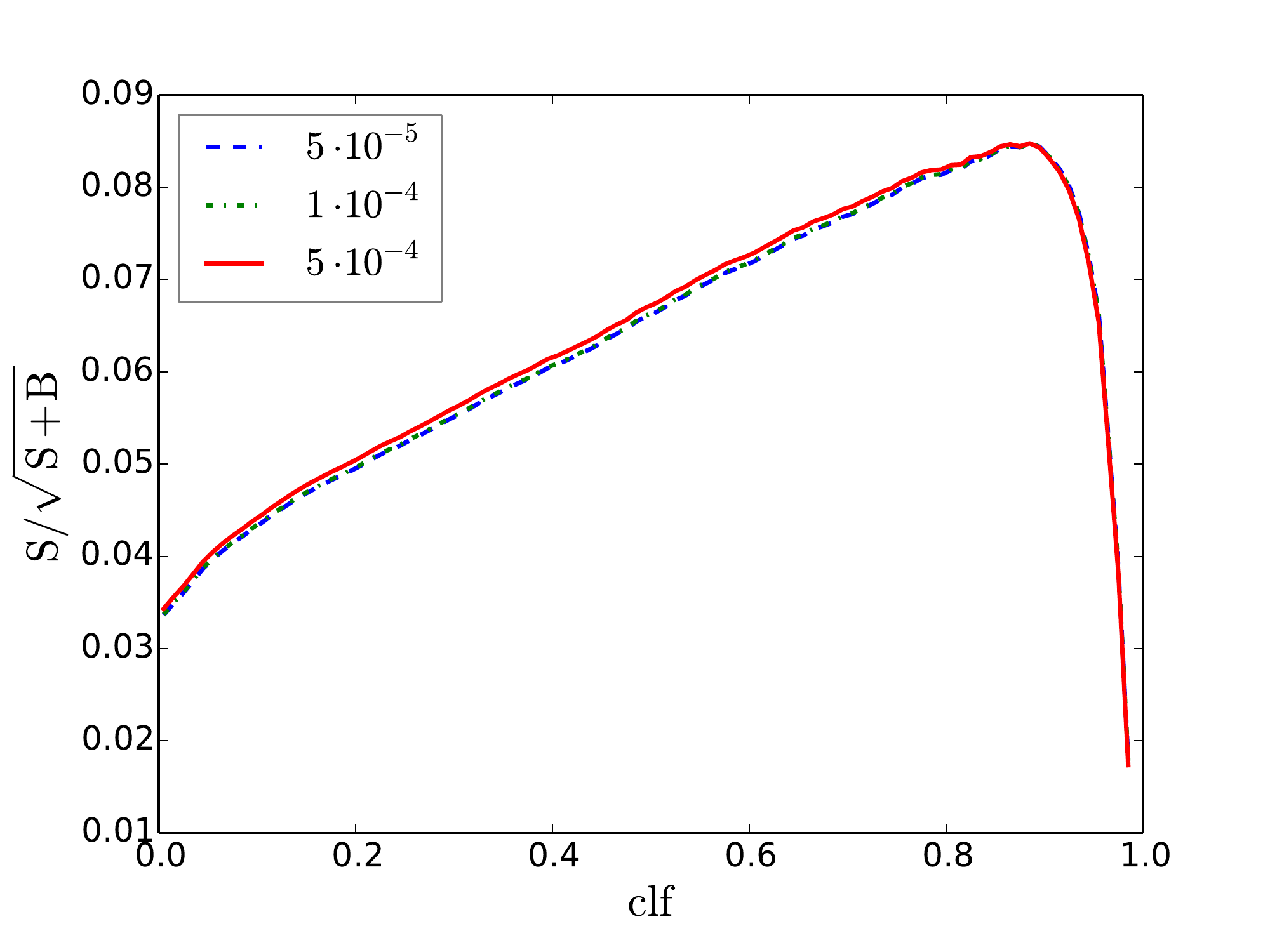}
    \label{fig:SoverSplusB:electron}
  }
  \subfigure[]{
    \includegraphics[width=.48\textwidth]{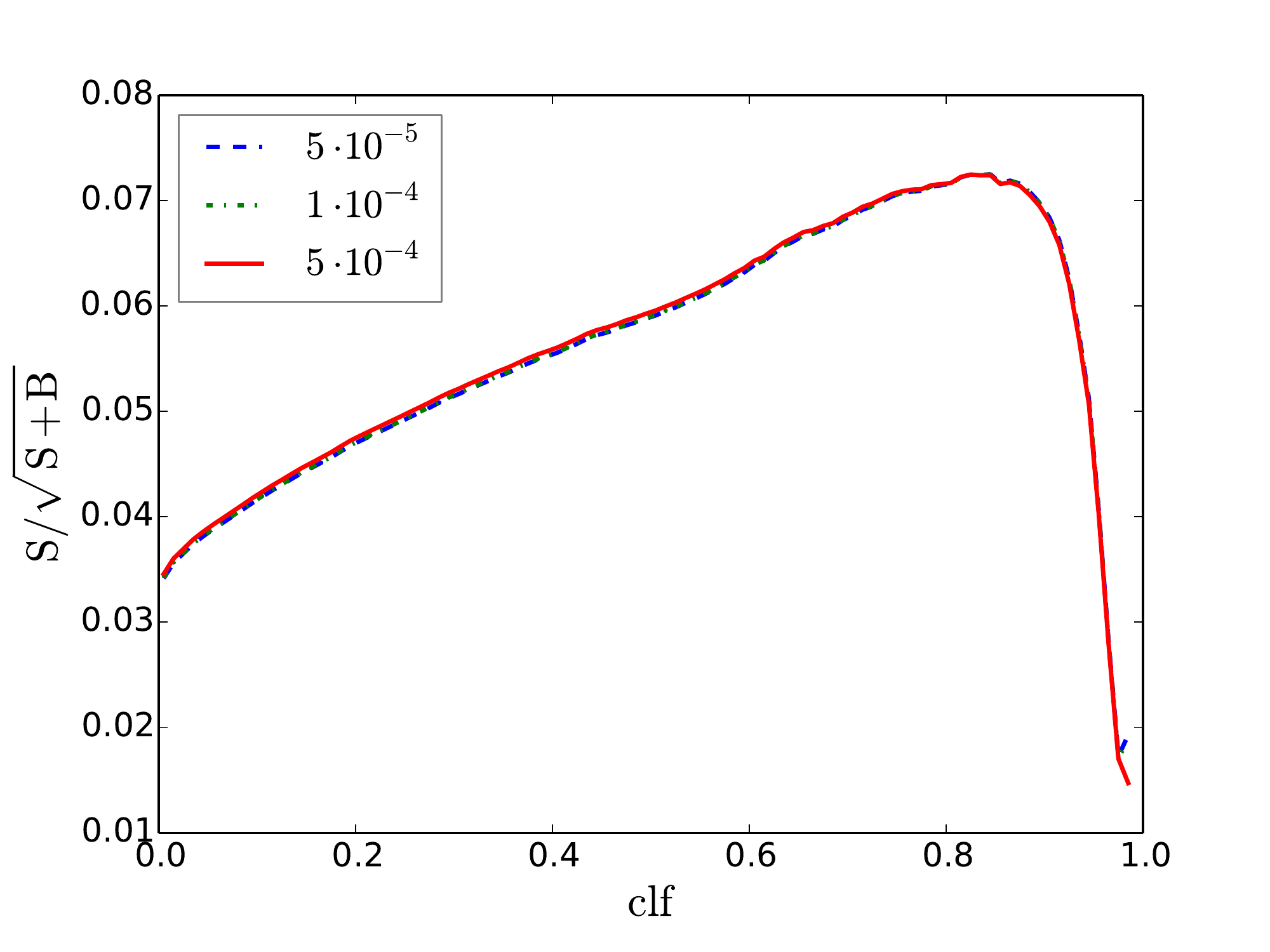}
    \label{fig:SoverSplusB:muon}
  }
  \caption{$S/\sqrt{S+B}$ versus the \BBbar classifier for the electron \subref{fig:SoverSplusB:electron} and muon \subref{fig:SoverSplusB:muon} channel.}
  \label{fig:SoverSplusB}
\end{figure}

For the further analysis of the decay we select events above a certain threshold in the \BBbar classifier. The optimal value of this threshold is determined by maximizing the statistical significance $S/\sqrt{S+B}$, where $S$ is the expected signal yield, and $B$ the background yield. Here, we use all events that passed the previously described selection criteria. For the background yield we scale the \qqbar and \BBbar Monte Carlo data sets to \texttt{OnPeak} luminosity. For the signal yield we vary the expected branching fraction between $0.5 \times 10^{-5}$ and $5 \times 10^{-4}$. Fig. \ref{fig:SoverSplusB} shows $S/\sqrt{S+B}$ versus the classifier cut for both signal channels. The maximal value for the electron channel is achieved for  $clf > 0.9$, reducing background to $2\%$ while preserving $36\%$ of signal. The maximum for the muon channel is at $>0.85$, reducing background to $3\%$ while preserving $42\%$ of signal. Both optimal cuts show no significant variation with the expected signal yield.

\begin{figure}[h]
  \subfigure[]{
    \includegraphics[width=.48\textwidth]{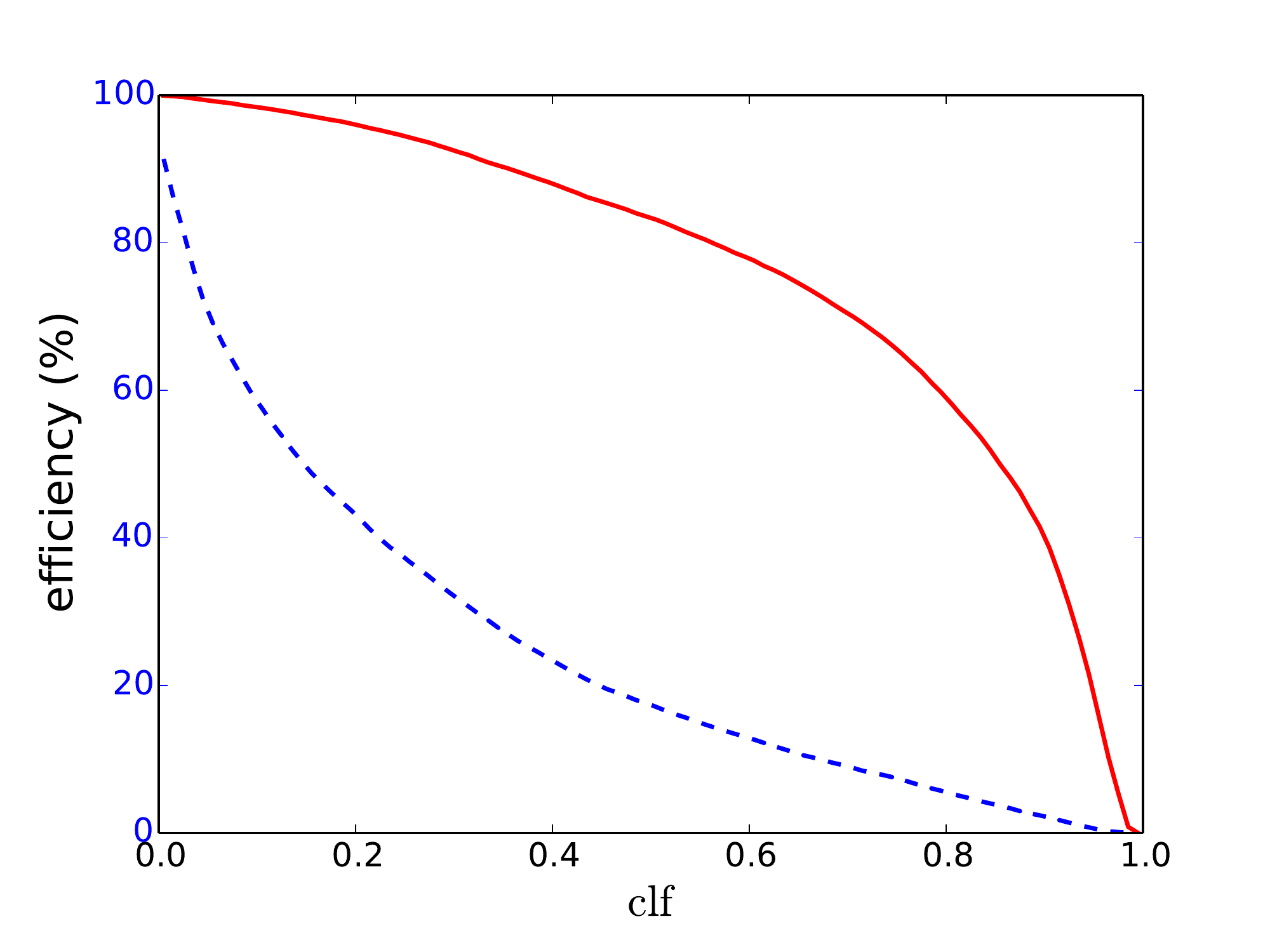}
    \label{subfig:Eff:BBbar:electron}
  }
  \subfigure[]{
    \includegraphics[width=.48\textwidth]{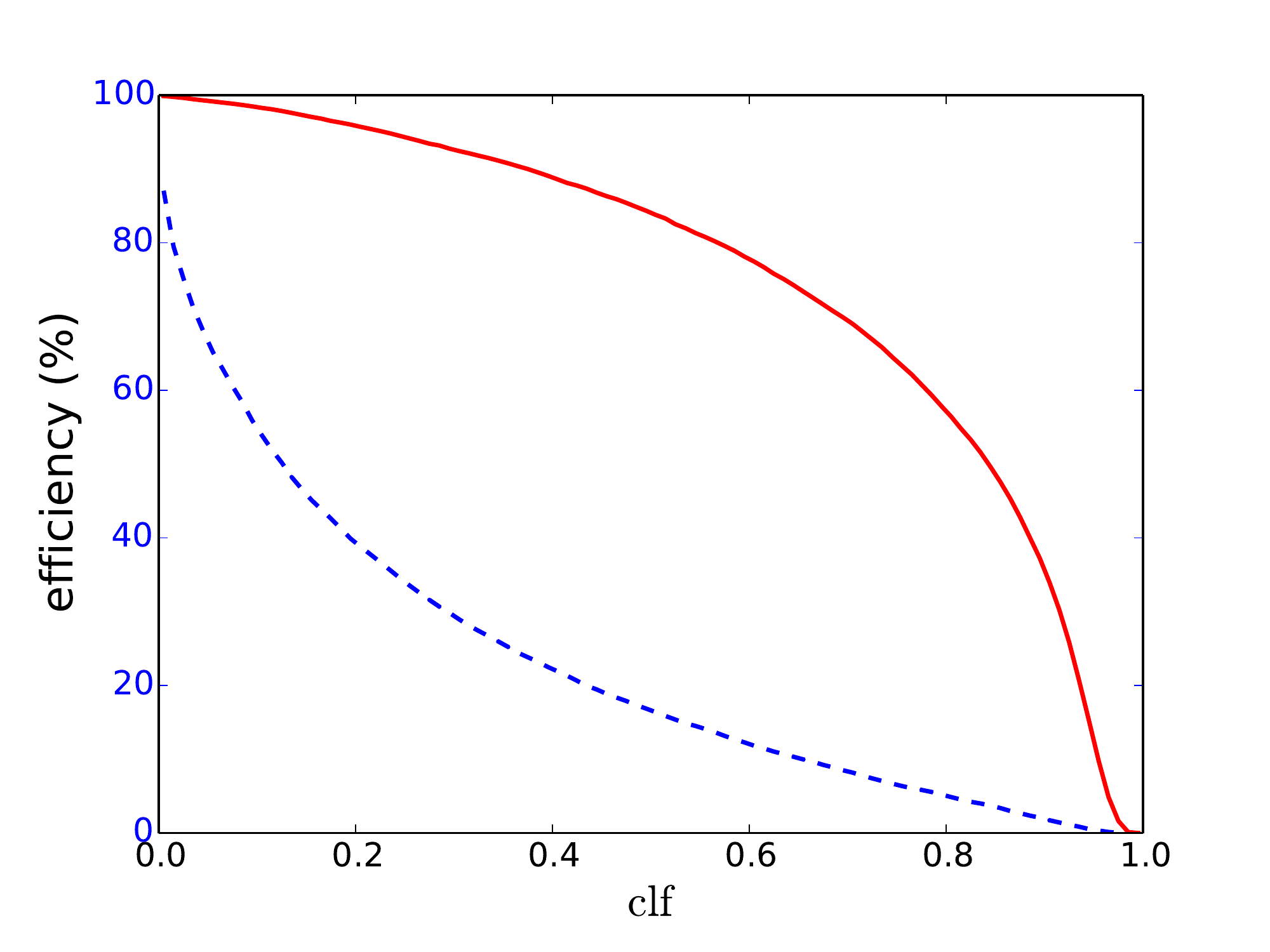}
    \label{subfig:Eff:BBbar:muon}
  }
  \caption{Signal (red, solid line) and background (blue, dashed line) efficiencies in dependence of the cut on the random forest classifier. \subref{subfig:clf:BBbar:electron} $\Bm \ra \LCp \antiproton \en \nueb$, \subref{subfig:clf:BBbar:muon} $\Bm \ra \LCp \antiproton \mun \numb$.}
  \label{fig:Eff:BBbar}
\end{figure}

Figure \ref{fig:Eff:BBbar} shows a comparison of background and signal efficiency in dependence of the classifier cut for both signal channels.

\section{Fit variable}

For signal extraction we have two possible sets of variables to determine the signal yield, first the classic neutrino variables
\begin{itemize}
  \item $m^2_{\rm miss} = E^2_{\rm miss} - \vec{p}^2_{\rm miss}$
  \item $m^2_{\rm miss}/2E_{\rm miss}$
  \item $m_{\nu}^2 = (E_{\rm beam}^* - E_Y^*)^2-\vec{p}^{*2}_Y$
\end{itemize}
where $E_{\rm beam}$, $E_Y^*$ and $\vec{p}_Y^*$ are measured in the CM system. For $m_{\nu}^2$ we neglect the small momentum of the \B meson in the CM system, in consequence the neutrino momentum and the momentum of the Y system have the same magnitude. The second set consists of
\begin{itemize}
  \item \DeltaE
  \item \mes,
\end{itemize}
as defined in eq. \eqref{eq:DeltaE} and \eqref{eq:mES}.

Crucial for both variable sets is a good description of the background shape by background Monte Carlo, and different shapes for background and signal.
The next sections show a comparison of background Monte Carlo with \texttt{OnPeak} data and with WI signal Monte Carlo to decide on the variables for signal extraction. If the background description from generic Monte Carlo is capable of describing data background we expect a good agreement between background Monte Carlo events and \texttt{OnPeak} data, with a possible variation due to a signal component in data.

\subsection{Neutrino Variables}\label{sect:nuVars}

All three neutrino variables work well for low-mass final states like $\B \ra \pi \ell \nu$, but might have problems with high-mass final states, as it can be seen in the analysis of $\Bm \ra D_s^+ \Km \ellm \nulb$ \cite{BAD2183}. In this analysis the signal yield was extracted in $m^2_{\rm miss}$, but in contrast to decays like $\B \ra \pi \ell \nu$ background in this variable is not flat. In consequence, the signal peak had to be fitted onto a steep slope. The situation might be even more difficult in the present analysis, where the final state is about $0.8\gevcc$ heavier.

\begin{figure}[h]
  \begin{center}
  \subfigure[]{
    \includegraphics[width=.42\textwidth]{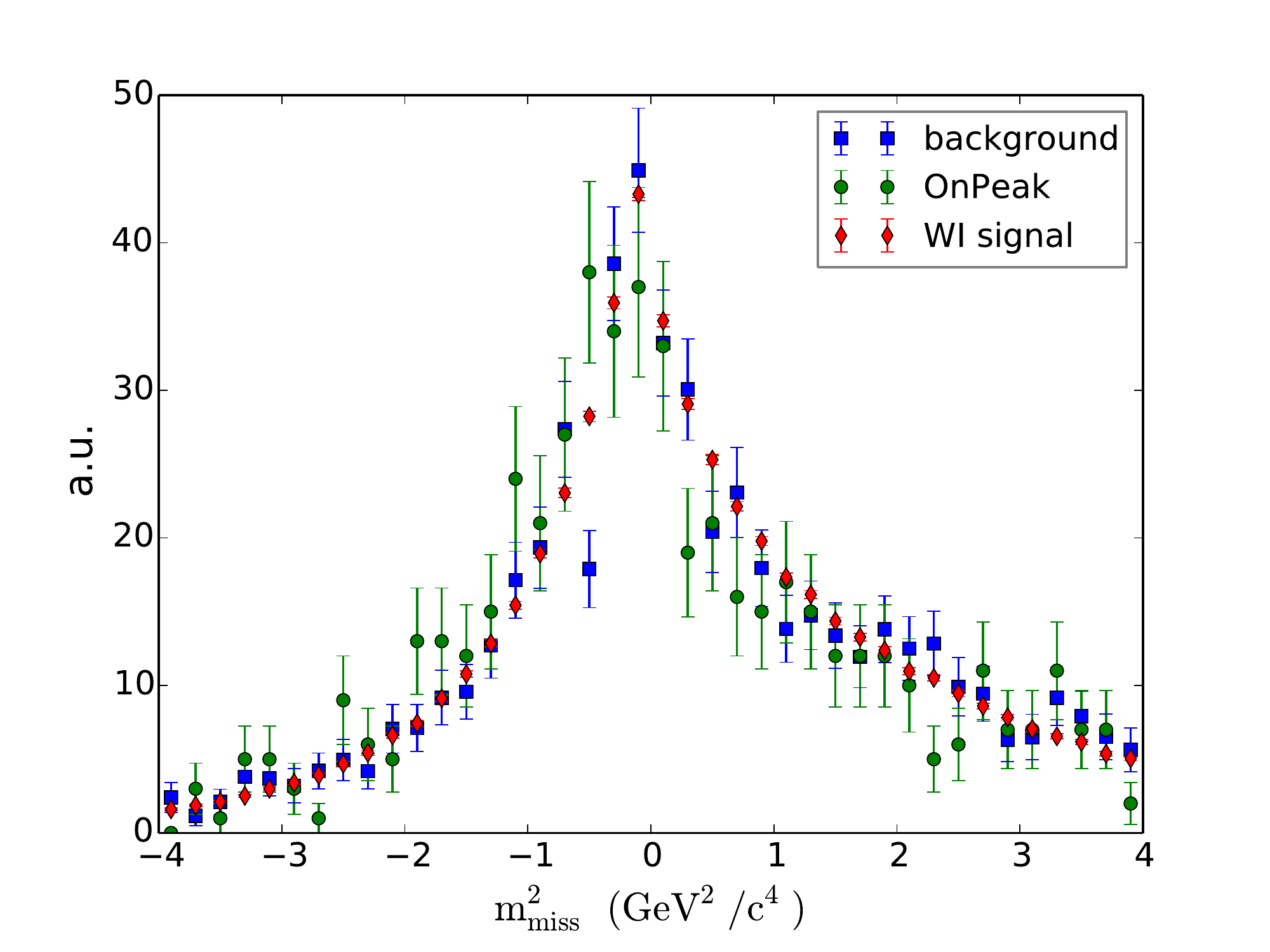}
    \label{subfig:target:NeutrinomM2}
  }
  \subfigure[]{
    \includegraphics[width=.42\textwidth]{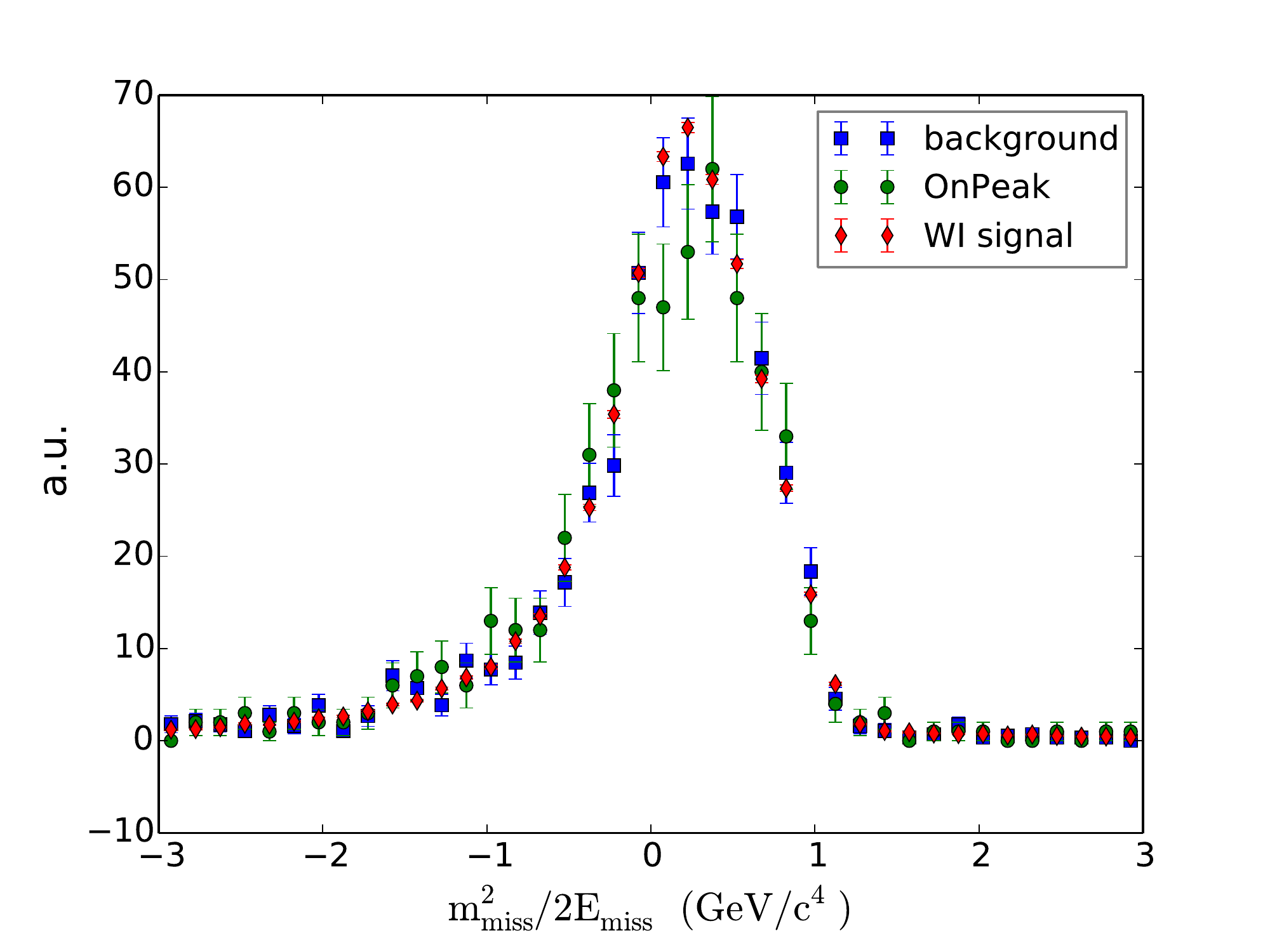}
    \label{subfig:target:NeutrinomM2o2mE}
  }
  \subfigure[]{
    \includegraphics[width=.42\textwidth]{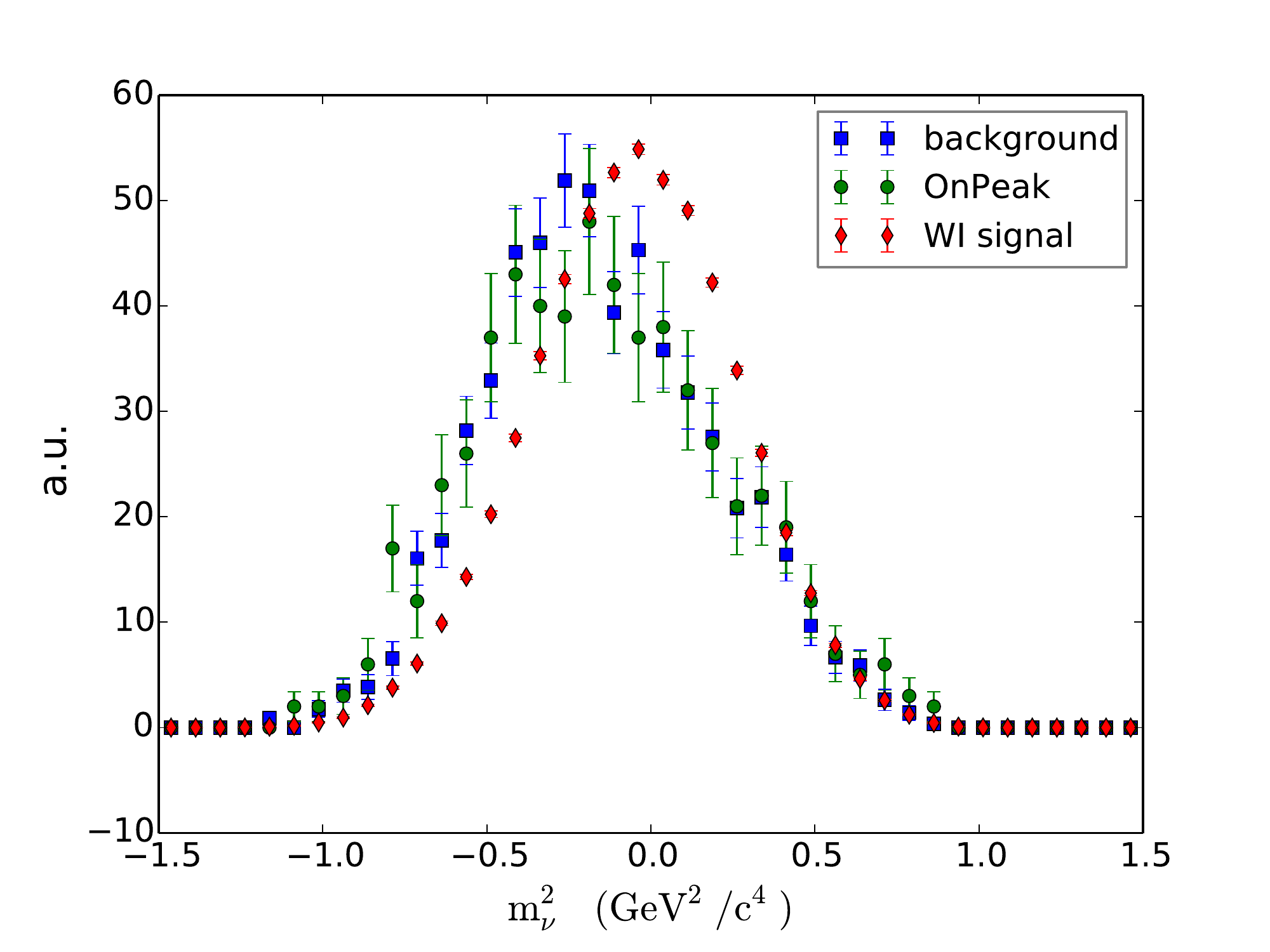}
    \label{subfig:target:NeutrinoMass}
  }
  \end{center}
  \caption{Comparison of background Monte Carlo (blue squares), OnPeak data (green dots) and WI signal Monte Carlo (red diamonds) for the three neutrino variables \subref{subfig:target:NeutrinomM2} $m^2_{\rm miss}$, \subref{subfig:target:NeutrinomM2o2mE} $m^2_{\rm miss}/2E_{\rm miss}$, and \subref{subfig:target:NeutrinoMass} $m_{\nu}^2$.}
  \label{fig:targetVarsComp}
\end{figure}

A comparison between background Monte Carlo events, OnPeak data, and WI signal Monte Carlo data for these three variables is shown in Fig. \ref{fig:targetVarsComp}. The comparison of background and signal Monte Carlo events for $m^2_{\rm miss}$ and $m^2_{\rm miss}/2E_{\rm miss}$ shows that the distributions equal each other. This can be explained in terms of the large invariant mass of the $Y$ system, leaving only a small fraction of energy and momentum for the neutrino, in consequence a neutrino signature can be easily faked by a wrong particle identification in the neutrino reconstruction, or a missing photon.

In contrast $m_{\nu}^2$ is independent of the particle identification in the neutrino reconstruction and offers a small difference between background and signal. In addition the agreement between the OnPeak and the background Monte Carlo distribution is reasonably good, making $m_{\nu}^2$ a possible fit variable. The down-side of $m_{\nu}^2$ as fit variable is that a fit in this quantity requires to fit the signal on the right slope of the background distribution.

\subsection{\B Variables}

\begin{figure}[h]
  \begin{center}
  \subfigure[]{
    \includegraphics[width=.48\textwidth]{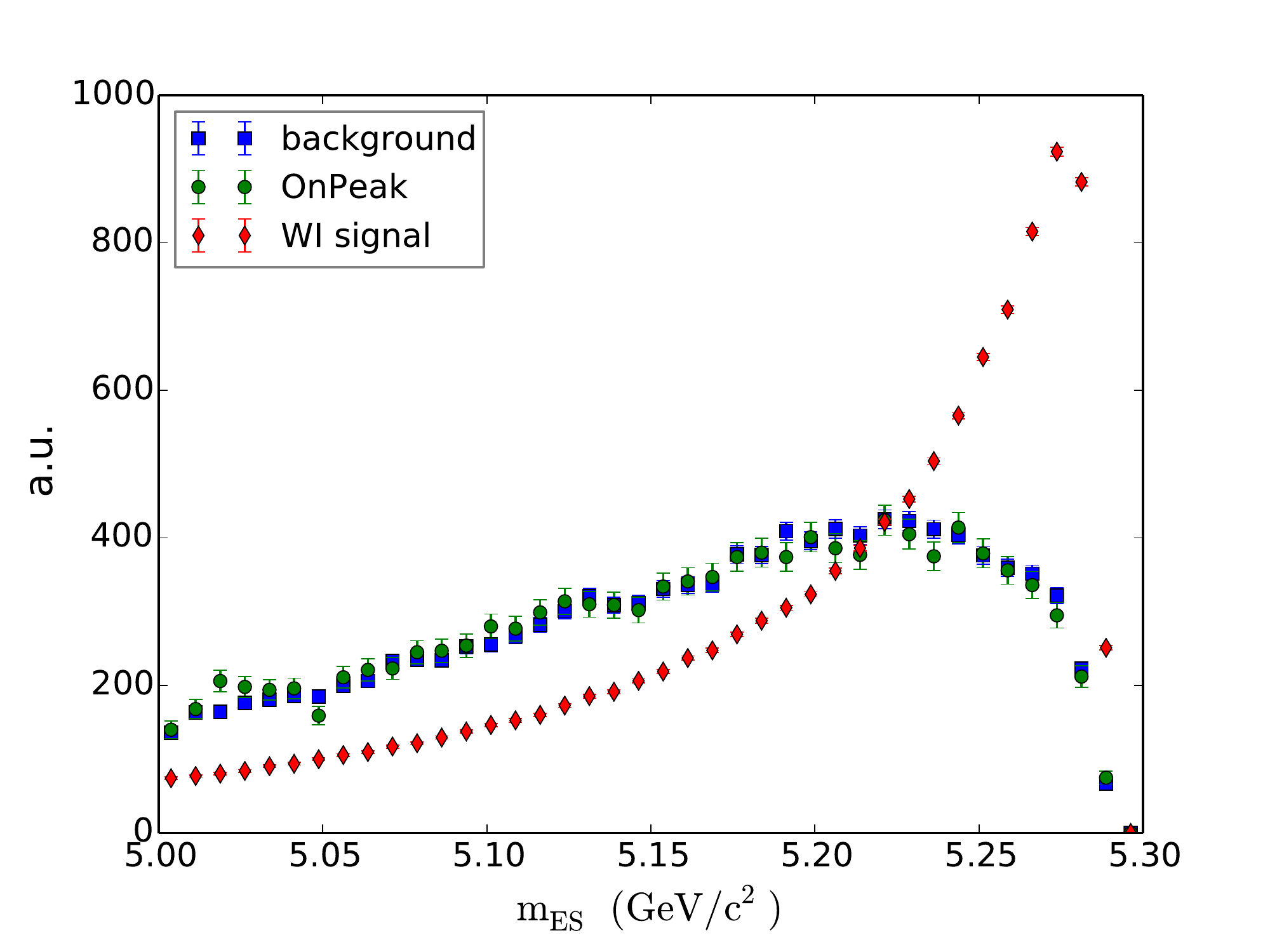}
    \label{subfig:target:mES}
  }
  \subfigure[]{
    \includegraphics[width=.48\textwidth]{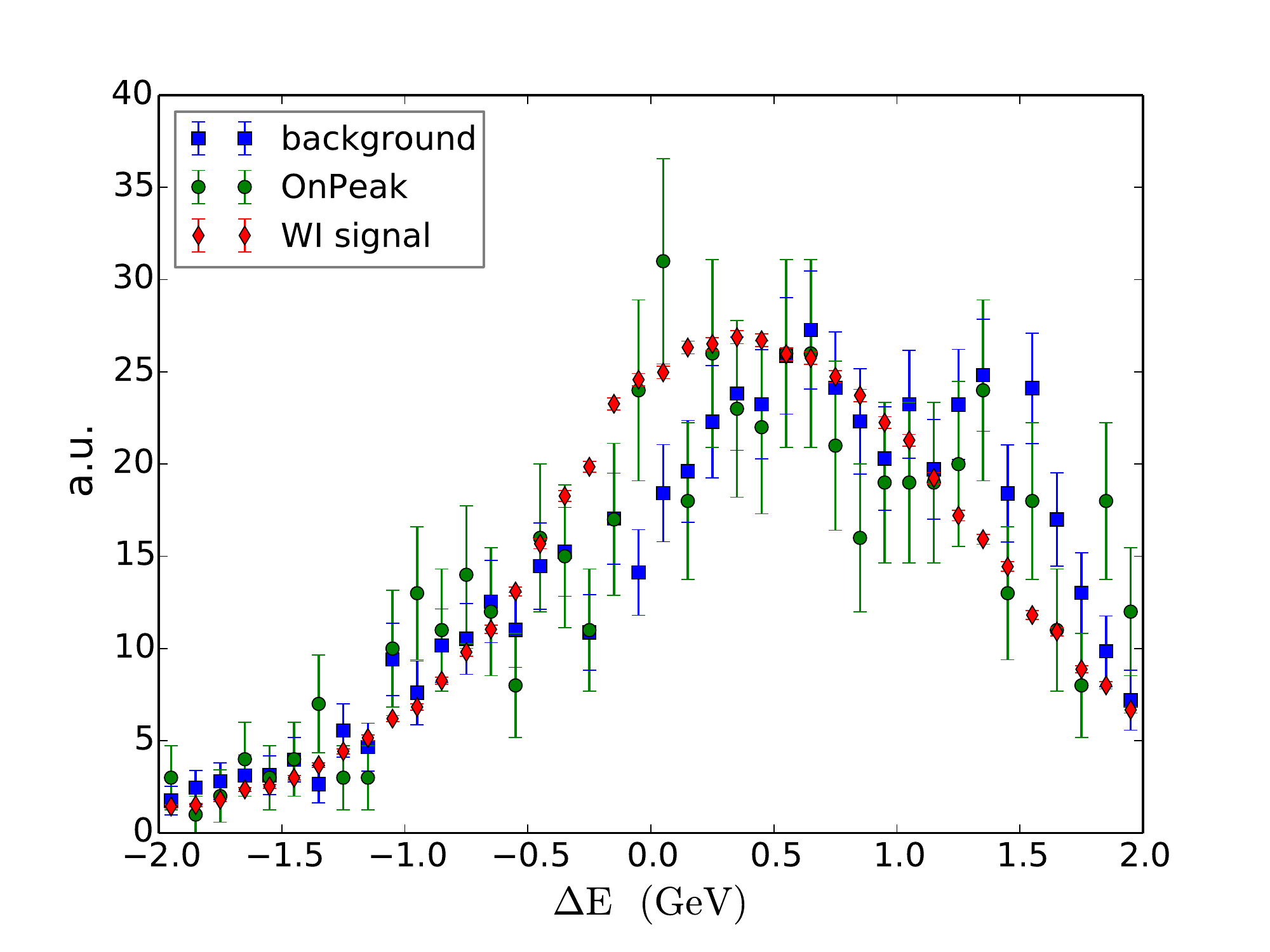}
    \label{subfig:target:DeltaE}
  }
  \end{center}
  \caption{Comparison of background Monte Carlo (blue squares), \texttt{OnPeak} data (green dots) and WI signal Monte Carlo (red diamonds) for \subref{subfig:target:mES} \mes, and \subref{subfig:target:DeltaE} \DeltaE.}
  \label{fig:targetBVarsComp}
\end{figure}

Two other possible target variables are \mes and \DeltaE for the reconstructed \B meson. Fig. \ref{fig:targetBVarsComp} shows a comparison between data and Monte Carlo for these two variables. Comparing the \mes distributions we see a good separation between background and signal, as well as a reasonable agreement between the background Monte Carlo and data distributions. This makes \mes a good choice as fit variable. In contrast the \DeltaE distribution shows a rather similar shape for background and signal Monte Carlo events.

\section{Signal extraction}

\begin{figure}[t]\centering
  \subfigure[]{
    \includegraphics[width=.42\textwidth]{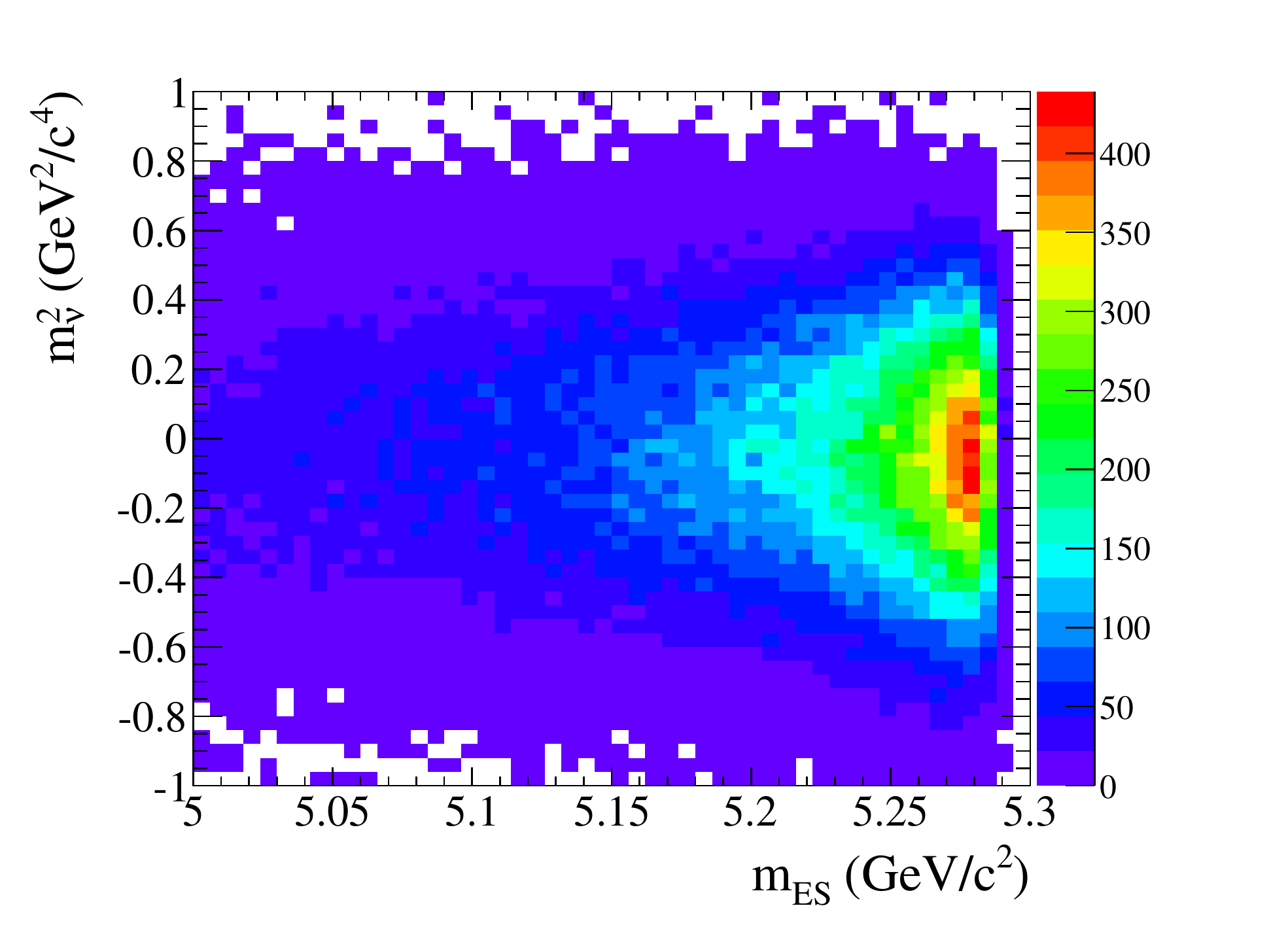}
    \label{fig:mESmNu2MC:sig:e}
  }
  \subfigure[]{
    \includegraphics[width=.42\textwidth]{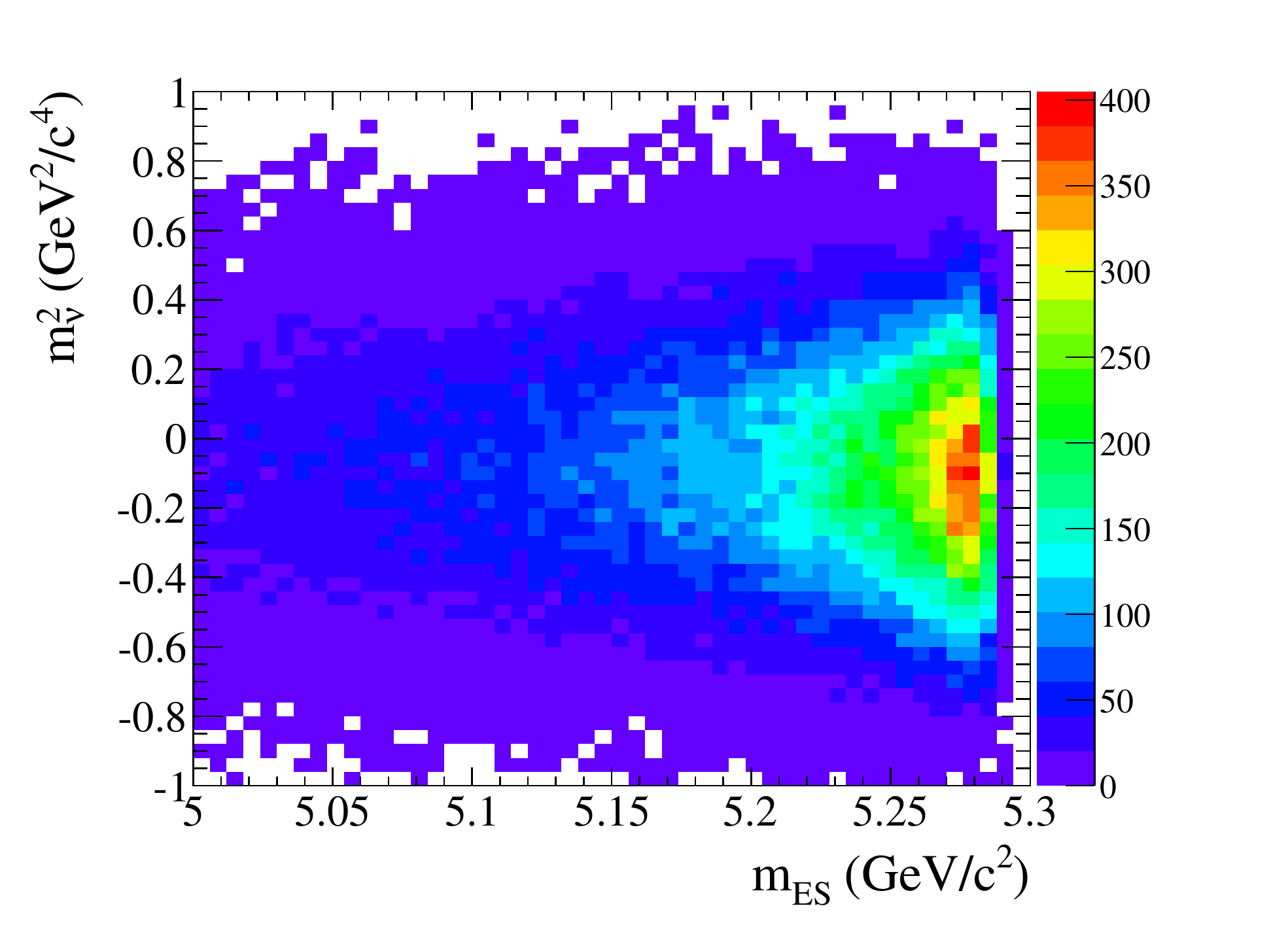}
    \label{fig:mESmNu2MC:sig:mu}
  }
  \subfigure[]{
    \includegraphics[width=.42\textwidth]{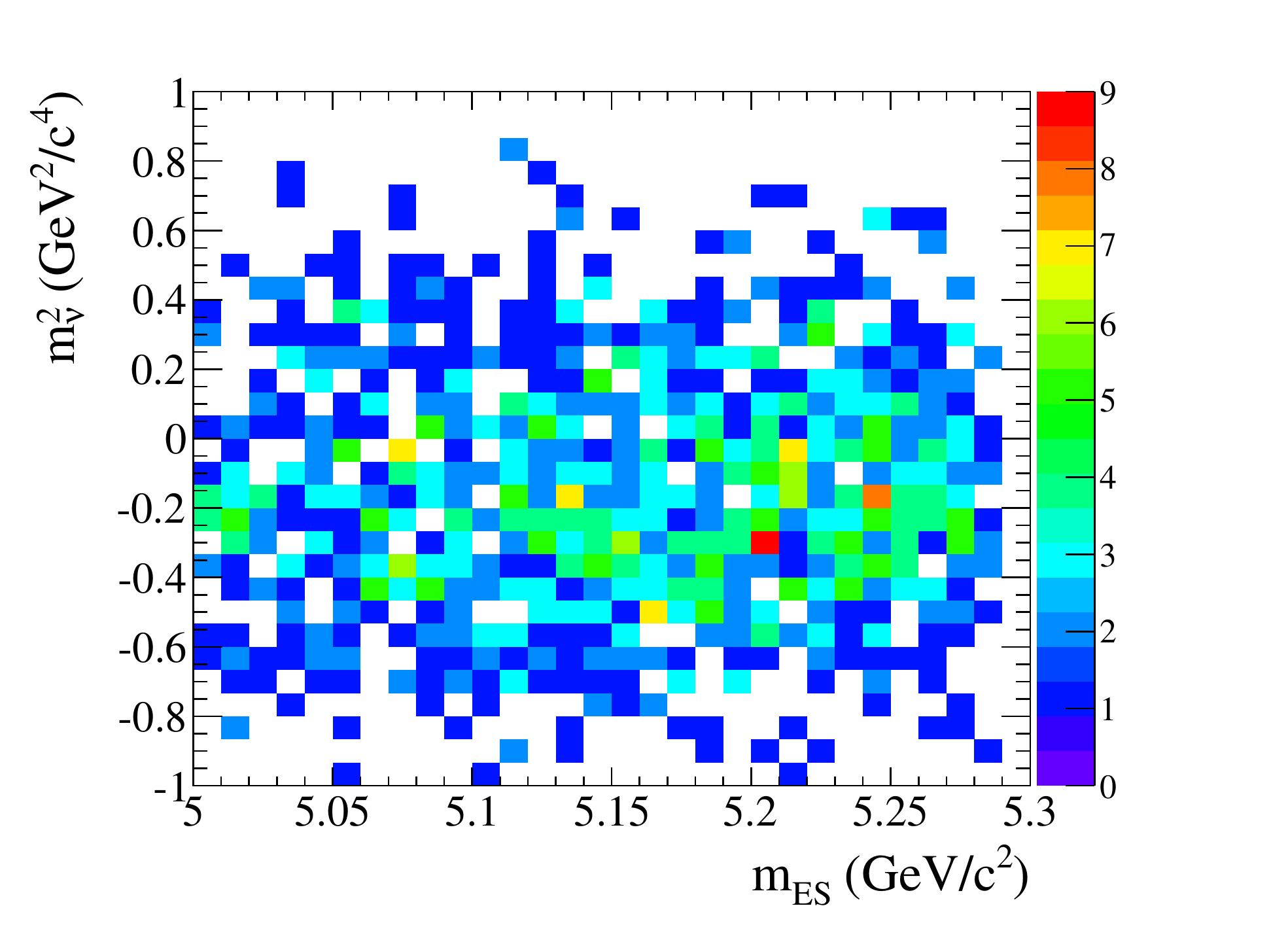}
    \label{fig:mESmNu2MC:bkg:e}
  }
  \subfigure[]{
    \includegraphics[width=.42\textwidth]{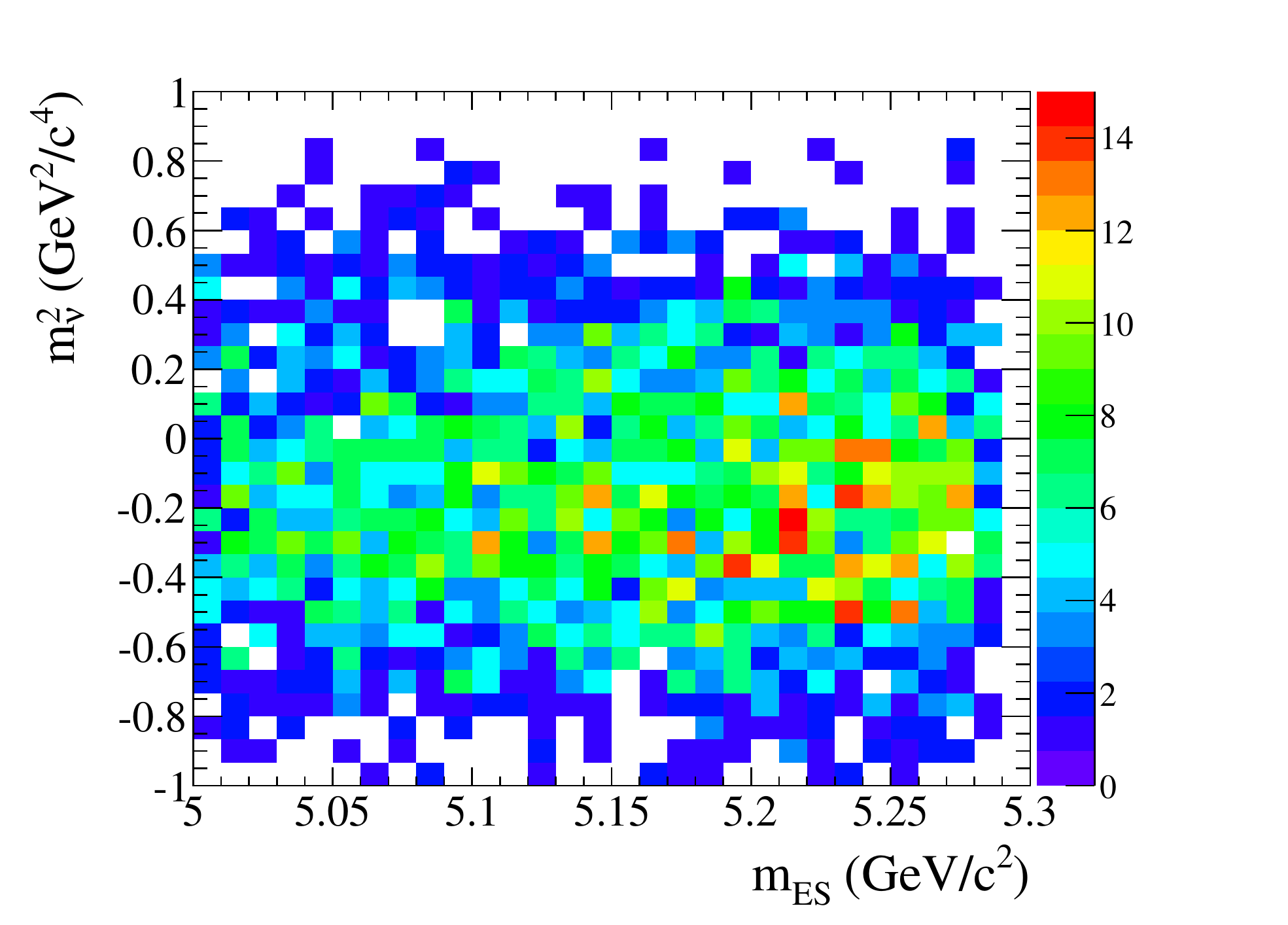}
    \label{fig:mESmNu2MC:bkg:mu}
  }
  \caption{$m_{\nu}^2:\mes$ plane for WI signal Monte Carlo \subref{fig:mESmNu2MC:sig:e},\subref{fig:mESmNu2MC:sig:mu} and generic background Monte Carlo data \subref{fig:mESmNu2MC:bkg:e}, \subref{fig:mESmNu2MC:bkg:mu}.  For the electron channel \subref{fig:mESmNu2MC:sig:e}, \subref{fig:mESmNu2MC:bkg:e}, and for the muon channel \subref{fig:mESmNu2MC:sig:mu}, \subref{fig:mESmNu2MC:bkg:mu}.}
  \label{fig:mESmNu2MC}
\end{figure}

For signal extraction we decide to perform a simultaneous maximum likelihood fit in \mes and $m_{\nu}^2$. Therefore, it is crucial to exclude a strong correlation between these two variables. Fig. \ref{fig:mESmNu2MC} shows no correlation between these two variables, neither for signal nor for background. 
To exclude the possibility that the shape in one variable varies in dependence of the second variable, we compare the projections in slices of the second variable as well. Fig. \ref{fig:mESmNu2Slices_signal} and \ref{fig:mESmNu2Slices_background} show these projections for WI signal and background Monte Carlo.
There seems to be no strong variation in shape parameters for neither signal, nor background. 
Thus, we can apply a factorization ansatz. In the following sections we determine the parameters of the background and signal probability density functions (pdf) by fitting Monte Carlo events.

\subsection{Signal parametrization}\label{sect:fit_procedure}

For signal description we decide to use for $m_{\nu}^2$ a Cruijff function $C(x; m_0, \sigma_L, \sigma_R, \alpha_L, \alpha_R)$ defined as
\begin{align}
  C(x; m_0, \sigma_L, \sigma_R, \alpha_L, \alpha_R) = 
  \begin{cases}
    \exp{\left(\frac{-(x-m_0)^2}{2\cdot\sigma_R^2+\alpha_R\cdot(x-m_0)^2}\right)} & \text{if } x-m_0>0 \\
    \\
    \exp{\left(\frac{-(x-m_0)^2}{2\cdot\sigma_L^2+\alpha_L\cdot(x-m_0)^2}\right)} & \text{if } x-m_0\leq 0,
  \end{cases}
\end{align}
and the sum of an ARGUS function ${\cal A}(x; c, p, m_0)$ and a Novosibirsk function ${\cal N}(x; x_0, \sigma, \Lambda)$ for \mes.
\begin{align}
  S(x; c, p, m_0, x_0, \sigma, \Lambda, f_1) &= f_1\cdot {\cal A}(x; c, p, ) + (1-f_1) \cdot {\cal N}(x; x_0, \sigma, \Lambda)
\end{align}
The ARGUS function ${\cal A}(x; c, p, m_0)$ is defined as
\begin{align}
  {\cal A}(x; c, p) &= x \cdot \left( 1-\left(\frac{x}{m_0}\right)^2\right)^p\cdot \exp{\left(c \cdot \left( 1-\left(\frac{x}{m_0}\right)^2\right)\right)} ,
\end{align}
where $m_0$ denotes the upper boundary which coincides with half the center of mass energy in \babar, and can thus be taken from data as a conditional observable. The parameters $p$ and $c$ determine the shape of the ARGUS function. In the definition \cite{ARGUSfunc} first used by the ARGUS collaboration $p$ is fixed to $0.5$. To reduce the number of free parameters we will do the same here.\\
The Novosibirsk distribution ${\cal N}(x; x_0, \sigma, \Lambda)$ is defined as
\begin{align}
  {\cal N}(x; x_0, \sigma, \Lambda) &= \exp{\left(-\frac{1}{2}\frac{(\ln q_y)^2}{\Lambda^2}+\Lambda^2\right)} \\
  q_y &= 1+\frac{\Lambda(x-x_0)}{\sigma}\cdot\frac{\sinh\left(\Lambda\sqrt{\ln 4}\right)}{\Lambda\sqrt{\ln 4}},
\end{align}
where $x_0$ represents the peak position, $\sigma$ the width of the peak, and $\Lambda$ the asymmetric tail.
For the fit only correctly reconstructed events are considered. Thus, we can determine the signal parameters, without a pollution from background events.

\begin{figure}[h]
  \subfigure[]{
    \includegraphics[width=.48\textwidth]{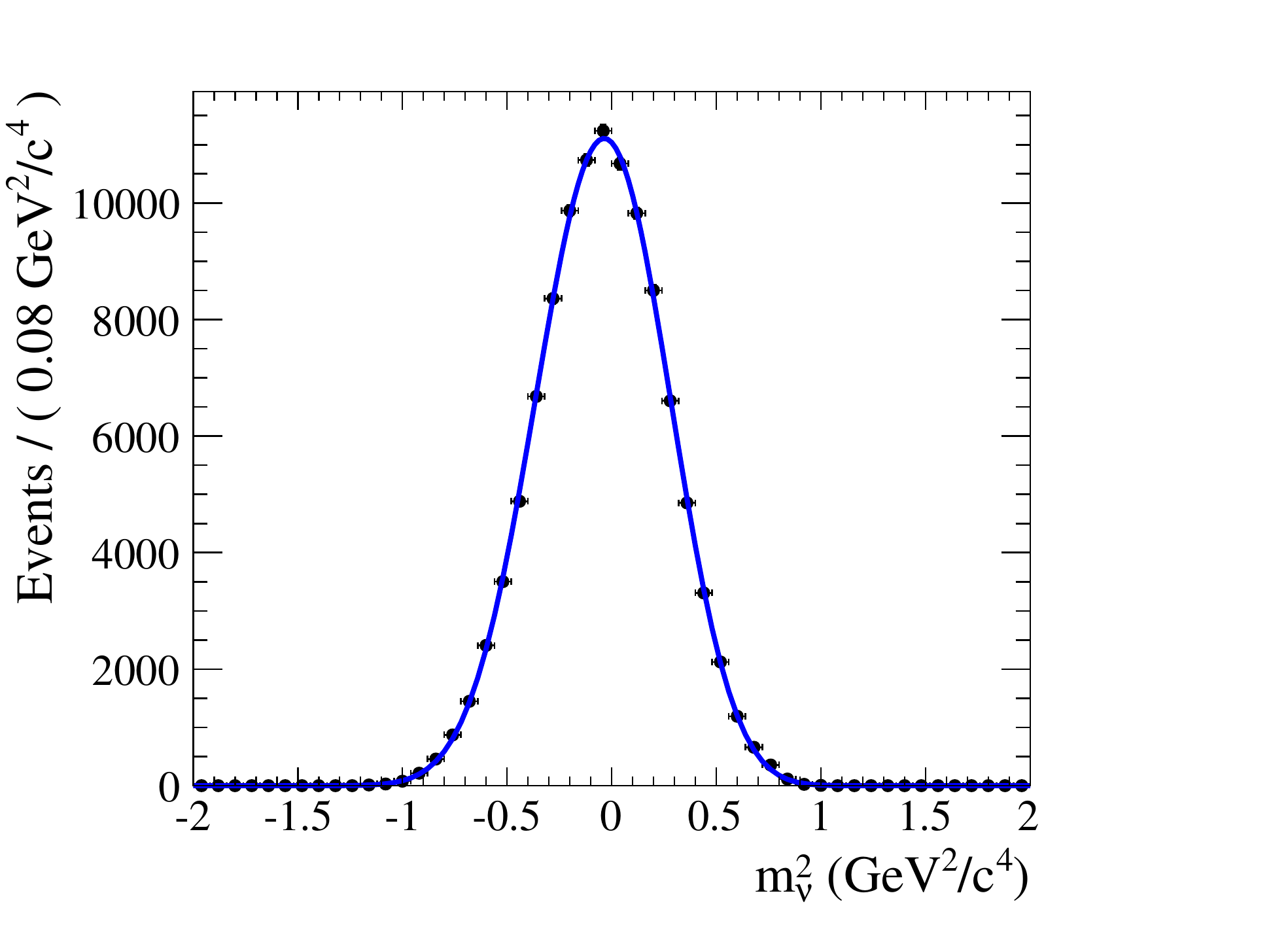}
    \label{subfig:signalParam_mNu2:e}
  }
  \subfigure[]{
    \includegraphics[width=.48\textwidth]{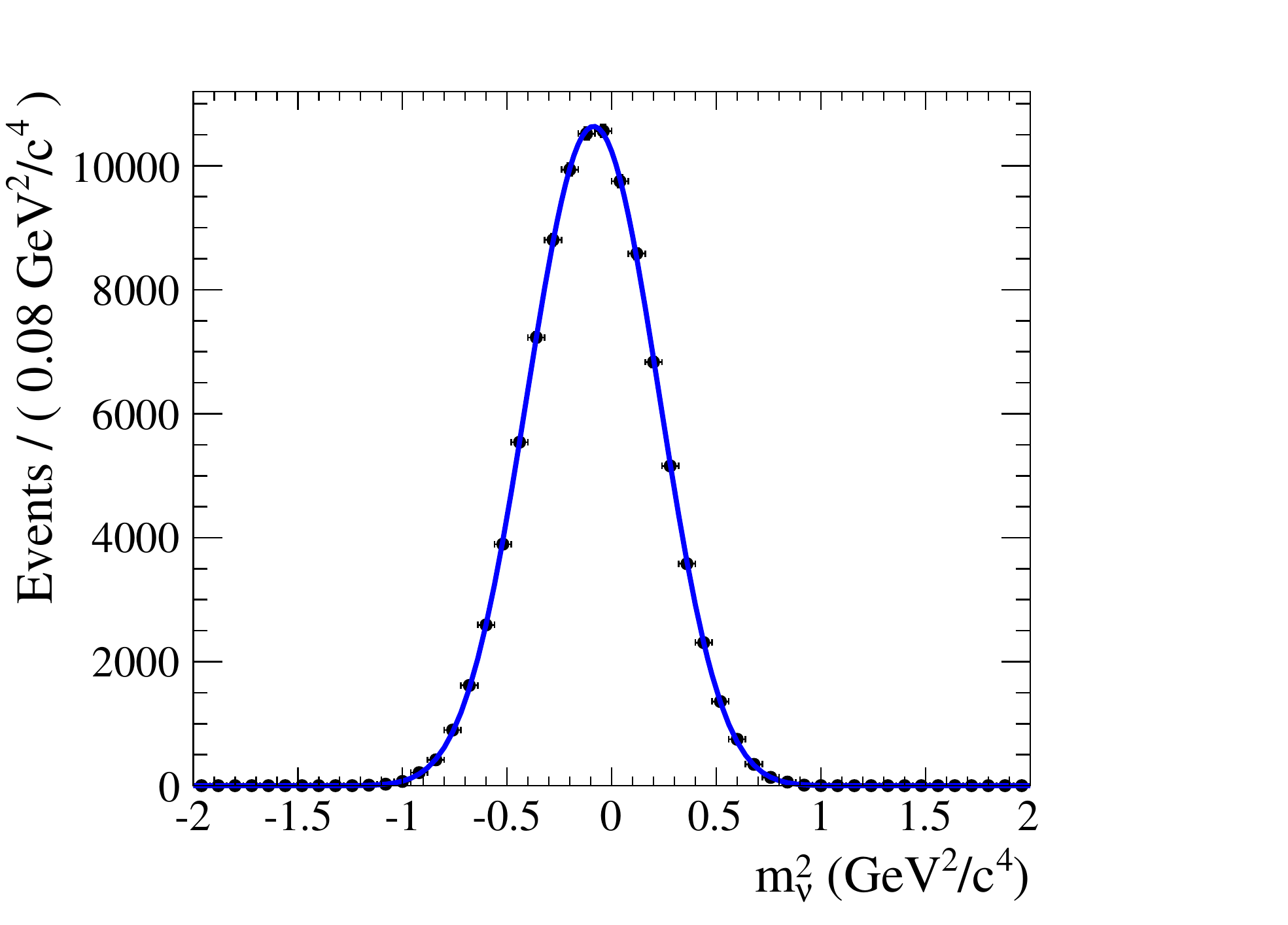}
    \label{subfig:signalParam_mNu2:mu}
  }
  \caption{Fitted $m_{\nu}^2$ distribution for WI signal Monte Carlo data: \subref{subfig:signalParam_mNu2:e} for $\Bm \ra \LCp \antiproton \en \nueb$, and \subref{subfig:signalParam_mNu2:mu} for $\Bm \ra \LCp \antiproton \mun \numb$.}
  \label{fig:signalParam_mNu2}
\end{figure}

\begin{figure}[h]
  \subfigure[]{
    \includegraphics[width=.48\textwidth]{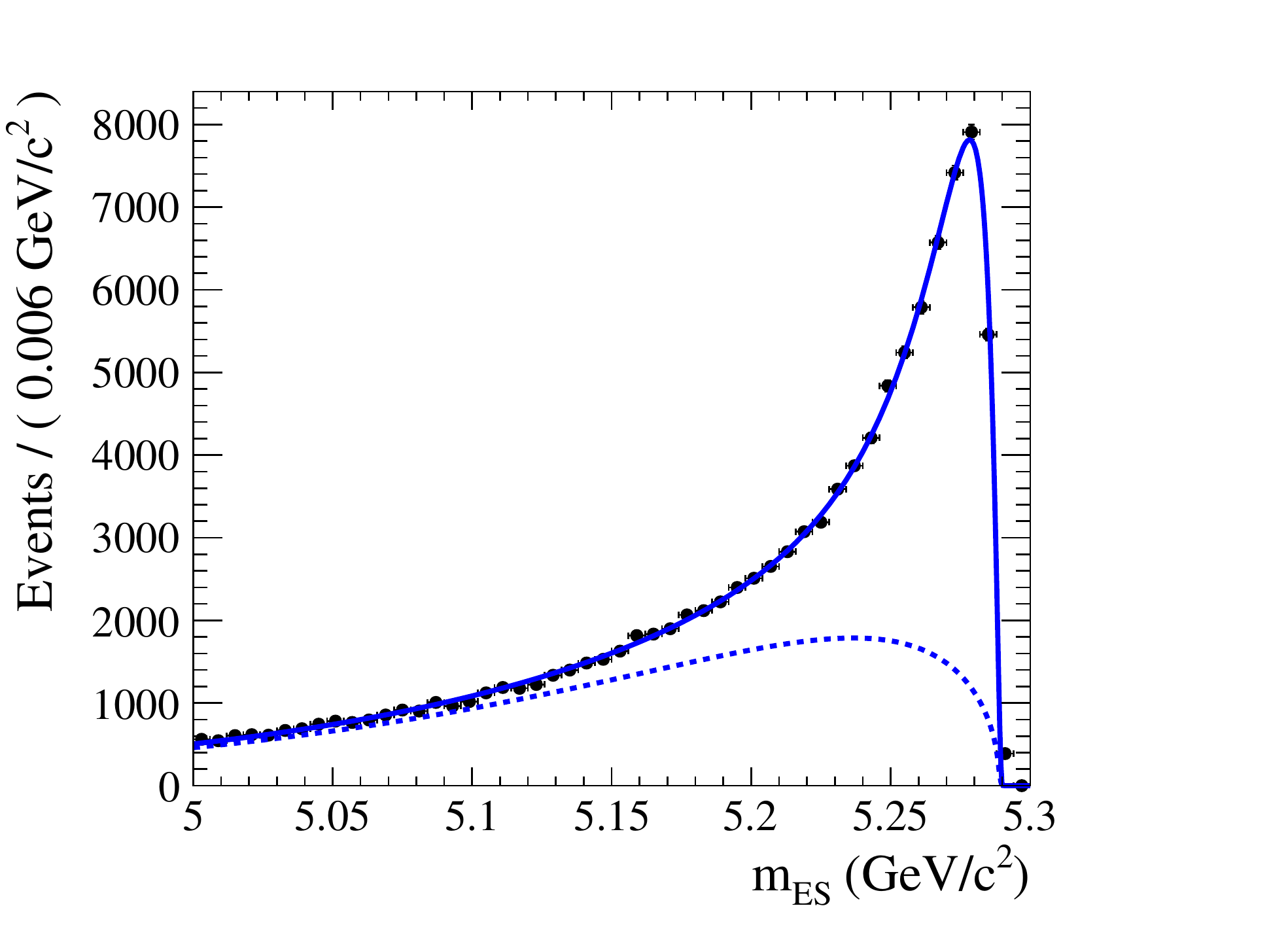}
    \label{subfig:signalParam_mES:e}
  }
  \subfigure[]{
    \includegraphics[width=.48\textwidth]{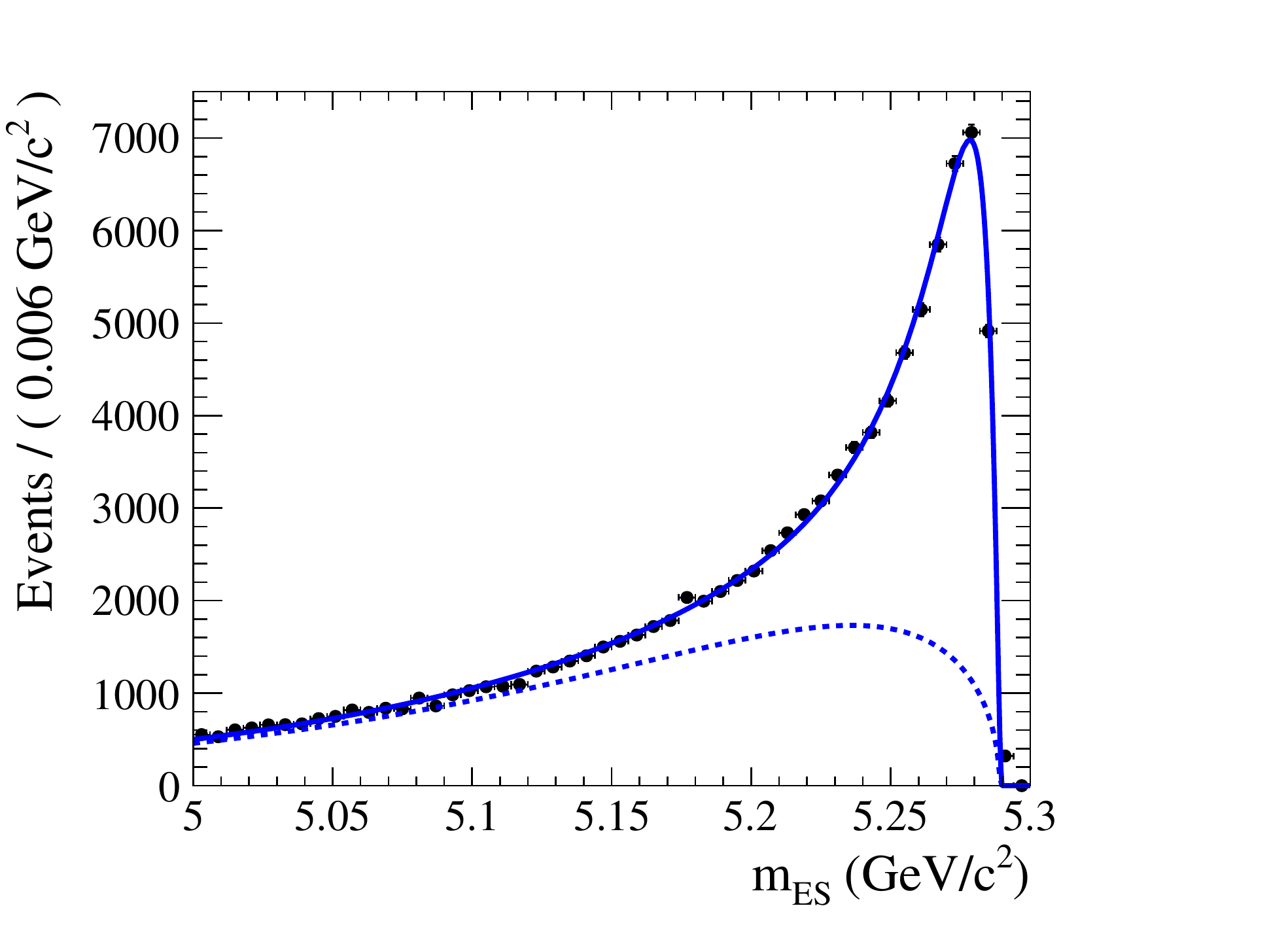}
    \label{subfig:signalParam_mES:mu}
  }
    \caption{Fitted $m_{ES}$ distribution for WI signal Monte Carlo data: \subref{subfig:signalParam_mES:e} for $\Bm \ra \LCp \antiproton \en \nueb$, and \subref{subfig:signalParam_mES:mu} for $\Bm \ra \LCp \antiproton \mun \numb$.}
  \label{fig:signalParam_mES}
\end{figure}

The fitted $m_{\nu}^2$ distributions are shown in Fig. \ref{fig:signalParam_mNu2} and the fit parameters in Table \ref{tab:signalParam_mNu2}. For the \mes fit the fitted signal distributions are shown in Fig. \ref{fig:signalParam_mES}, and the fit parameters in Tab. \ref{tab:signalParam_mES}.
\begin{table}[h]
  \caption{Parameters for the fit to the $m_{\nu}^2$ distribution for WI signal Monte Carlo events: \subref{subtab:signalParam_mNu2:e} for $\Bm \ra \LCp \antiproton \en \nueb$, and \subref{subtab:signalParam_mNu2:mu} for $\Bm \ra \LCp \antiproton \mun \numb$.}
  \centering
  \subtable[]{
    \begin{tabular}{cr@{ $\pm$ }l}
      \toprule
      parameter & \multicolumn{2}{c}{value}	\\\midrule
      $m_0$	& $-0.035$ 	& $0.005$ 	\\
      $\sigma_L$& $0.326$	& $0.004$	\\
      $\sigma_R$& $0.316$ 	& $0.004$	\\
      $\alpha_L$& $-0.023$ 	& $0.005$	\\
      $\alpha_R$& $-0.044$ 	& $0.006$	\\
      \bottomrule
    \end{tabular}
    \label{subtab:signalParam_mNu2:e}
  }\hspace{.5cm}
  \subtable[]{
    \begin{tabular}{cr@{ $\pm$ }l}
      \toprule
      parameter & \multicolumn{2}{c}{value}	\\\midrule
      $m_0$	& $-0.086$	& $0.004$ 	\\
      $\sigma_L$& $0.314$	& $0.004$	\\
      $\sigma_R$& $0.310$ 	& $0.004$	\\
      $\alpha_L$& $-0.035$ 	& $0.006$	\\
      $\alpha_R$& $-0.039$ 	& $0.006$	\\
      \bottomrule
    \end{tabular}
    \label{subtab:signalParam_mNu2:mu}
  }
  \label{tab:signalParam_mNu2}
\end{table}

\begin{figure}[h]
  \subfigure[]{
    \includegraphics[width=.48\textwidth]{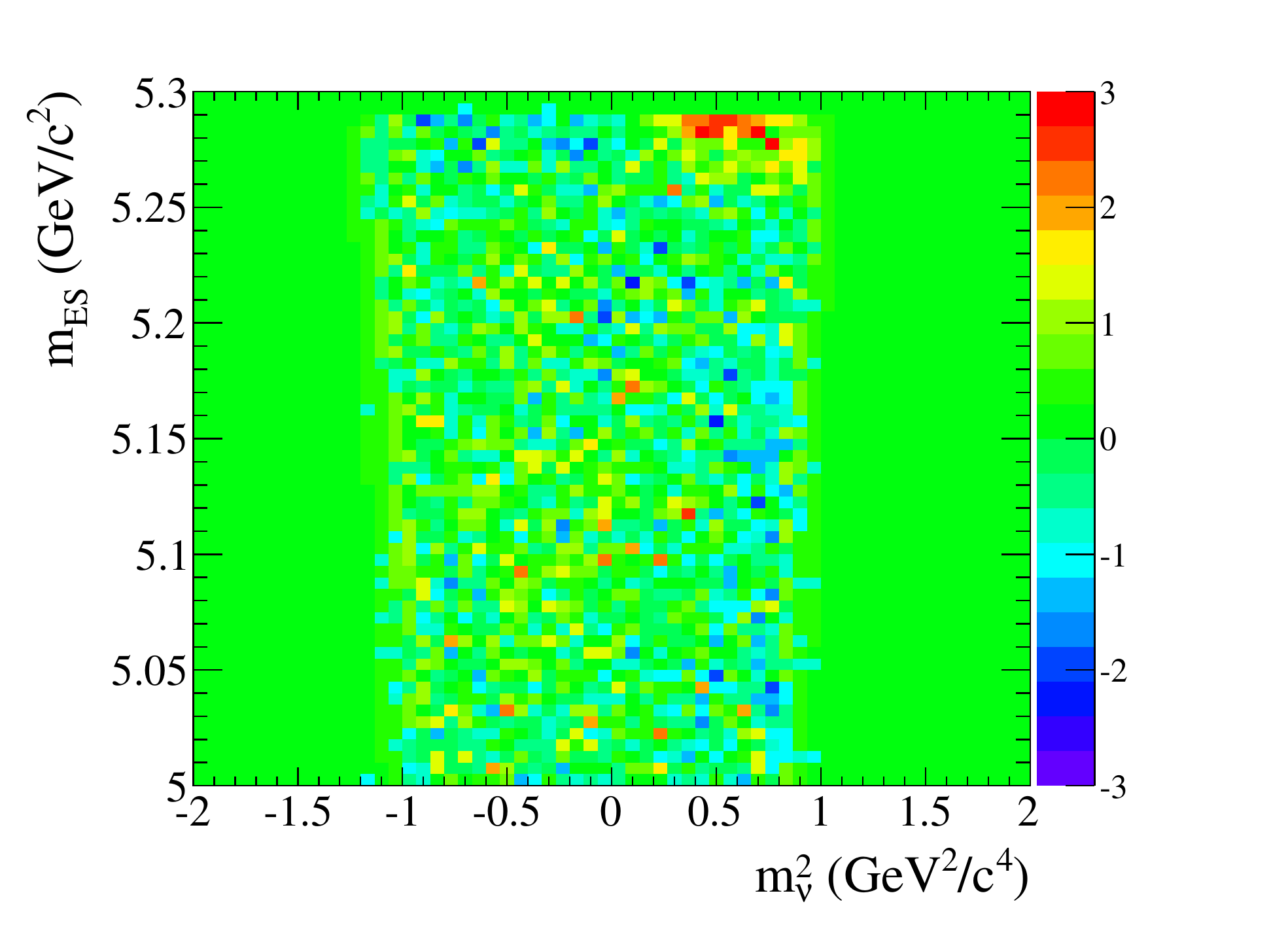}
    \label{subfig:signalPull:e}
  }
  \subfigure[]{
    \includegraphics[width=.48\textwidth]{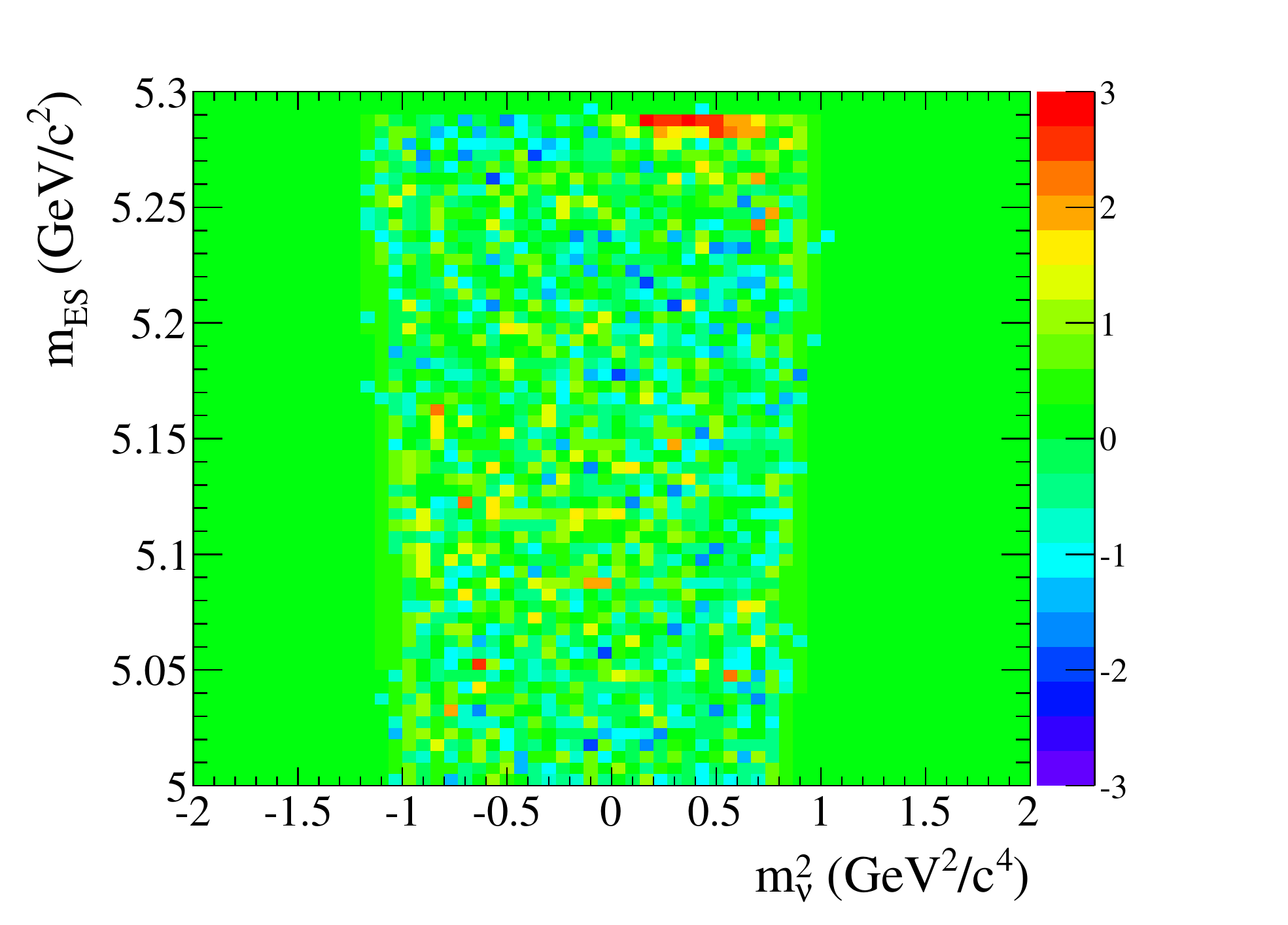}
    \label{subfig:signalPull:mu}
  }
    \caption{Pull plots for the fit to WI signal Monte Carlo data: \subref{subfig:signalPull:e} for $\Bm \ra \LCp \antiproton \en \nueb$, and \subref{subfig:signalPull:mu} for $\Bm \ra \LCp \antiproton \mun \numb$.}
  \label{fig:signalParam_pull}
\end{figure}

\begin{table}[h]
  \caption{Parameters for the fit to the $m_{ES}$ distribution for WI signal Monte Carlo events. \subref{subtab:signalParam_mES:e} for $\Bm \ra \LCp \antiproton \en \nueb$, and \subref{subtab:signalParam_mES:mu} for $\Bm \ra \LCp \antiproton \mun \numb$.}
  \centering
  \subtable[]{
    \begin{tabular}{cr@{ $\pm$ }l}
      \toprule
      parameter & \multicolumn{2}{c}{value}	\\\midrule
      $c$	& $-24.9$ 	& $0.6$ 	\\
      $f_1$	& $0.514$	& $0.020$	\\
      $x_0$	& $5.27930$ 	& $0.00015$	\\
      $\sigma$	& $0.0144$ 	& $0.0004$	\\
      $\Lambda$	& $1.32$ 	& $0.05$	\\
      \bottomrule
    \end{tabular}
    \label{subtab:signalParam_mES:e}
  }\hspace{.5cm}
  \subtable[]{
    \begin{tabular}{cr@{ $\pm$ }l}
      \toprule
      parameter & \multicolumn{2}{c}{value}	\\\midrule
      $c$	& $-24.5$ 	& $0.6$ 	\\
      $f_1$	& $0.542$	& $0.021$	\\
      $x_0$	& $5.27950$ 	& $0.00016$	\\
      $\sigma$	& $0.0142$ 	& $0.0005$	\\
      $\Lambda$	& $1.34$ 	& $0.05$	\\
      \bottomrule
    \end{tabular}
    \label{subtab:signalParam_mES:mu}
  }
  \label{tab:signalParam_mES}
\end{table}
In order to assess the quality of the fit we can check the binned two-dimensional pull $p$
\begin{align}
  p_i =\frac{f_i(m_{\nu}^2, \mes) - N_i(m_{\nu}^2, \mes)}{\sigma_i},
\end{align}
where $N_i(m_{\nu}^2, \mes)$ is the number of entries in the $i$-th bin, and $\sigma$ the uncertainty on $N_i$. The pull distributions are shown in Fig. \ref{fig:signalParam_pull}. The pull plots show a systematic deviation at large \mes values and $m_{\nu}^2 \approx 0.4\gev^2/c^4$. This deviation is not accounted for in the simultaneous fit, since an unbinned two-dimensional fit without correlations equals a simultaneous fit in the $m_{\nu}^2$ and \mes distribution. As for a simultaneous fit only the agreement in the projections is relevant, which is good. The pull plots are only shown for completeness.
\clearpage
\subsection{Background parametrization}\label{sect:bkgParam}

\begin{figure}[h]
  \subfigure[]{
    \includegraphics[width=.48\textwidth]{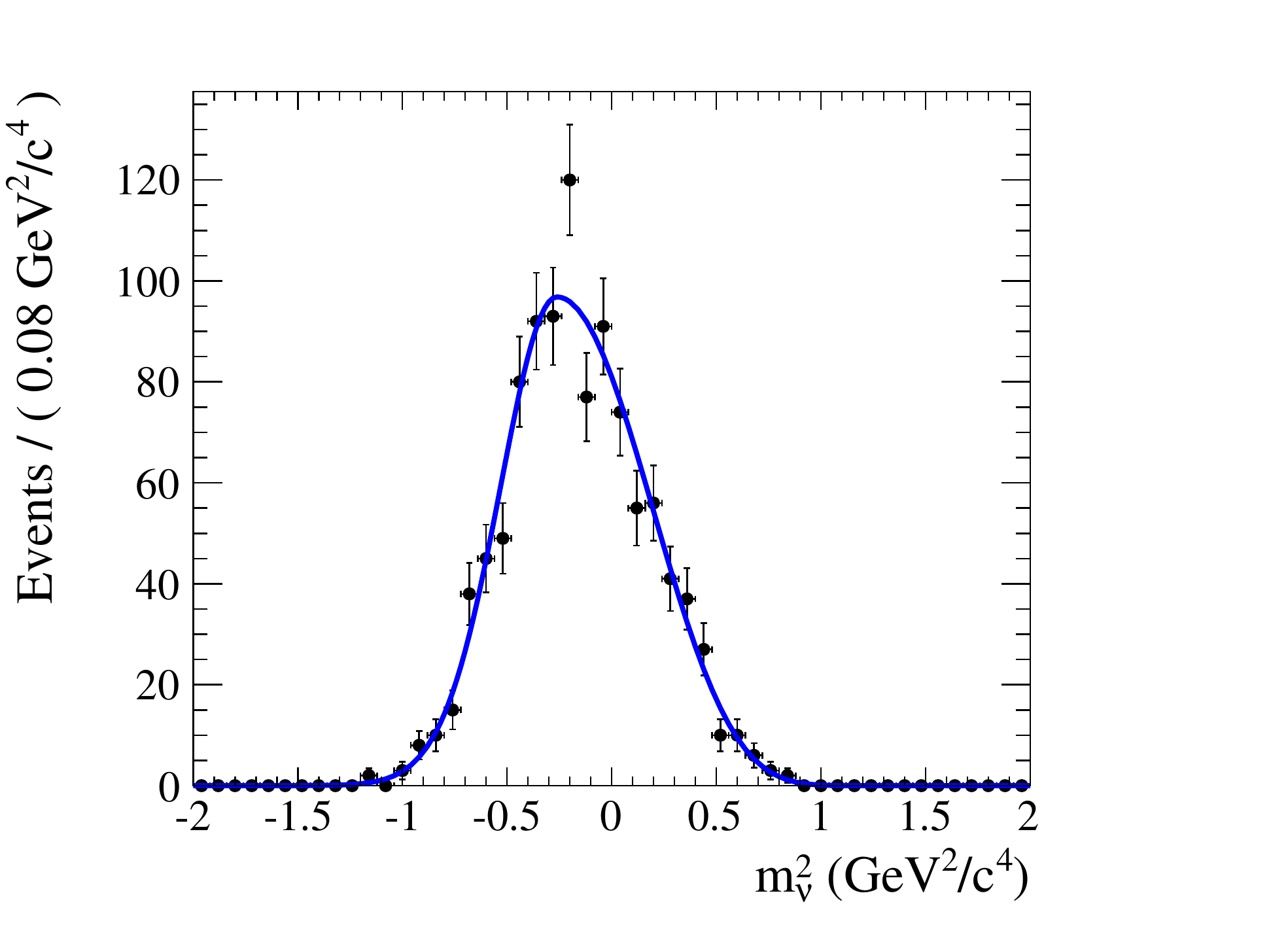}
    \label{subfig:backgroundParam_mNu2:e}
  }
  \subfigure[]{
    \includegraphics[width=.48\textwidth]{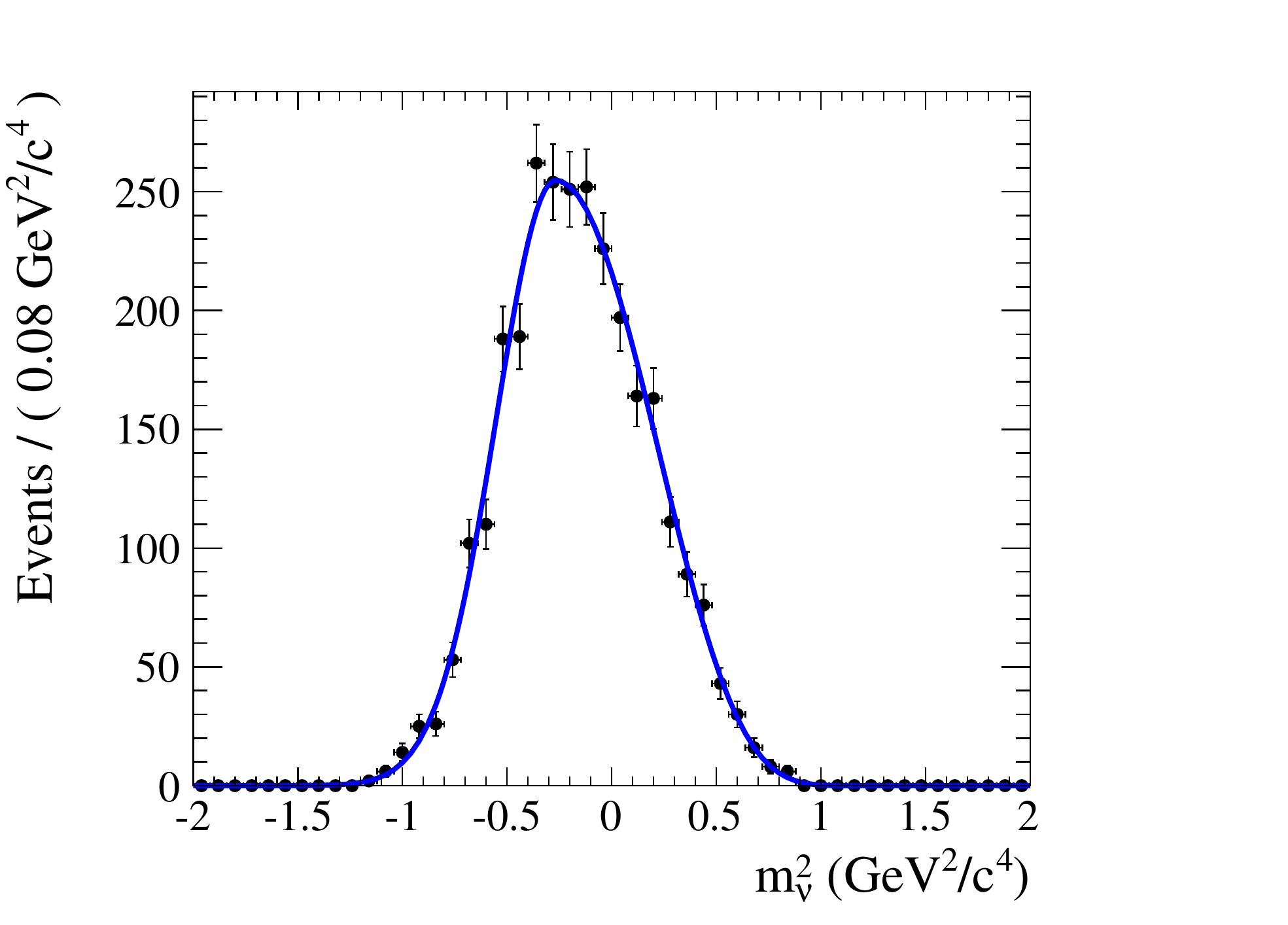}
    \label{subfig:backgroundParam_mNu2:mu}
  }
  \caption{Fitted $m_{\nu}^2$ distribution for background Monte Carlo data: \subref{subfig:backgroundParam_mNu2:e} for $\Bm \ra \LCp \antiproton \en \nueb$, and \subref{subfig:backgroundParam_mNu2:mu} for $\Bm \ra \LCp \antiproton \mun \numb$.}
  \label{fig:backgroundParam_mNu2}
\end{figure}

\begin{figure}[h]
  \subfigure[]{
    \includegraphics[width=.48\textwidth]{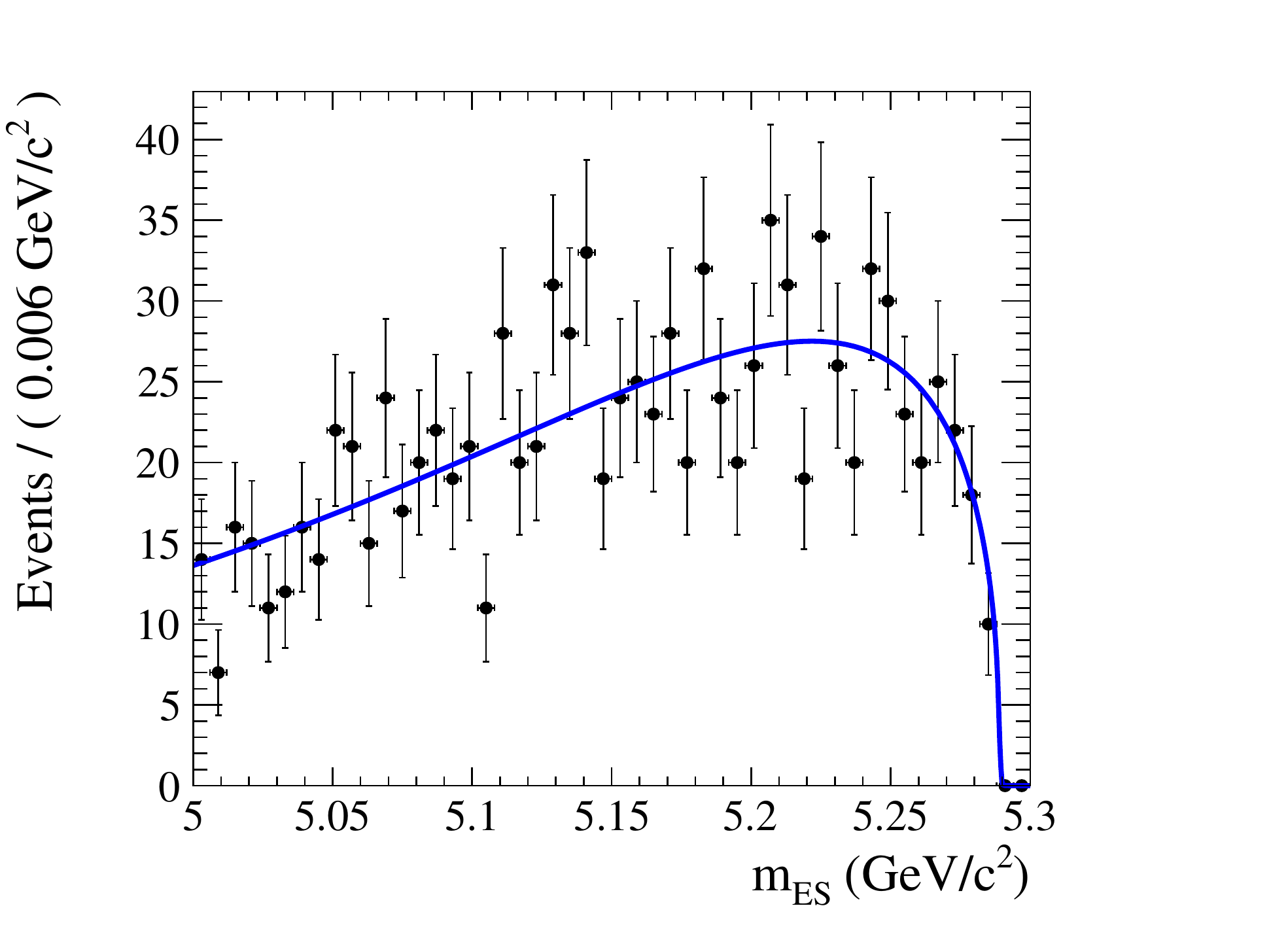}
    \label{subfig:backgroundParam_mES:e}
  }
  \subfigure[]{
    \includegraphics[width=.48\textwidth]{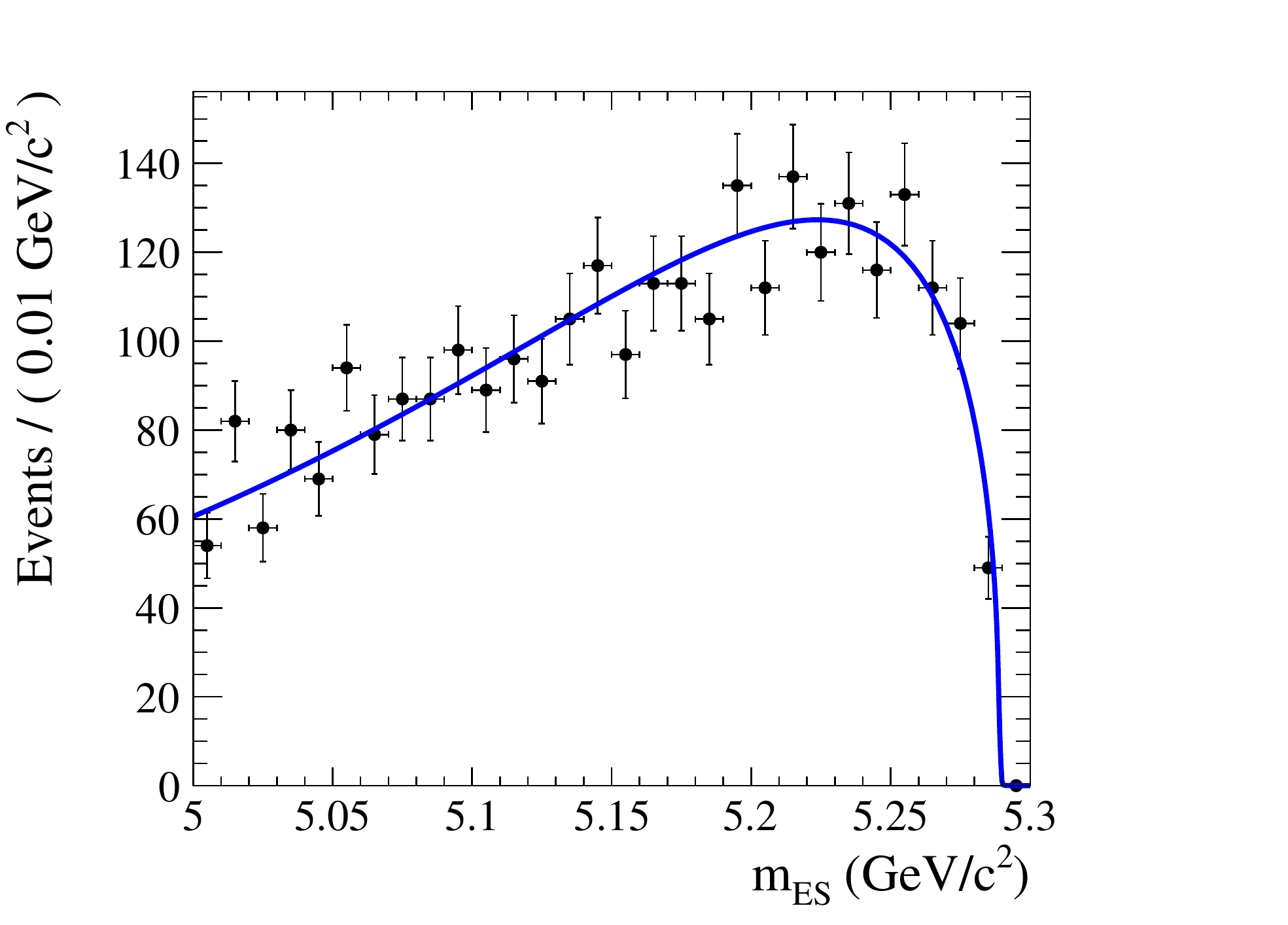}
    \label{subfig:backgroundParam_mES:mu}
  }
    \caption{Fitted $m_{ES}$ distribution for background Monte Carlo data. \subref{subfig:backgroundParam_mES:e} for $\Bm \ra \LCp \antiproton \en \nueb$, and \subref{subfig:backgroundParam_mES:mu} for $\Bm \ra \LCp \antiproton \mun \numb$.}
  \label{fig:backgroundParam_mES}
\end{figure}

\begin{figure}[h]
  \subfigure[]{
    \includegraphics[width=.48\textwidth]{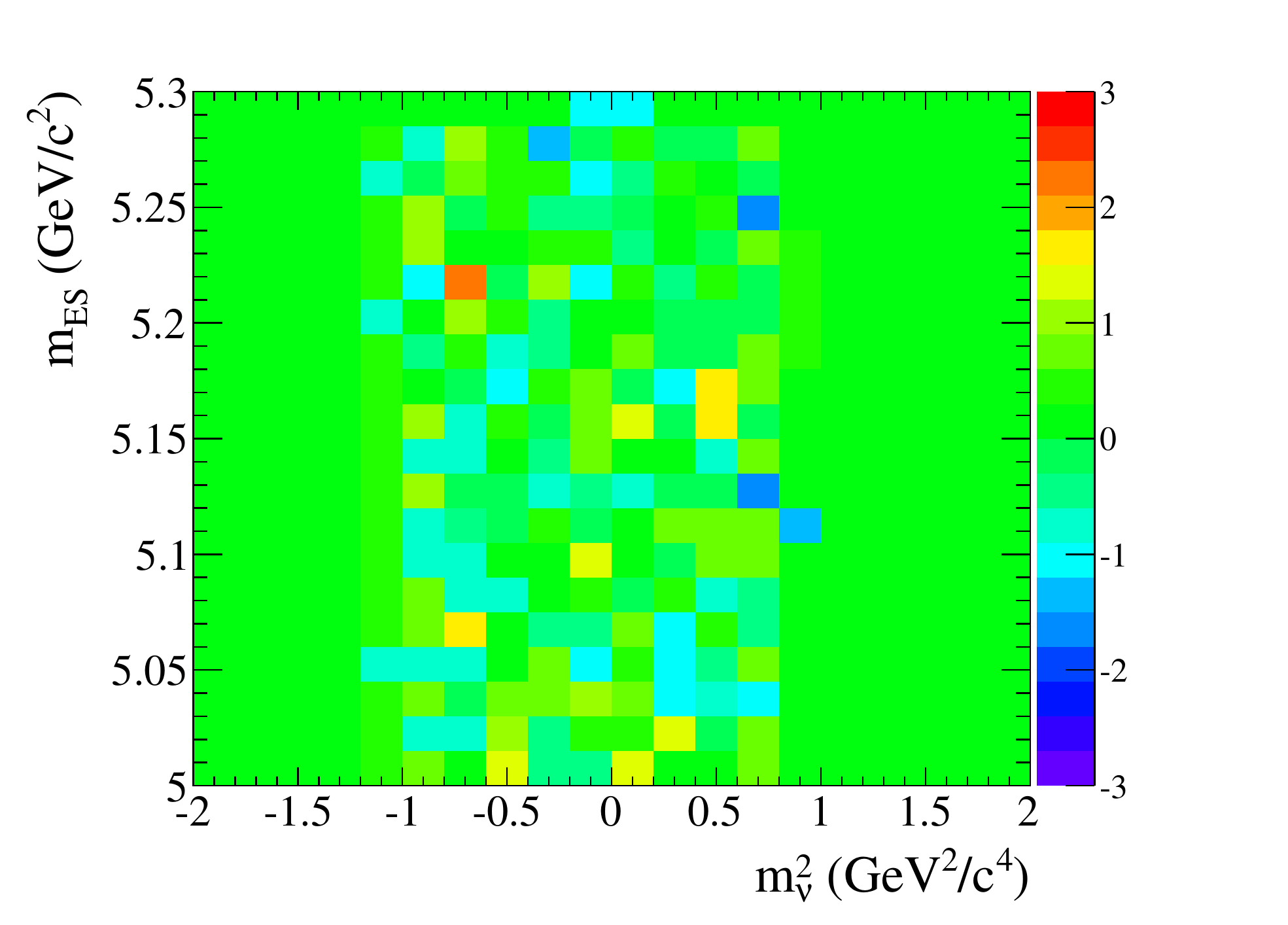}
    \label{subfig:backgroundPull:e}
  }
  \subfigure[]{
    \includegraphics[width=.48\textwidth]{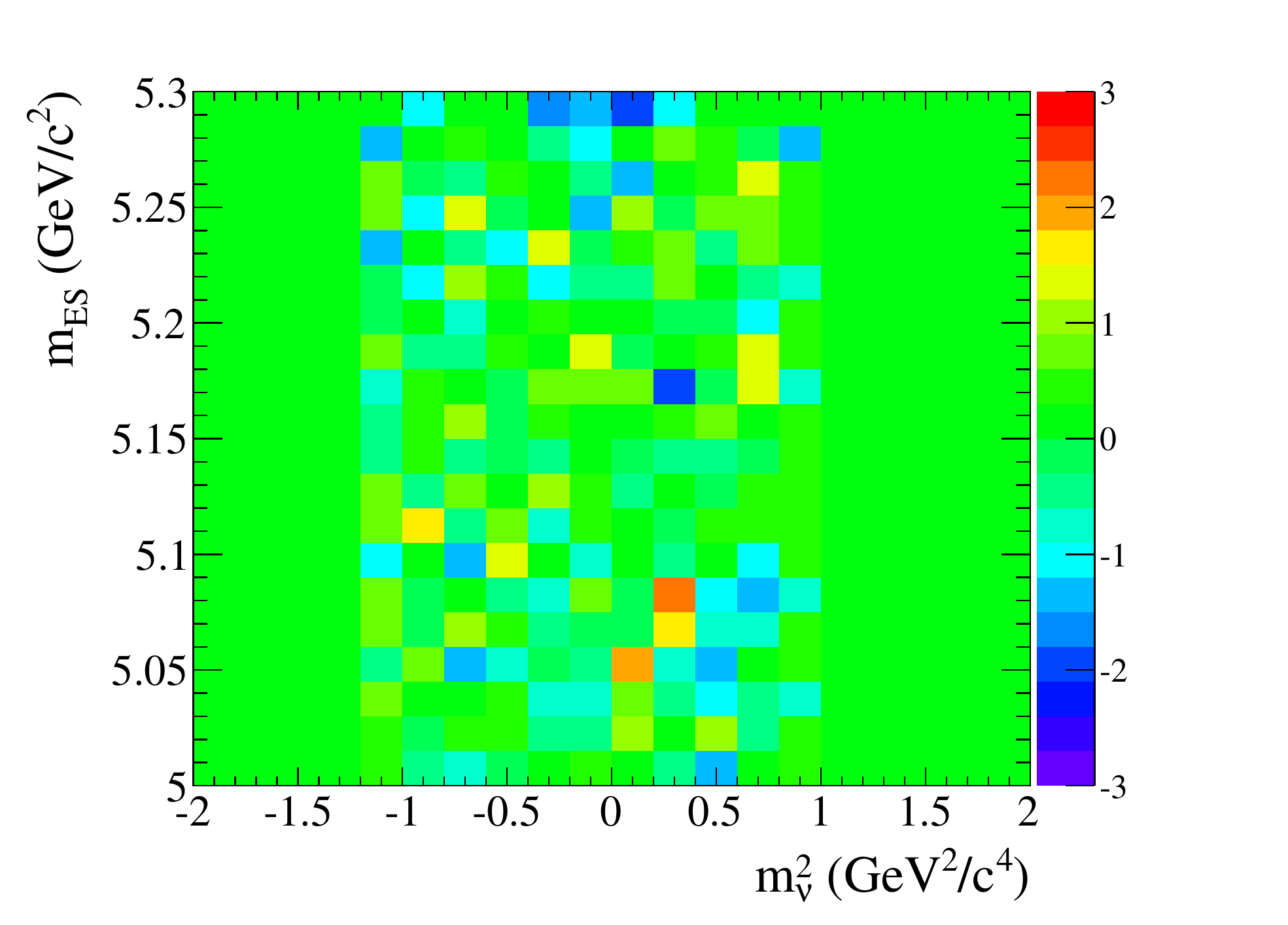}
    \label{subfig:backgroundPull:mu}
  }
    \caption{Pull plots for the fit to background Monte Carlo data: \subref{subfig:backgroundPull:e} for $\Bm \ra \LCp \antiproton \en \nueb$, and \subref{subfig:backgroundPull:mu} for $\Bm \ra \LCp \antiproton \mun \numb$.}
  \label{fig:backgroundParam_pull}
\end{figure}

For background we use a Cruijff function for $m_{\nu}^2$ as well, and for \mes an ARGUS function. For the ARGUS we leave the exponent $p$ floating to allow for a steeper distribution. The fitted projections are shown in Fig. \ref{fig:backgroundParam_mNu2} and \ref{fig:backgroundParam_mES}, and the parameters are given in Tab. \ref{tab:backgroundParam_mNu2} and \ref{tab:backgroundParam_mES}. The input data set for the fit consists of a luminosity scaled mixture of \qqbar and \BBbar Monte Carlo events, i.e. the relative contributions of \qqbar and \BBbar events to the complete input set are equivalent to the expected ratios in OnPeak data.
The corresponding pull plots, shown in Fig. \ref{fig:backgroundParam_pull}, show no systematic deviations between the fitted function and the background Monte Carlo data.
\begin{table}[h]
  \caption{Fit parameters for the fit to the $m_{\nu}^2$ distribution for background Monte Carlo events: \subref{subtab:backgroundParam_mNu2:e} for $\Bm \ra \LCp \antiproton \en \nueb$, and \subref{subtab:backgroundParam_mNu2:mu} for $\Bm \ra \LCp \antiproton \mun \numb$.}
  \centering
  \subtable[]{
    \begin{tabular}{cr@{ $\pm$ }l}
      \toprule
      parameter & \multicolumn{2}{c}{value}	\\\midrule
      $m_0$	& $-0.26$ 	& $0.04$ 	\\
      $\sigma_L$& $0.271$	& $0.030$	\\
      $\sigma_R$& $0.44$ 	& $0.04$	\\
      $\alpha_L$& $0.01$ 	& $0.04$	\\
      $\alpha_R$& $-0.10$ 	& $0.08$	\\
      \bottomrule
    \end{tabular}
    \label{subtab:backgroundParam_mNu2:e}
  }\hspace{.5cm}
  \subtable[]{
    \begin{tabular}{cr@{ $\pm$ }l}
      \toprule
      parameter & \multicolumn{2}{c}{value}	\\\midrule
      $m_0$	& $-0.268$	& $0.025$ 	\\
      $\sigma_L$& $0.283$	& $0.022$	\\
      $\sigma_R$& $0.469$ 	& $0.029$	\\
      $\alpha_L$& $0.007$ 	& $0.028$	\\
      $\alpha_R$& $-0.13$ 	& $0.04$	\\
      \bottomrule
    \end{tabular}
    \label{subtab:backgroundParam_mNu2:mu}
  }
  \label{tab:backgroundParam_mNu2}
\end{table}
\begin{table}[h]
  \caption{Fit parameters for the fit to the $m_{ES}$ distribution for background Monte Carlo events: \subref{subtab:backgroundParam_mES:e} for $\Bm \ra \LCp \antiproton \en \nueb$, and \subref{subtab:backgroundParam_mES:mu} for $\Bm \ra \LCp \antiproton \mun \numb$.}
  \centering
  \subtable[]{
    \begin{tabular}{cr@{ $\pm$ }l}
      \toprule
      parameter & \multicolumn{2}{c}{value}	\\\midrule
      $c$	& $-16.5$ 	& $1.1$ 	\\
      $p$	& $0.45$	& $0.08$	\\
      \bottomrule
    \end{tabular}
    \label{subtab:backgroundParam_mES:e}
  }\hspace{.5cm}
  \subtable[]{
    \begin{tabular}{cr@{ $\pm$ }l}
      \toprule
      parameter & \multicolumn{2}{c}{value}	\\\midrule
      $c$	& $-15.7$ 	& $0.07$ 	\\
      $p$	& $0.3976$	& $0.0010$	\\
      \bottomrule
    \end{tabular}
    \label{subtab:backgroundParam_mES:mu}
  }
  \label{tab:backgroundParam_mES}
\end{table}
\clearpage
\subsection{Fit validation}

In order to validate the fit procedure, and exclude any significant systematic uncertainties in the extracted signal yield, we prepare different mixtures of background and signal Monte Carlo data. While the used background set stays the same we vary the signal fraction from $0$ to $100$ events and compare the fitted signal yields with the size of the input data sets. Table \ref{tab:FitValid}, listing the fit result, shows no significant deviations between the size of the input data set and the fitted signal yield. Thus, we conclude that the fit procedure does not lead to a significant over- or underestimation of the signal yield and is save to use for signal extraction. The fits are shown in Fig. \ref{fig:FitValidation:electron} and \ref{fig:FitValidation:muon}.
\begin{table}[h]
  \begin{center}
  \caption{Size of the signal input data set, and the fitted signal yield for the electron and muon channel.}
  \begin{tabular}{ccc}\toprule
    signal events & $\en$ fit result & $\mun$ fit result \\\midrule
    $100$ & $80 \pm 40$ & $100 \pm 70$ \\
    $50$  & $30 \pm 40$ & $60 \pm 70$  \\
    $10$  & $10 \pm 40$ & $20 \pm 70$  \\
    $0$   & $ -1 \pm 40$& $30 \pm 70$  \\\bottomrule
  \end{tabular}
  \label{tab:FitValid}
  \end{center}
\end{table}

\subsection{Fit to data}

\begin{figure}[h]
  \begin{center}
  \subfigure[]{
    \includegraphics[width=.48\textwidth]{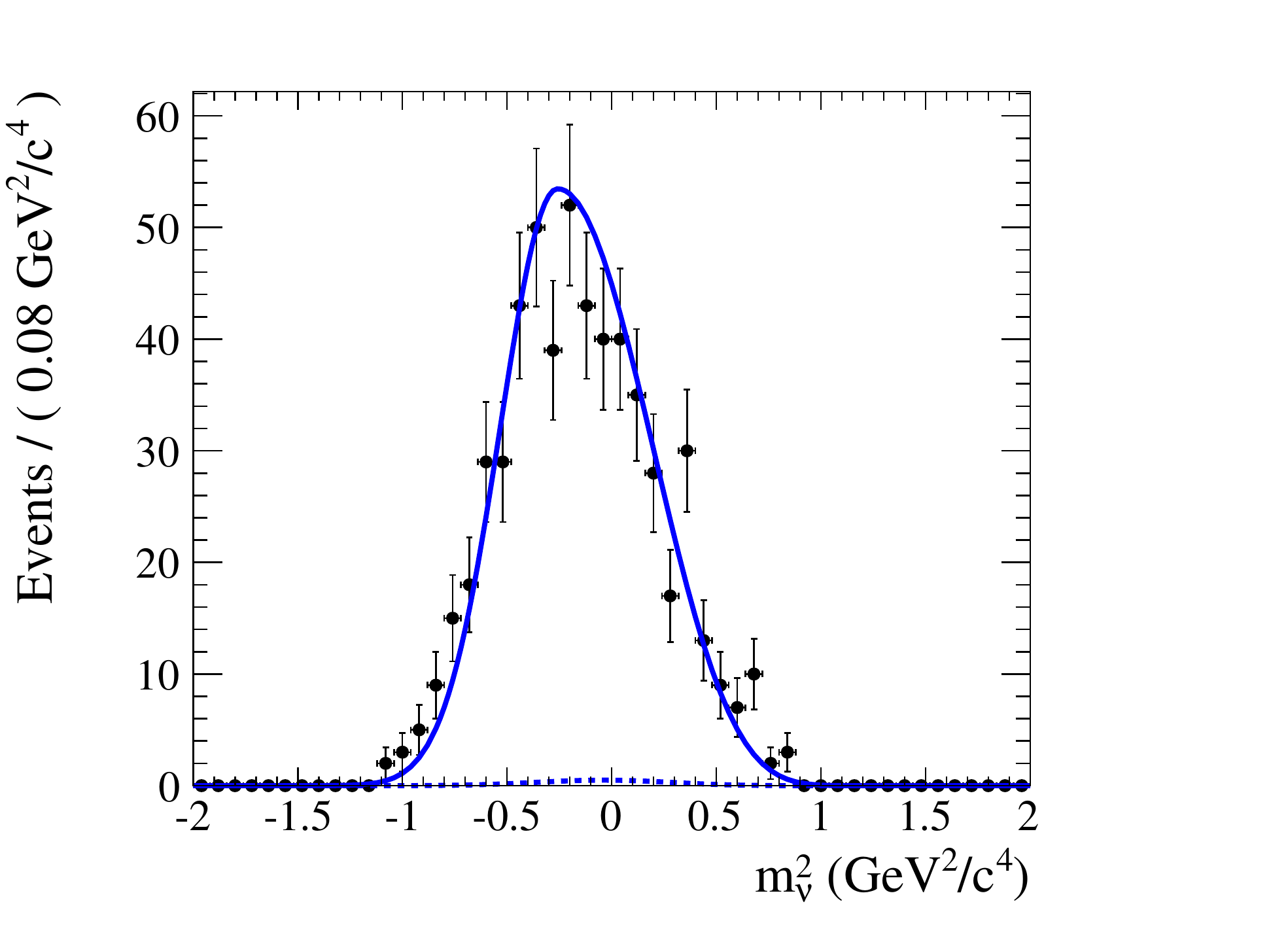}
  }
  \subfigure[]{
    \includegraphics[width=.48\textwidth]{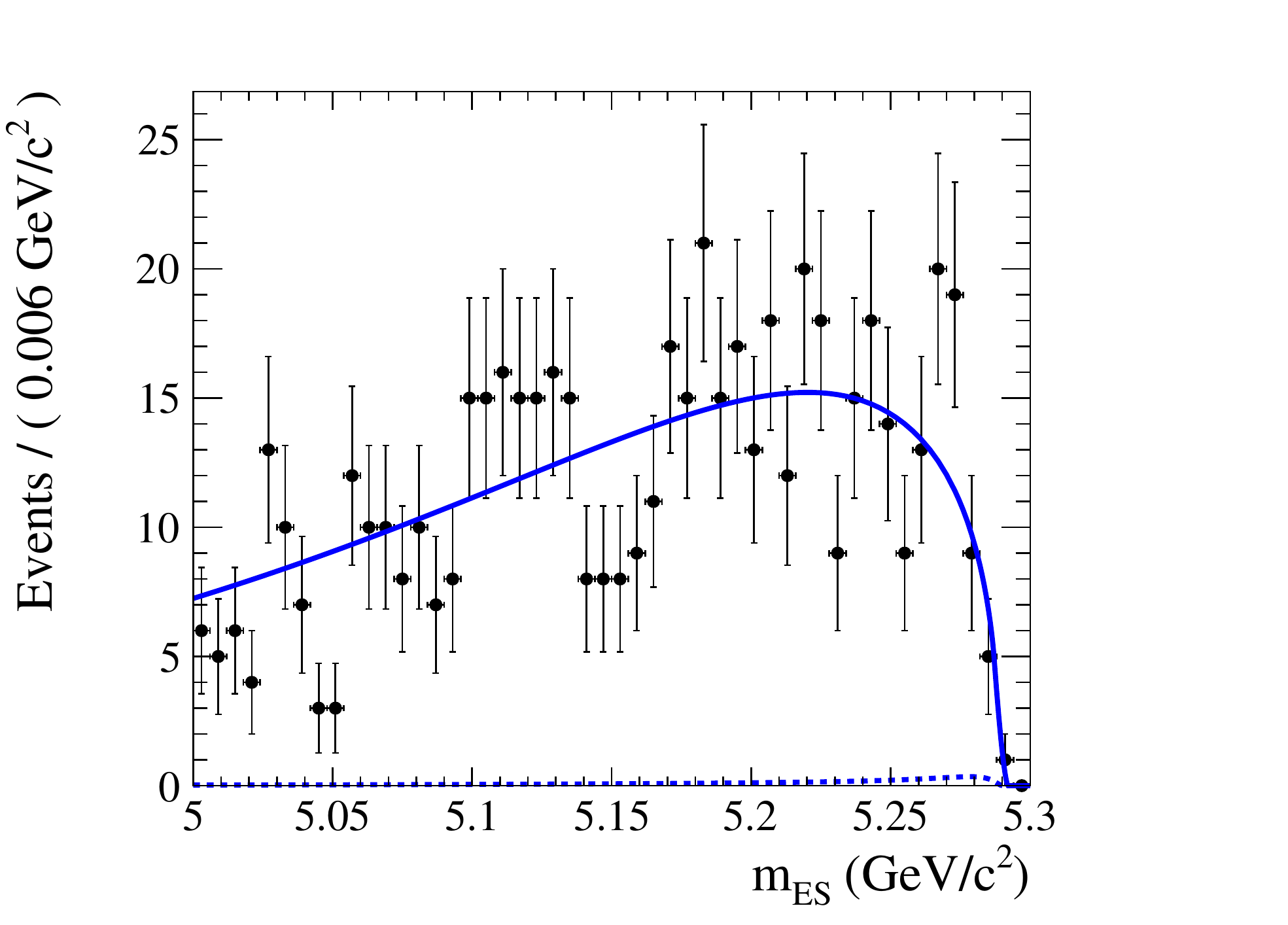}
  }
  \end{center}
  \caption{Fitted $m_{\nu}^2$ and \mes distributions of the fit to \texttt{OnPeak} data for the electron channel.}
  \label{fig:OnPeakFit:e}
\end{figure}

\begin{figure}[h]
  \begin{center}
  \subfigure[]{
    \includegraphics[width=.48\textwidth]{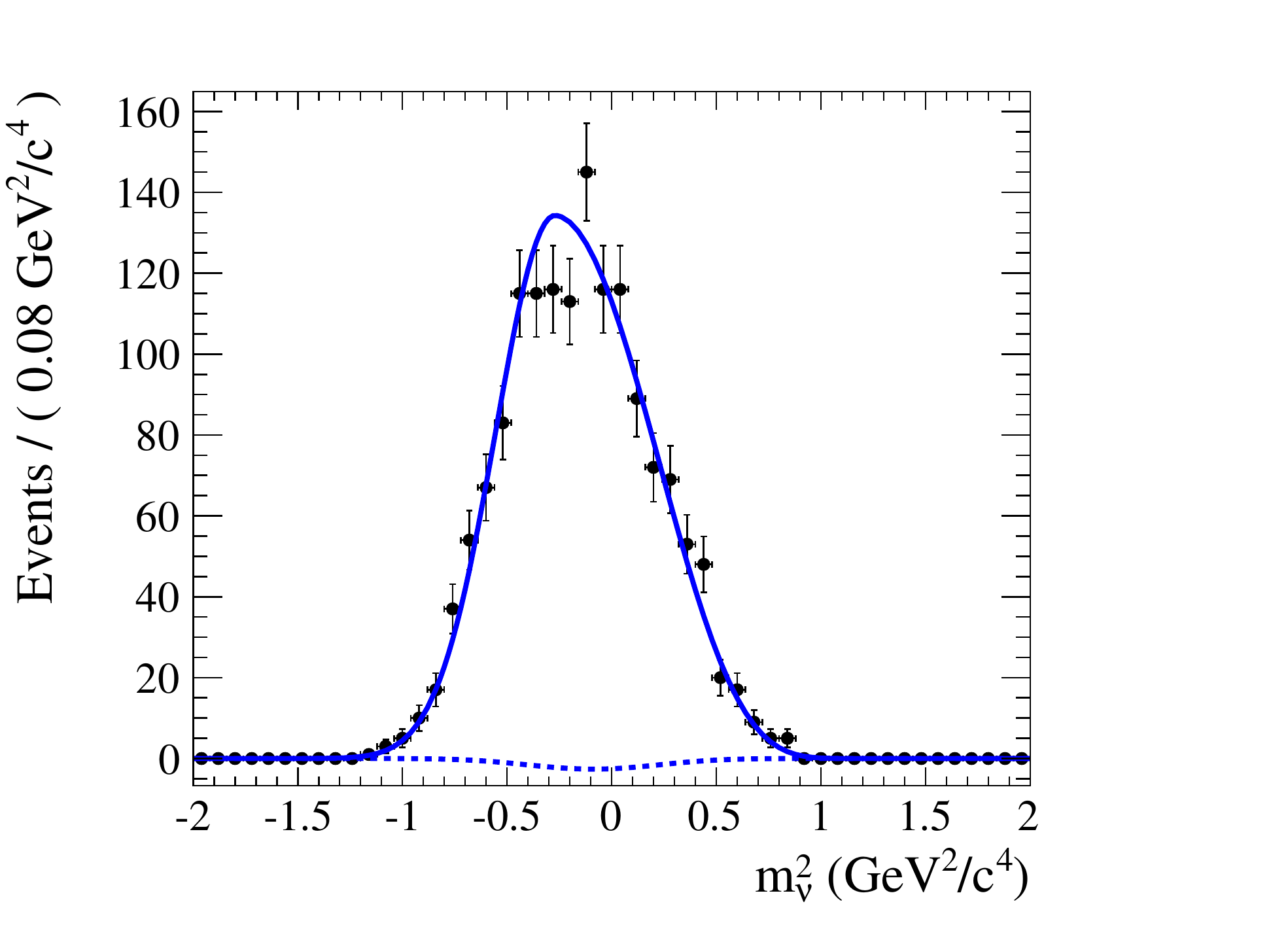}
  }
  \subfigure[]{
    \includegraphics[width=.48\textwidth]{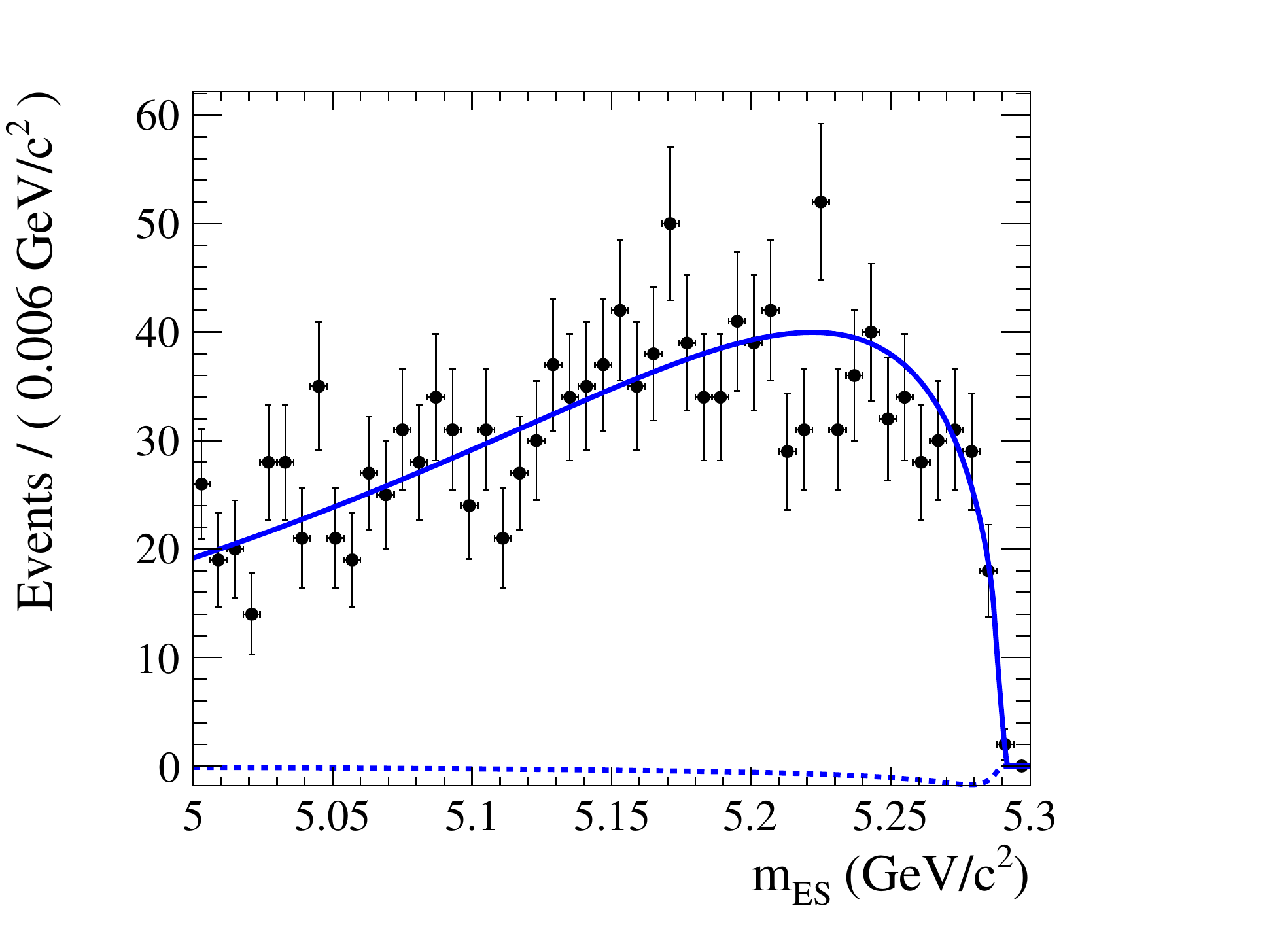}
  }
  \end{center}
  \caption{Fitted $m_{\nu}^2$ and \mes distributions of the fit to \texttt{OnPeak} data for the muon channel.}
  \label{fig:OnPeakFit:mu}
\end{figure}

\begin{figure}[h]
  \begin{center}
  \subfigure[]{
    \includegraphics[width=.48\textwidth]{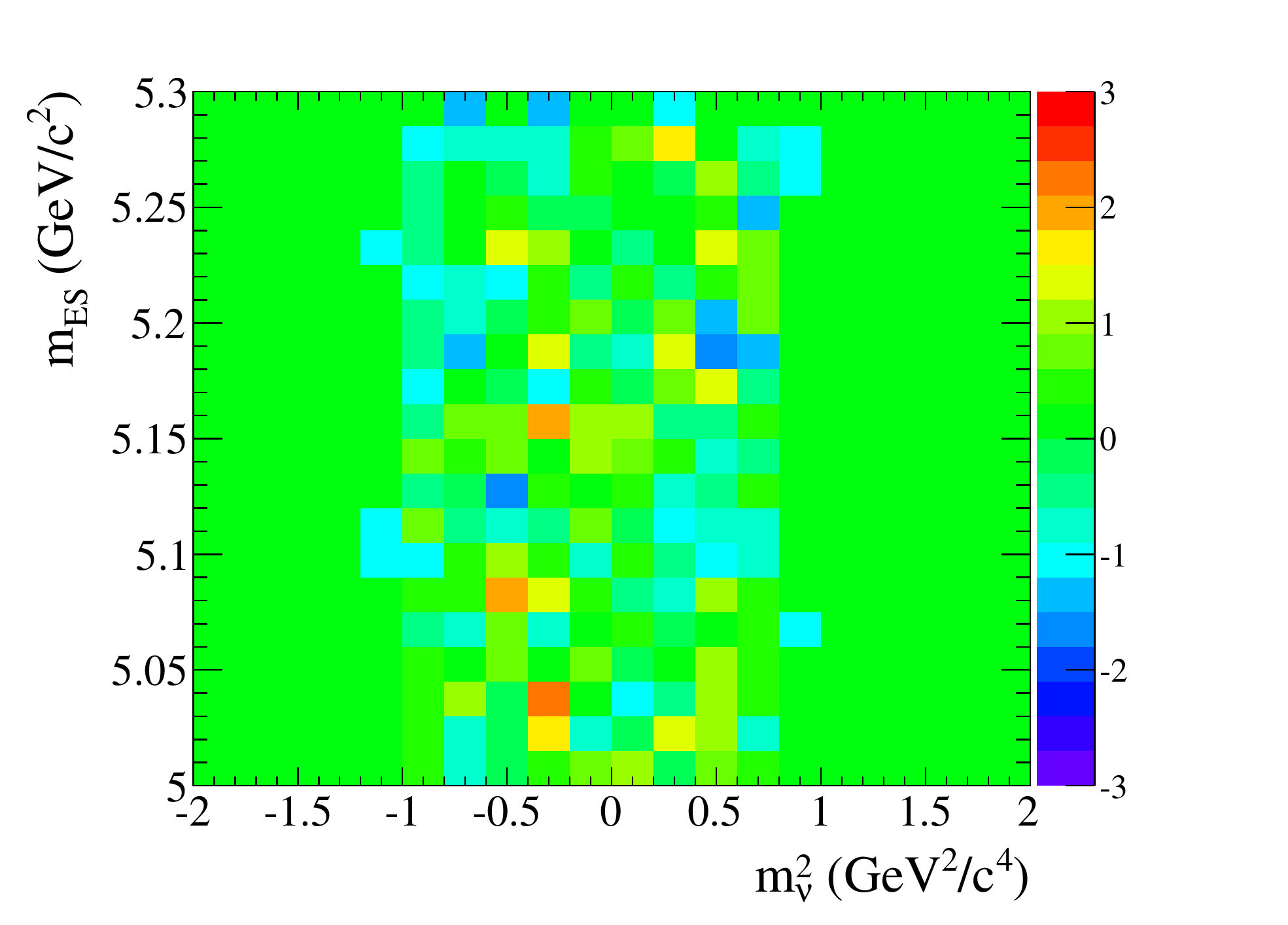}
    \label{subfig:e:OnPeak:pull}
  }
  \subfigure[]{
    \includegraphics[width=.48\textwidth]{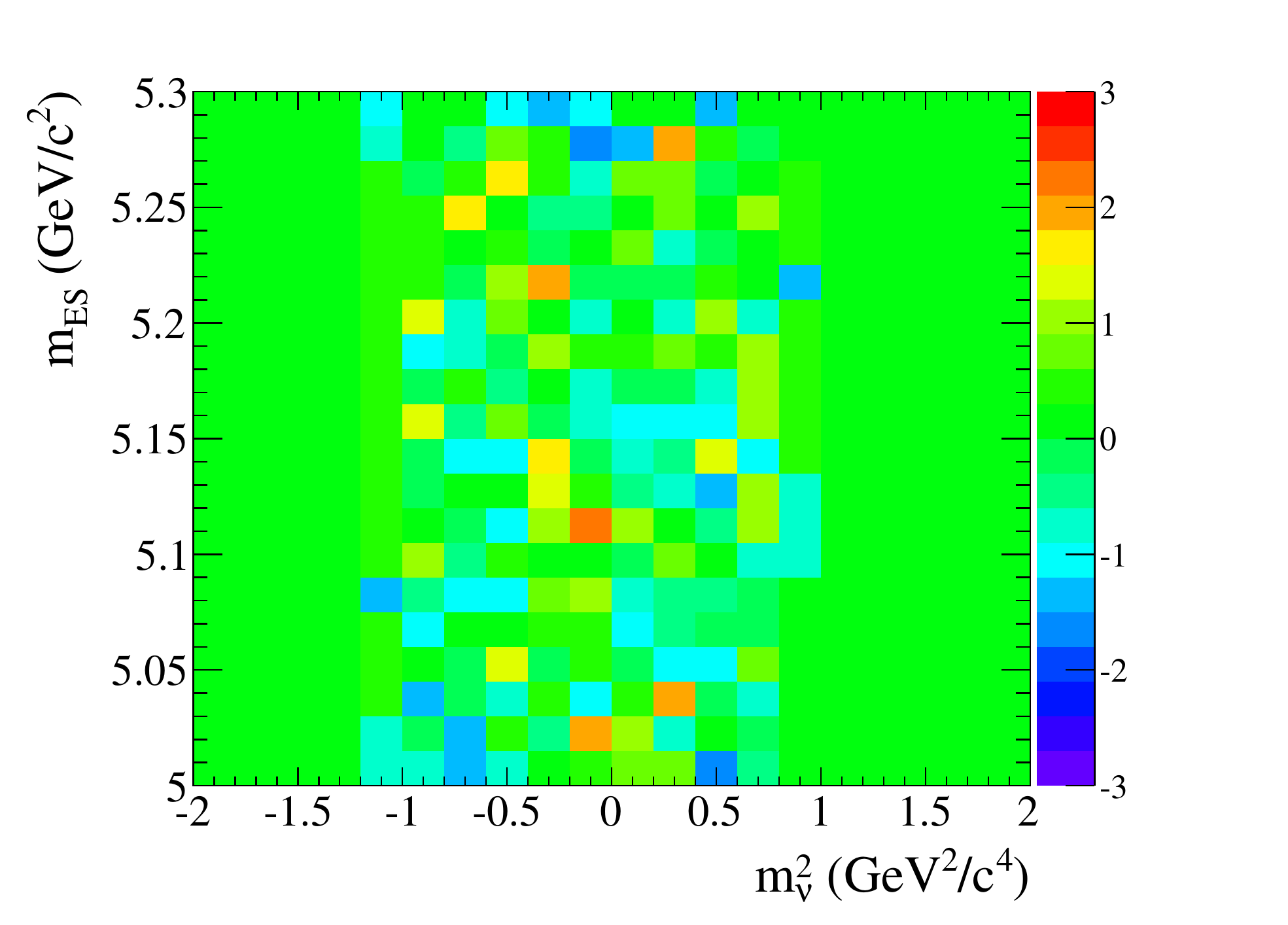}
    \label{subfig:mu:OnPeak:pull}
  }
  \end{center}
  \caption{The two-dimensional pull of the fit to \texttt{OnPeak} data for the electron \subref{subfig:e:OnPeak:pull} and muon \subref{subfig:mu:OnPeak:pull} channel.}
  \label{fig:OnPeakFit:pull}
\end{figure}

We extract the signal yield for both signal channels on data separately. Therefore, all parameters of the background and signal pdfs are fixed to their optimal values, determined in section \ref{sect:fit_procedure} and \ref{sect:bkgParam}. The fitted distributions and the pull distribution are shown in Fig. \ref{fig:OnPeakFit:e}, \ref{fig:OnPeakFit:mu}, and \ref{fig:OnPeakFit:pull}. The obtained signal yields are
\begin{align}
  N_{\rm sig}(\Bm \ra \LCp \antiproton \en \nueb) &= 5 \pm 27,\\
  N_{\rm sig}(\Bm \ra \LCp \antiproton \mun \numb) &= -30 \pm 50.
\end{align}
Although the yield for $\Bm \ra \LCp \antiproton \mun \numb$ is negative both yield are compatible with zero. 

\subsection{Statistical Upper Limit}\label{sect:UL:stat}

\begin{figure}[h]
  \subfigure[]{
    \includegraphics[width=.48\textwidth]{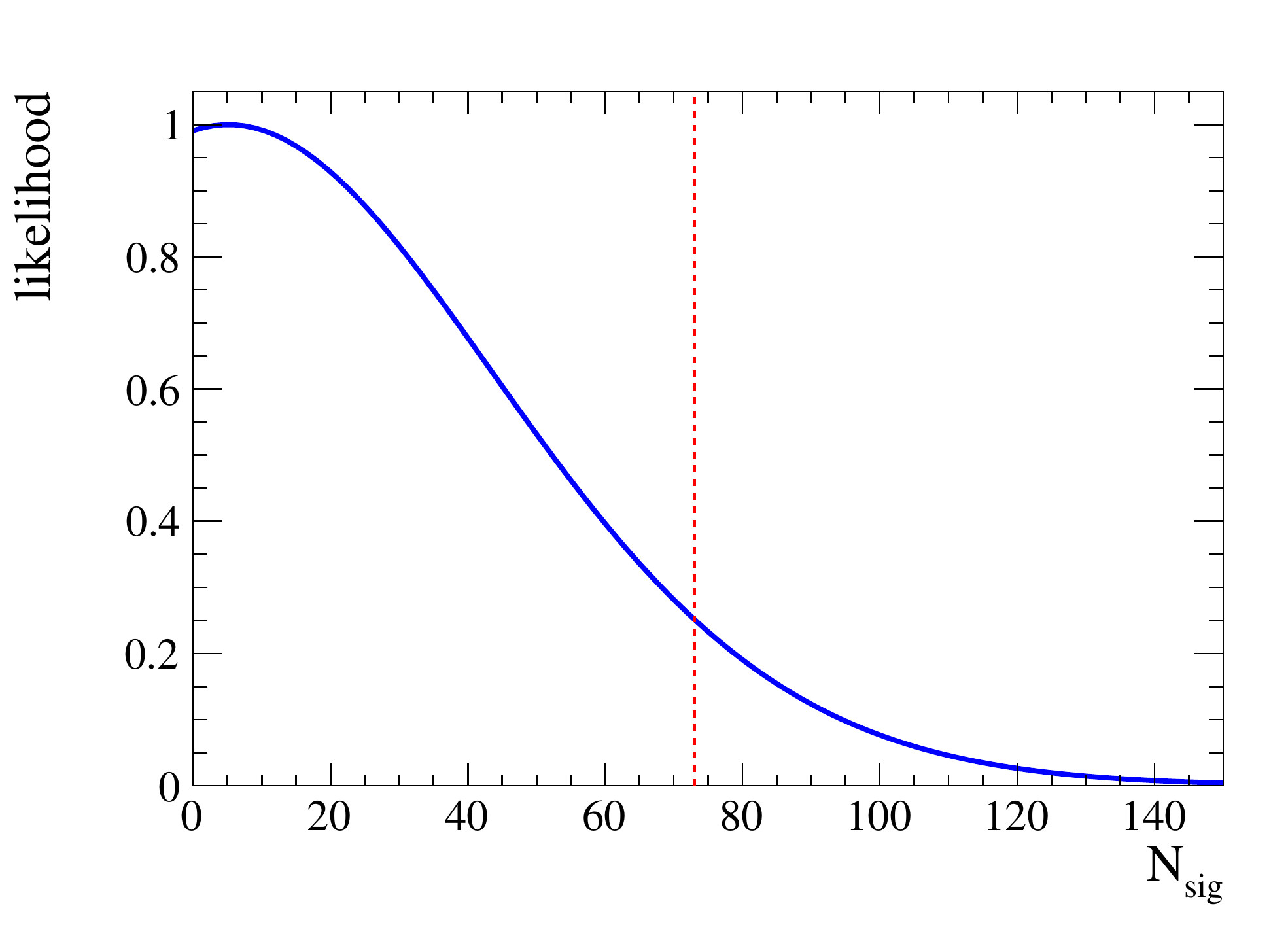}
    \label{subfig:LL:e}
  }
  \subfigure[]{
    \includegraphics[width=.48\textwidth]{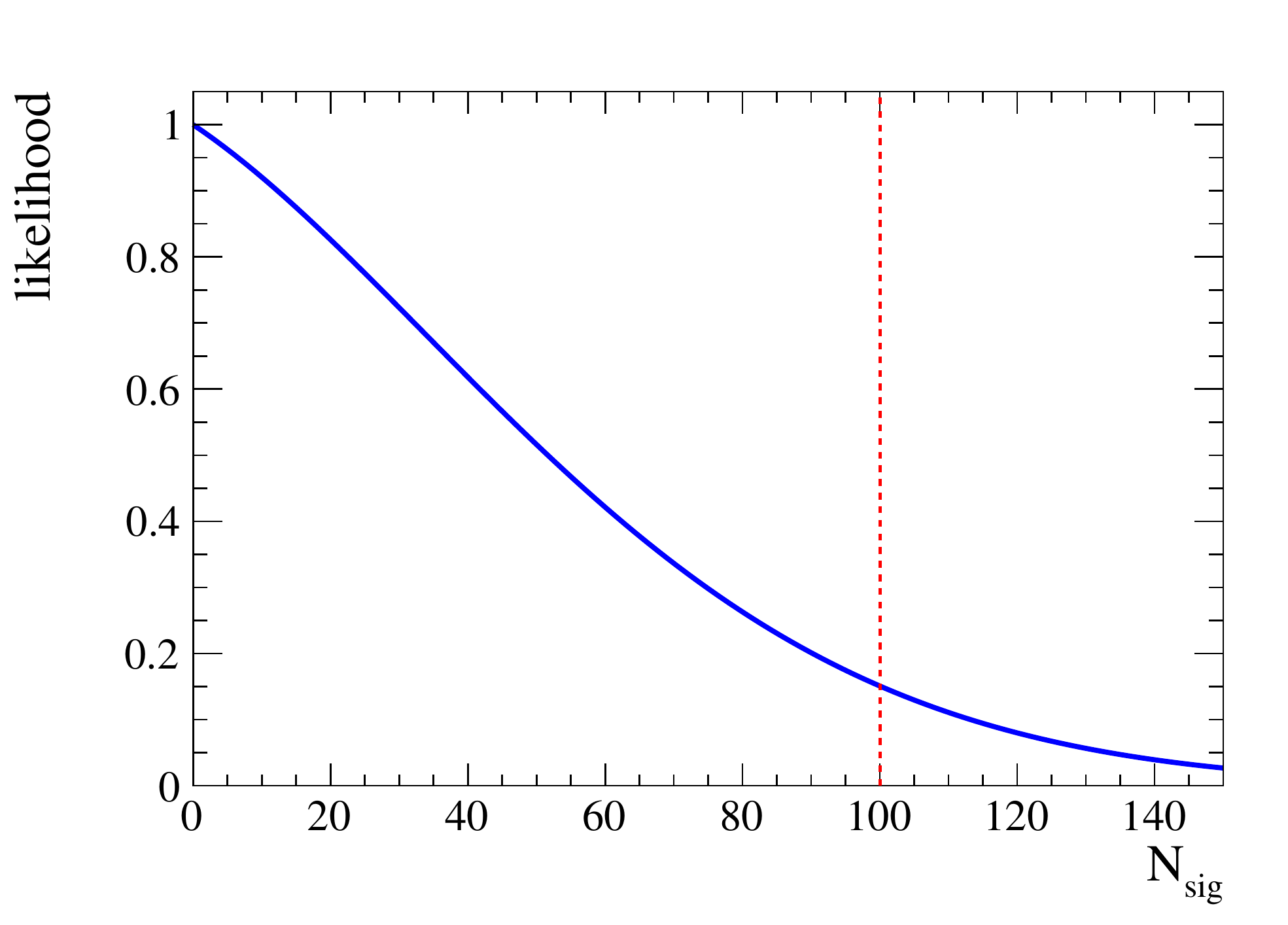}
    \label{subfig:LL:mu}
  }
  \caption{Log-likelihood projections on $N_{\rm sig}$ in \texttt{OnPeak} data, for the electron \subref{subfig:LL:e} and muon \subref{subfig:LL:mu} channel. The dashed line shows the $90\%$ CL limit.}
  \label{fig:LL}
\end{figure}

The significance for both decay channels is well below the threshold for an observation, and hence we can only give an upper limit for their branching fraction.
The pure statistical upper limit is obtained using a Bayesian approach by integrating the likelihood obtained in the previous section from $0$ to $N$, where $N$ denotes the number of signal events at which the integral coincides with $90\%$ of the integral from $0$ to $\infty$. The obtained upper limits for the signal yield at $90\%$ confidence level are
\begin{align}
  N_{\rm sig}(\Bm \ra \LCp \antiproton \en \nueb) &< 73 \\
  N_{\rm sig}(\Bm \ra \LCp \antiproton \mun \numb) &< 100.
\end{align}
The projection of the likelihood onto the signal yield $N_{\rm sig}$ is shown in Fig. \ref{fig:LL}.

\section{Efficiency calculation}\label{sect:eff}

\begin{figure}[H]
  \begin{center}
  \subfigure[]{
    \includegraphics[width=.31\textwidth]{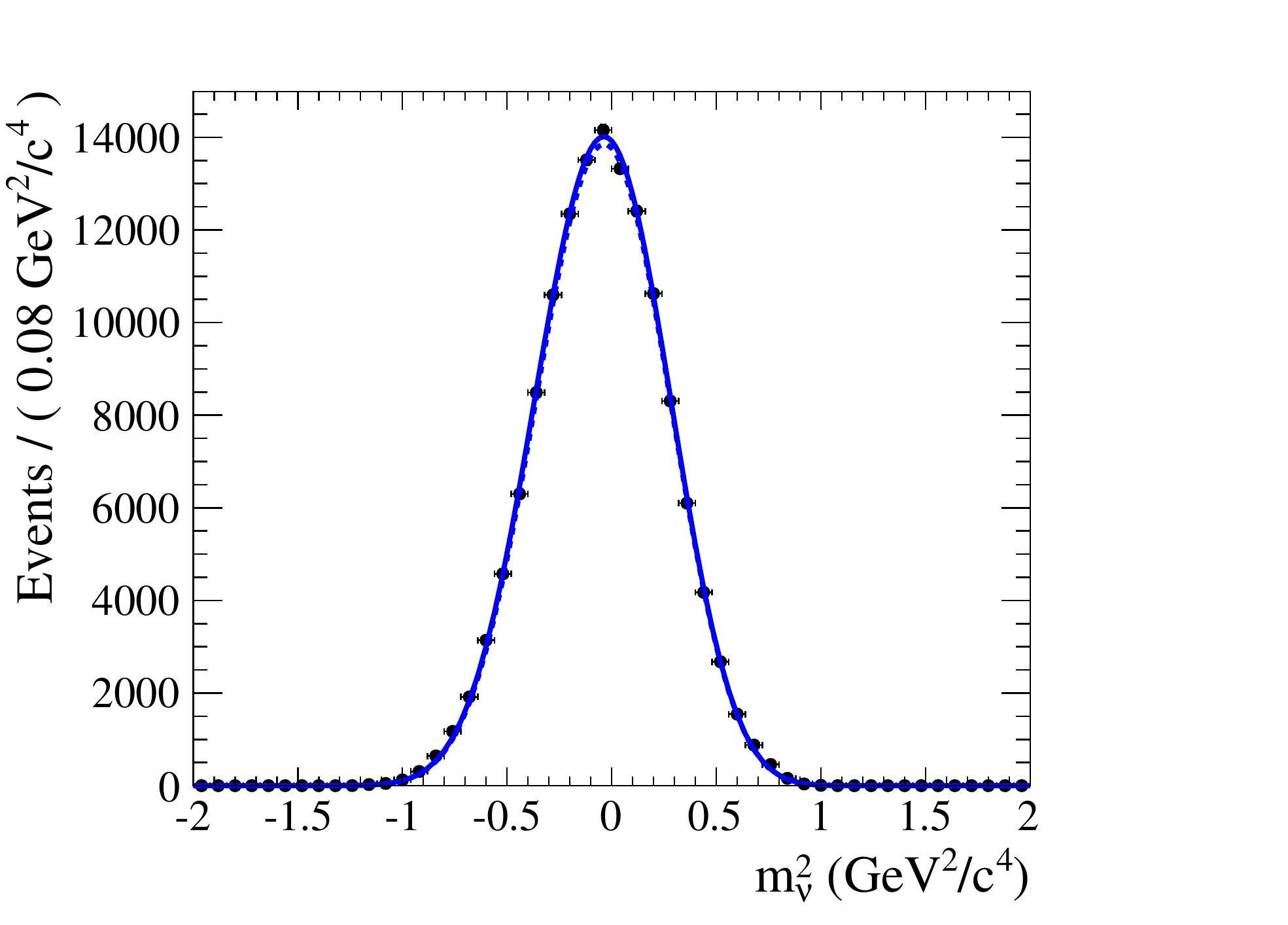}
  }
  \subfigure[]{
    \includegraphics[width=.31\textwidth]{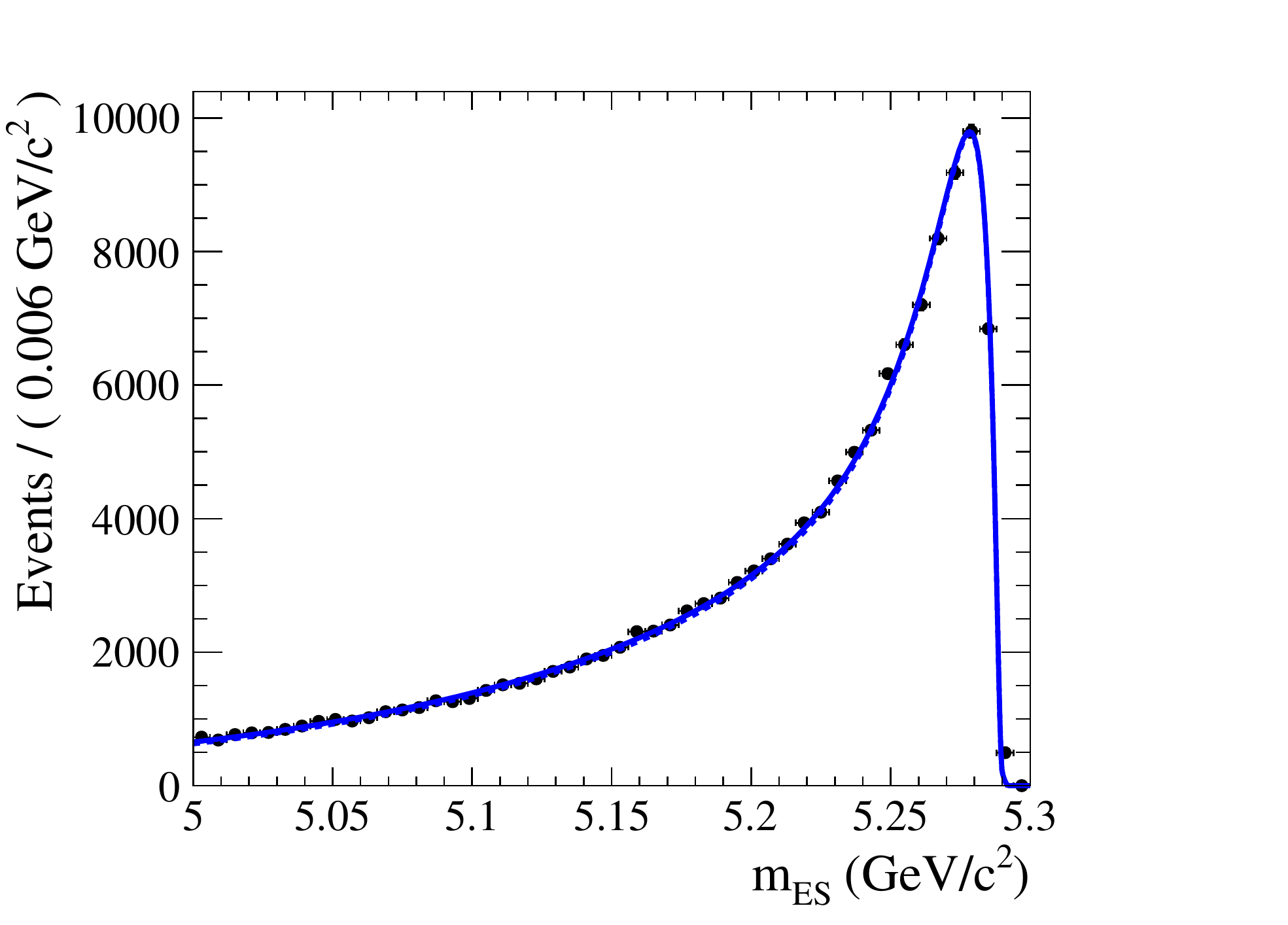}
  }
  \subfigure[]{
    \includegraphics[width=.31\textwidth]{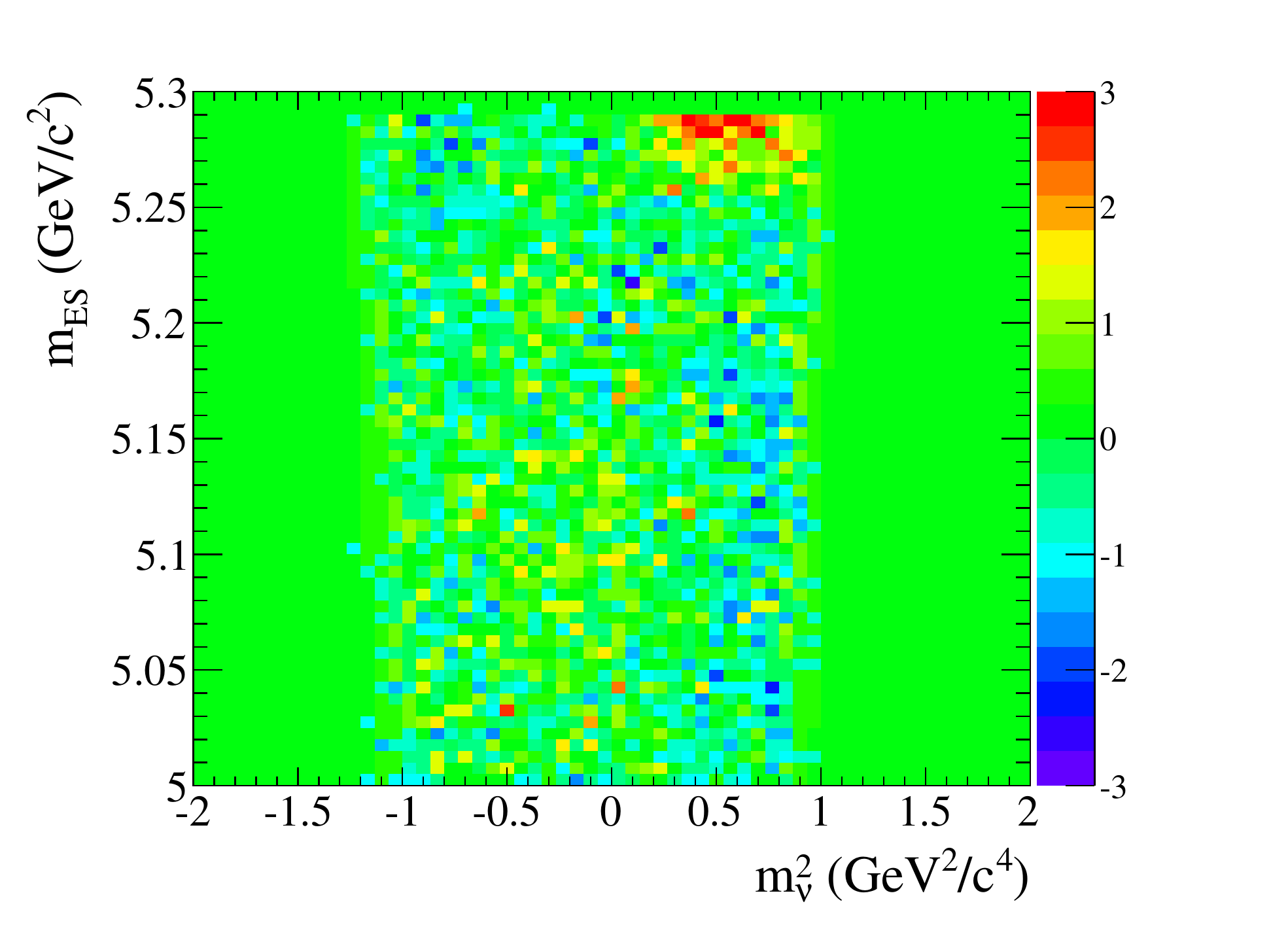}
    \label{subfig:e:Eff:pull}
  }
  \end{center}
  \caption{Fitted $m_{\nu}^2$ and \mes distributions, as well as the two-dimensional pull of the fit to WI signal Monte Carlo data for the electron channel. The dashed line represents the signal contribution.}
  \label{fig:Efficiency:e}
\end{figure}

\begin{figure}[H]
  \begin{center}
  \subfigure[]{
    \includegraphics[width=.31\textwidth]{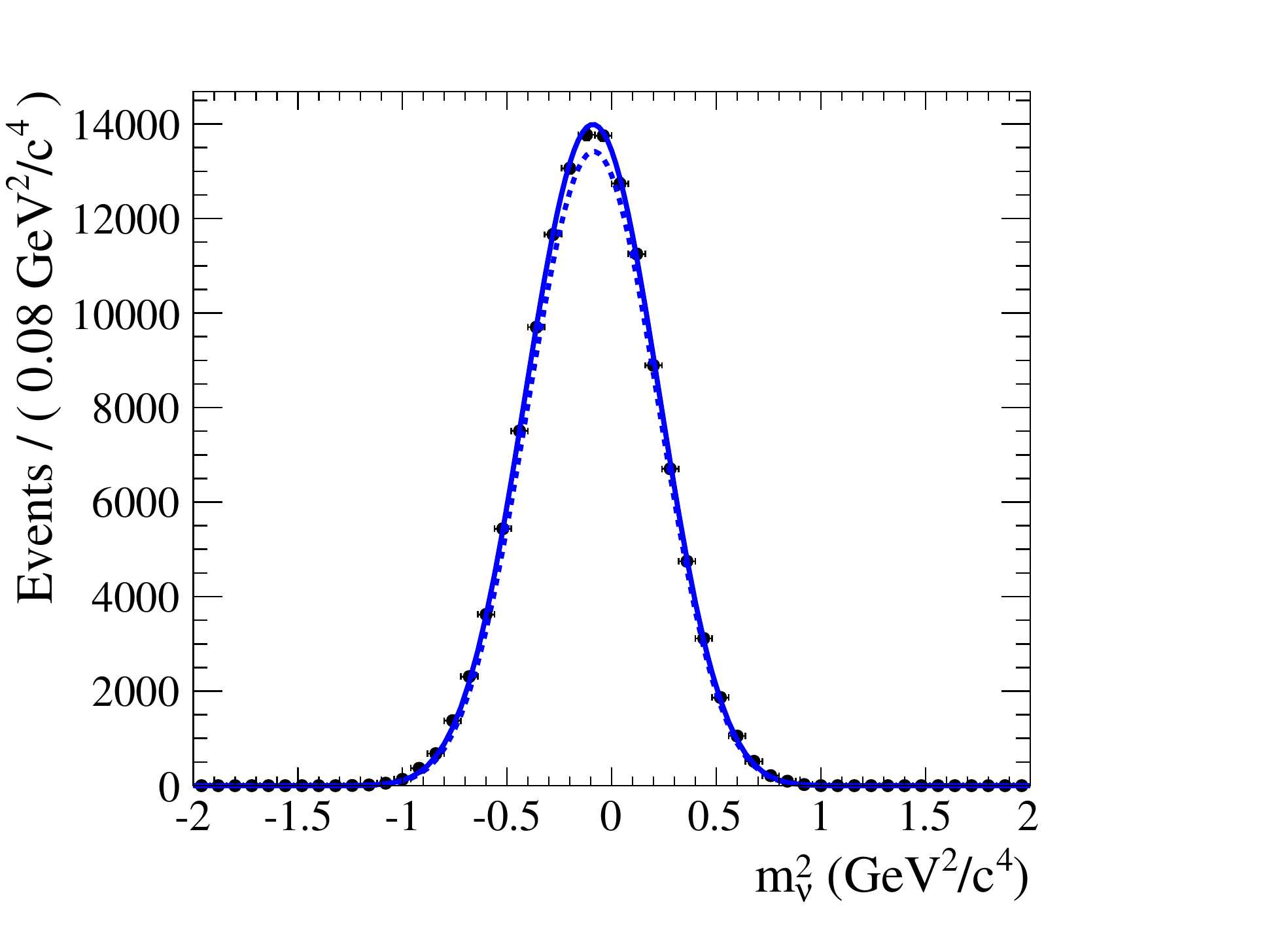}
  }
  \subfigure[]{
    \includegraphics[width=.31\textwidth]{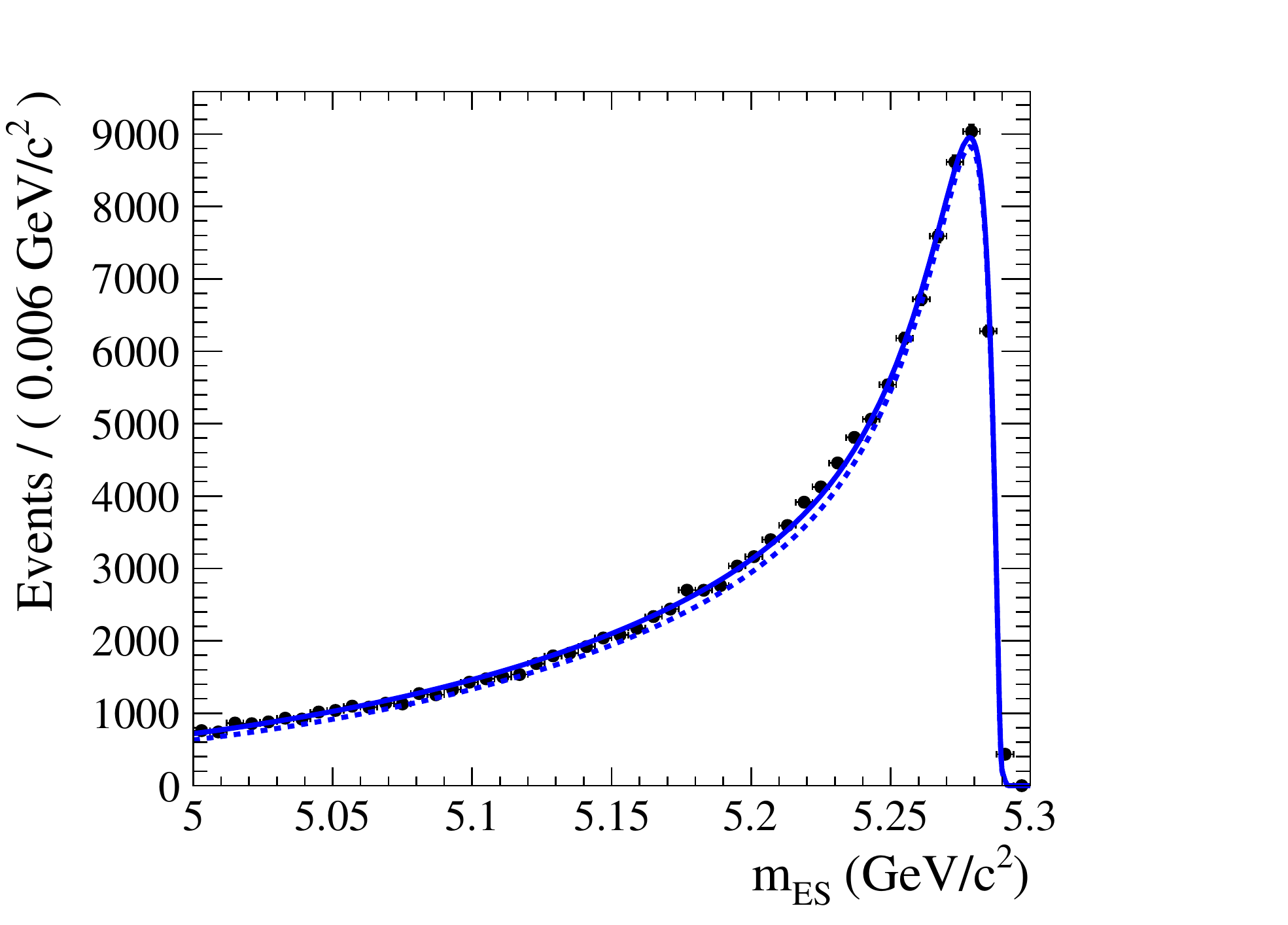}
  }
  \subfigure[]{
    \includegraphics[width=.31\textwidth]{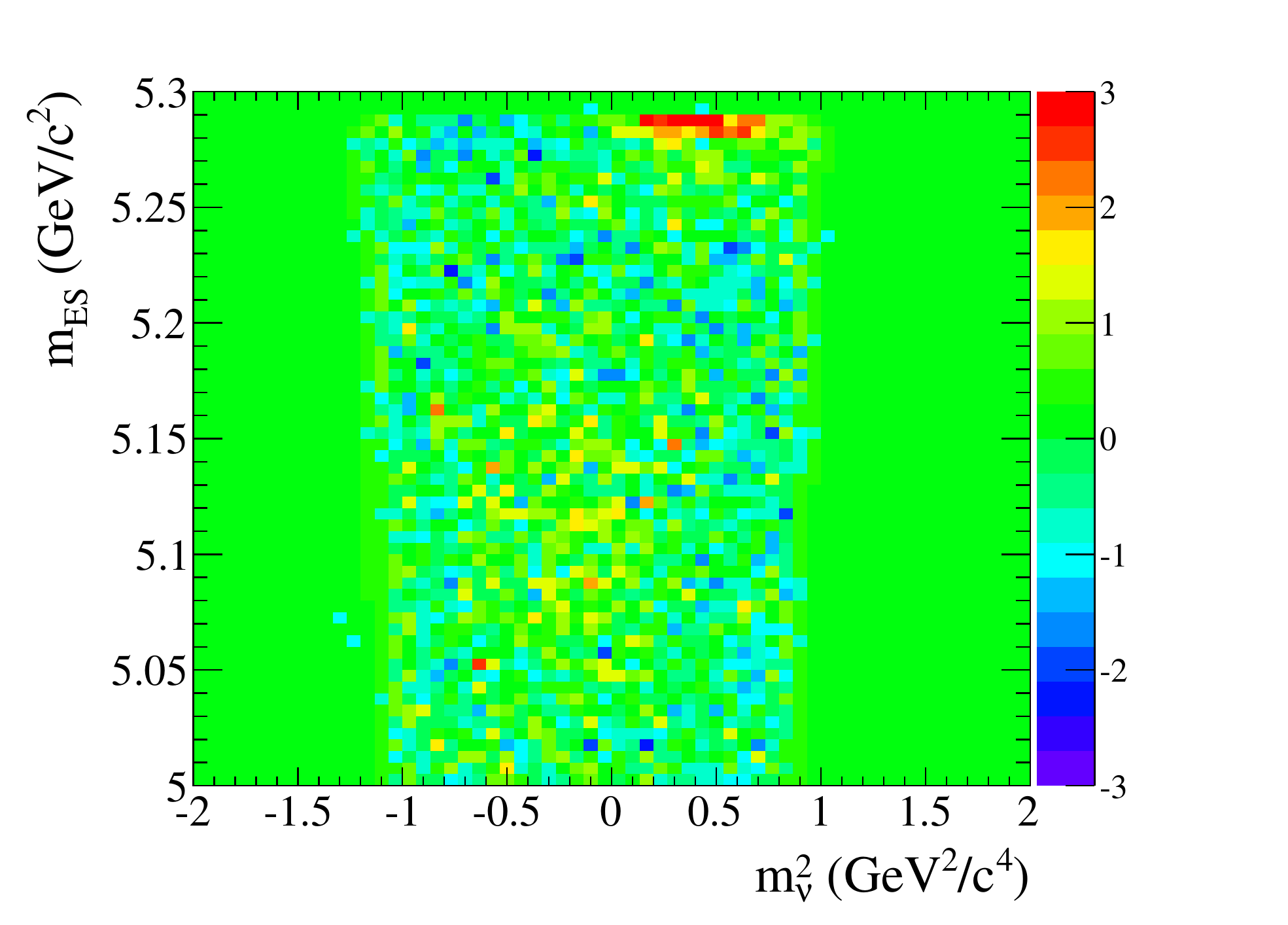}
    \label{subfig:mu:Eff:pull}
  }
  \end{center}
  \caption{Fitted $m_{\nu}^2$ and \mes distributions, as well as the two-dimensional pull of the fit to WI signal Monte Carlo data for the muon channel. The dashed line represents the signal contribution.}
  \label{fig:Efficiency:mu}
\end{figure}

We determine the reconstruction efficiency by fitting WI signal Monte Carlo with the signal and background pdfs described before. Note, that in the determination of the signal pdf only Monte Carlo events with a correctly reconstructed signal decay were used. The fitted $m_{\nu}^2$ and \mes distributions, as well as the two-dimensional pull plot are shown in Fig. \ref{fig:Efficiency:e} and \ref{fig:Efficiency:mu}. As expected only a small background fraction is contained in signal Monte Carlo, and the agreement between Monte Carlo data and the fit function is adequate, as shown in the pull plot.
From the fit we extract a signal yield of $136000 \pm 600$ for the electron channel, and $127800 \pm 600$ for the muon channel. Together with the number of generated events (given in Table \ref{tab:SPmodesSignal}) we obtain reconstruction efficiencies of
\begin{align}
  \begin{split}
  \epsilon (\Bm \ra \LCp \antiproton \en \nueb)  &= (4.7422 \pm 0.0021)\% \\
  \epsilon (\Bm \ra \LCp \antiproton \mun \numb) &= (4.1248 \pm 0.0020)\%.
  \end{split}
  \label{eq:eff}
\end{align}

\section{Systematic Uncertainties}

Since the previously determined signal yields are compatible with $0$ the use of relative systematic uncertainties is not possible. Instead, we have to distinguish two types of systematic uncertainties and treat them separately. The first type are systematic uncertainties  affecting dominantly the signal yield in \texttt{OnPeak} data, but represent a second order effect on the efficiency, like the used fit function. The second type of systematic uncertainties affect the reconstruction efficiency.

\subsection{Fit function}\label{sys:fitFunc}

The used fit function influences mainly the signal yield obtained in the fit to experimental data. On the efficiency the effect is expected to be small due to the negligible background contribution. Thus, we concentrate on \texttt{OnPeak} data for this systematic.
Therefore, we decide to change the signal description in $m_{\nu}^2$ from a Cruijff to a Gaussian and for \mes we leave the ARGUS shape parameter $c$ floating. The parameters of the Gaussian are
\begin{alignat*}{3}
    \mu &= -&&0.0477 \pm 0.0010 \\
    \sigma &= &&0.3065 \pm 0.0007
\end{alignat*} 
for the electron, and
\begin{alignat*}{3}
    \mu &= -&&0.0899 \pm 0.0010 \\
    \sigma &= &&0.2963 \pm 0.0007
\end{alignat*}
for the muon channel. Fixing these parameters to the obtained values, while leaving $c$ floating we obtain
\begin{alignat*}{3}
    c &= -&&30\pm 12 \\
    N_{\rm sig} &= &&10 \pm 26
\end{alignat*} 
for the electron, and
\begin{alignat*}{3}
    c &= -&&13 \pm 15 \\
    N_{\rm sig} &= -&&20\pm 50
\end{alignat*}
for the muon channel. Integrating the log-likelihood as before we obtain as upper limits for the signal yield
\begin{align}
  N_{\rm sig}(\Bm \ra \LCp \antiproton \en \nueb) &< 81 \\
  N_{\rm sig}(\Bm \ra \LCp \antiproton \mun \numb) &< 130.
\end{align}
We decide to use these values for the upper limit, rather than those obtained in section \ref{sect:UL:stat}, as a conservative limit.

\subsection{\B counting, tracking and particle identification}

Three standard sources of systematic uncertainties are the \B counting, the charged track reconstruction and the particle identification. The uncertainty from the \B counting can be directly taken from the number of \BBb pairs, given in eq. \eqref{eq:Nbb}, and accounts for an uncertainty of $0.6\%$.

The tracking uncertainty is given by the tracking group \cite{track_eff}, and amounts to a relative uncertainty of $0.128\%$ per track for the used tracking list \texttt{Charged Tracks}. Adding up these uncertainties in quadrature leads to an overall tracking uncertainty of $0.3\%$.

For the uncertainty from particle identification we take advantage of the correction tables applied to Monte Carlo events. These correction tables try to balance the differences between data and Monte Carlo events. A standard method to obtain the systematic uncertainties is to turn off these correction tables and compare the number of reconstructed events with each other. The difference in these numbers can be taken as systematic uncertainty from particle identification. With the correction tables enabled, we obtain signal yields of
\begin{align}
  N_{\rm sig}(\Bm \ra \LCp \antiproton \en \nueb) &= 136000 \pm 600,\\
  N_{\rm sig}(\Bm \ra \LCp \antiproton \mun \numb) &= 127800 \pm 600
\end{align}
on WI signal Monte Carlo, as described in section \ref{sect:eff} for the efficiency calculation. Repeating this fit on WI signal Monte Carlo events without the correction tables return signal yields of
\begin{align}
  N_{\rm sig}(\Bm \ra \LCp \antiproton \en \nueb) &= 141800 \pm 600,\\
  N_{\rm sig}(\Bm \ra \LCp \antiproton \mun \numb) &= 131500 \pm 700.
\end{align}
The differences in the signal yield lead to systematic uncertainties of $4.3\%$ for the electron and $2.9\%$ for the muon channel. These uncertainties are mainly due to the use of the $\texttt{Tight}$ particle ID lists in the \B and the neutrino reconstruction. While the influence on the \B reconstruction directly affect $m_{\nu}^2$ since it depends on the energy of the $Y$ system, the neutrino reconstruction affects the loose \DeltaE selection cut. For the latter upper limit we have to neglect this uncertainty, since it leads to a larger efficiency, and in turn would reduce the upper limit.

\subsection{Model dependence}

For the model dependence we have two possibilities, either we determine the efficiency for phase space Monte Carlo, or for a WI Monte Carlo with altered properties of the pole at threshold.

\begin{figure}[h]
  \subfigure[]{
    \includegraphics[width=.32\textwidth]{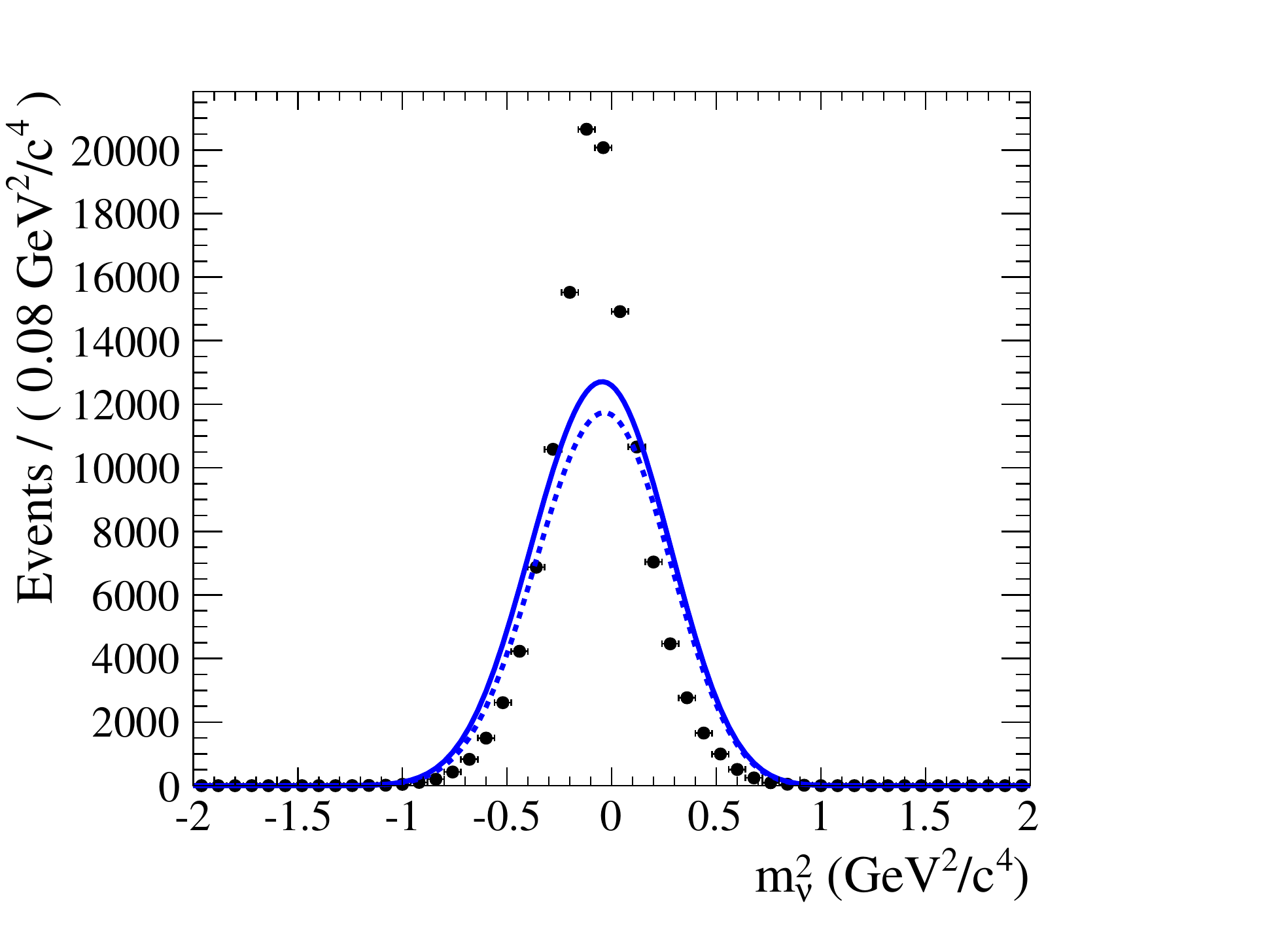}
    \includegraphics[width=.32\textwidth]{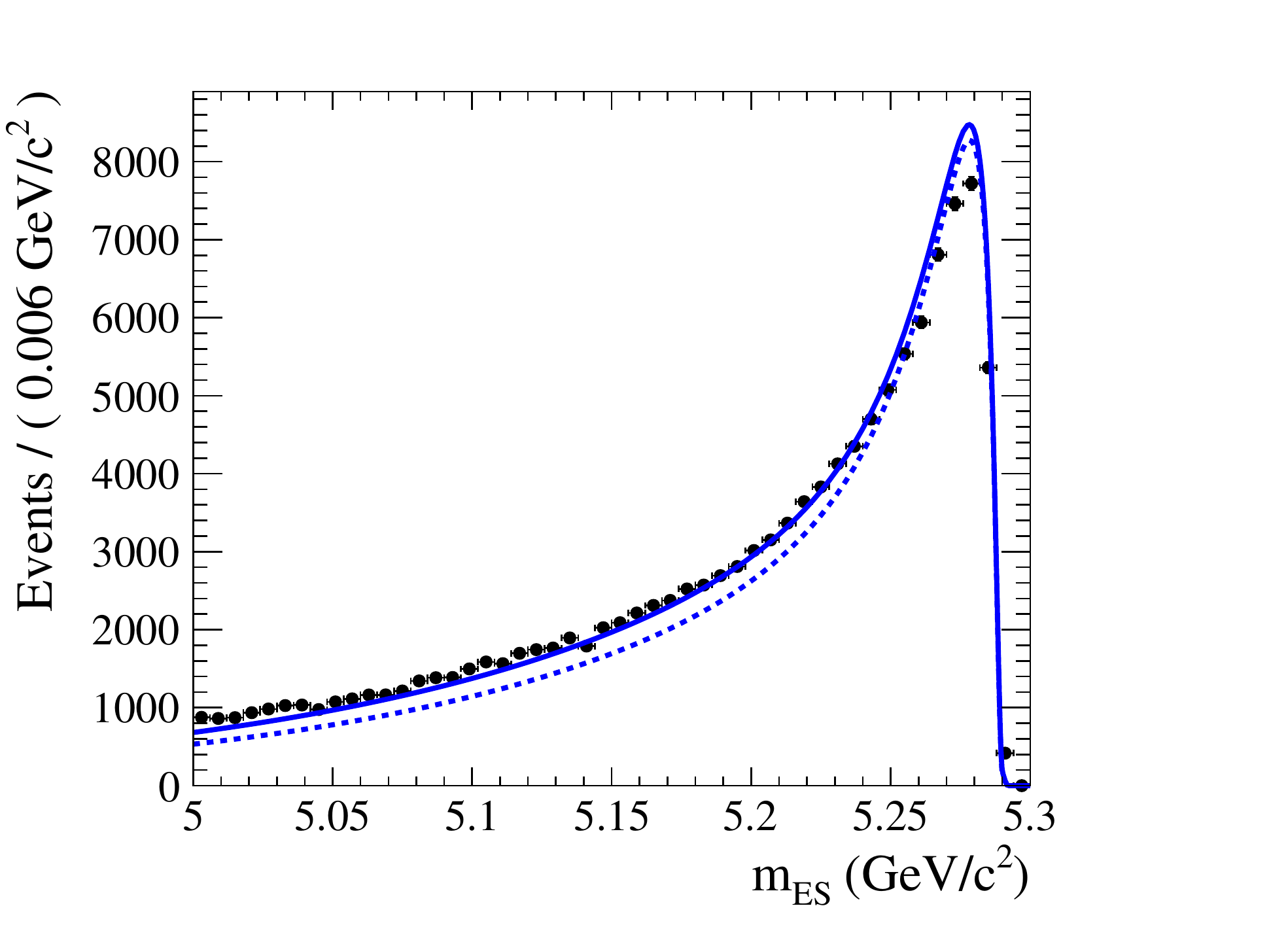}
    \includegraphics[width=.32\textwidth]{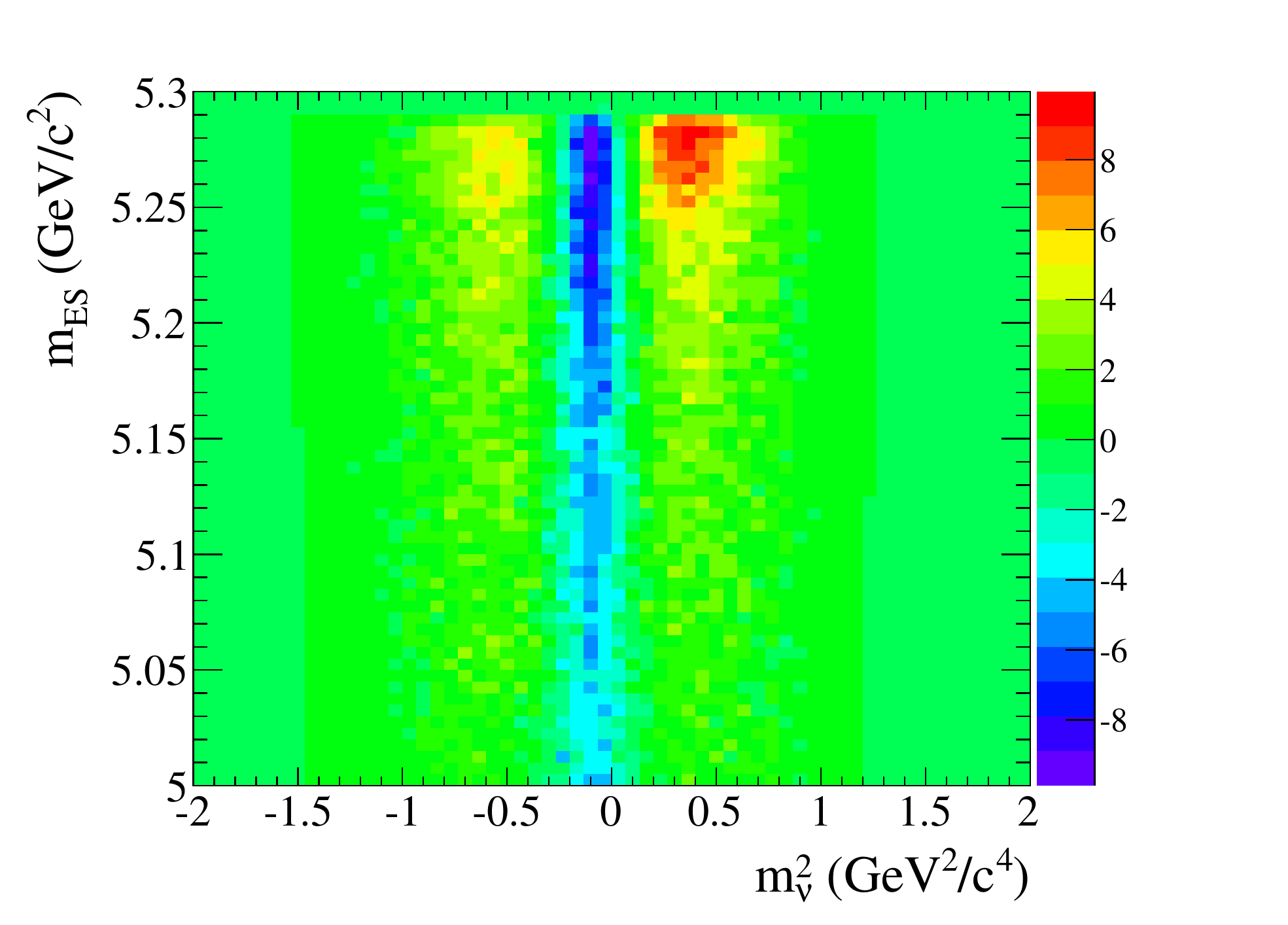}
    \label{subfig:PhSpElectron} 
  }
  \subfigure[]{
    \includegraphics[width=.32\textwidth]{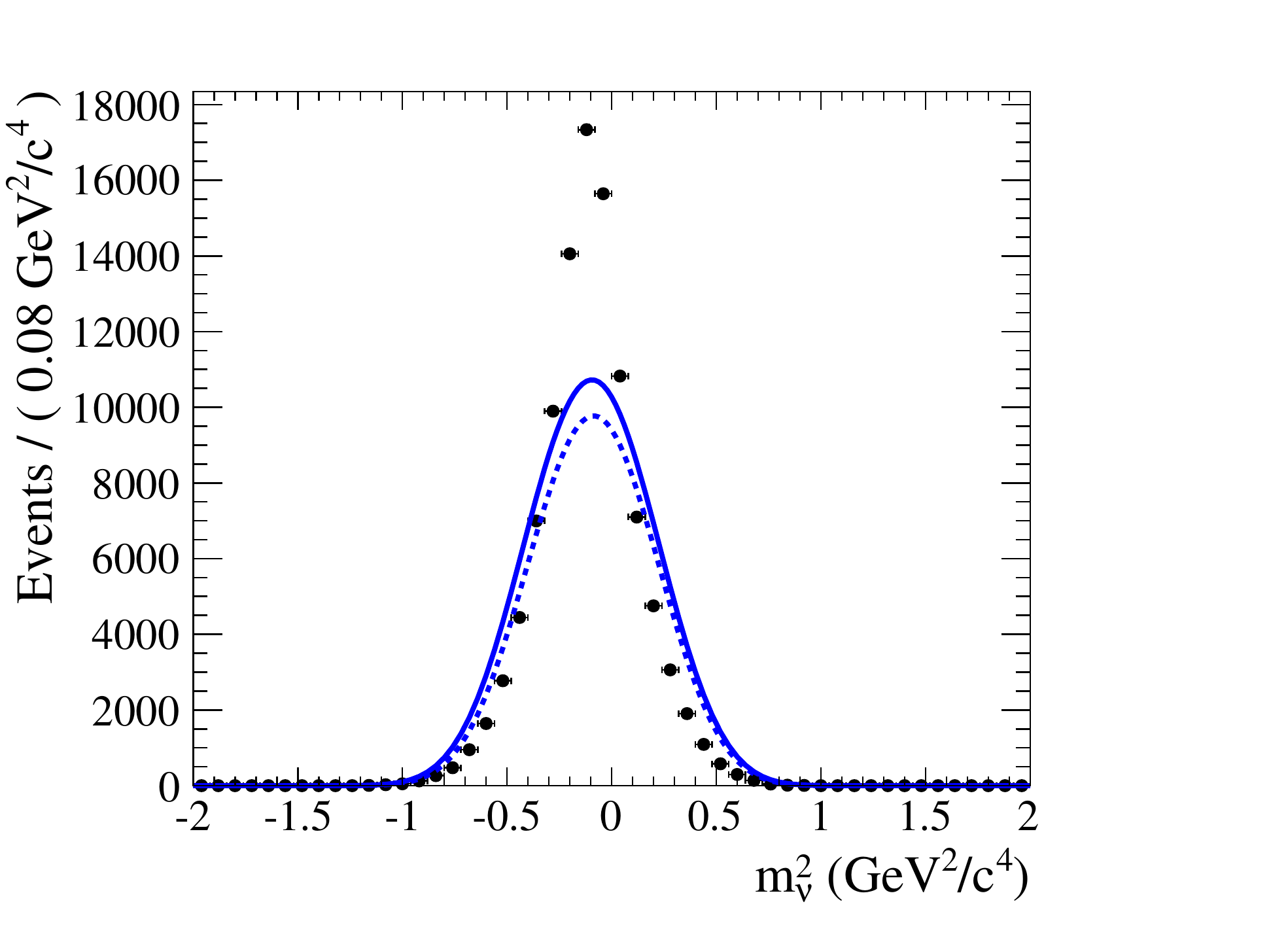}
    \includegraphics[width=.32\textwidth]{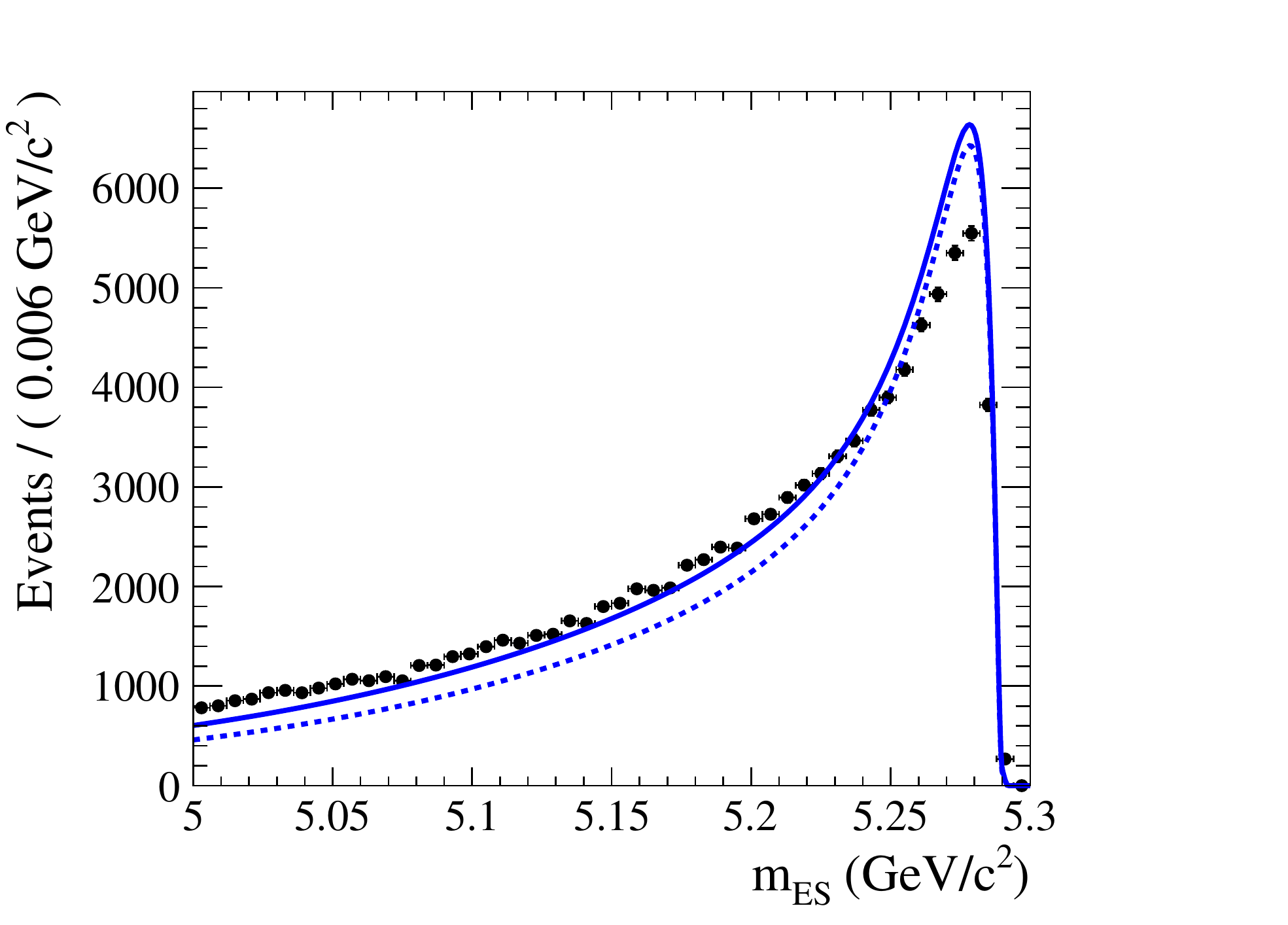}
    \includegraphics[width=.32\textwidth]{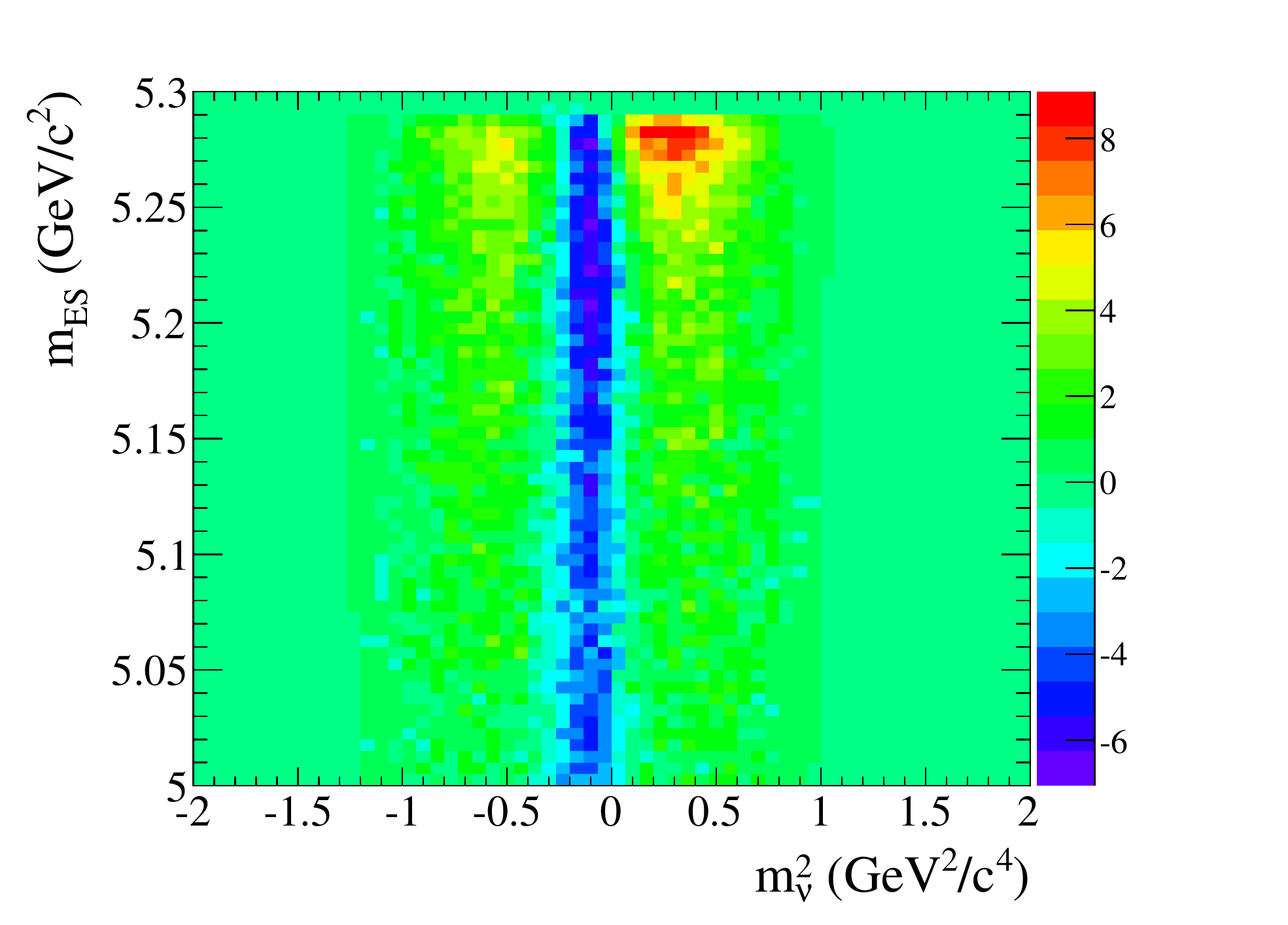}
    \label{subfig:PhSpMuon} 
  }
  \caption{$m_{\nu}^2$ and \mes projections as well as the two-dimensional pull distribution for the fit to Monte Carlo events distributed according to phase space. \subref{subfig:PhSpElectron} for the electron and \subref{subfig:PhSpMuon} for the muon channel.}
  \label{fig:PhSpFits}
\end{figure}

Using phase-space Monte Carlo instead of our WI model for efficiency calculation we obtain signal yields of $115300 \pm 600$ for the electron and $93100 \pm 600$ for the muon channel. This translates into efficiencies of $2.8\%$ and $2.2\%$, respectively. A comparison with the efficiencies given in section \ref{sect:eff}, this translates into systematic uncertainties of
\begin{align}
  u_{\rm model} (\Bm \ra \LCp \antiproton \en \nueb)  &= 41\% \\
  u_{\rm model} (\Bm \ra \LCp \antiproton \mun \numb) &= 47\%.
\end{align}
The fits are shown in Fig. \ref{fig:PhSpFits}.

\begin{figure}[h]
  \begin{center}
    \includegraphics[width=.6\textwidth]{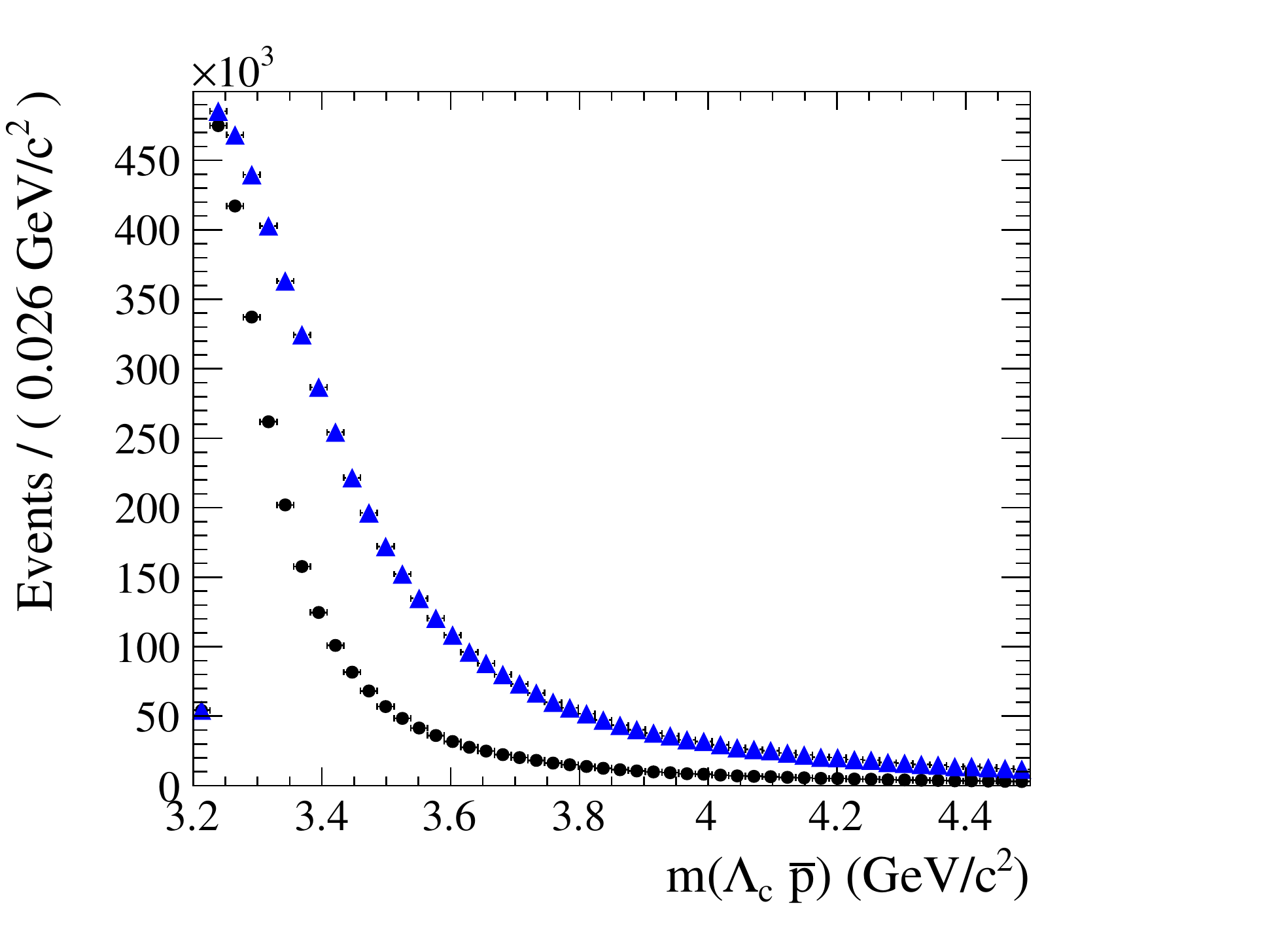}
  \end{center}
  \caption{Invariant $m(\LCp \antiproton)$ mass for WI signal Monte Carlo on generator level. The black dots are the original enhancement with a width of $200\mevcc$, while the blue triangles show the reweighted events with a width of $400\mevcc$.}
  \label{fig:ModesSysTLcomp}
\end{figure}

\begin{figure}[h]
  \subfigure[]{
    \includegraphics[width=.32\textwidth]{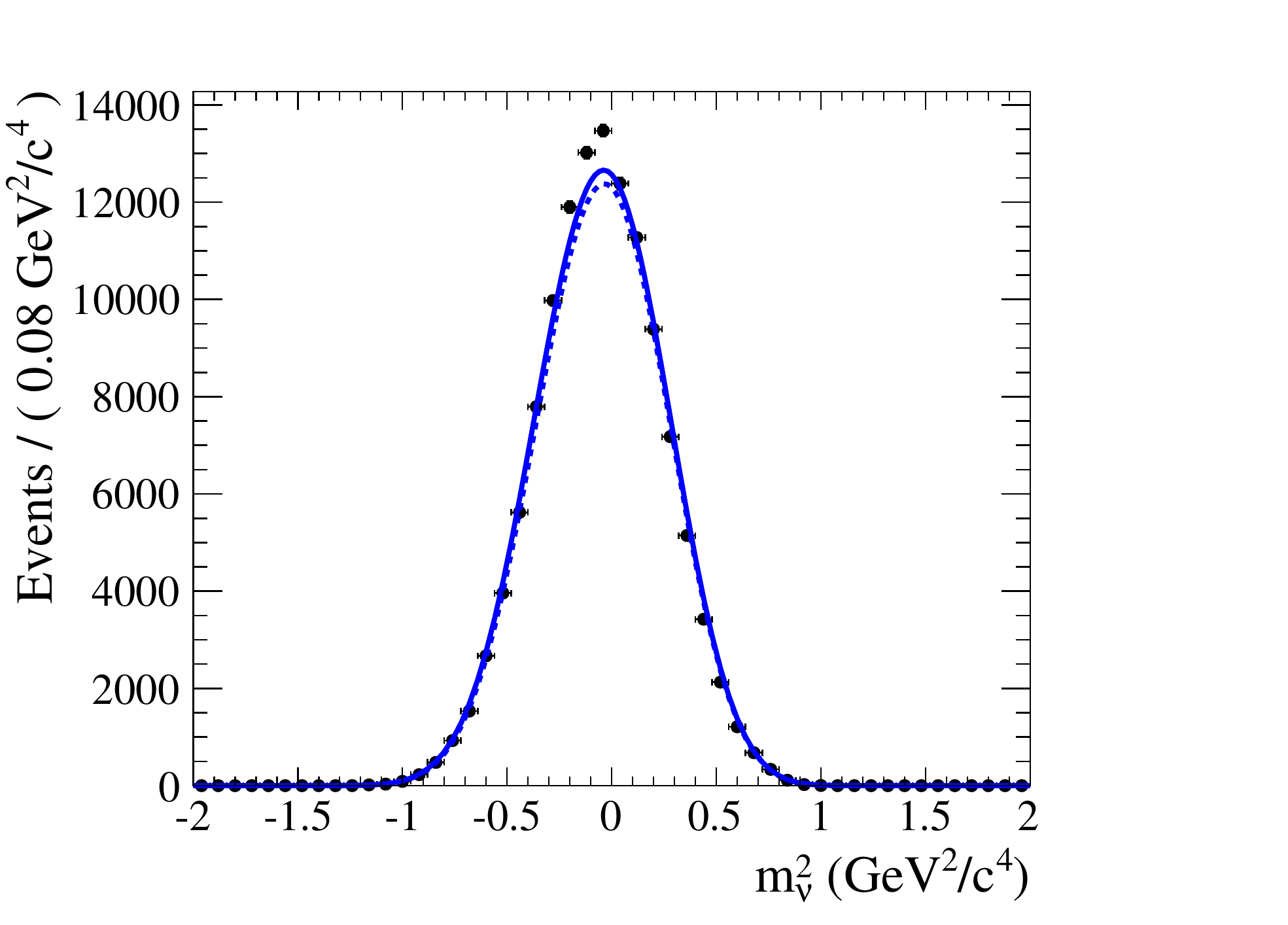}
    \includegraphics[width=.32\textwidth]{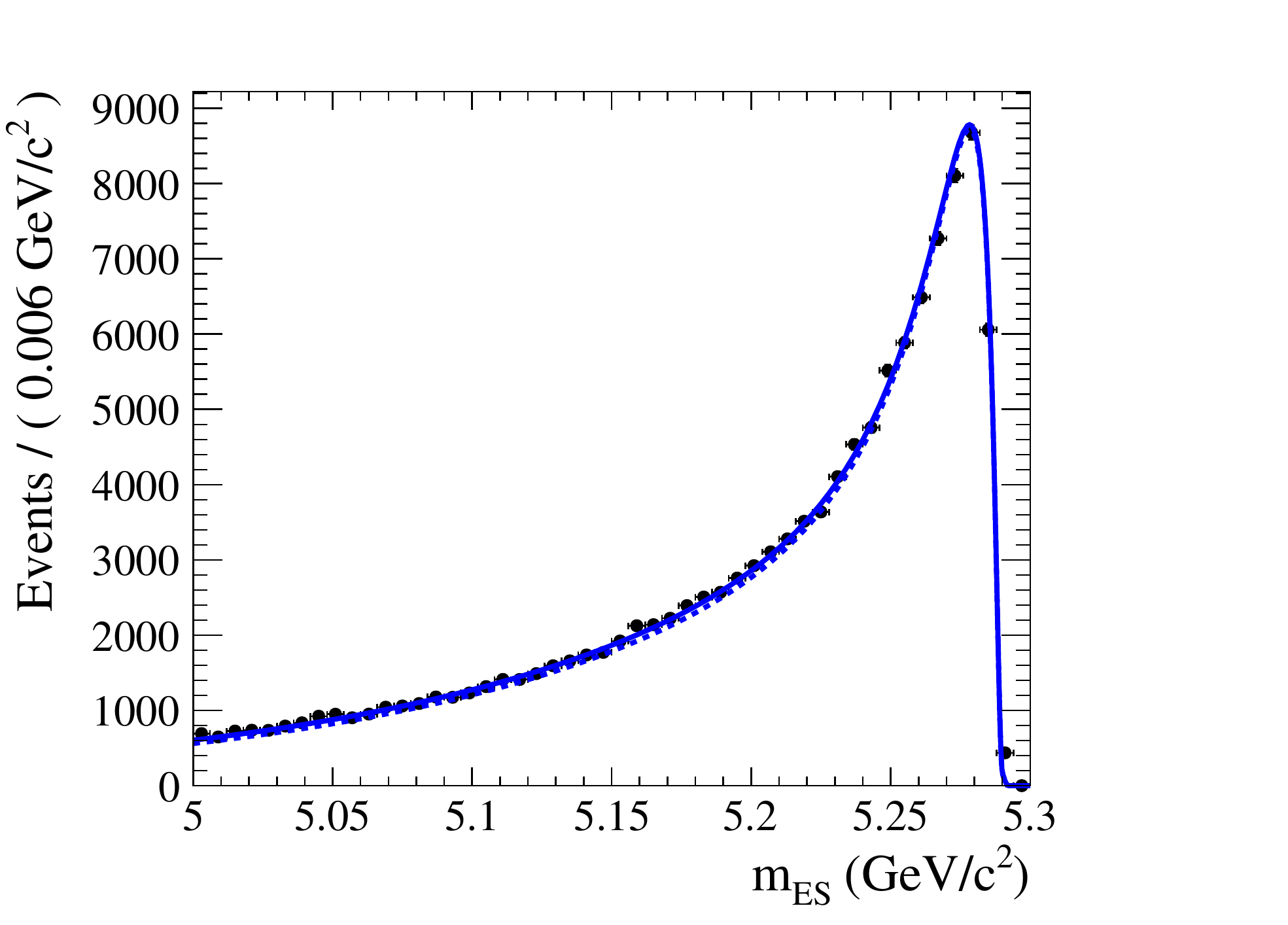}
    \includegraphics[width=.32\textwidth]{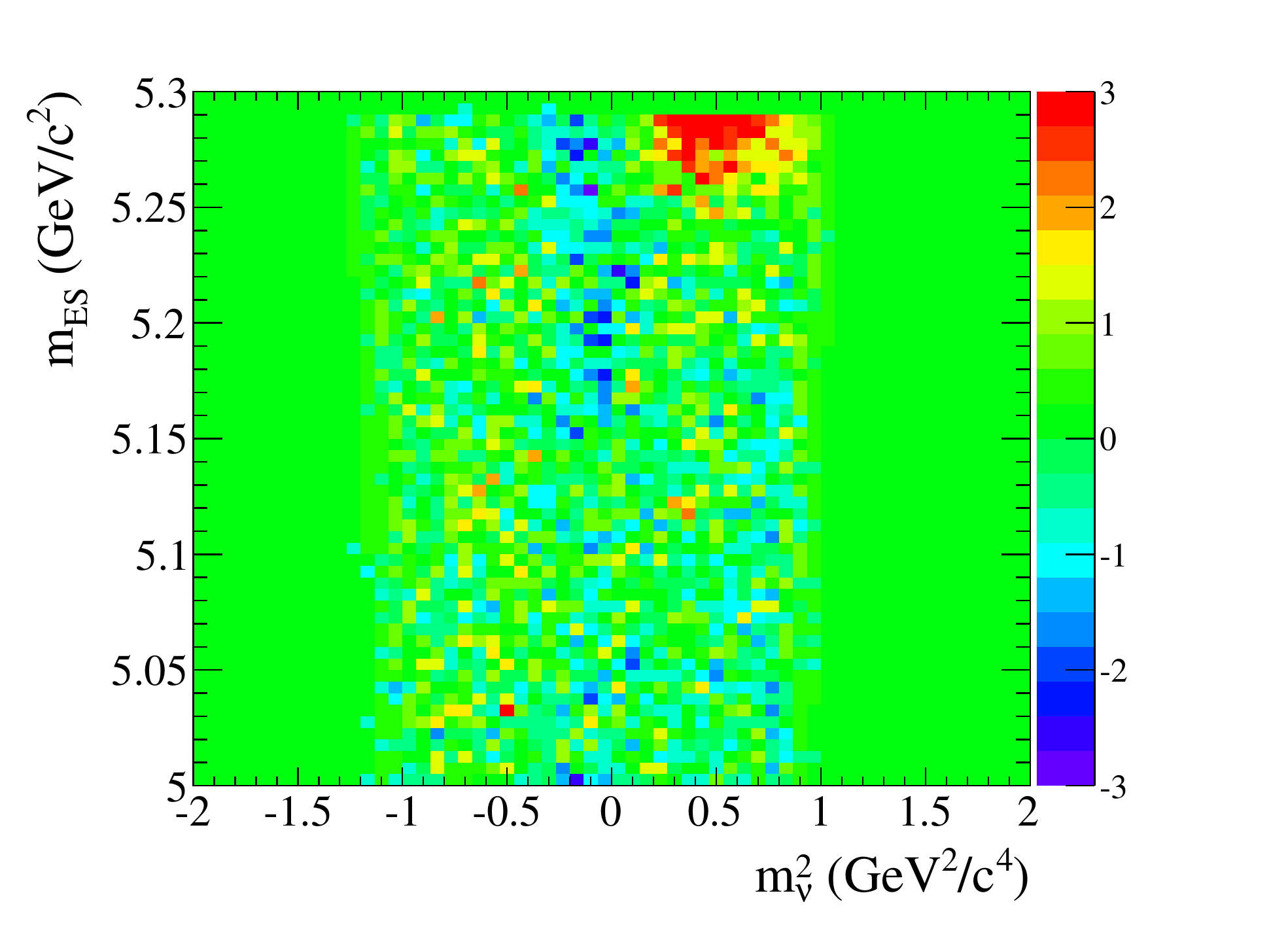}
    \label{subfig:ModelElectron} 
  }
  \subfigure[]{
    \includegraphics[width=.32\textwidth]{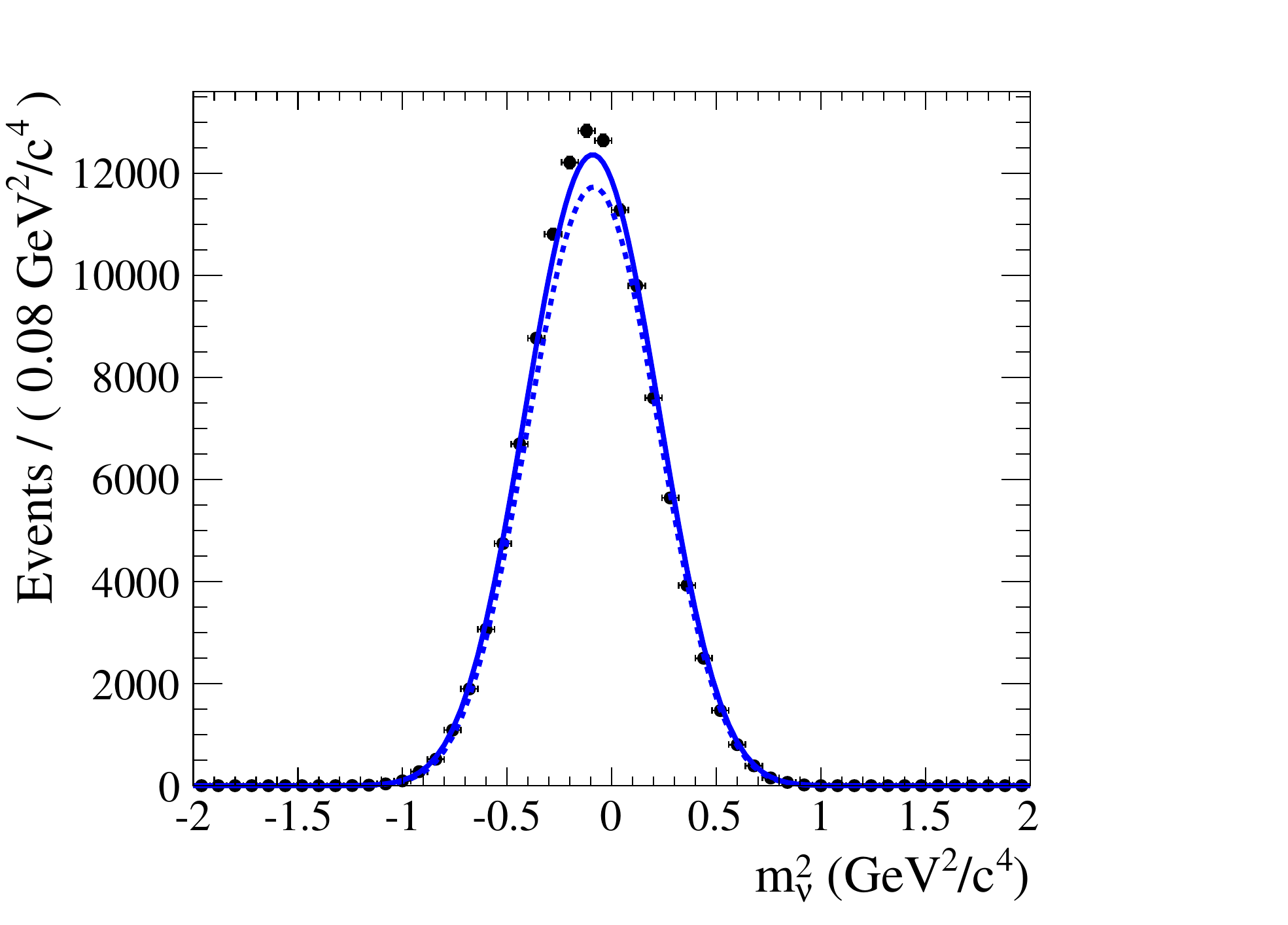}
    \includegraphics[width=.32\textwidth]{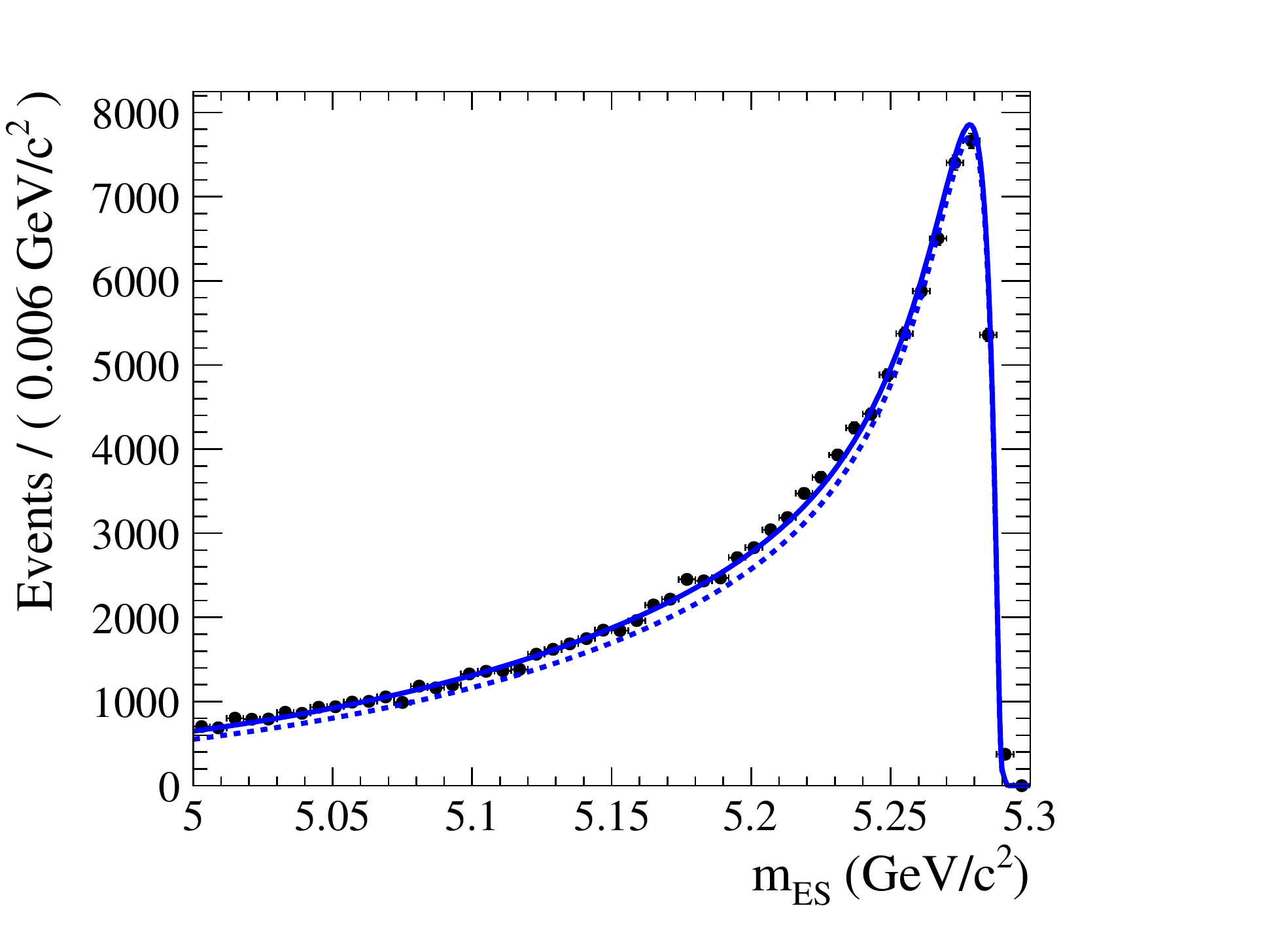}
    \includegraphics[width=.32\textwidth]{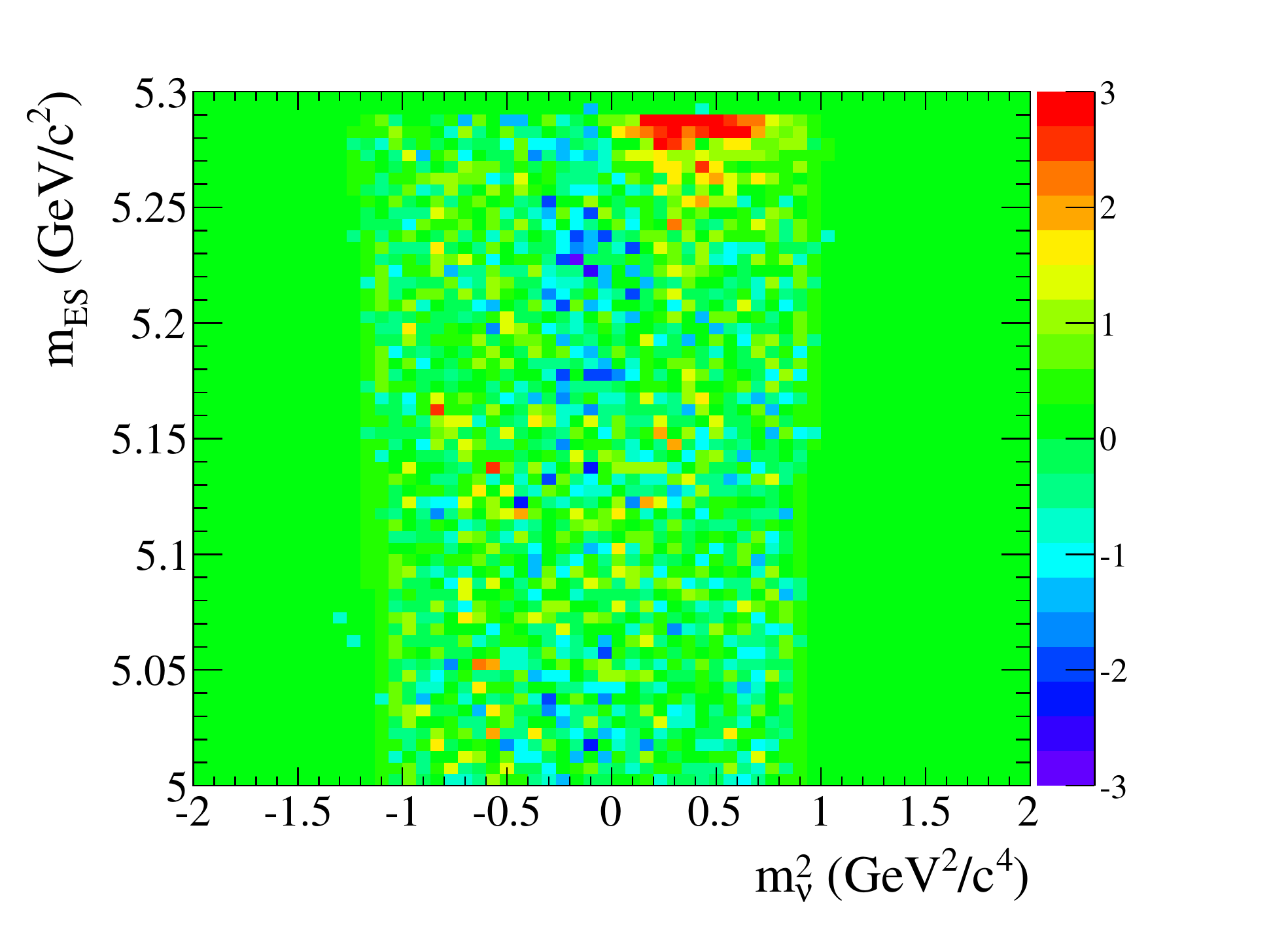}
    \label{subfig:ModelMuon} 
  }
  \caption{$m_{\nu}^2$ and \mes projections as well as the two-dimensional pull distribution for the fit to reweighted Monte Carlo events. \subref{subfig:ModelElectron} for the electron and \subref{subfig:ModelMuon} for the muon channel.}
  \label{fig:ModelFits}
\end{figure}

For the second option we re-weight WI signal Monte Carlo to a threshold enhancement at the same peak position ($3.225\gevcc$), but with a doubled width of $400\mevcc$. A  comparison of the $200\mevcc$ and $400\mevcc$ broad $m(\LCp \antiproton)$ enhancement on generator level is shown in Fig. \ref{fig:ModesSysTLcomp}.
The reweighted number of generated events is $2797130$ for the electron and $3018900$ for the muon channel. To determine the analysis efficiency for these reweighted events we repeat the fit described in section \ref{sect:eff}. Here, we obtain a signal yield of $121600 \pm 500$ for the electron channel, and $1306300 \pm 600$ for the muon channel. This translates into the efficiencies
\begin{align}
  \epsilon_W (\Bm \ra \LCp \antiproton \en \nueb)  &= 4.34\% \\
  \epsilon_W (\Bm \ra \LCp \antiproton \mun \numb) &= 4.32\%.
\end{align}
Taking the difference to the non-weighted efficiencies, given in sect. \ref{sect:eff}, as systematic uncertainties we obtain as model-dependent systematic uncertainties
\begin{align}
  u_{\rm model} (\Bm \ra \LCp \antiproton \en \nueb)  &= 8.4\% \\
  u_{\rm model} (\Bm \ra \LCp \antiproton \mun \numb) &= 4.9\%.
\end{align}
The fitted $m_{\nu}^2$ and \mes projections as well as the two-dimensional pull plots are shown in Fig. \ref{fig:ModelFits}. As can be seen from the distributions the signal pdf describes the weighted signal well.

In conclusion we have two sets of systematic uncertainties to choose from. The most conservative comes from phase space Monte Carlo, since this represents an absolute border of the model dependence. The re-weighted Monte Carlo can be seen as more realistic, but at the cost of an unknown confidence level for this uncertainty. In order to preserve a $90\%$ confidence level we decide to use the systematic uncertainties determined via phase space Monte Carlo events.

\section{Upper Limit}

With the upper limit for the signal yield, determined in sect. \ref{sys:fitFunc} we can determine the upper limit for the product branching fraction $\BR(\Bm \ra \LCp \antiproton \ellm \nulb)\cdot \BR(\LCp \ra \proton \Km \pip)$ as
\begin{align}
  \frac{N_{\rm sig}}{\epsilon \cdot N_{\BBb}},
\end{align}
with the efficiencies $\epsilon$, given in eq. \eqref{eq:eff}, and the  number of \BBb pairs, given in eq. \eqref{eq:Nbb}. This leads to
\begin{align}
  \begin{split}
    \BR(\Bm \ra \LCp \antiproton \en \nueb)\cdot \BR(\LCp \ra \proton \Km \pip) &< 3.6 \times 10^{-6}\\
    \BR(\Bm \ra \LCp \antiproton \mun \numb)\cdot \BR(\LCp \ra \proton \Km \pip) &< 6.7 \times 10^{-6},
  \end{split}
\end{align}
neglecting the small statistical uncertainties on the efficiency and number of \BBb pairs.
These values for the upper limit are calculated at $90\%$ confidence level, but without the systematic uncertainties affecting the efficiency. Here, we only have to consider the model uncertainties which reduce the efficiency, and lead in turn to a larger upper limit. This uncertainty amounts to $41\%$ for the electron, and $47\%$ for the muon channel. Due to these large uncertainties we have to work with asymmetric error ranges and modify the efficiency with a factor $1-u_{\epsilon}$ of $0.59$ for the electron, and $0.53$ for the muon case.
\begin{align}
  \begin{split}
    \BR(\Bm \ra \LCp \antiproton \en \nueb)\cdot \BR(\LCp \ra \proton \Km \pip) &< 6.1 \times 10^{-6}\\
    \BR(\Bm \ra \LCp \antiproton \mun \numb)\cdot \BR(\LCp \ra \proton \Km \pip) &< 12.6 \times 10^{-6}.
  \end{split}
\end{align}
Correcting for the \LCp branching fraction, by dividing by the world average value of $5\%$, given in \cite{PDG:2012}, we obtain
\begin{align}
  \begin{split}
    \BR(\Bm \ra \LCp \antiproton \en \nueb)\cdot \frac{\BR(\LCp \ra \proton \Km \pip)}{5\%} &< 1.2 \times 10^{-4}\\
    \BR(\Bm \ra \LCp \antiproton \mun \numb)\cdot \frac{\BR(\LCp \ra \proton \Km \pip)}{5\%} &< 2.5 \times 10^{-4}.
  \end{split}
\end{align}
We decide to use this notation, to allow for a later correction of the upper limit for a changed \LCp branching fraction, since the current value has a large model-dependent uncertainty. If we would take the recent Belle measurement on the \LCp branching fraction \cite{Zupanc:2013iki},
\begin{equation}
  \BR(\LCp \ra \proton \Km \pip) = (6.84 \pm 0.24^{+0.21}_{-0.27})\%
\end{equation}
the upper limits would be
\begin{align}
  \begin{split}
    \BR(\Bm \ra \LCp \antiproton \en \nueb)\cdot \frac{\BR(\LCp \ra \proton \Km \pip)}{6.84\%} &< 0.9 \times 10^{-4}\\
    \BR(\Bm \ra \LCp \antiproton \mun \numb)\cdot \frac{\BR(\LCp \ra \proton \Km \pip)}{6.84\%} &< 1.8 \times 10^{-4}.
  \end{split}
\end{align}

\begin{figure}[t]
  \subfigure[]{
    \includegraphics[width=.48\textwidth]{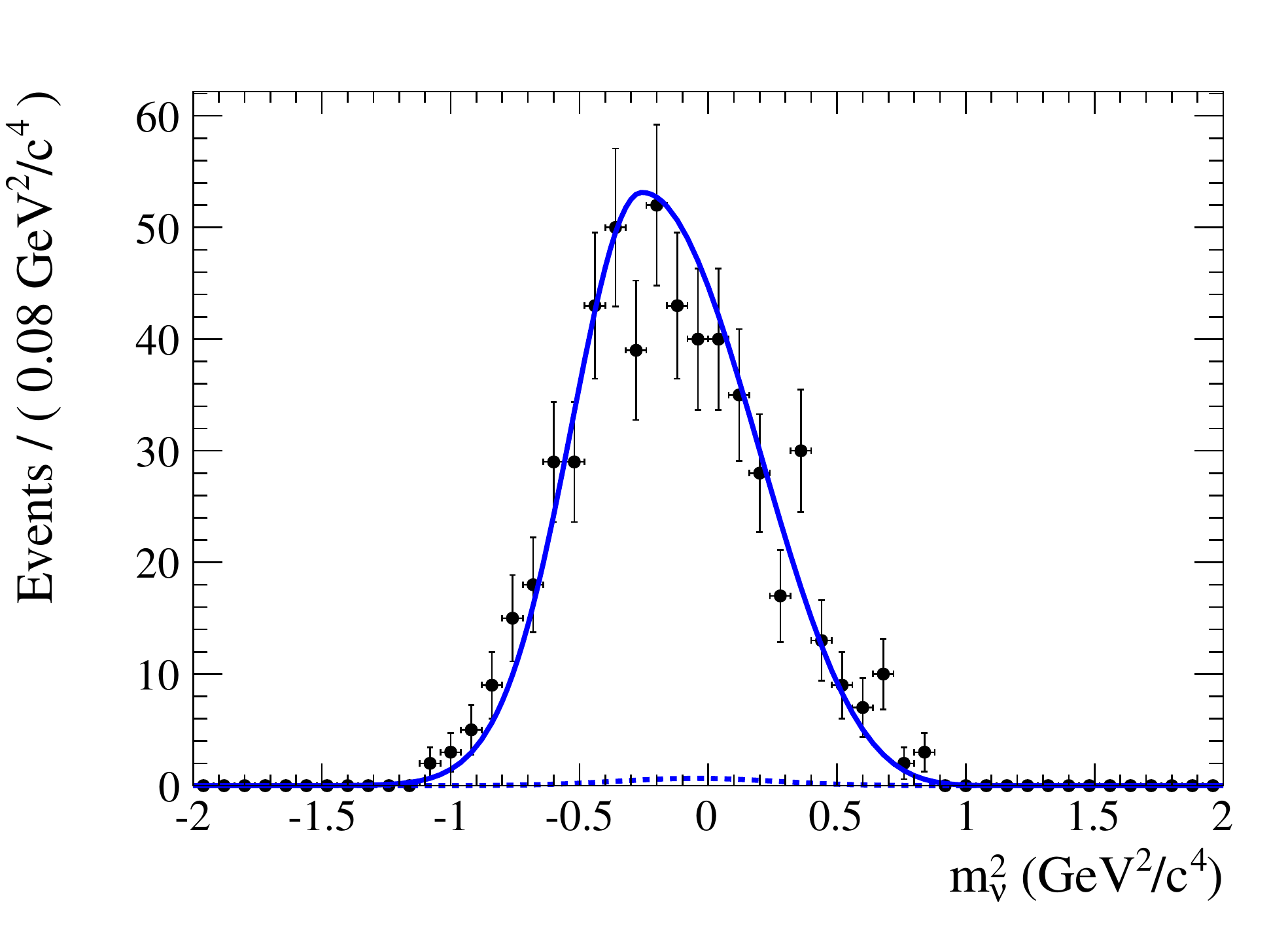}
    \label{subfig:combFit:e:mNu2}
  }
  \subfigure[]{
    \includegraphics[width=.48\textwidth]{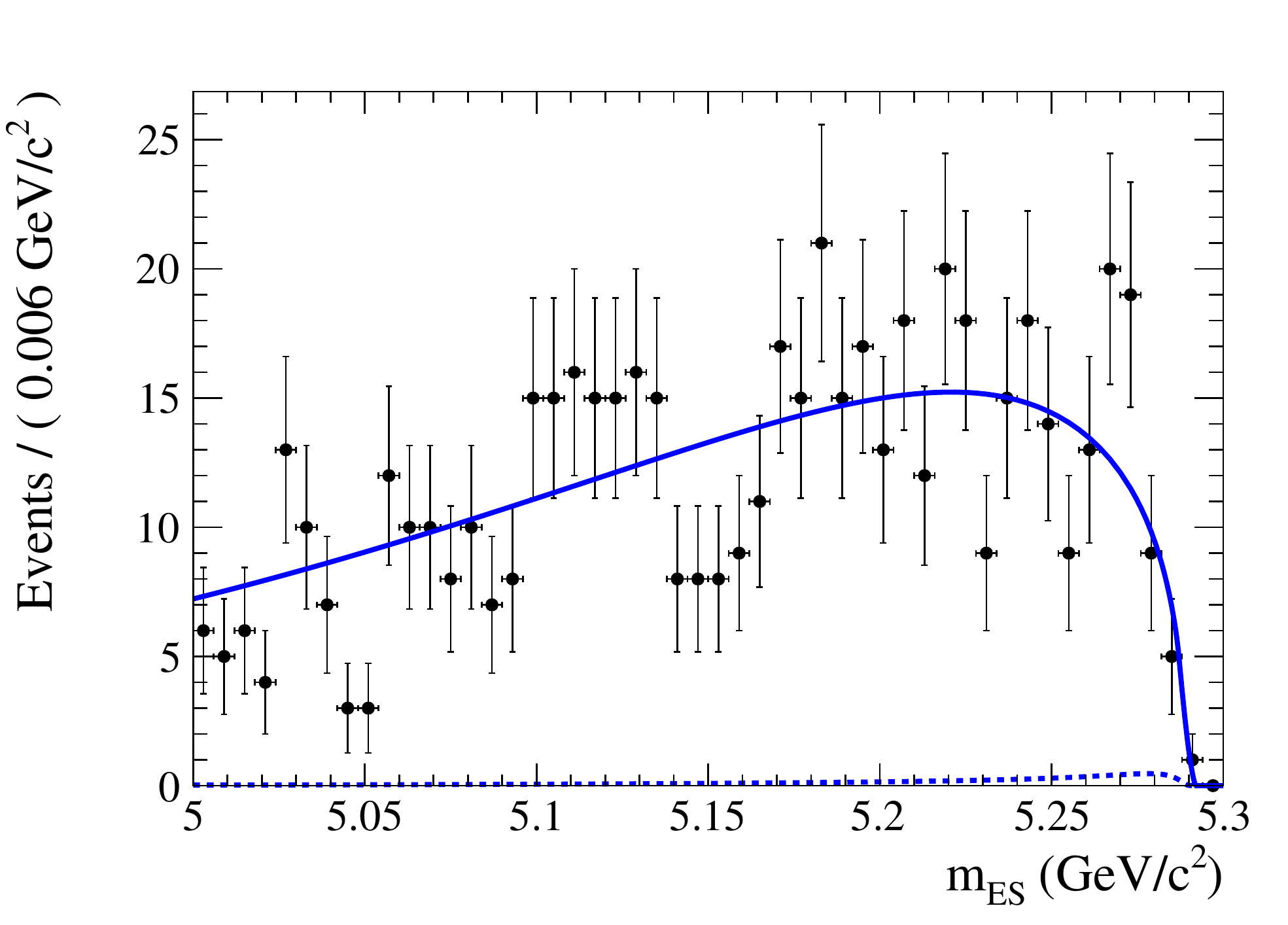}
    \label{subfig:combFit:e:mES}
  }
  \subfigure[]{
    \includegraphics[width=.48\textwidth]{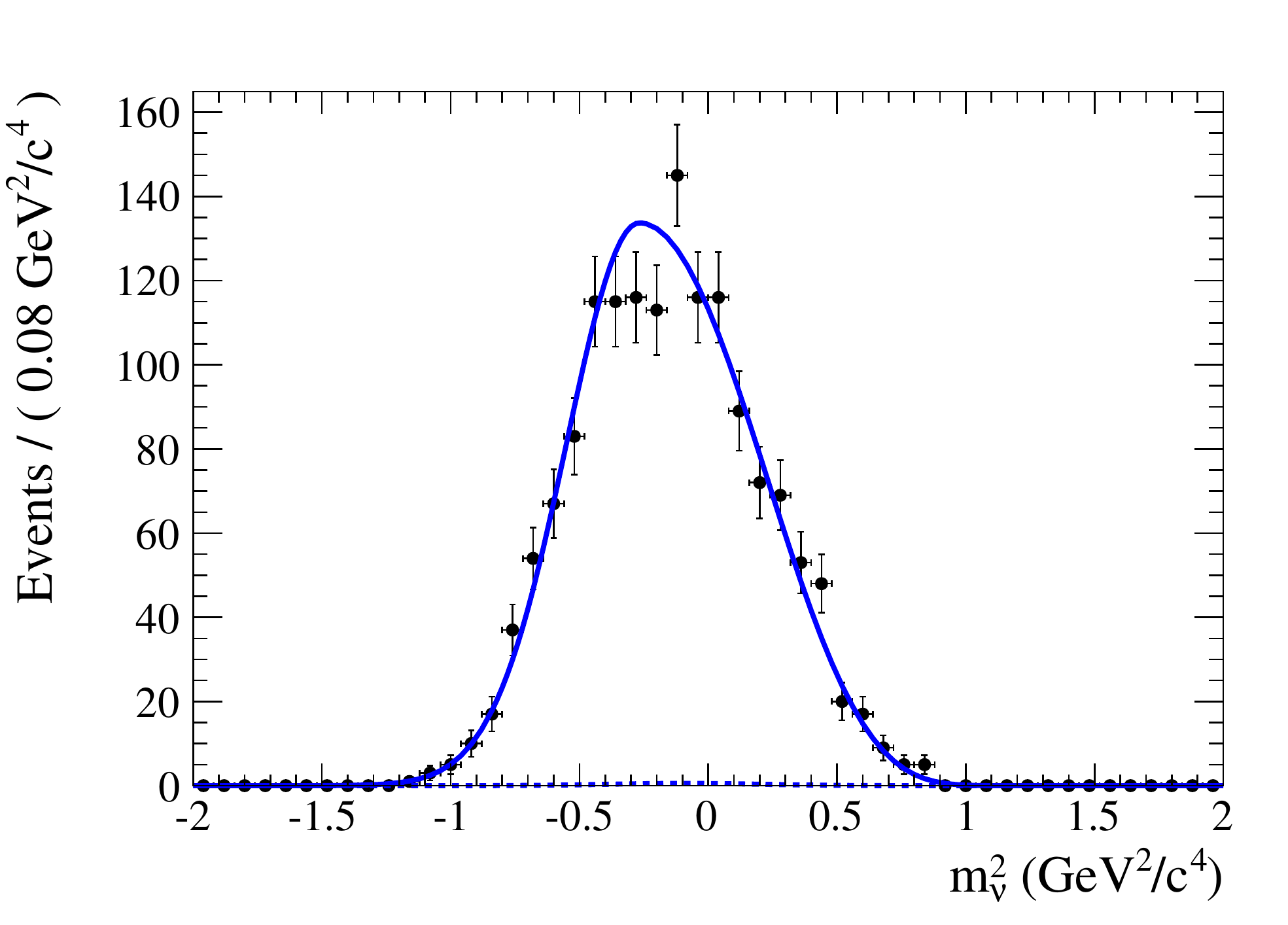}
    \label{subfig:combFit:mu:mNu2}
  }
  \subfigure[]{
    \includegraphics[width=.48\textwidth]{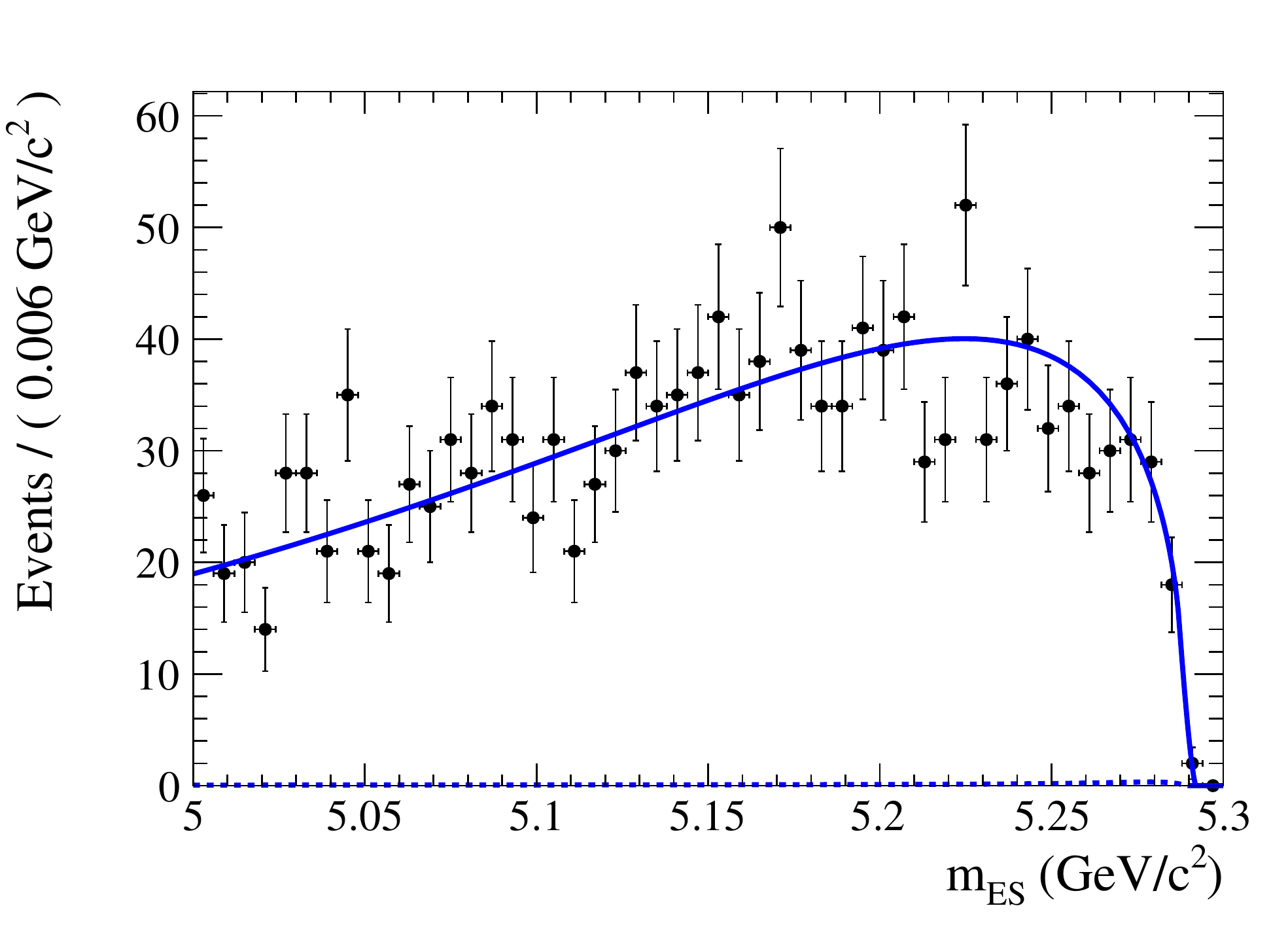}
    \label{subfig:combFit:mu:mES}
  }
  \caption{Fit projections for the electron (\subref{subfig:combFit:e:mNu2}, \subref{subfig:combFit:e:mES}) and muon (\subref{subfig:combFit:mu:mNu2}, \subref{subfig:combFit:mu:mES}) channel from the simultaneous fit. The dashed line shows the signal component.}
  \label{fig:combFit}
\end{figure}
\begin{figure}[h]
  \begin{center}
    \includegraphics[width=.6\textwidth]{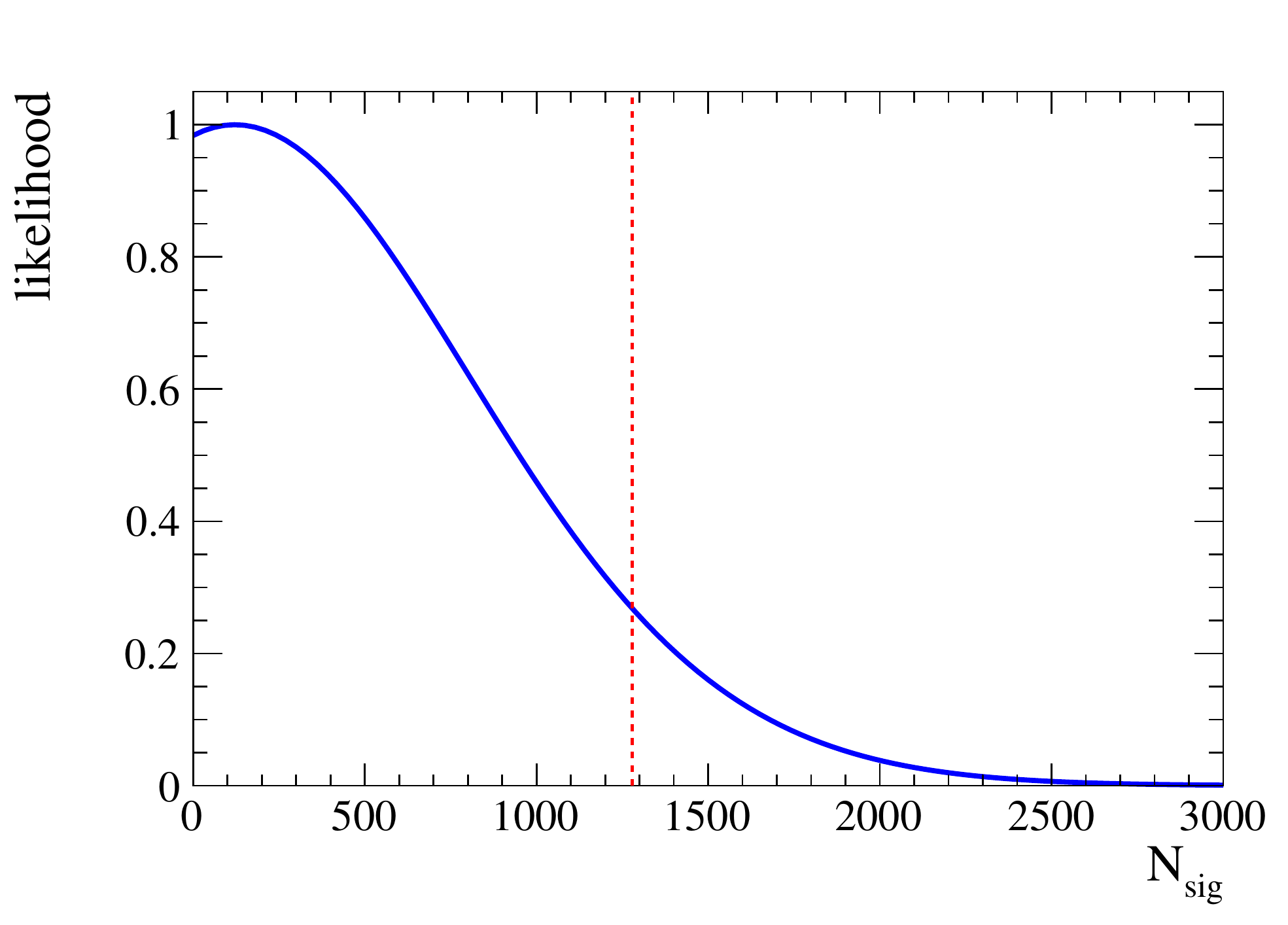}
  \end{center}
  \caption{Likelihood for the simultaneous fit for the electron and the muon channel.}
  \label{fig:UL:comb}
\end{figure}

For the combined upper limit for $\Bm \ra \LCp \antiproton \ellm \nulb$ we perform a simultaneous, unbinned maximum likelihood fit in $\mes$ and $m_{\nu}^2$ for the electron and the muon channel. The common variable is the efficiency corrected signal yield. For the fit we use the same functions as in section \ref{sys:fitFunc}, where we determined the systematic uncertainty arising from the used fit function. Here, we fix all parameters to the obtained values in section \ref{sys:fitFunc}. For the efficiency correction of the signal yields we use the values given in eq. \eqref{eq:eff}. The \mes and $m_{\nu}^2$ projections for both decay modes are shown in Fig. \ref{fig:combFit}.
The likelihood in dependence of the signal yield is shown in Fig. \ref{fig:UL:comb}. The Bayesian upper limit for the signal yield at $90\%$ confidence level is $1279$ events.
With the number of \BBb pairs, given in eq. \eqref{eq:Nbb}, this translates into
\begin{align}
    \BR(\Bm \ra \LCp \antiproton \ellm \nulb)\cdot \BR(\LCp \ra \proton \Km \pip) &< 2.7 \times 10^{-6}.
\end{align}
For the systematic uncertainty arising from the model dependence of the efficiency we take the average of the correction factors given in the previous section, which is $0.56$, and obtain after the correction for the \LCp branching fraction an upper limit of
\begin{align}
    \BR(\Bm \ra \LCp \antiproton \ellm \nulb)\cdot \frac{\BR(\LCp \ra \proton \Km \pip)}{5\%} &< 1.0 \times 10^{-4},
\end{align}
for the world average value of $5\%$ for $\BR(\LCp \ra \proton \Km \pip)$, and for the recent Belle result of $6.84\%$ an upper limit of
\begin{align}
    \BR(\Bm \ra \LCp \antiproton \ellm \nulb)\cdot \frac{\BR(\LCp \ra \proton \Km \pip)}{6.84\%} &< 0.7 \times 10^{-4}.
\end{align}

\clearpage
\chapter{Conclusion}

The present analysis is a search for the semi-leptonic \B decay $\Bm \ra \LCp \antiproton \ellm \nulb$. Neither for the electron, nor the muon mode a signal could be established. This led to the $90\%$ CL upper limits of
\begin{align}
  \begin{split}
    \BR(\Bm \ra \LCp \antiproton \en \nueb)\cdot \frac{\BR(\LCp \ra \proton \Km \pip)}{5\%} &< 1.2 \times 10^{-4},\\
    \BR(\Bm \ra \LCp \antiproton \mun \numb)\cdot \frac{\BR(\LCp \ra \proton \Km \pip)}{5\%} &< 2.5 \times 10^{-4},\\
    \BR(\Bm \ra \LCp \antiproton \ellm \nulb)\cdot \frac{\BR(\LCp \ra \proton \Km \pip)}{5\%} &< 1.0 \times 10^{-4}.
  \end{split}
\end{align}
All three limits are in a slight tension with two of the predictions given in sect. \ref{semileptBdecay}, which are of the order $5\times 10^{-4}$. 
There are three possible explanations for this tension:
\begin{enumerate}
  \item An unknown suppression factor reduces this branching fraction.
  \item The predictions in sect. \ref{semileptBdecay} rely on inaccurate assumptions.
  \item The decay model used in this analysis for signal simulation is not accurate.
\end{enumerate}
Although the first one is the most interesting one, it is the most unlikely explanation as well. For the second explanation we have to take a closer look on the two predictions again. The first one is derived from the branching fraction for $\Bm \ra \proton\antiproton \ellm \nulb$ by assuming a simple CKM suppression between the two decays. This ignores completely the different size of the available phase space, as well as different form factors for the fragmentation into the hadronic part of the final state. These form factors can depend on the invariant mass of the final state particles, and thus might affect the prediction substantially. The second prediction comes from the decay $\Bm \ra \LCp \antiproton \pim$, assuming that $100\%$ of this decay proceeds via the external $W$ Feynman graph, which is probably not accurate. A possible solution for this problem comes from the isospin analysis given in \cite{ebertPHD}, where the contribution of the external graph is determined based on simple isospin arguments. The upper limit we obtained here is compatible with the predicted limit in \cite{ebertPHD} and might corroborate the isospin argumentation.
Last but not least, we have to consider the used Monte Carlo model in this analysis. We modelled the semi-leptonic decay according to the weak matrix element, assuming a meson-like pole near threshold decaying into $\LCp\antiproton$. Despite being well motivated from the analysis of full hadronic baryonic \B-decays this pole model might prove to be wrong. Another factor here might be the quark hadron duality, since the weak matrix element is only valid on quark-level, neglecting the influence of the hadronization.

On the experimental side the analysis suffered from the high mass of the final state particles, rendering the classical signal variables used in semi-leptonic \B decays useless, as was shown in sect. \ref{sect:nuVars}. A similar problem was already observed in the analysis of $\Bm \ra D_s^+ K^- \ellm \nulb$ \cite{BAD2183}. To circumvent this problem, a tagged analysis approach, where the second \B is fully reconstructed in a pure hadronic mode, is advisable. Belle II, which will have a multitude of the \babar dataset available for analysis, should be able to perform a tagged search for the decay $\Bm \ra \LCp \antiproton \ellm \nulb$. LHCb should be able to study this decay mode as well. The large boost and the resulting displaced vertex could enable them to obtain a rather clean signal for $\Bm \ra \LCp \antiproton \mun \numb$ without the need for a neutrino reconstruction. 

The measurement of the semileptonic decay $\Bm \ra \LCp \antiproton \ellm \nulb$ stays crucial for the understanding of baryonic \B-decays. A direct observation will enable us to quantify the impact of the external Feynman graph on the overall branching fraction of fully hadronic baryonic \B decays. In addition, it will provide valuable insight on the threshold enhancement described in sect. \ref{sect:thresh}, especially on the absence of such an enhancement in the decay $\Bzb \ra \Sigma_c(2455)^{0} \antiproton \pim$. Consequently it will help to decide if the model given in \cite{hartmann2011study} is correct.

\appendix

\chapter{Random Forest input variables}

\section{$\qqbar$ random forest}

\begin{figure}[H]
  \begin{center}
  \subfigure[]{
    \includegraphics[width=.48\textwidth]{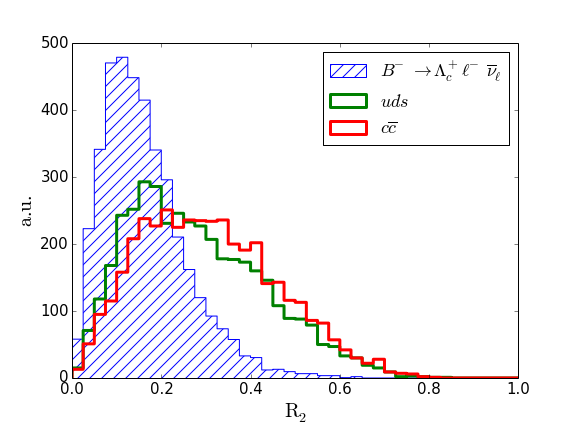}
    \label{subfig:mu:qqbar:R2}
  }
  \subfigure[]{
    \includegraphics[width=.48\textwidth]{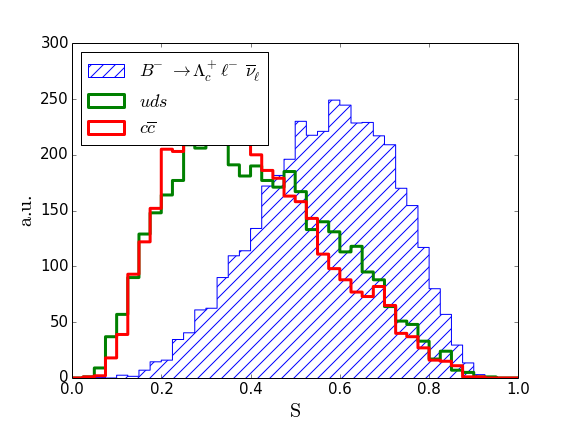}
    \label{subfig:mu:qqbar:S}
  }
  \subfigure[]{
    \includegraphics[width=.48\textwidth]{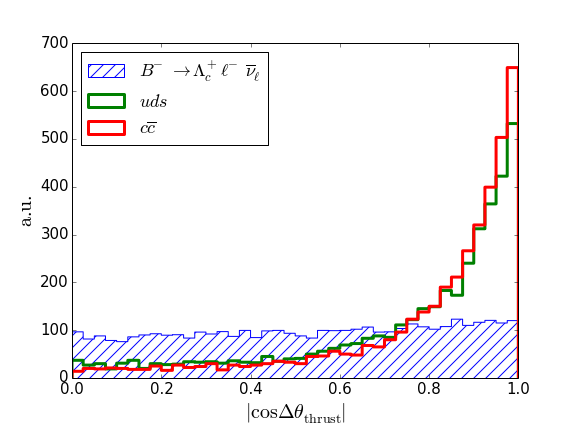}
    \label{subfig:mu:qqbar:cosDTh}
  }
  \end{center}
  \caption{Comparison of WI signal Monte Carlo (blue histogram) and $\qqbar$ background Monte Carlo (green and red lines) for the three Random Forest input variables, \subref{subfig:mu:qqbar:R2} $R_2$, \subref{subfig:mu:qqbar:S} $S$, and \subref{subfig:mu:qqbar:cosDTh} $cos\Delta\theta_{\rm thrust}$.}
  \label{fig:qqbarRF_input:mu}
\end{figure}

\section{\BBbar random forest}

  \begin{figure}[H]
  \subfigure[]{
    \includegraphics[width=.48\textwidth]{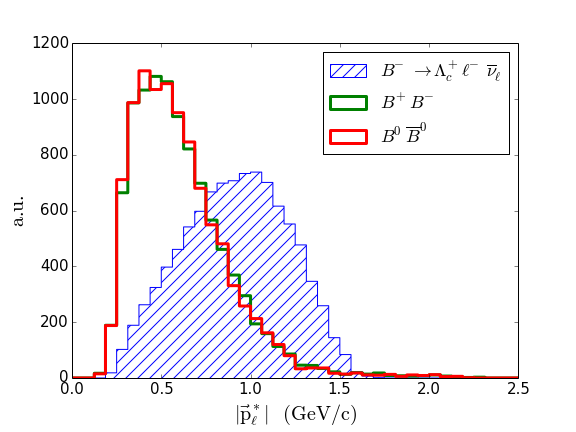}
    \label{fig:mu:BBbar_RF_vars:p_lT}
  }
  \subfigure[]{
    \includegraphics[width=.48\textwidth]{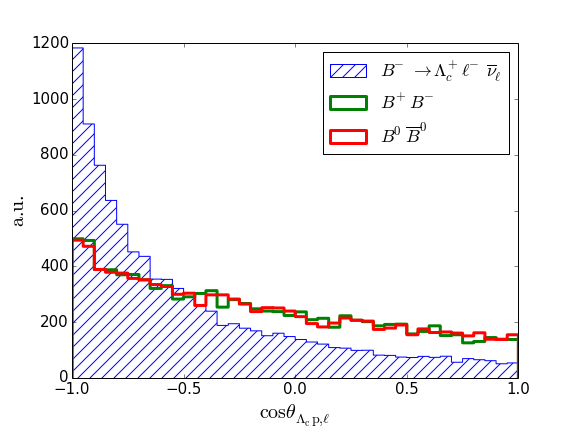}
    \label{fig:mu:BBbar_RF_vars:cosLcpl}
  }
  \subfigure[]{
    \includegraphics[width=.48\textwidth]{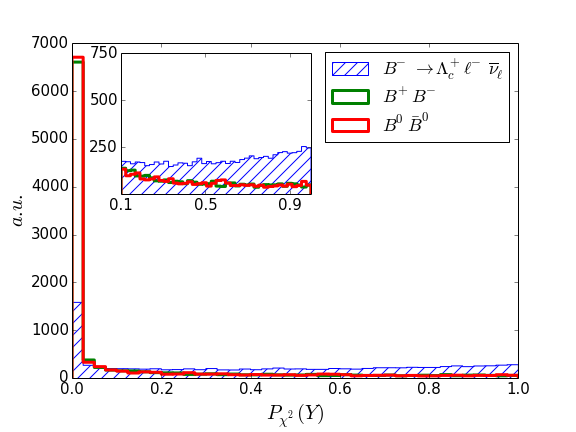}
    \label{fig:mu:BBbar_RF_vars:PY}
  }
  \subfigure[]{                                
    \includegraphics[width=.48\textwidth]{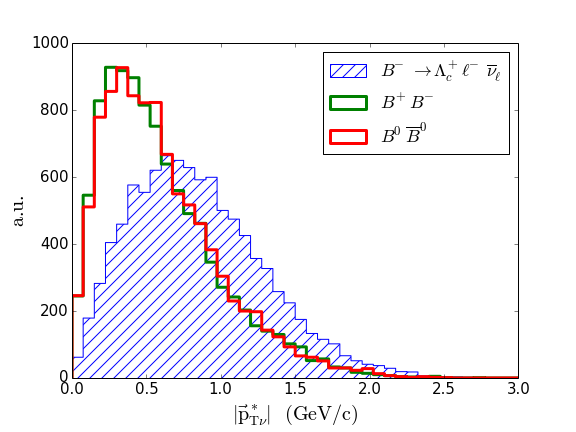}    
    \label{fig:mu:BBbar_RF_vars:pNu}
  }
  \caption{Comparison of WI signal simulation events and generic \B decays. \subref{fig:mu:BBbar_RF_vars:p_lT} $|\vec{p}_{\ell}^{*}|$, \subref{fig:mu:BBbar_RF_vars:cosLcpl} $\cos\theta_{\Lambda_c p, \ell}$, \subref{fig:mu:BBbar_RF_vars:PY} $P_{\chi^2}(Y)$, and \subref{fig:mu:BBbar_RF_vars:pNu} $p_{T\nu}$.}
  \label{fig:mu:BBbar_RF_vars}
\end{figure}


\chapter{Muon Channel target variables}
\begin{figure}[h]
  \begin{center}
  \subfigure[]{
    \includegraphics[width=.48\textwidth]{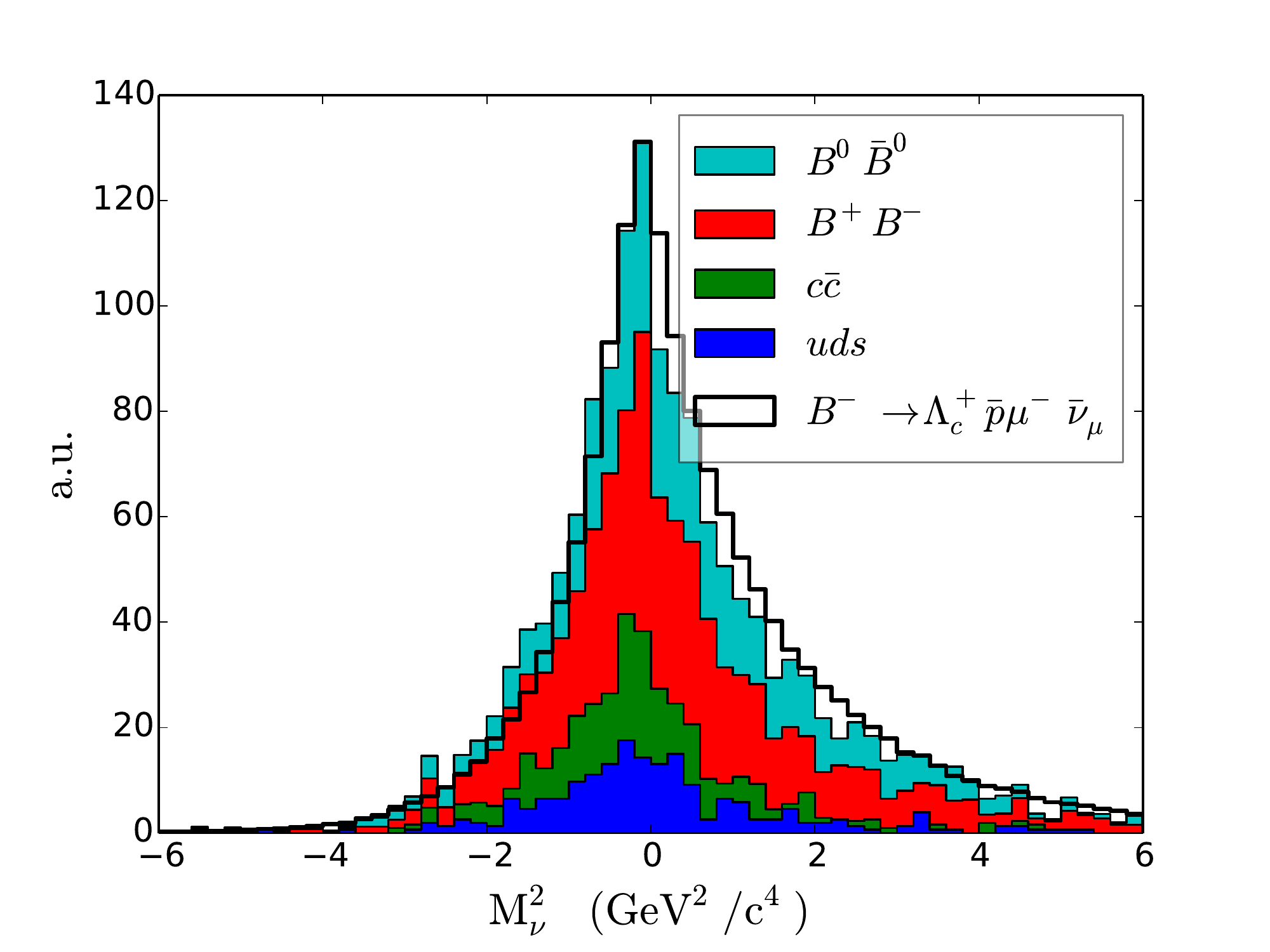}
    \label{subfig:target_Mnu2_mu}
  }
  \subfigure[]{
    \includegraphics[width=.48\textwidth]{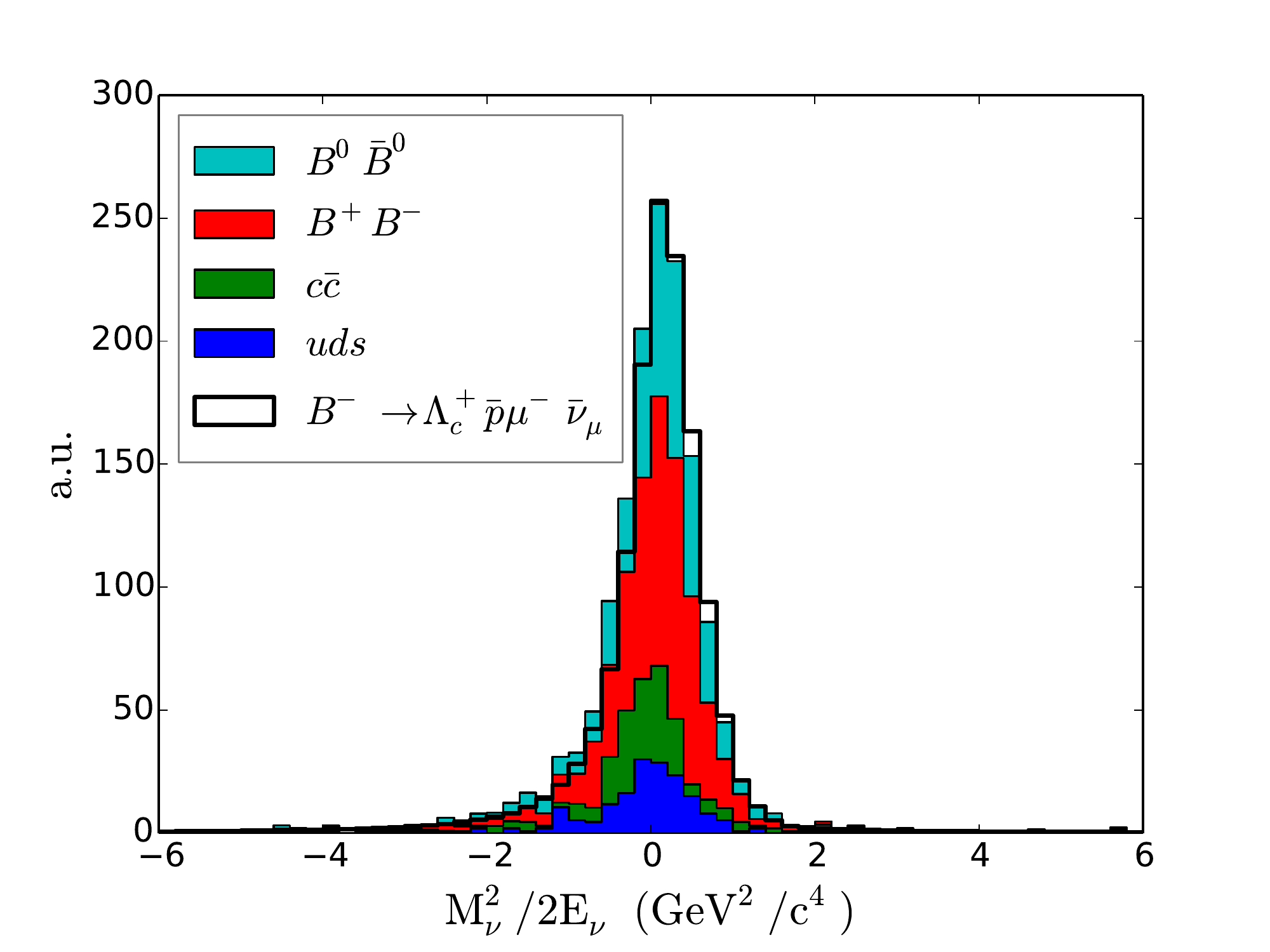}
    \label{subfig:target_Mnu2o2Enu_mu}
  }
  \subfigure[]{
    \includegraphics[width=.48\textwidth]{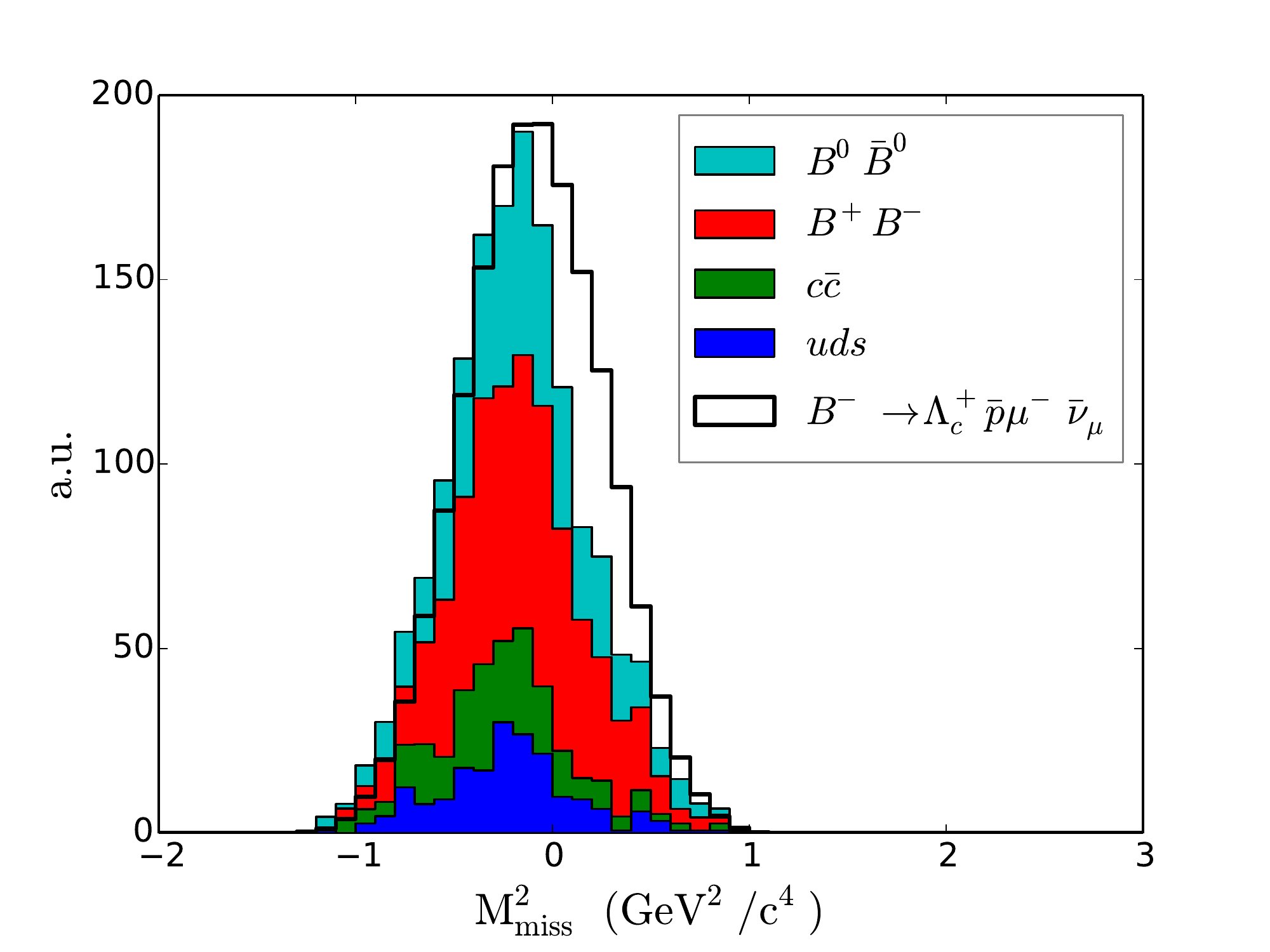}
    \label{subfig:target_Mmiss2_mu}
  }
  \subfigure[]{
    \includegraphics[width=.48\textwidth]{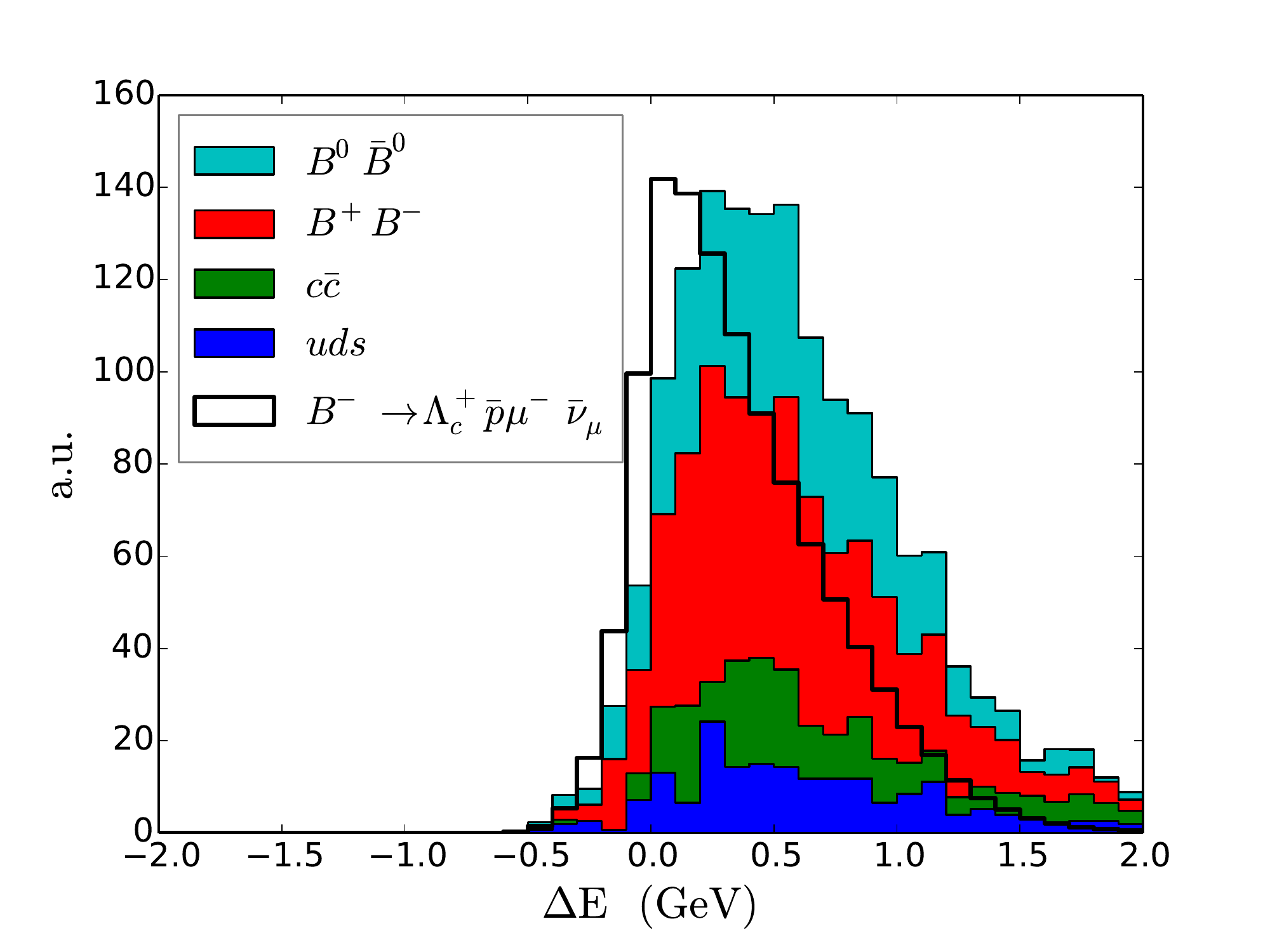}
    \label{subfig:target_DeltaE_mu}
  }
  \end{center}
  \caption{Comparison of WI signal Monte Carlo and generic background Monte Carlo for \subref{subfig:target_Mnu2_mu} $M_{\nu}^2$, \subref{subfig:target_Mnu2o2Enu_mu} $M_{\nu}^2/E_{\nu}$, \subref{subfig:target_Mmiss2_mu} $M_{\rm miss}^2$, and\subref{subfig:target_DeltaE_mu} \DeltaE. The background Monte Carlo is scaled to \texttt{OnPeak} luminosity and stacked. The WI Monte Carlo is scaled to match the height of the background peak.}
  \label{fig:targetVars_comp_muChannel}
\end{figure}


\chapter{Fit procedure}

\section{Correlation checks}

\begin{figure}[H]
  \centering
  \subfigure[]{
    \includegraphics[width=.4\textwidth]{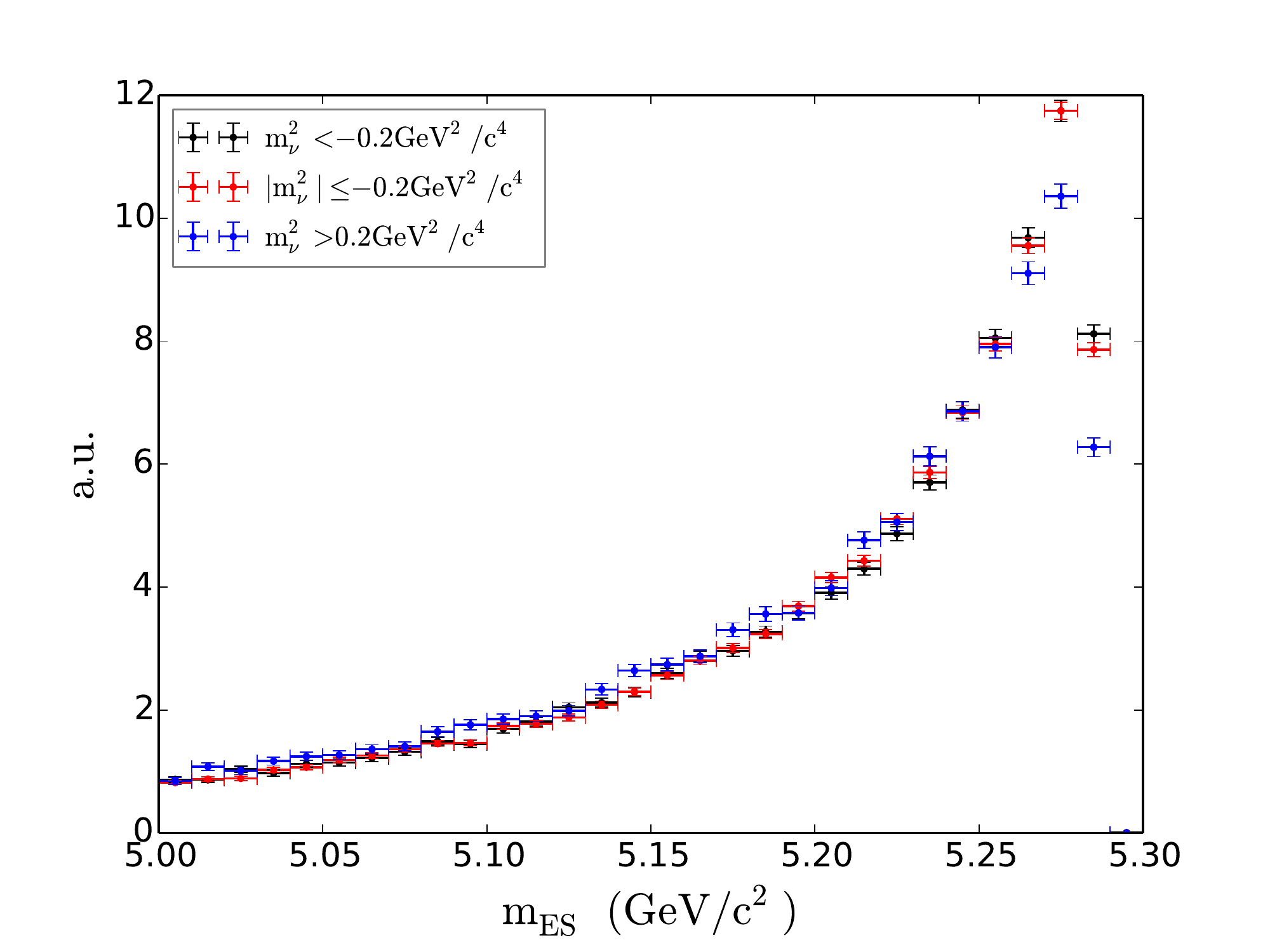}

  }
  \subfigure[]{
    \includegraphics[width=.4\textwidth]{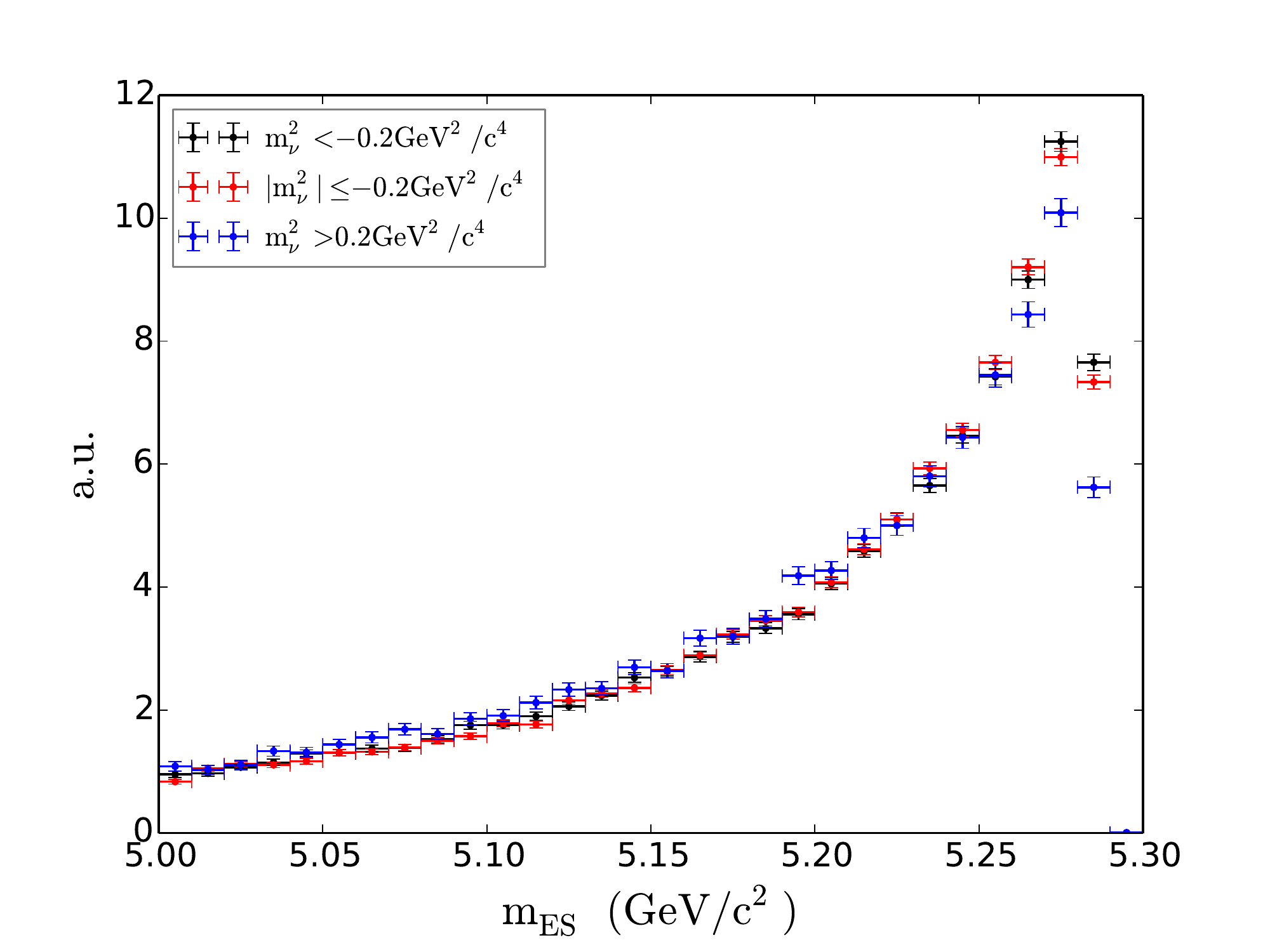}

  }
  \subfigure[]{
    \includegraphics[width=.4\textwidth]{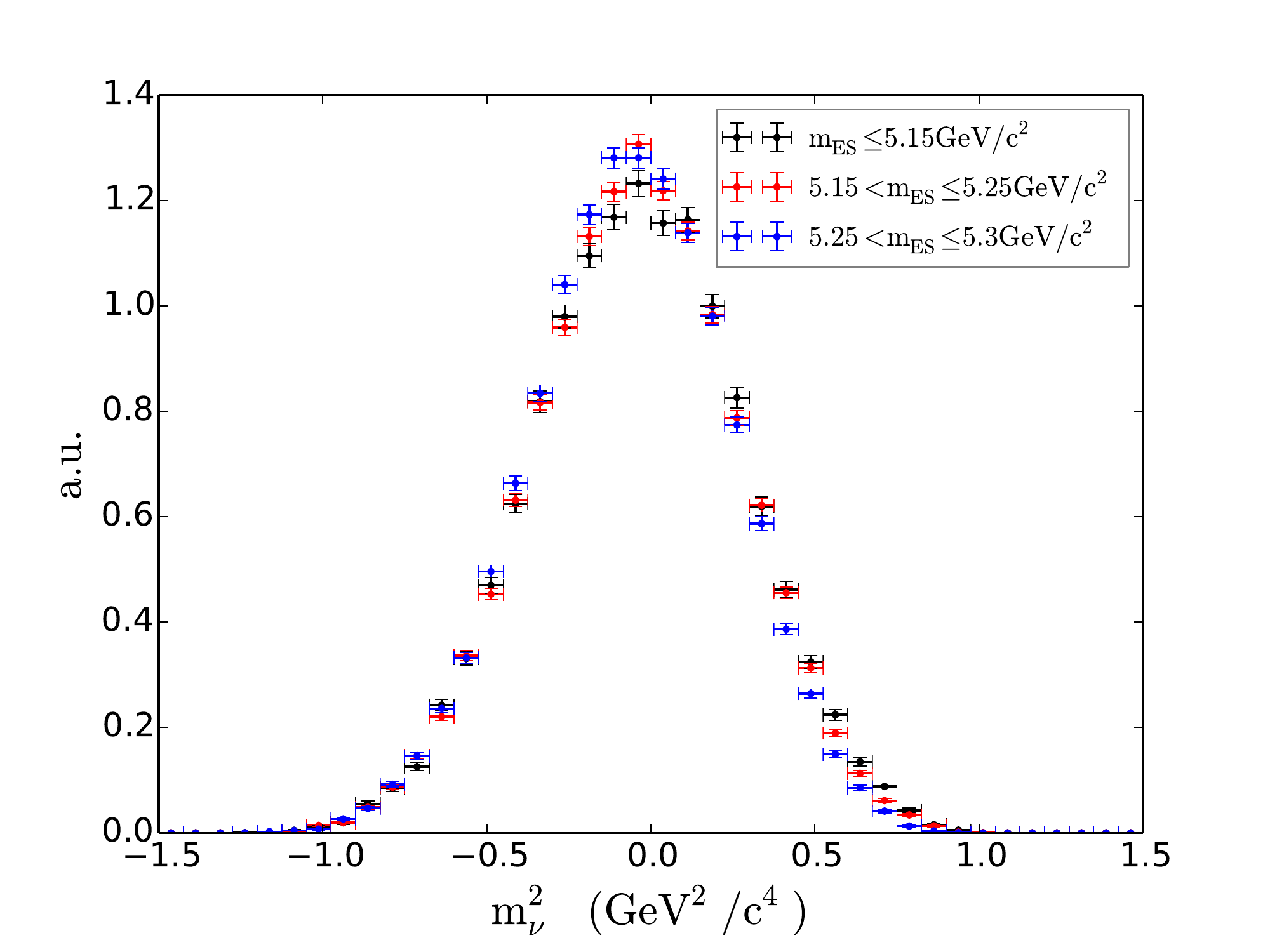}

  }
  \subfigure[]{
    \includegraphics[width=.4\textwidth]{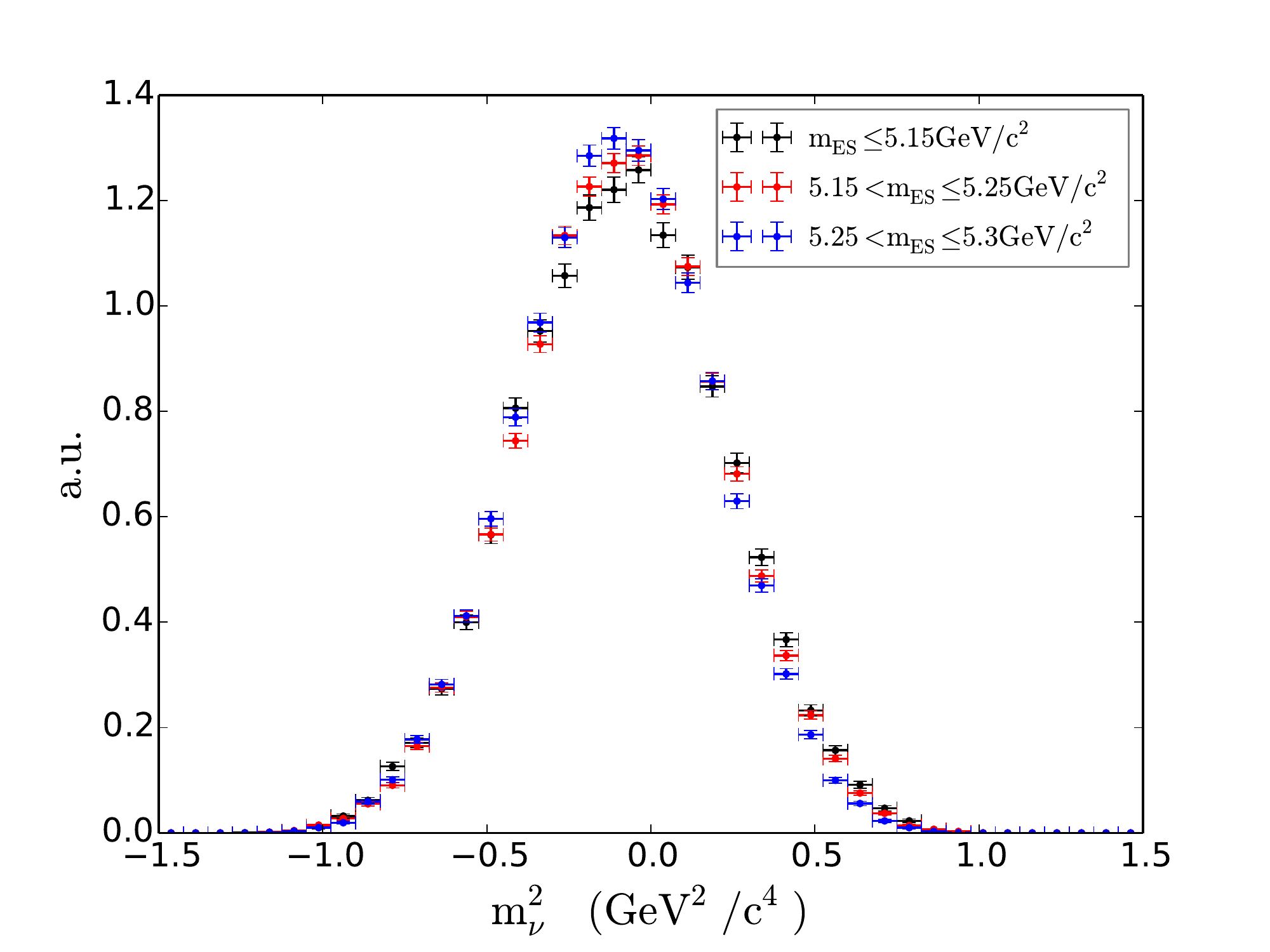}

  }
  \caption{\mes and $m_{\nu}^2$ in slices of each other. On the left hand side for electron, and on the right hand side for muon WI signal Monte Carlo events.}
  \label{fig:mESmNu2Slices_signal}
\end{figure}

\begin{figure}[H]\centering
  \subfigure[]{
    \includegraphics[width=.4\textwidth]{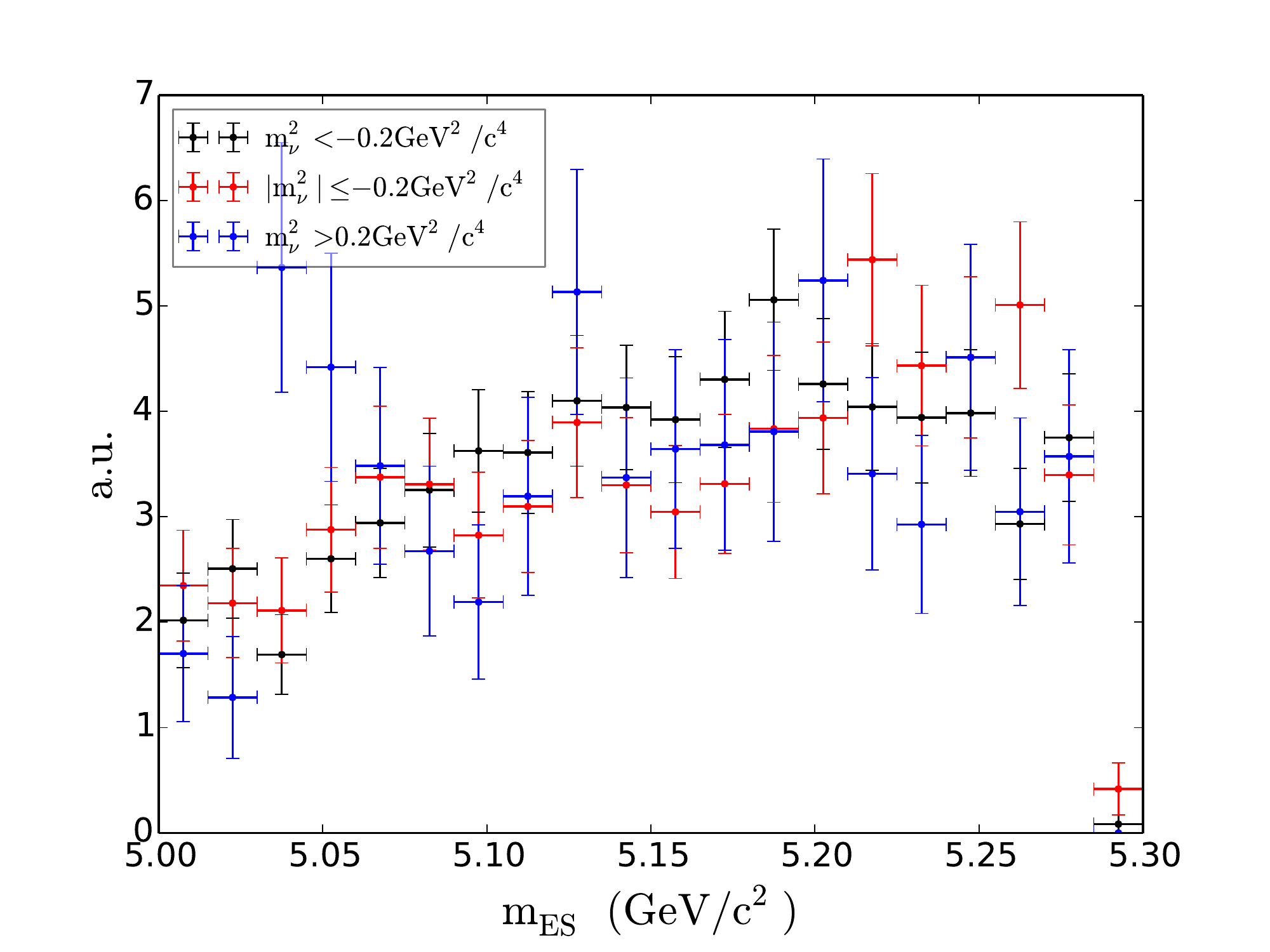}

  }
  \subfigure[]{
    \includegraphics[width=.4\textwidth]{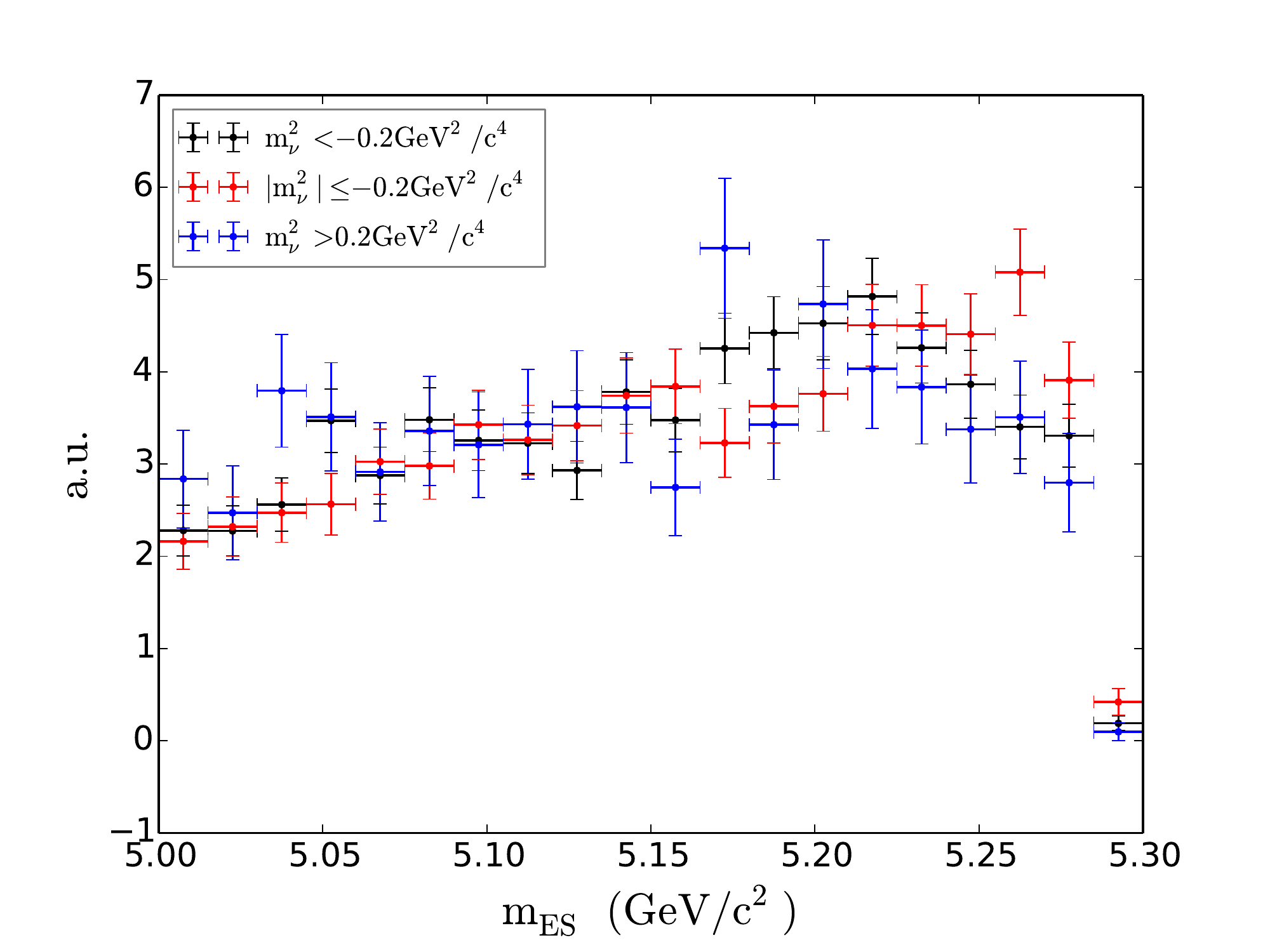}

  }
  \subfigure[]{
    \includegraphics[width=.4\textwidth]{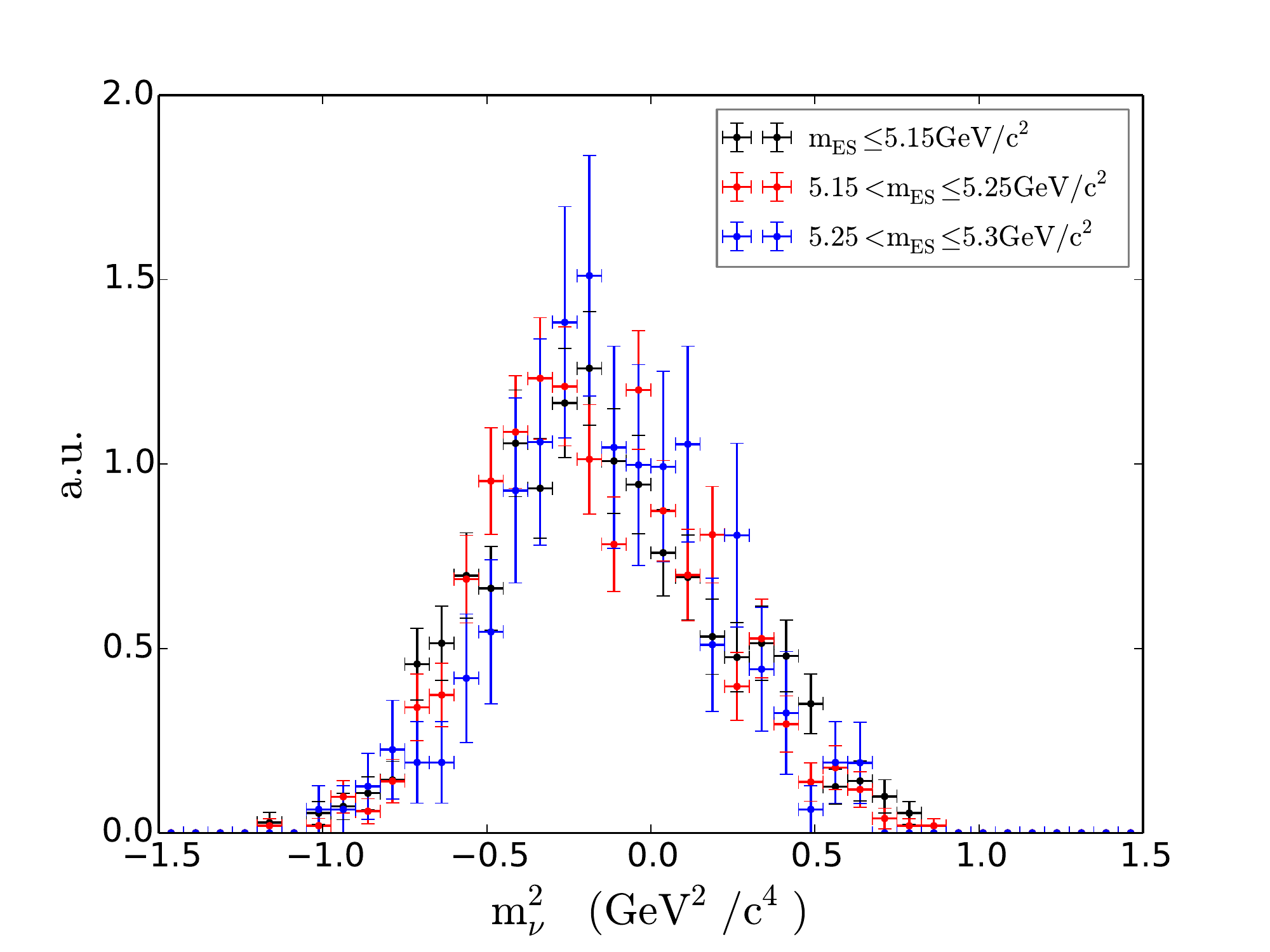}

  }
  \subfigure[]{
    \includegraphics[width=.4\textwidth]{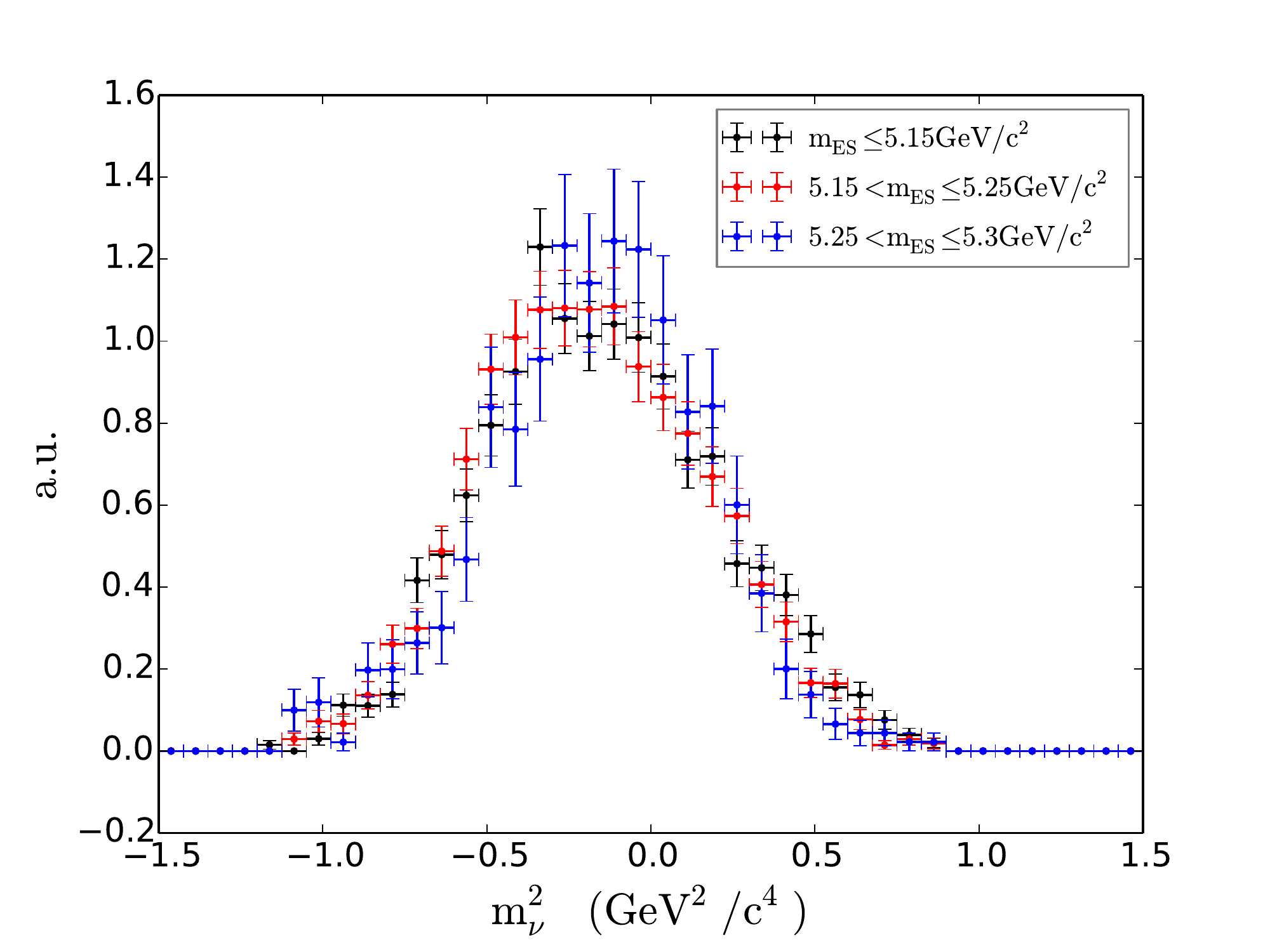}

  }
  \caption{\mes and $m_{\nu}^2$ in slices of each other. On the left hand side for electron, and on the right hand side for muon background Monte Carlo events.}
  \label{fig:mESmNu2Slices_background}
\end{figure}
\clearpage
\section{Fit validation}
\subfigtopskip=0pt
\subfigbottomskip=0pt
\subfigcapskip=0pt
\begin{figure}[h]
  \begin{center}
  \subfigure[100 signal events]{
    \includegraphics[width=.36\textwidth]{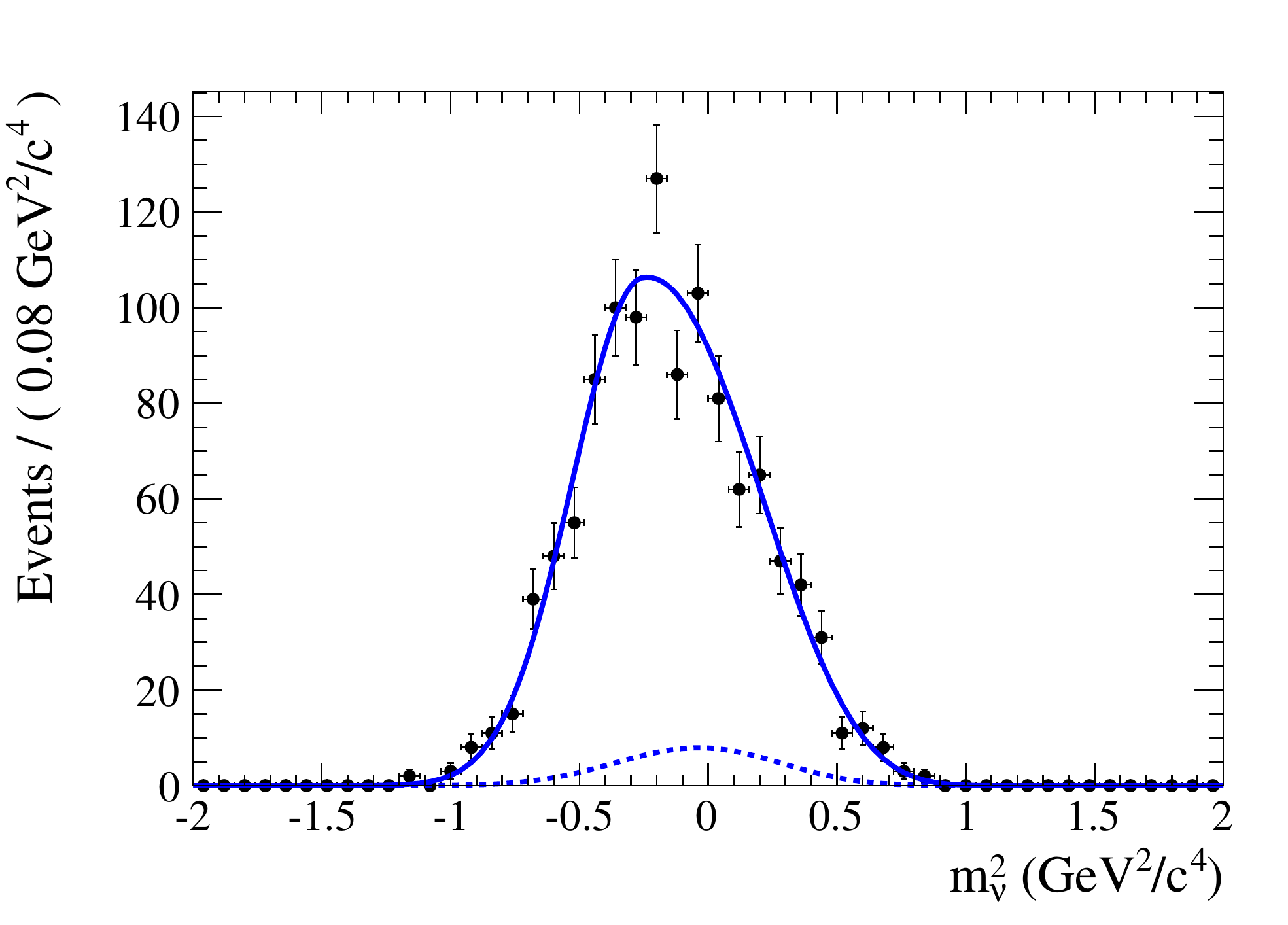}
  }
  \subfigure[100 signal events]{
    \includegraphics[width=.36\textwidth]{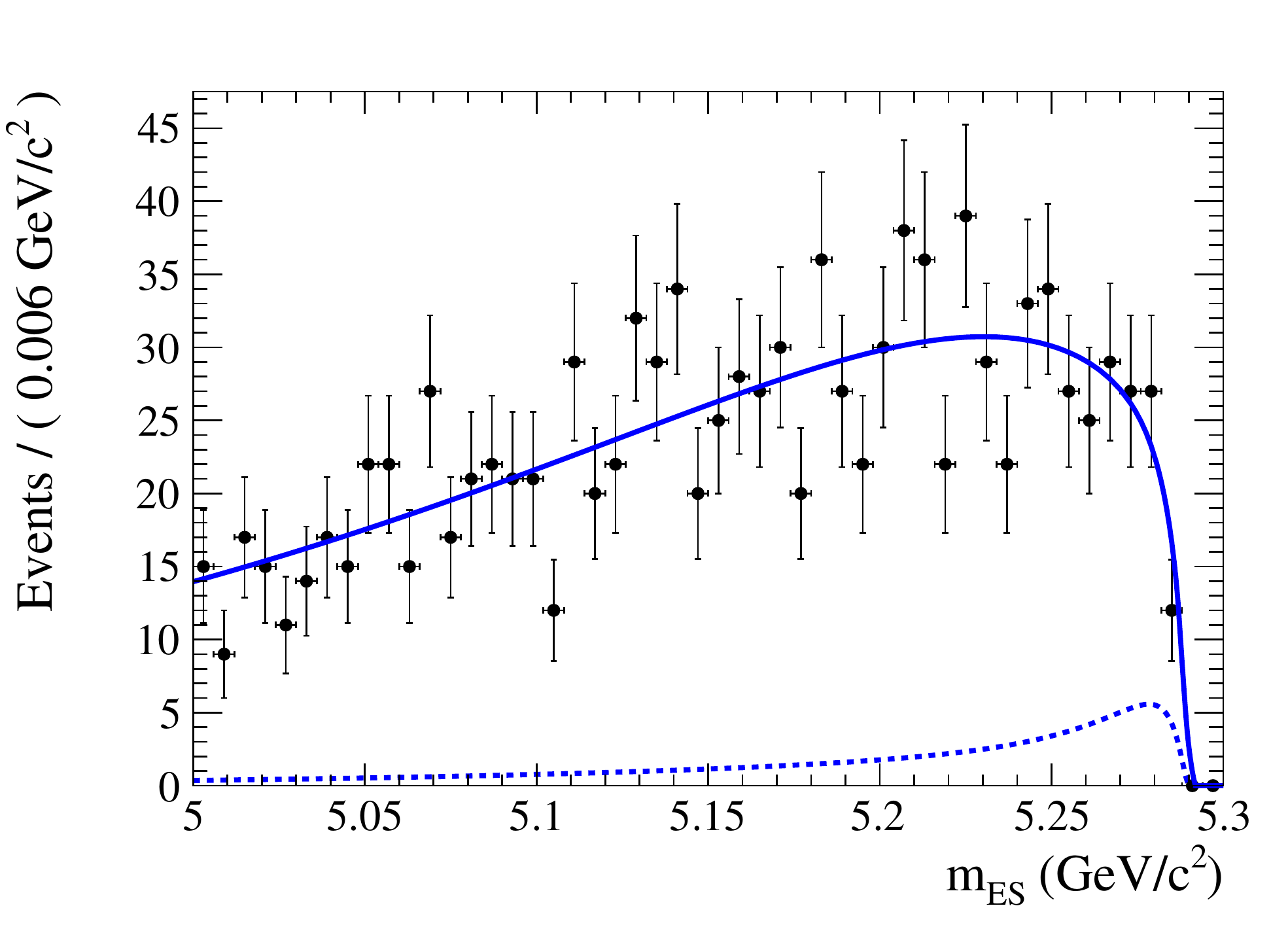}
  }
  \subfigure[50 signal events]{
    \includegraphics[width=.36\textwidth]{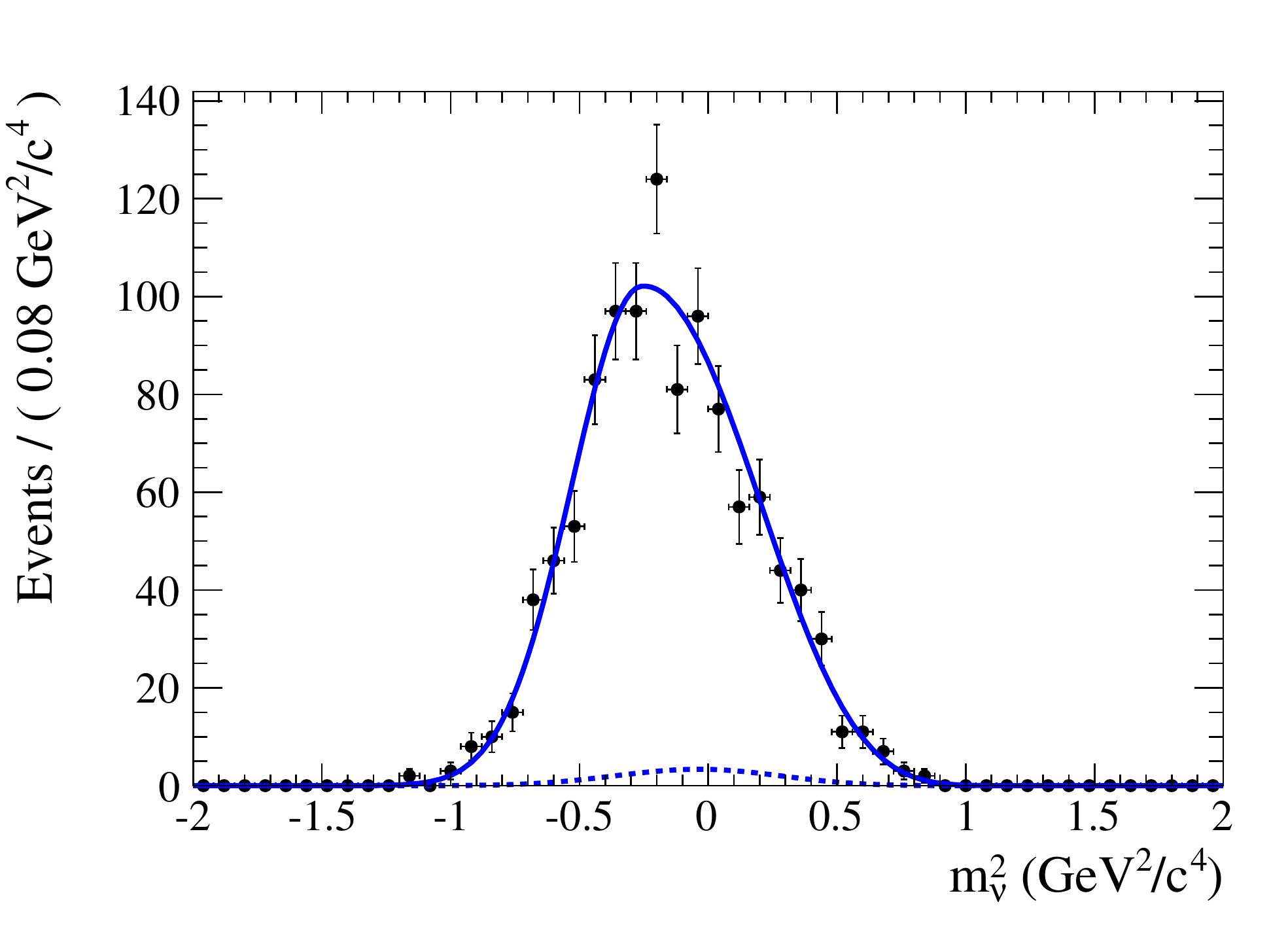}
  }
  \subfigure[50 signal events]{
    \includegraphics[width=.36\textwidth]{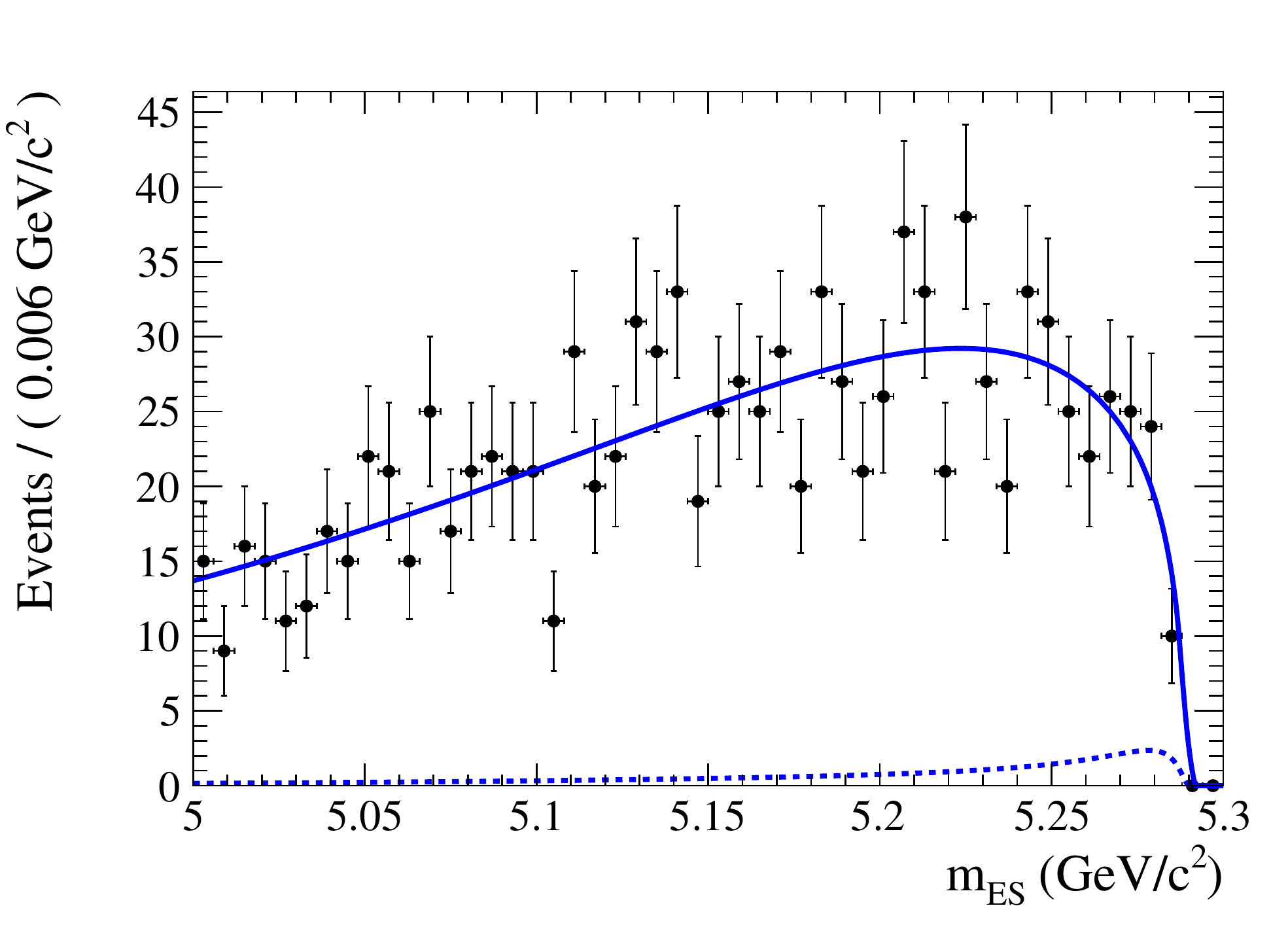}
  }
  \subfigure[10 signal events]{
    \includegraphics[width=.36\textwidth]{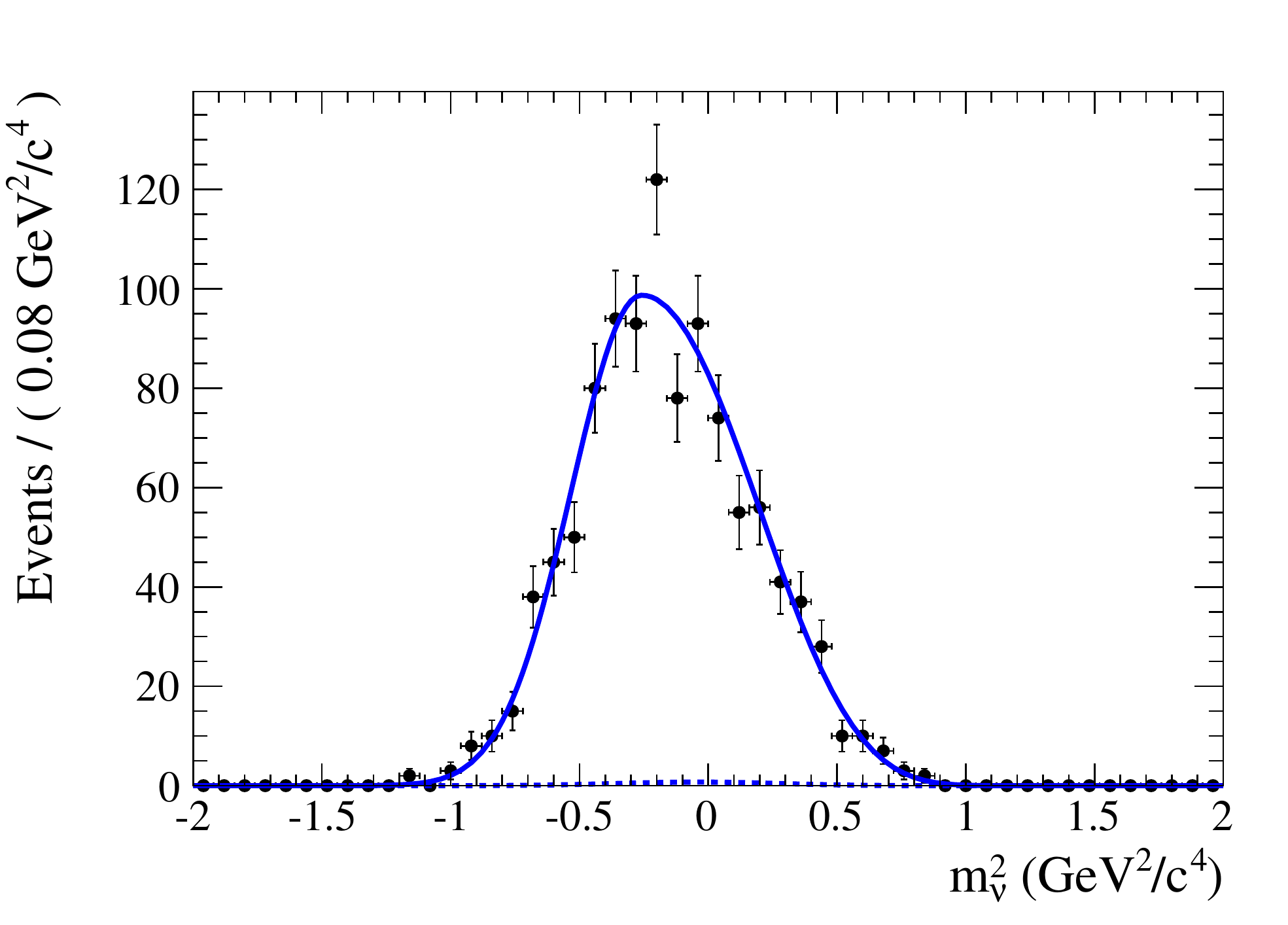}
  }
  \subfigure[10 signal events]{
    \includegraphics[width=.36\textwidth]{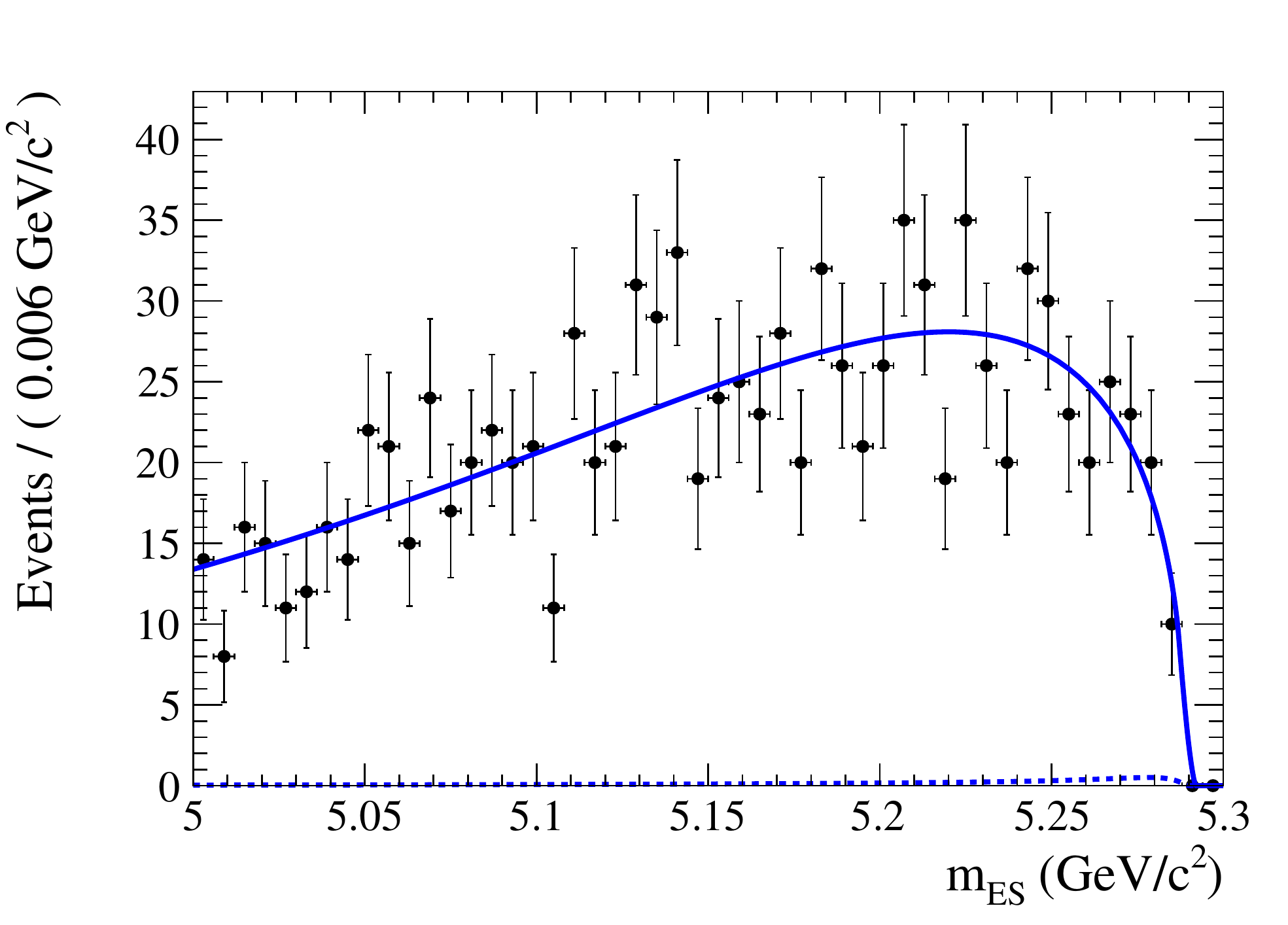}
  }
  \subfigure[0 signal events]{
    \includegraphics[width=.36\textwidth]{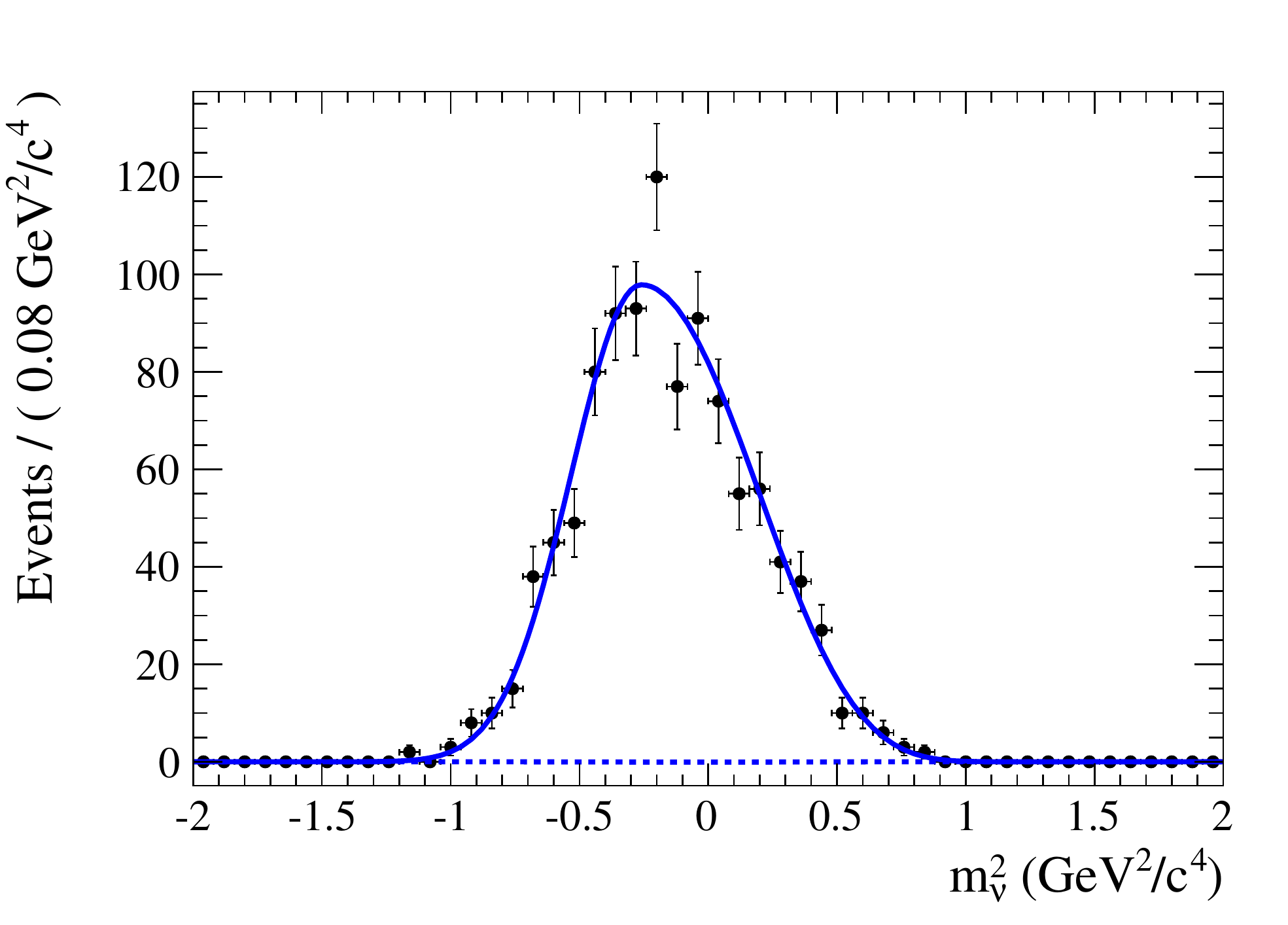}
  }
  \subfigure[0 signal events]{
    \includegraphics[width=.36\textwidth]{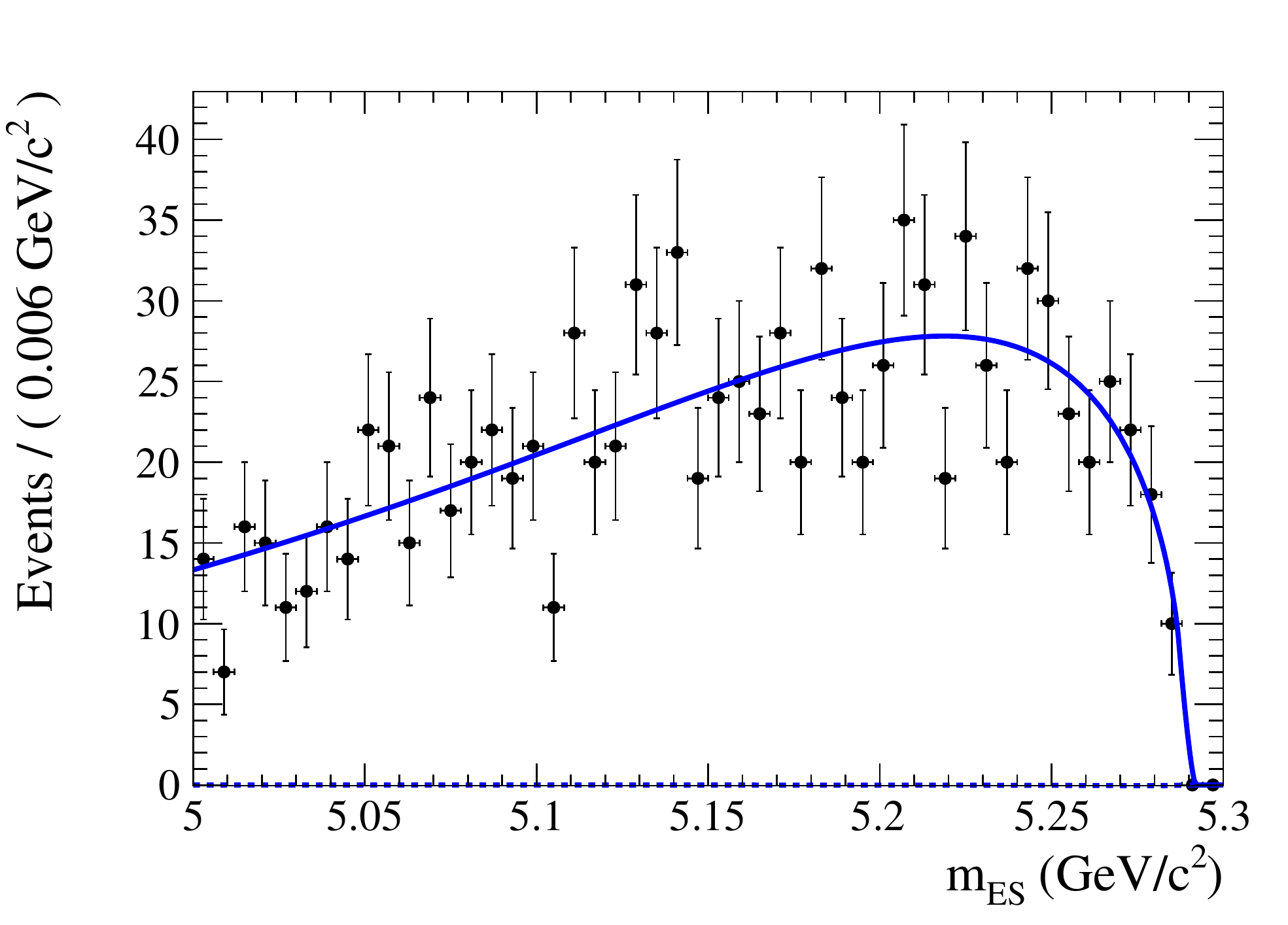}
  }
  \end{center}
  \vspace{-0.5cm}
  \caption{$m_{\nu}^2$ and \mes projections for the fit validation for the electron channel. Shown are the fits for $100, 50, 10$ and $0$ input signal events.}
  \label{fig:FitValidation:electron}
\end{figure}
\begin{figure}[h]
  \begin{center}
  \subfigure[100 signal events]{
    \includegraphics[width=.36\textwidth]{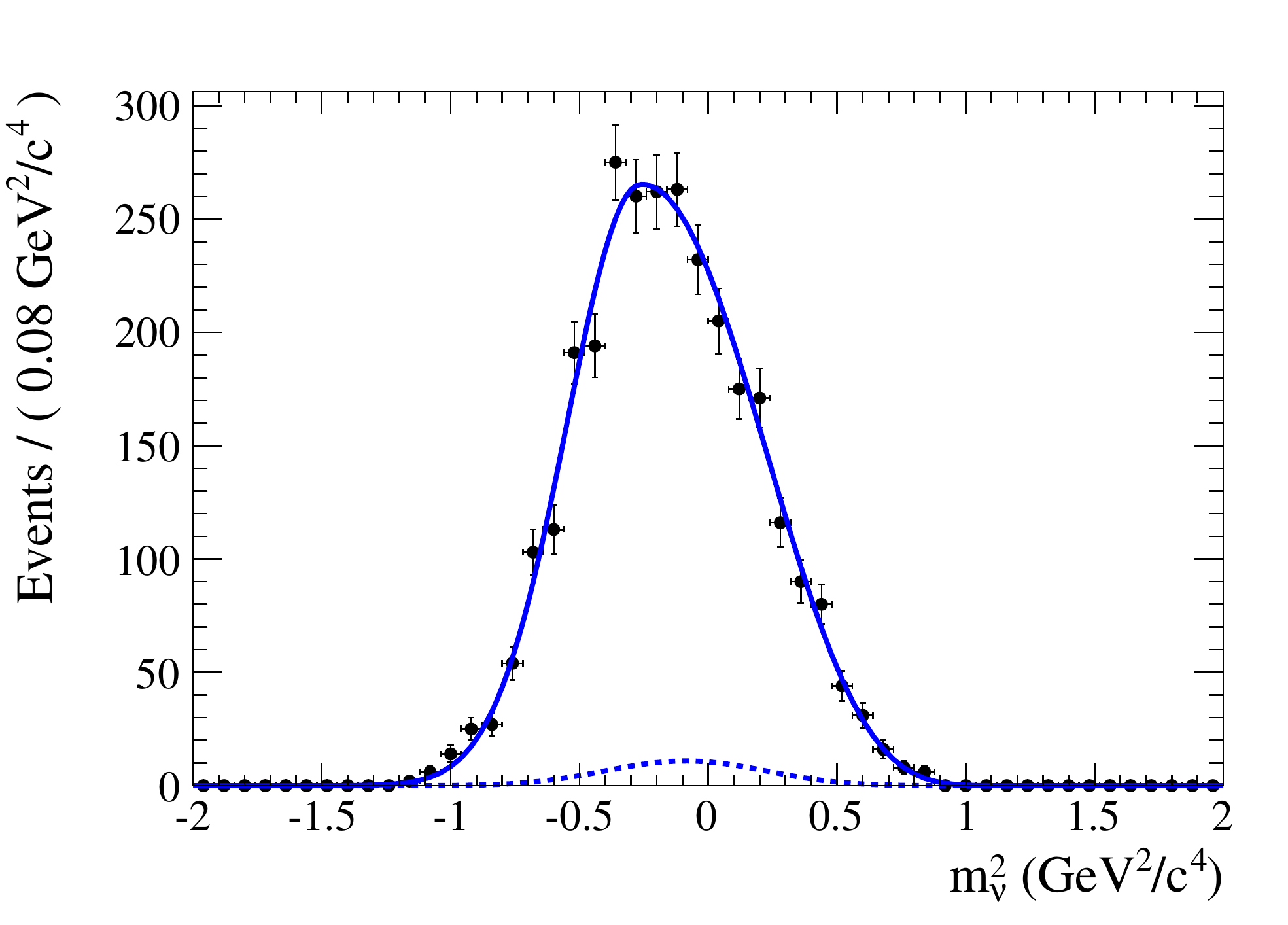}
  }
  \subfigure[100 signal events]{
    \includegraphics[width=.36\textwidth]{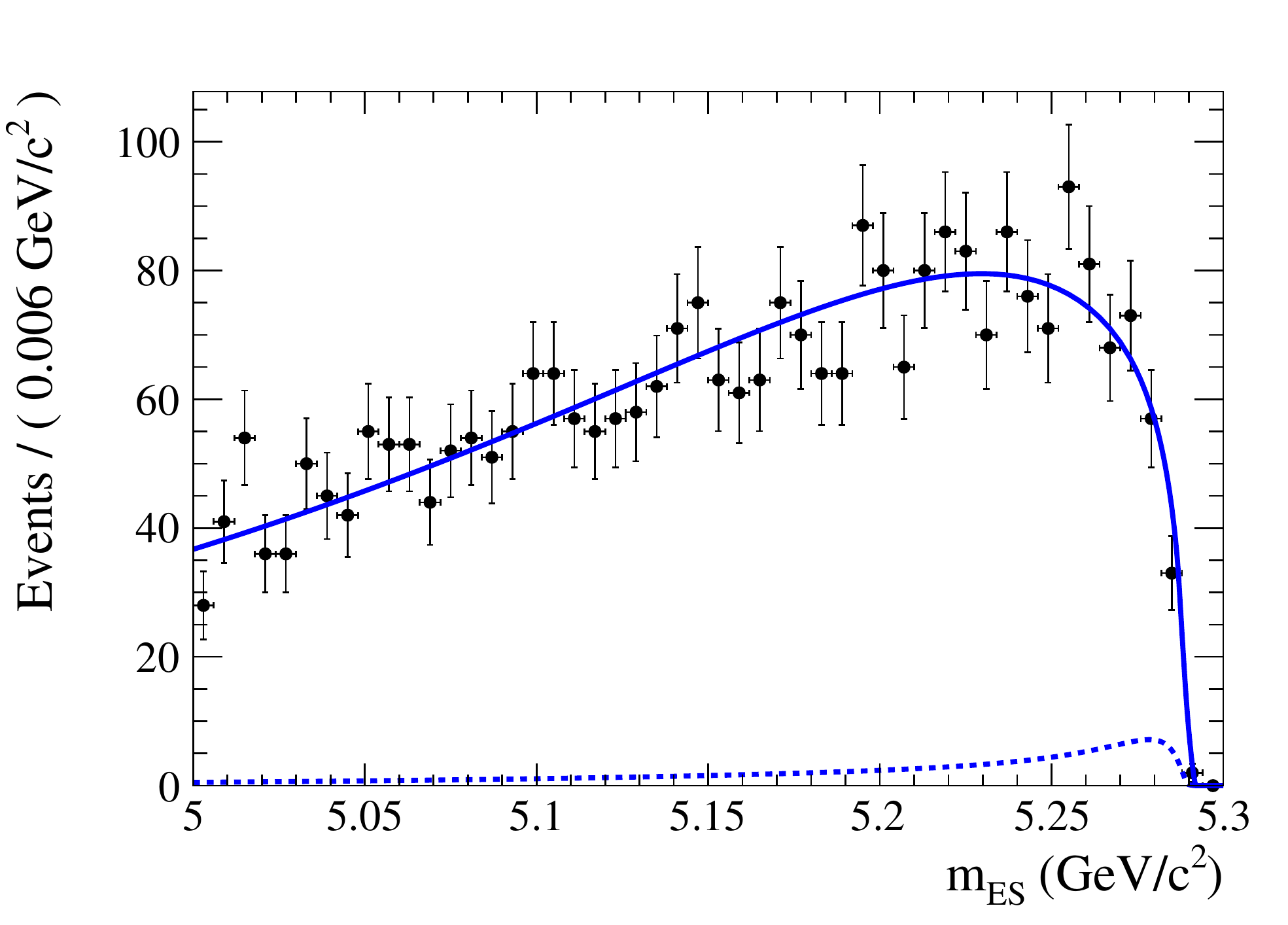}
  }
  \subfigure[50 signal events]{
    \includegraphics[width=.36\textwidth]{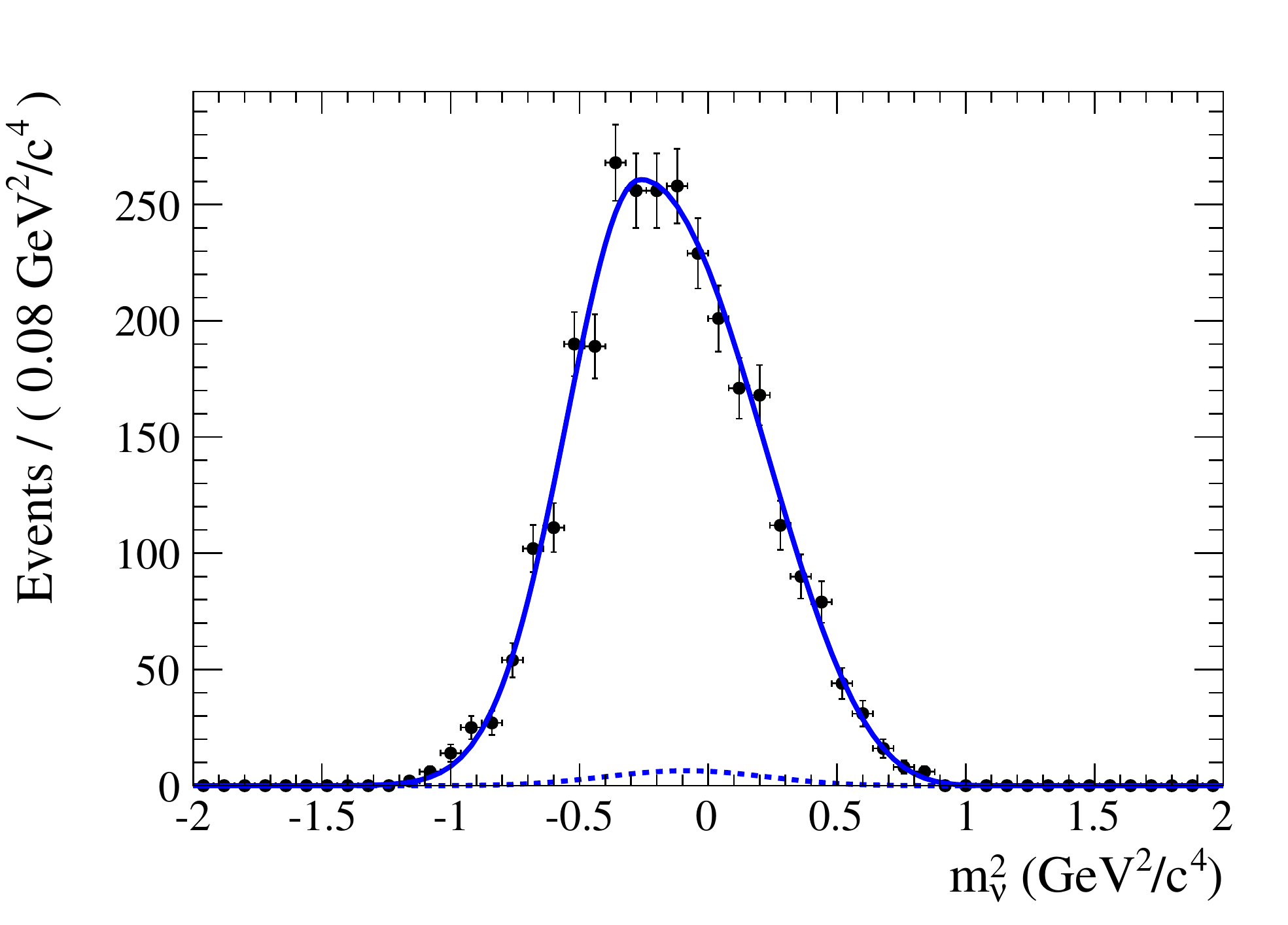}
  }
  \subfigure[50 signal events]{
    \includegraphics[width=.36\textwidth]{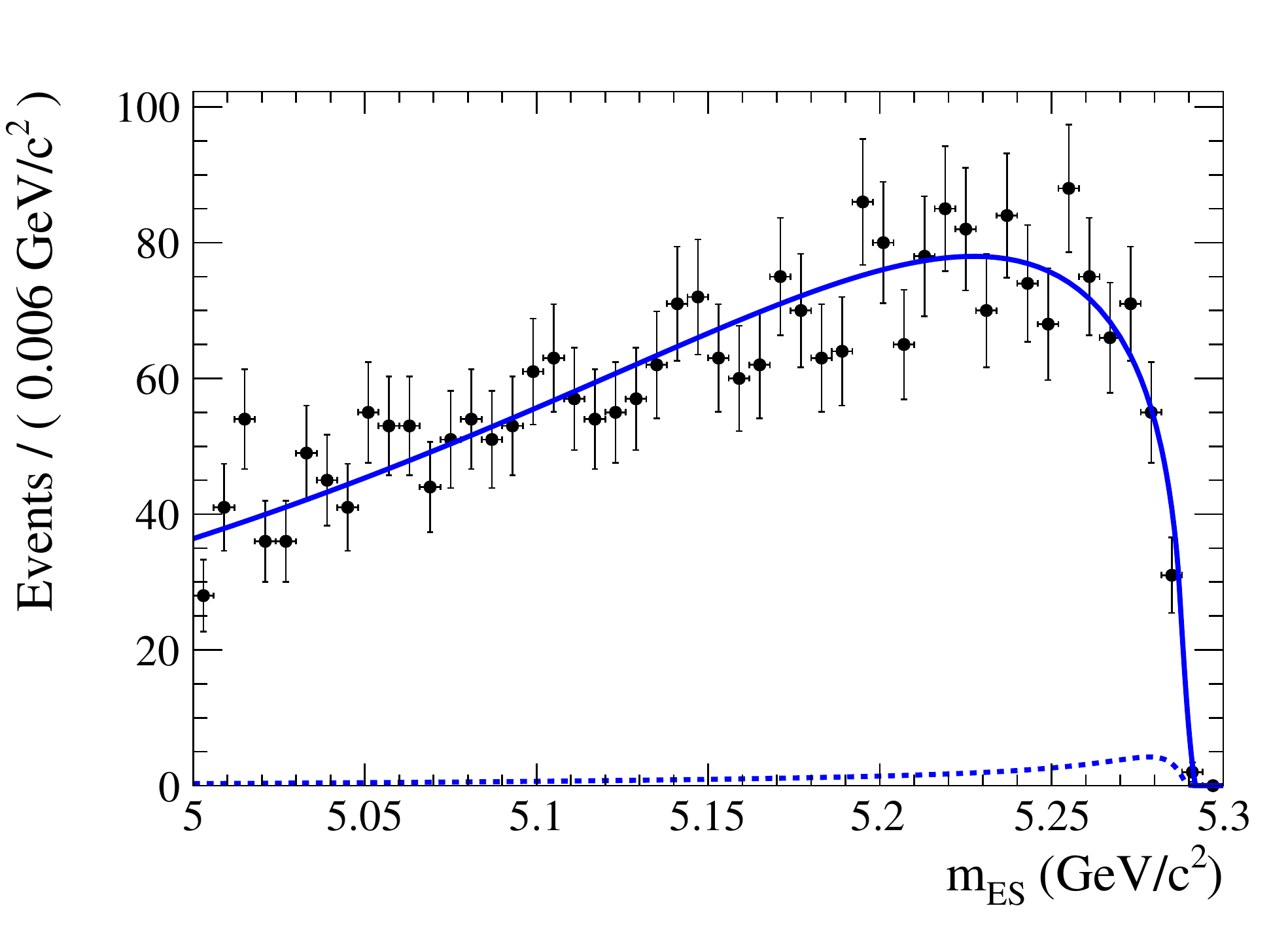}
  }
  \subfigure[10 signal events]{
    \includegraphics[width=.36\textwidth]{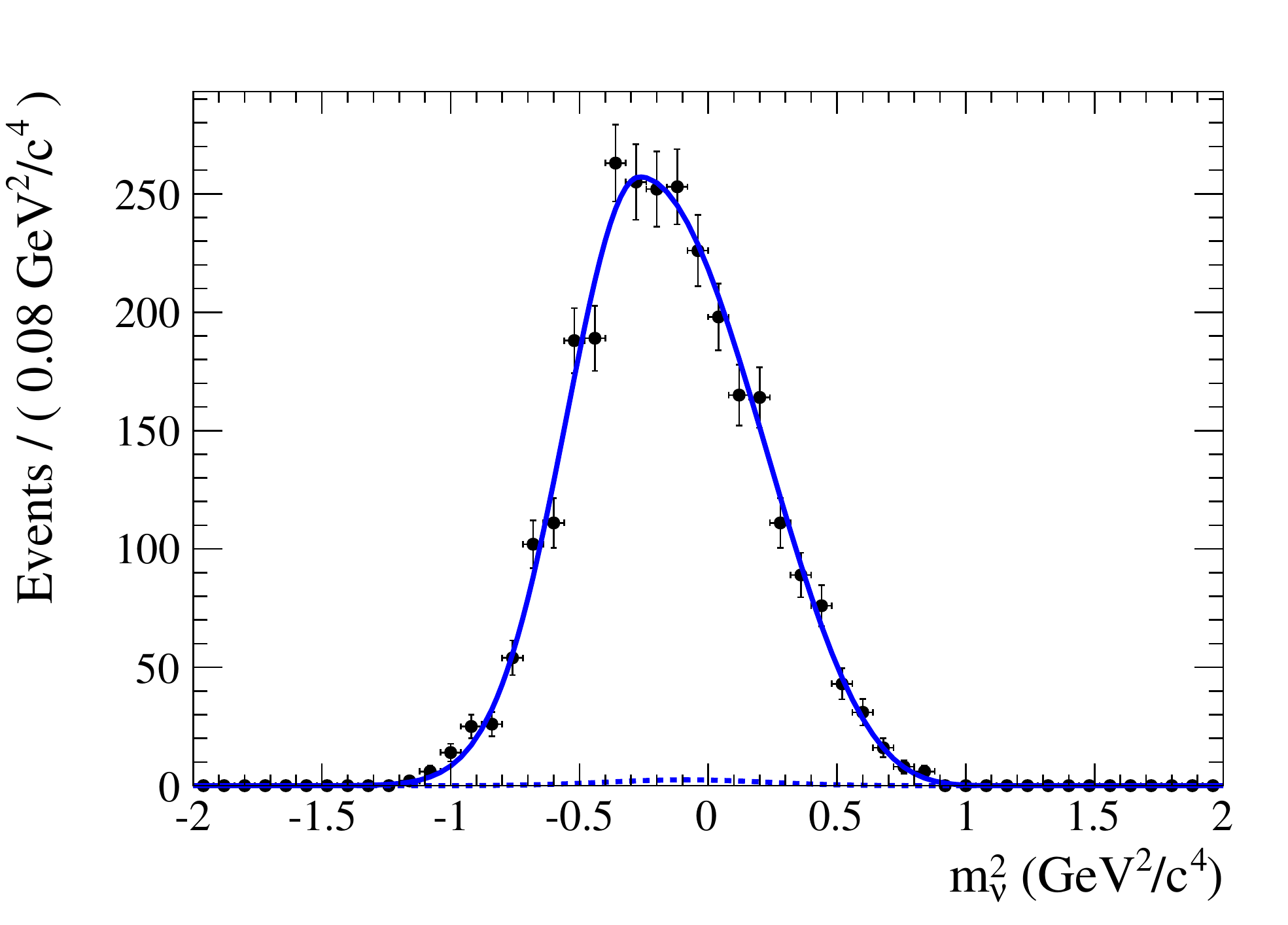}
  }
  \subfigure[10 signal events]{
    \includegraphics[width=.36\textwidth]{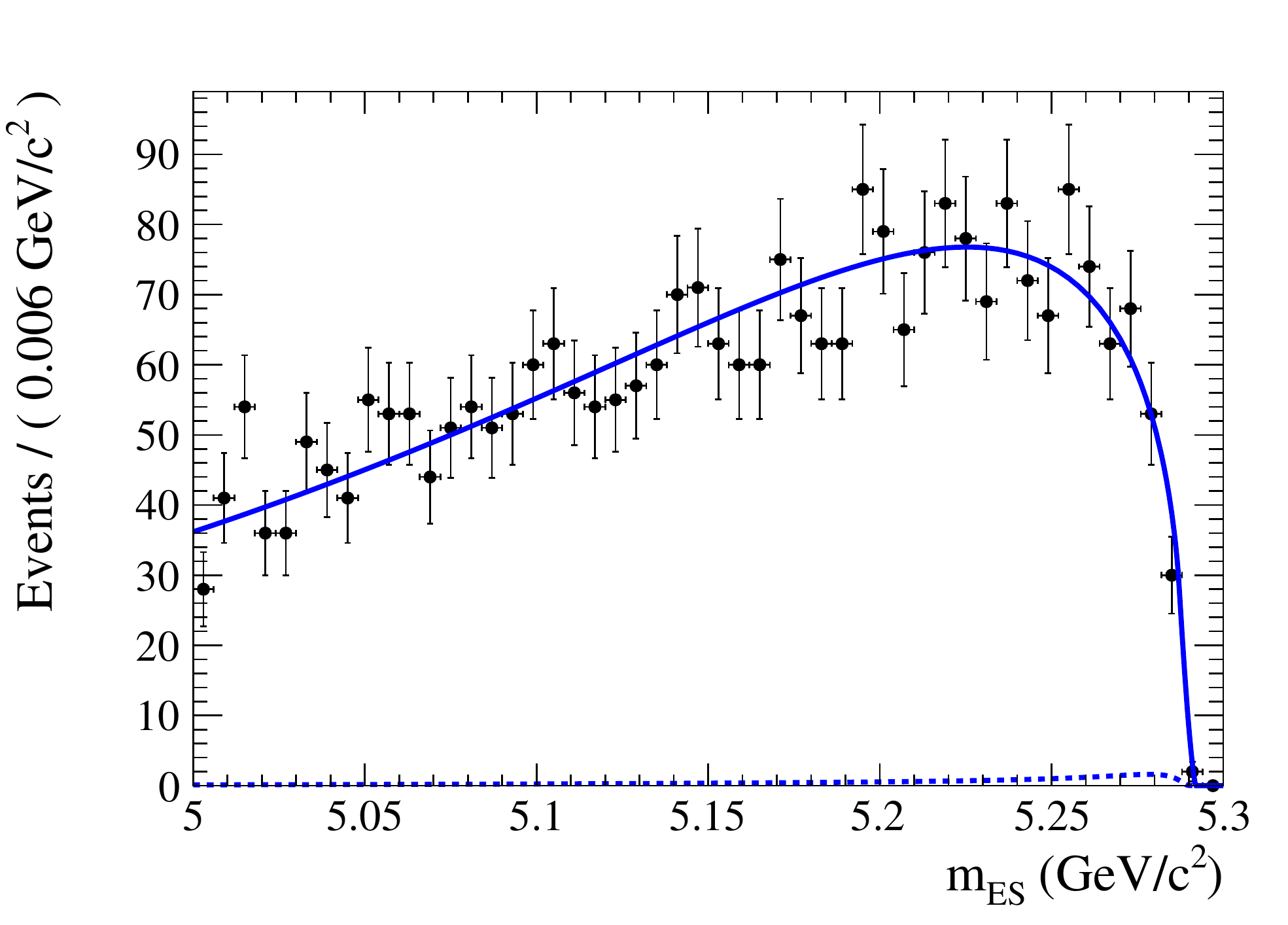}
  }
  \subfigure[0 signal events]{
    \includegraphics[width=.36\textwidth]{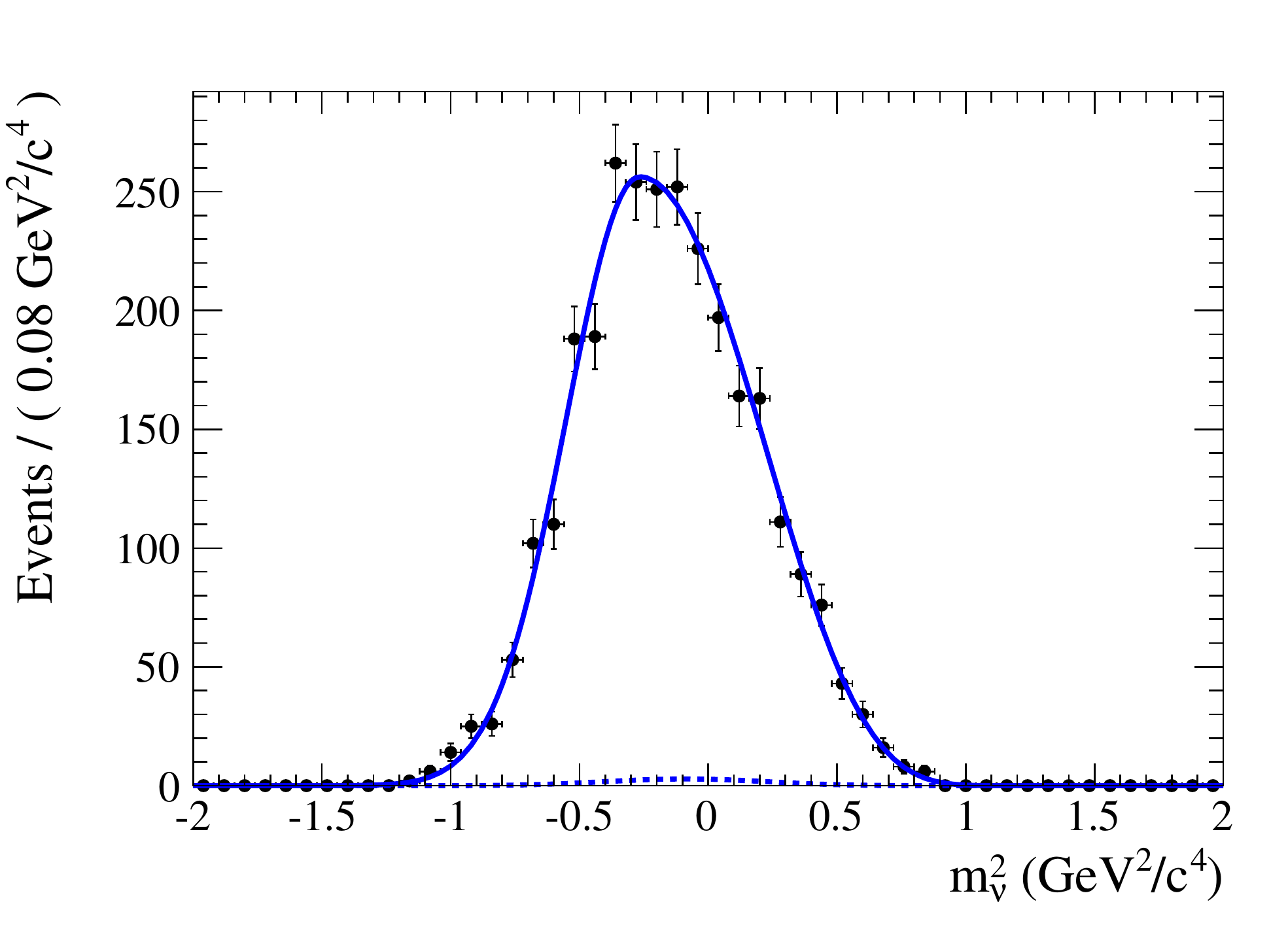}
  }
  \subfigure[0 signal events]{
    \includegraphics[width=.36\textwidth]{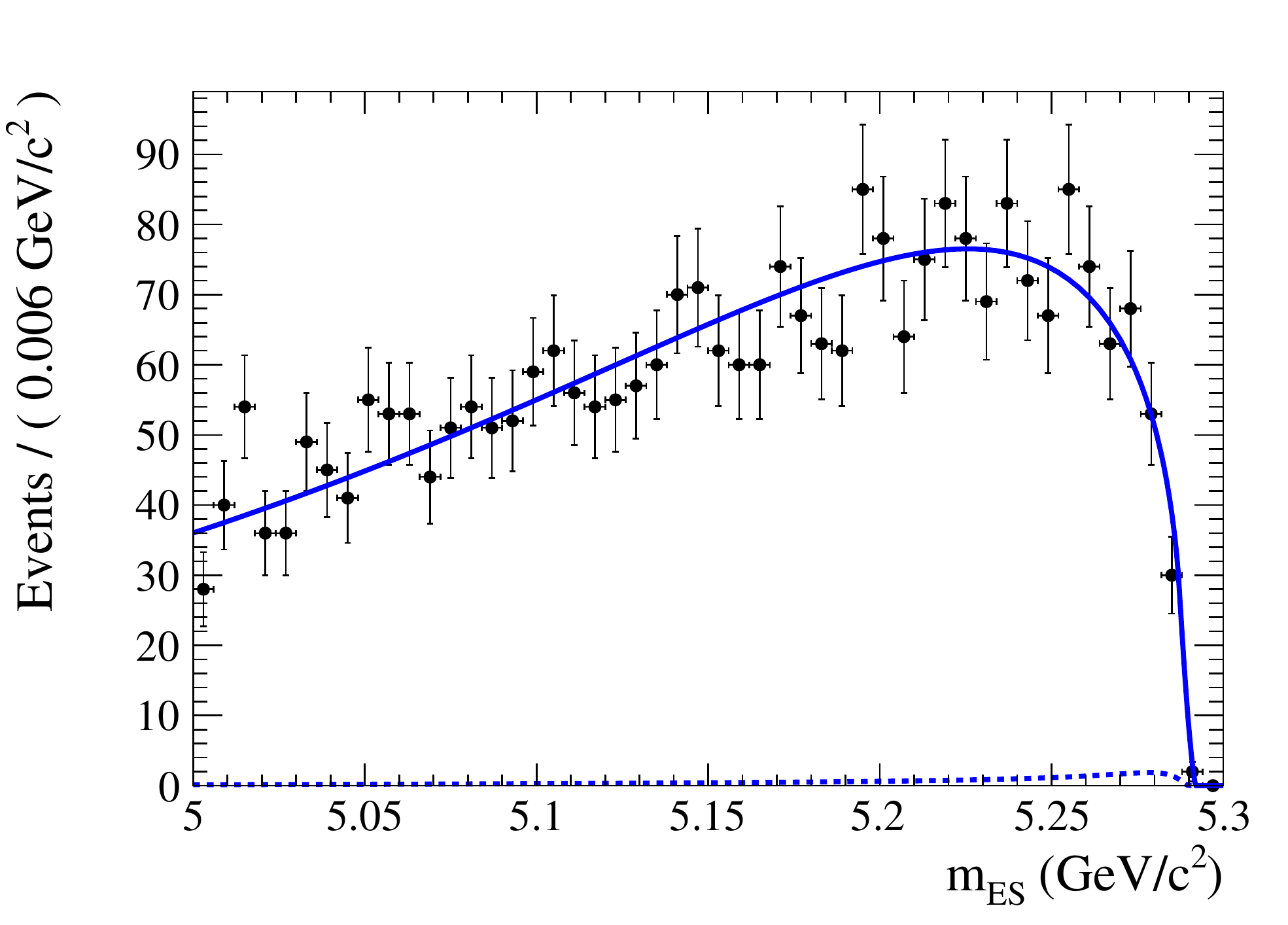}
  }
  \end{center}
  \vspace{-0.5cm}
  \caption{$m_{\nu}^2$ and \mes projections for the fit validation for the muon channel. Shown are the fits for $100, 50, 10$ and $0$ input signal events.}
  \label{fig:FitValidation:muon}
\end{figure}

\backmatter
\phantomsection
\nocite{*}
\cleardoublepage
\phantomsection 
\addcontentsline{toc}{chapter}{\bibname}
\bibliographystyle{utphys}
\bibliography{thesis}

%
%
%
%

\end{document}